\documentclass[11pt]{article}

\usepackage{graphicx}
\usepackage{color}
\usepackage{rotating}
\usepackage{makeidx}
\usepackage{alltt}
\usepackage{upquote}
\usepackage{longtable}

\usepackage{url}

\usepackage[pdfpagelabels,colorlinks,urlcolor=black,linkcolor=black,citecolor=blue]{hyperref}

\usepackage{amsmath}
\usepackage{amssymb}
\usepackage{dsfont}
\usepackage{amsthm}

\usepackage{bookmark}

\theoremstyle{remark}
\newtheorem{remark}{Remark}
\newtheorem{example}{Example}
\theoremstyle{definition}
\newtheorem{definition}{Definition}
\newtheorem{theorem}{Theorem}

\newtheorem{proposition}{Proposition}

\newcommand{\rank}{\mathop{\mathrm{rank}}}

\newcommand{\be}{\begin{equation}}
\newcommand{\ee}{\end{equation}}
\newcommand{\ba}{\left [ \begin{array}}
\newcommand{\ea}{\end{array} \right ]}
\newcommand{\bea}{\begin{eqnarray}}
\newcommand{\eea}{\end{eqnarray}}
\newcommand{\C}{{\mbox{\rm $\scriptscriptstyle ^\mid$\hspace{-0.40em}C}}}

\newcommand{\FF}{{{\rm I \kern -0.2em F}}}
\newcommand{\RR}{{{\rm I \kern -0.2em R}}}
\newcommand{\DD}{{{\rm I \kern -0.2em D}}}
\newcommand{\CC}{{{\mbox{\rm \hspace*{0.05ex}
\rule[.18ex]{.18ex}{1.24ex} \kern -.65em C}}}}

\newcommand{\finr}{{\ \hfill $\Box$}}

\newcommand{\ii}[1]{{\it #1}}
\newcommand{\diag}{\mathop{\mathrm{diag}}}
\renewcommand{\ker}{\mathcal{N}}
\newcommand{\wno}{\mathop{\mathrm{wno}}}
\newcommand{\dist}{\mathop{\mathrm{dist}}}

\makeindex             

\begin{document}
%
\title{\vspace*{-0cm}\LARGE{{\sc Fault Detection and Isolation Tools (FDITOOLS)} } \newline\newline {\LARGE \textsc{User's Guide} } \vspace*{4cm}
}
\author{\Large \textbf{Andreas~Varga}\footnote{Andreas Varga lives in Gilching, Germany. \emph{E-mail address}: \url{varga.andreas@gmail.com}, \emph{URL}: \url{https://sites.google.com/site/andreasvargacontact/}}
}
\date{November 30, 2018}

\maketitle
\vspace*{2cm}
\begin{abstract}
The Fault Detection and Isolation Tools (\textbf{FDITOOLS}) is a collection of MATLAB functions for the analysis and solution of fault detection and model detection problems.
The implemented functions are based on the computational procedures described in the Chapters 5, 6 and 7 of the book: "A. Varga, Solving Fault Diagnosis Problems -- Linear Synthesis Techniques, Springer, 2017". This document is the User's Guide for the version V1.0 of \textbf{FDITOOLS}. First, we present the mathematical background for solving several basic exact and approximate synthesis problems of fault detection filters and model detection filters. Then, we give in-depth information on the command syntax of the main analysis and synthesis functions. Several examples illustrate the use of the main functions of \textbf{FDITOOLS}.
\end{abstract}
\hypersetup{pageanchor=true}

\vspace*{1.5cm}

\newpage
\pdfbookmark[1]{Contents}{Cont}

\tableofcontents
\hypersetup{linkcolor=blue}

\newpage
\vspace*{-18mm}
\section*{Notations and Symbols}
\addcontentsline{toc}{section}{Notations and Symbols}

%
\textbf{General notations}
\begin{longtable}{lp{13cm}}
$\mathds{C}$ & {field of complex numbers} \\
$\mathds{R}$ & {field of real numbers} \\
$\mathds{C}_s$ & {stability domain (i.e., open left complex half-plane in continuous-time or open unit disk centered in the origin in discrete-time)}\\
$\partial\mathds{C}_s$ & {boundary of stability domain (i.e., extended imaginary axis with infinity included in continuous-time, or unit circle centered in the origin in discrete-time)} \\
$\overline{\mathds{C}}_s$ & {closure of $\mathds{C}_s$: $\overline{\mathds{C}}_s = \mathds{C}_s \cup \partial\mathds{C}_s$} \\
${\mathds{C}}_u$ & {open instability domain: $\mathds{C}_u := \mathds{C} \setminus \overline{\mathds{C}}_s$}
\\ $\overline{\mathds{C}}_u$ & {closure of ${\mathds{C}}_u$: $\overline{\mathds{C}}_u := \mathds{C}_u \cup \partial\mathds{C}_s$}
\\ $\mathds{C}_g$ & {``good'' domain of $\mathds{C}$}
\\ $\mathds{C}_b$ & {``bad'' domain of $\mathds{C}$: $\mathds{C}_b = \mathds{C} \setminus \mathds{C}_g$}
\\ $s$ & {complex frequency variable in the Laplace transform: $s = \sigma+\mathrm{i}\omega$}
\\ $z$ & {complex frequency variable in the Z-transform: $z = \mathrm{e}^{\,sT}$, $T$ -- sampling time}
\\ $\lambda$ & {complex  frequency variable: $\lambda = s$ in continuous-time or $\lambda = z$ in discrete-time}
\\ $\bar\lambda$ & {complex conjugate of the complex number $\lambda$ }
\\ $\mathds{R}(\lambda)$ & {set of rational matrices  in  indeterminate $\lambda$ with real coefficients  }
\\ $\mathds{R}(\lambda)^{p\times m}$ & {set of $p\times m$ rational matrices  in  indeterminate $\lambda$ with real coefficients}
\\ $\delta(G(\lambda))$ & {McMillan degree of the rational matrix $G(\lambda)$}
\\ $G^\sim(\lambda)$ & {Conjugate of $G(\lambda) \in \mathds{R}(\lambda)$: $G^\sim(s) = G^T(-s)$ in continuous-time and $G^\sim(z) = G^T(1/z)$ in discrete-time}
\\ $\ell_2$ & {Banach-space of square-summable sequences}
\\ $\mathcal{L}_2$ & {Lebesgue-space of square-integrable functions}
\\  $\mathcal{L}_\infty$ & {Space of complex-valued functions bounded and analytic in ${\partial\mathds{C}}_s$}
\\ $\mathcal{H}_\infty$ & {Hardy-space of complex-valued functions bounded and analytic in ${\mathds{C}}_u$}
\\ $\|G\|_{2}$ & {$\mathcal{H}_2$- or $\mathcal{L}_2$-norm of the transfer function matrix $G(\lambda)$ or 2-norm of a matrix $G$}
\\ $\|G\|_{\infty}$ & {$\mathcal{H}_\infty$- or $\mathcal{L}_\infty$-norm of the transfer function matrix $G(\lambda)$}
\\ $\|G\|_{\infty/2}$ & {either the $\mathcal{H}_\infty$- or $\mathcal{H}_2$-norm of the transfer function matrix $G(\lambda)$}
\\ $\|G\|_{\infty-}$ & {$\mathcal{H}_{\infty-}$-index of the transfer function matrix $G(\lambda)$}
\\ $\|G\|_{\Omega -}$ & $\mathcal{H}_-$-index over a frequency domain $\Omega$ of the transfer function matrix $G(\lambda)$
\\ $\delta_\nu(G_1,G_2)$ & $\nu$-gap distance between the transfer function matrices  $G_1(\lambda)$ and $G_2(\lambda)$
\\ $M^T$ & {transpose of the matrix $M$}
\\ $M^{-1}$ & {inverse of the matrix $M$}
\\ $M^{-L}$ & {left inverse of the matrix $M$}
\\ $\overline\sigma(M)$ & {largest singular value of the matrix $M$}
\\ $\underline\sigma(M)$ & {least singular value of the  matrix $M$}
\\ $\ker(M)$ & {kernel (or right nullspace) of the matrix $M$}
\\ $\ker_L(G(\lambda))$ & {left kernel (or left nullspace) of $G(\lambda) \in \mathds{R}(\lambda)$}
\\ $\ker_R(G(\lambda))$ & {right kernel (or right nullspace) of $G(\lambda) \in \mathds{R}(\lambda)$}
\\ $\mathcal{R}(M)$ & {range (or image space) of the matrix $M$}
\\ $I_n$ or $I$ & {identity matrix of order $n$ or of an order resulting from context}
\\ $e_i$ & {the $i$-th column of the (known size) identity matrix }
\\ $0_{m\times n}$ or $0$ & {zero matrix of size ${m\times n}$ or of a size resulting from context}
 \end{longtable}

\newpage
\noindent\textbf{Fault diagnosis related notations}
\begin{longtable}{lp{13cm}}
$y(t)$ & {measured output vector: $y(t) \in \mathds{R}^{p}$ }
\\$\mathbf{y}(\lambda)$ & {Laplace- or $\mathcal{Z}$-transformed measured output  vector}
\\$u(t)$ & {control input vector: $u(t) \in \mathds{R}^{m_u}$ }
\\$\mathbf{u}(\lambda)$ & {Laplace- or $\mathcal{Z}$-transformed control input vector}
\\$d(t)$ & {disturbance input vector: $d(t) \in \mathds{R}^{m_d}$ }
\\$\mathbf{d}(\lambda)$ & {Laplace- or $\mathcal{Z}$-transformed disturbance input vector}
\\$w(t)$ & {noise input vector: $w(t) \in \mathds{R}^{m_w}$ }
\\$\mathbf{w}(\lambda)$ & {Laplace- or $\mathcal{Z}$-transformed noise input vector}
\\$f(t)$ & {fault input vector: $f(t) \in \mathds{R}^{m_f}$ }
\\$\mathbf{f}(\lambda)$ & {Laplace- or $\mathcal{Z}$-transformed fault input vector}
\\$x(t)$ & {state vector: $x(t) \in \mathds{R}^n$}
\\$G_u(\lambda)$ & {transfer function matrix from  $u$ to $y$}
\\$G_d(\lambda)$ & {transfer function matrix from  $d$ to $y$}
\\$G_w(\lambda)$ & {transfer function matrix from  $w$ to $y$}
\\$G_f(\lambda)$ & {transfer function matrix from  $f$ to $y$}
\\$G_{f_j}(\lambda)$ & {transfer function matrix from the $j$-th fault input $f_j$ to $y$}
\\$A$ & {system state matrix}
\\$E$ & {system descriptor matrix}
\\$B_u$, $B_d$, $B_w$, $B_f$ & {system input matrices from $u$, $d$, $w$, $f$}
\\$C$ & {system output matrix}
\\$D_u$, $D_d$, $D_w$, $D_f$ & {system feedthrough matrices from $u$, $d$, $w$, $f$}
\\$r(t)$ & {residual vector: $r(t) \in \mathds{R}^{q}$}
\\$\mathbf{r}(\lambda)$ & {Laplace- or $\mathcal{Z}$-transformed residual  vector}
\\$n_b$ & {number of components of residual vector $r$}
\\$r^{(i)}(t)$ & {$i$-th residual vector component: $r^{(i)}(t) \in \mathds{R}^{q_i}$}
\\$\mathbf{r}^{(i)}(\lambda)$ & {Laplace- or $\mathcal{Z}$-transformed $i$-th residual  vector component}
\\$Q(\lambda)$ & {transfer function matrix of the implementation form of the residual generator from  $y$ and $u$ to $r$}
\\$Q_y(\lambda)$ & {transfer function matrix of residual generator from  $y$ to $r$}
\\$Q_u(\lambda)$ & {transfer function matrix of residual generator from  $u$ to $r$}
\\$Q^{(i)}(\lambda)$ & {transfer function matrix of the implementation form of the $i$-th residual generator from  $y$ and $u$ to $r^{(i)}$}
\\$R(\lambda)$ & {transfer function matrix of the internal form of the residual generator from  $u$, $d$, $w$ and $f$ to $r$}
\\$R_u(\lambda)$ & {transfer function matrix from  $u$ to $r$}
\\$R_d(\lambda)$ & {transfer function matrix from  $d$ to $r$}
\\$R_w(\lambda)$ & {transfer function matrix from  $w$ to $r$}
\\$R_f(\lambda)$ & {transfer function matrix from  $f$ to $r$}
\\$R_{f_j}(\lambda)$ & {transfer function matrix from the $j$-th fault input $f_j$ to $r$}
\\$R^{(i)}_{f_j}(\lambda)$ & {transfer function matrix from the $j$-th fault input $f_j$ to $r^{(i)}$}
\\$S$ & {binary structure matrix}
\\$S_{R_f}$ & {binary structure matrix corresponding to $R_f(\lambda)$}
\\$M_r(\lambda)$ & {transfer function matrix of a reference model from  $f$ to $r$}
\\$\theta(t)$ & {residual evaluation vector}
\\$\iota(t)$ & {binary decision vector}
\\$\tau$, $\tau_i$ & {decision thresholds}
\end{longtable}

\noindent\textbf{Model detection related notations}
\begin{longtable}{lp{13cm}}
$N$ & {number of component models of the multiple model}
\\$y(t)$ & {measured output vector: $y(t) \in \mathds{R}^{p}$ }
\\$\mathbf{y}(\lambda)$ & {Laplace- or $\mathcal{Z}$-transformed measured output  vector}
\\$u(t)$ & {control input vector: $u(t) \in \mathds{R}^{m_u}$ }
\\$\mathbf{u}(\lambda)$ & {Laplace- or $\mathcal{Z}$-transformed control input vector}
\\$u^{(j)}(t)$ & {control input vector of $j$-th model: $u^{(j)}(t) := u(t) \in \mathds{R}^{m_u}$}
\\$\mathbf{u}^{(j)}(\lambda)$ & {Laplace- or $\mathcal{Z}$-transformed control input vector of $j$-th model}
\\$d^{(j)}(t)$ & {disturbance input vector of $j$-th model: $d^{(j)}(t) \in \mathds{R}^{m_d^{(j)}}$ }
\\$\mathbf{d}^{(j)}(\lambda)$ & {Laplace- or $\mathcal{Z}$-transformed disturbance input vector of $j$-th model}
\\$w^{(j)}(t)$ & {noise input vector of $j$-th model: $w^{(j)}(t) \in \mathds{R}^{m_w^{(j)}}$ }
\\$\mathbf{w}^{(j)}(\lambda)$ & {Laplace- or $\mathcal{Z}$-transformed noise input vector of $j$-th model}
\\$y^{(j)}(t)$ & {output vector of $j$-th model: $y^{(j)}(t) \in \mathds{R}^{p}$ }
\\$\mathbf{y}^{(j)}(\lambda)$ & {Laplace- or $\mathcal{Z}$-transformed output  vector of $j$-th model}
\\$x^{(j)}(t)$ & {state vector of $j$-th model: $x^{(j)}(t) \in \mathds{R}^{n_i}$}
\\$G_u^{(j)}(\lambda)$ & {transfer function matrix  of $j$-th model from  $u^{(j)}$ to $y^{(j)}$}
\\$G_d^{(j)}(\lambda)$ & {transfer function matrix  of $j$-th model from  $d^{(j)}$ to $y^{(j)}$}
\\$G_w^{(j)}(\lambda)$ & {transfer function matrix  of $j$-th model from  $w^{(j)}$ to $y^{(j)}$}
\\$A^{(j)}$ & {system state matrix of $j$-th model }
\\$E^{(j)}$ & {system descriptor matrix of $j$-th model }
\\$B_u^{(j)}$, $B_d^{(j)}$, $B_w^{(j)}$ & {system input matrices  of $j$-th model from $u^{(j)}\!$, $d^{(j)}\!$, $w^{(j)}$}
\\$C^{(j)}$ & {system output matrix of $j$-th model }
\\$D_u^{(j)}$, $D_d^{(j)}$, $D_w^{(j)}$ & {system feedthrough matrices  of $j$-th model from $u^{(j)}\!$, $d^{(j)}\!$, $w^{(j)}$}
\\$r^{(i)}(t)$ & {$i$-th residual vector component: $r^{(i)}(t) \in \mathds{R}^{q_i}$}
\\$\mathbf{r}^{(i)}(\lambda)$ & {Laplace- or $\mathcal{Z}$-transformed $i$-th residual  vector component}
\\$r(t)$ & {overall residual vector: $r(t) \in \mathds{R}^{q}$, $q = \sum_{i=1}^N q_i$}
\\$\mathbf{r}(\lambda)$ & {Laplace- or $\mathcal{Z}$-transformed overall residual  vector}
\\$Q^{(i)}(\lambda)$ & {transfer function matrix of the implementation form of the $i$-th residual generator from  $y$ and $u$ to $r^{(i)}$}
\\$Q_y^{(i)}(\lambda)$ & {transfer function matrix of residual generator from  $y$ to $r^{(i)}$}
\\$Q_u^{(i)}(\lambda)$ & {transfer function matrix of residual generator from  $u$ to $r^{(i)}$}
\\$Q(\lambda)$ & {transfer function matrix of the implementation form of the overall residual generator from  $y$ and $u$ to $r$}
\\$R^{(i,j)}(\lambda)$ & {the transfer function matrix of the internal form of the overall residual generator from $(u^{(j)}\!, d^{(j)}\!, w^{(j)})$ to $r^{(i)}$}
\\$R_u^{(i,j)}(\lambda)$ & {the transfer function matrix of the internal form of the overall residual generator from  $u^{(j)}$ to $r^{(i)}$}
\\$R_d^{(i,j)}(\lambda)$ & {the transfer function matrix of the internal form of the overall residual generator from  $d^{(j)}$ to $r^{(i)}$}
\\$R_w^{(i,j)}(\lambda)$ & {the transfer function matrix of the internal form of the overall residual generator from  $w^{(j)}$ to $r^{(i)}$}
\\$\theta(t)$ & {$N$-dimensional residual evaluation vector}
\\$\iota(t)$ & {$N$-dimensional binary decision vector}
\\$\tau_i$ & {decision threshold for $i$-th component of the residual vector}
\end{longtable}

\newpage
\section*{Acronyms}
\addcontentsline{toc}{section}{Acronyms}

\begin{tabular}{lp{13cm}}
AFDP & Approximate fault detection problem\\
AFDIP & Approximate fault detection and isolation problem\\
AMDP & Approximate model detection problem\\
AMMP & Approximate model matching problem\\
EFDP & Exact fault detection problem\\
EFEP & Exact fault estimation problem\\
EFDIP & Exact fault detection and isolation problem\\
EMDP & Exact model detection problem\\
EMMP & Exact model matching problem\\
FDD & Fault detection and diagnosis\\
FDI & Fault detection and isolation\\
LTI & Linear time-invariant\\
LFT & Linear fractional transformation\\
LPV & Linear parameter-varying\\
MIMO & Multiple-input multiple-output\\
MMP & Model-matching problem\\
TFM & Transfer function matrix
 \end{tabular}

\newpage

\section{Introduction}\label{sec:intro}

The {\sc Fault Detection and Isolation Tools} (\textbf{FDITOOLS}) is a collection of {MATLAB} functions for the analysis and solution of fault detection problems. \textbf{FDITOOLS} supports various synthesis approaches of linear residual generation filters for continuous- or discrete-time linear systems. The underlying synthesis techniques rely on reliable numerical algorithms developed by the author and described in the Chapters 5, 6 and 7 of the author's book \cite{Varg17}: \\

\begin{tabular}{l}Andreas Varga, \emph{Solving Fault Diagnosis Problems - Linear Synthesis Techniques}, \\
vol. 84 of Studies in Systems, Decision and Control, Springer International Publishing, \\
xxviii+394, 2017.
\end{tabular}\\

The functions of the \textbf{FDITOOLS} collection rely on the \emph{Control System Toolbox} \cite{MLCO15} and the \emph{Descriptor System Tools} (\textbf{DSTOOLS}) V0.71 \cite{Varg17a}.
The current release of \textbf{FDITOOLS} is version V1.0, dated November 30, 2018. \textbf{FDITOOLS} is distributed as a free software via the Bitbucket repository.\footnote{\url{https://bitbucket.org/DSVarga/fditools}} The codes have been developed under MATLAB 2015b and have been also tested with MATLAB 2016a through 2018b. To use the functions of \textbf{FDITOOLS}, the \emph{Control System Toolbox} and the \textbf{DSTOOLS} collection must be installed in MATLAB running under 64-bit Windows 7, 8, 8.1 or 10.

This document describes version V1.0 of the \textbf{FDITOOLS} collection. This version covers all synthesis procedures described in the book \cite{Varg17} and, additionally, includes a comprehensive collection of analysis functions,  as well as functions for an easy setup of synthesis models.  The book \cite{Varg17} represents an important complementary documentation for the \textbf{FDITOOLS} collection: it describes the mathematical background of solving synthesis problems of fault detection and model detection filters and gives detailed descriptions of the underlying synthesis procedures. Additionally, the M-files of the functions are self-documenting and a detailed documentation can be obtained online by typing help with the M-file name.
Please cite \textbf{FDITOOLS} as follows: \\

\begin{tabular}{l} A. Varga. FDITOOLS -- The Fault Detection and Isolation Tools for MATLAB, 2018. \\ \url{https://sites.google.com/site/andreasvargacontact/home/software/fditools}.
\end{tabular}\\

The implementation of the functions included in the \textbf{FDITOOLS} collection follows several principles, which have been consequently enforced when implementing these functions. These principles are listed below and partly consists of the requirements for robust software implementation, but also include several requirements which are specific  to the field of fault detection:
\begin{itemize}
\item \emph{Using general, numerically reliable and computationally efficient numerical approaches as basis for the implementation of all computational functions, to guarantee the solvability of problems under the most general existence conditions of the solutions.} Consequently, the implemented methods provide a solution whenever a solution exists. These methods are extensively described in the book \cite{Varg17}, which forms the methodological and computational basis of all implemented analysis and synthesis functions.
\item \emph{Support for the most general model representation of linear time-invariant systems in form of generalized state-space representation, also known as descriptor systems.} All analysis and synthesis functions are applicable to both continuous- and discrete-time systems. The basis for implementation of all functions is the \emph{Descriptor System Tools} (\textbf{DSTOOLS}) \cite{Varg17a}, a collection of functions to handle  rational transfer function matrices (proper or improper), via their equivalent descriptor system representations. The initial version of this collection  has been implemented in conjunction with the book \cite{Varg17}.
\item \emph{Providing simple user interface to all synthesis functions. } All functions rely on default settings of problem parameters and synthesis options, which allow to easily obtain preliminary synthesis results. Also, all functions to solve a class of problems (e.g., fault detection), are applicable to the same input models. Therefore,  the synthesis functions to solve approximate synthesis problems are applicable to solve the exact synthesis problems as well. On the other side, the solution of an exact problem for a system with noise inputs, represents a first approximation to the solution of the approximate synthesis problem.
\item \emph{Providing an exhaustive set of options to ensure the complete freedom in choosing problem specific parameter and synthesis options.} Among the frequently used synthesis options are: the number of residual signal outputs or the numbers of outputs of the components of structured residual signals; stability degree for the poles of the resulting filters or the location of their poles; frequency values to enforce strong fault detectability; type of the employed nullspace basis (e.g., proper, proper and simple, full-order observer); performing least-order synthesis, etc.
\item \emph{Guaranteeing the reproducibility of results.  }This feature is enforced by employing the so-called \emph{design matrices}. These matrices are internally used to build linear combinations of left nullspace basis vectors and are frequently randomly generated (if not explicitly provided). The values of the employed design matrices are returned as additional information by all synthesis functions. The use of design matrices also represents a convenient mean to perform an optimization-based tuning of these matrices to achieve specific performance characteristics  for the resulting filters.
\end{itemize}

\newpage
\section{Fault Detection Basics}\label{sec:Basics}

In this section we describe first the basic fault monitoring tasks, such as fault detection and  fault isolation, and then introduce and characterize the concepts of fault detectability and fault isolability. Six ``canonical'' fault detection problems are formulated in the book \cite{Varg17} for the class of \emph{linear time-invariant} (LTI) systems with additive faults.  Of the formulated six problems, three  involve the exact synthesis and three involve the approximate synthesis of fault detection filters. The current release of \textbf{FDITOOLS} covers all synthesis techniques described in \cite{Varg17}.  Jointly with the formulation of the fault detection problems, general solvability conditions are given for each problem in terms of ranks of certain transfer function matrices. More details and the proofs of the  results are available in  Chapters 2 and 3 of \cite{Varg17}.

\subsection{Basic Fault Monitoring Tasks}
\index{f@FDD|see{fault detection and diagnosis}}
\index{f@FDI|see{fault detection and isolation}}

A fault represents a deviation from the normal behaviour of a system
due to an unexpected event (e.g., physical component failure or
supply breakdown). The occurrence of faults
must be detected as early as possible to prevent any
serious consequence. For this purpose, {fault diagnosis} techniques
are used to allow the {detection} of occurrence of faults (fault detection) and the
localization of detected faults (fault isolation). The term
\emph{fault detection and diagnosis} (FDD) includes the requirements for \emph{fault detection and isolation }(FDI). \index{fault detection and isolation (FDI)}

\begin{figure}[thpb]
\begin{center}
\includegraphics[height=4.8cm]{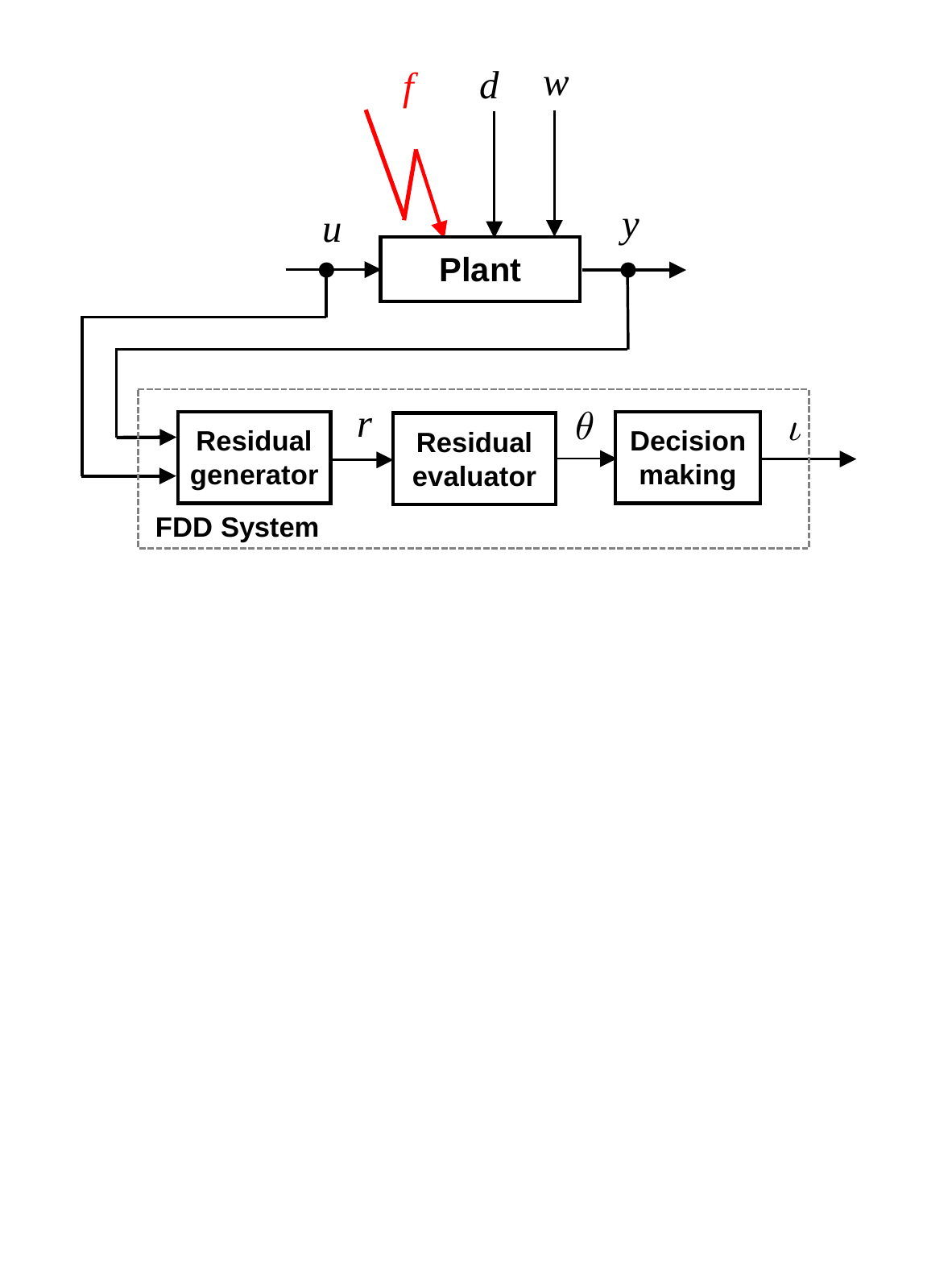}
\vspace*{-2mm}   
\caption{Basic fault diagnosis setup.}  
\label{Fig_FDDSystem}                                 
\end{center}                                 
\end{figure}

A FDD system is a device (usually based on a collection of real-time processing algorithms)
suitably set-up to fulfill the above tasks. The minimal functionality of any FDD system is illustrated in Fig.~\ref{Fig_FDDSystem}.

The main plant variables are the control
inputs $u$, the unknown disturbance inputs $d$,  the noise inputs $w$, and the
output measurements $y$.
The output $y$ and control input $u$ are the only measurable signals which can be used for fault monitoring purposes.
The disturbance inputs $d$ and noise inputs  $w$ are non-measurable ``unknown'' input signals, which  act adversely on the system performance. For example, the unknown disturbance inputs $d$ may represent physical disturbance inputs, as for example, wind turbulence acting on an aircraft or external loads acting on a plant. Typical noise inputs are sensor noise signals as well as
process input noise.  However, fictive noise inputs can also account for the cumulative effects of unmodelled
system dynamics or for the effects of parametric uncertainties. In general, there is no clear-cut separation between disturbances and noise, and therefore, the appropriate definition of the disturbance and noise
inputs is a challenging aspect when modelling systems for solving fault detection problems.
A \emph{fault} is any unexpected variation of some physical
parameters or variables of a plant causing an unacceptable violation of
certain specification limits for normal operation. Frequently, a fault input $f$ is defined to account for any anomalous behaviour of the plant.

The main component of any FDD system (as that in Fig.~\ref{Fig_FDDSystem}) is the \emph{residual
generator} (or \emph{fault detection filter}, or simply \emph{fault detector}),
which produces residual signals grouped in a $q$-dimensional vector $r$ by processing the available
measurements $y$ and the known values of control inputs $u$. The role of the residual signals
is to indicate the presence  or absence of faults, and therefore the residual $r$ must be equal (or
close) to zero in the absence of faults and significantly different
from zero after a fault occurs.
For decision-making, suitable measures of the residual magnitudes  (e.g., signal norms) are generated in a vector $\theta$, which is then used to produce the corresponding  decision vector $\iota$. In what follows, two basic fault monitoring tasks are formulated and discussed.

\index{fault detection and diagnosis (FDD)!fault detection}
\emph{Fault detection} is simply a binary
decision on the presence of any fault ($f\not = 0$) or the absence of all
faults ($f = 0$). Typically, $\theta(t)$ is scalar evaluation signal, which approximates $\|r\|_2$, the $\mathcal{L}_2$- or $\ell_2$-norms of signal $r$, while $\iota(t)$ is a scalar decision making signal defined as $\iota(t) = 1$ if $\theta(t) > \tau$ (fault occurrence) or $\iota(t) = 0$ if $\theta(t) \leq \tau$ (no fault), where $\tau$ is a suitable threshold quantifying the gap between the ``small'' and ``large'' magnitudes of the residual. The decision on the occurrence or absence of faults must be done in the presence of
arbitrary control inputs $u$, disturbance inputs $d$, and noise inputs $w$ acting simultaneously on the system. The effects of the control inputs on the residual can be always decoupled by a suitable choice of the residual generation filter.  In the ideal case, when no noise inputs are present ($w \equiv 0$), the residual generation filter must additionally be able to \emph{exactly}
decouple the effects of the disturbances inputs in the
residual and ensure, simultaneously, the sensitivity of the residual
to all faults (i.e., \emph{complete fault detectability}, see Section~\ref{sec:fdetectability}).
\index{fault detectability} In this case, $\tau = 0$ can be (ideally) used.
However, in the general case when $w \not\equiv 0$, only an \emph{approximate} decoupling of $w$ can be achieved (at best) and a sufficient gap must exist between the magnitudes of residuals in fault-free and faulty situations. Therefore, an appropriate choice of $\tau > 0$ must avoid false alarms and missed detections.

\index{fault detection and diagnosis (FDD)!fault isolation}
\emph{Fault isolation} concerns with the exact localization of occurred
faults and involves for each component $f_j$ of the fault vector $f$ the decision on the presence of $j$-th fault ($f_j\not = 0$) or its absence ($f_j = 0$).  Ideally, this must be achieved regardless the faults occur one at a time or several faults occur
simultaneously. Therefore, the fault isolation task is
significantly more difficult than the simpler fault detection. For fault isolation purposes, we will assume a partitioning of the $q$-dimensional residual vector $r$ in $n_b$ stacked $q_i$-dimensional subvectors $r^{(i)}$, $i = 1, \ldots, n_b$, in the form
\be\label{rstruct} r = \ba{c} r^{(1)}\\ \vdots \\ r^{(n_b)} \ea \, ,\ee
where $q = \sum_{i=1}^{n_b}q_i$. A typical fault evaluation setup used for fault isolation is  to define $\theta_i(t)$, the $i$-th component of $\theta(t)$,  as a real-time computable approximation of $\|r^{(i)}\|_2$. The $i$-th component of $\iota(t)$ is set to $\iota_i(t) = 1$ if $\theta_i(t) > \tau_i$ ($i$-th residual fired) or $\iota_i(t) = 0$ if $\theta_i(t) \leq \tau_i$ ($i$-th residual not fired), where $\tau_i$ is a suitable threshold for the $i$-th subvector $r^{(i)}(t)$.
If a sufficiently large number of measurements are available, then  it can be aimed that $r^{(i)}$  is influenced only by the $i$-th fault signal $f_i$. This setting, with $n_b$ chosen equal to the actual number of fault components, allows \emph{strong fault isolation},\index{fault detection and diagnosis (FDD)!strong fault isolation}
where an arbitrary number of simultaneous faults can be isolated. The isolation of the $i$-th fault is achieved if $\iota_i(t) = 1$, while for $\iota_i(t) = 0$ the $i$-th fault is not present. In many practical applications, the  lack of a sufficiently large number
of measurements impedes strong isolation of simultaneous faults. Therefore, often only  \emph{weak fault isolation}\index{fault detection and diagnosis (FDD)!weak fault isolation}
 can be performed under simplifying assumptions as, for example, that the faults occur one at a time or no more than two faults may occur simultaneously. The fault isolation schemes providing weak fault isolation compare the resulting $n_b$-dimensional binary decision vector $\iota(t)$, with a predefined set of binary fault signatures. If each individual fault $f_j$ has associated a distinct signature $s_j$, then the $j$-th fault can be isolated by simply checking that $\iota(t)$ matches the associated signature $s_j$. Similarly to fault detection, besides the decoupling of the control inputs $u$ from the residual $r$ (always possible), the exact decoupling of the disturbance inputs $d$ from $r$ can be strived in the case when $w\equiv 0$. However, in the general case when $w\not\equiv 0$, only approximate decoupling of $w$ can be achieved (at best) and a careful selection of tolerances $\tau_i$ is necessary to perform fault isolation without false alarms and missed detections.

\subsection{Plant Models with Additive Faults}\label{sec:additive_faults}

\index{faulty system model!additive}
The following input-output
representation  is used to describe LTI systems with additive faults
\be\label{systemw} {\mathbf{y}}(\lambda) =
G_u(\lambda){\mathbf{u}}(\lambda) +
G_d(\lambda){\mathbf{d}}(\lambda) +
G_f(\lambda){\mathbf{f}}(\lambda) +
G_w(\lambda){\mathbf{w}}(\lambda)
,\ee
where  ${\mathbf{y}}(\lambda)$, ${\mathbf{u}}(\lambda)$,
${\mathbf{d}}(\lambda)$, ${\mathbf{f}}(\lambda)$, and
${\mathbf{w}}(\lambda)$,  with boldface notation,  denote the Laplace-transformed (in the continuous-time case) or Z-transformed (in the discrete-time case) time-dependent vectors, namely,
the $p$-dimensional system output vector $y(t)$,
$m_u$-dimensional control input vector $u(t)$,
$m_d$-dimensional disturbance vector $d(t)$, $m_f$-dimensional fault vector $f(t)$, and
$m_w$-dimensional noise vector $w(t)$
respectively. $G_u(\lambda)$, $G_d(\lambda)$,
$G_f(\lambda)$   and $G_w(\lambda)$ are the {transfer-function matrices} (TFMs) from the control
inputs $u$,  disturbance inputs $d$, fault inputs $f$, and noise inputs $w$  to the outputs $y$, respectively.
 According to the system type, $\lambda = s$, the complex variable in the Laplace-transform
  in the case of a continuous-time system or $\lambda = z$,
the complex variable in the Z-transform in the case of a
discrete-time system.
For most of practical applications, the TFMs $G_u(\lambda)$, $G_d(\lambda)$, $G_f(\lambda)$, and $G_w(\lambda)$  are proper rational matrices.
However, for complete generality of our
problem settings, we will allow that these TFMs are general improper rational matrices for which
we will not \emph{a priori} assume any further properties  (e.g., stability, full
rank, etc.).
\index{faulty system model!additive!input-output}

The main difference
between the disturbance input $d(t)$ and noise input $w(t)$ arises from
the formulation of the fault monitoring goals. In this respect, when synthesizing devices to serve for fault diagnosis purposes, we will generally target the \emph{exact}
decoupling of the effects of disturbance inputs.
Since generally
the exact decoupling of effects of noise inputs is  not achievable, we will simultaneously try to attenuate their effects, to achieve an \emph{approximate} decoupling. Consequently, we will try to solve synthesis problems exactly or approximately, in accordance with the absence or presence of noise inputs  in the underlying plant model, respectively.

An equivalent
\emph{descriptor} state-space realization of the input-output model (\ref{systemw}) has the form
\be\label{ssystemw} \begin{aligned}E\lambda x(t) &= Ax(t) + B_u u(t) + B_d d(t) + B_f f(t) + B_w w(t) \, , \\
y(t) &= Cx(t) + D_u u(t) + D_d d(t)  + D_f f(t) + D_w w(t)\, ,
\end{aligned} \ee
with the $n$-dimensional state vector $x(t)$, where $\lambda
x(t) = \dot{x}(t)$ or $\lambda x(t) = x(t+1)$ depending on the
type of the system, continuous- or discrete-time, respectively. In
general, the square matrix $E$ can be singular, but we
will assume that the linear pencil $A-\lambda E$ is regular. For systems with
proper TFMs in (\ref{systemw}), we can always choose a \emph{standard}
state-space realization where $E = I$. In general, it is advantageous to choose
the representation (\ref{ssystemw}) minimal, with the
pair $(A-\lambda E, C)$ \textit{observable} and the pair
$(A-\lambda E, [\,B_u \; B_d \; B_f\; B_w \,])$
\textit{controllable}.
The corresponding TFMs of the model in (\ref{systemw}) are
\be\label{tfms}
\begin{aligned} G_u(\lambda) &= C(\lambda E-A)^{-1}B_u+D_u ,\\
G_d(\lambda) &= C(\lambda E-A)^{-1}B_d+D_d ,\\
G_f(\lambda) &= C(\lambda E-A)^{-1}B_f + D_f ,\\
G_w(\lambda) &= C(\lambda E-A)^{-1}B_w + D_w
\end{aligned}\ee
or in an equivalent notation
\[ \ba{cccc} G_u(\lambda) & G_d(\lambda) & G_f(\lambda)& G_w(\lambda) \ea :=
{\arraycolsep=1mm\ba{c|cccc} A-\lambda E & B_u & B_d & B_f & B_w \\ \hline
C & D_u & D_d & D_f& D_w \ea }\, .\]
\index{faulty system model!additive!state-space}


\subsection{Residual Generation} \label{fdi_res}
\index{residual generation!for fault detection and isolation}

\index{residual generator!implementation form}
A linear residual generator (or fault detection filter)
processes the measurable system outputs $y(t)$ and known control
inputs $u(t)$ and generates the residual signals $r(t)$ which
serve for decision-making on the presence or absence of faults.
The input-output form of this filter  is
 \be\label{detec}
{\mathbf{r}}(\lambda) = Q(\lambda)\ba{c}
{\mathbf{y}}(\lambda)\\{\mathbf{u}}(\lambda)\ea = Q_y(\lambda){\mathbf{y}}(\lambda)+ Q_u(\lambda) {\mathbf{u}}(\lambda)  \, , \ee
with $Q(\lambda) = [\, Q_y(\lambda)\; Q_u(\lambda)\,]$, and is called the \emph{implementation form}. The TFM $Q(\lambda)$ for a physically
realizable filter must be \emph{proper} (i.e.,
only with finite poles) and \emph{stable} (i.e., only with
poles having negative real parts for a continuous-time system
or magnitudes less than one for a discrete-time system).
The
dimension $q$ of the residual vector $r(t)$ depends on the
fault detection problem to be addressed.

\index{residual generator!internal form}
The residual signal $r(t)$ in (\ref{detec}) generally depends on
all system inputs $u(t)$,
$d(t)$, $f(t)$ and $w(t)$ via the system output $y(t)$. The \emph{internal form} of the filter is obtained by replacing in (\ref{detec})
${\mathbf{y}}(\lambda)$ by its expression in (\ref{systemw}), and is given by
\be\label{resys} \hspace*{-5mm}{\mathbf{r}}(\lambda) = R(\lambda)\!{\arraycolsep=0mm \ba{c}{\mathbf{u}}(\lambda)\\
{\mathbf{d}}(\lambda) \\ {\mathbf{f}}(\lambda) \\ {\mathbf{w}}(\lambda)\ea }\! =
R_u(\lambda){\mathbf{u}}(\lambda) \!+\!
R_d(\lambda){\mathbf{d}}(\lambda) \!+\!
R_f(\lambda){\mathbf{f}}(\lambda) \!+\! R_w(\lambda){\mathbf{w}}(\lambda)  \, , \hspace*{-4mm}\ee
with $R(\lambda) = [\, R_u(\lambda)\; R_d(\lambda)\; R_f(\lambda)\; R_w(\lambda\,]$ defined as
\be\label{resys1} \ba{c|c|c|c} R_u(\lambda) & R_d(\lambda) & R_f(\lambda) &  R_w(\lambda)\ea :=
{\arraycolsep=1mm Q(\lambda)  \ba{c|c|c|c} G_u(\lambda) & G_d(\lambda) & G_f(\lambda)  & G_w(\lambda) \\
         I_{m_u} & 0 & 0 & 0 \ea } \, .
         \ee
For a properly designed filter $Q(\lambda)$, the
corresponding internal representation $R(\lambda)$ is also a proper and stable system, and  additionally fulfills  specific fault detection and isolation requirements.

\subsection{Fault Detectability} \label{sec:fdetectability}

\index{fault detectability}
The concepts of fault detectability  and complete fault detectability deal with the sensitivity of the residual to an individual fault and to all faults, respectively. For the discussion of these concepts we will assume that no noise input is present in the system model (\ref{systemw}) ($w \equiv 0$).

\begin{definition}\label{fdetectability}
For the system (\ref{systemw}), the $j$-th fault $f_j$ is \emph{detectable} if there exists a fault detection filter $Q(\lambda)$ such that for all control inputs $u$ and all disturbance inputs $d$, the residual $r \not=0$ if $f_j \not= 0$ and $f_k = 0$ for all $k\not = j$.
\end{definition}

\index{fault detectability!complete}
\begin{definition}\label{cfdetectability}
The system (\ref{systemw}) is \emph{completely fault detectable} if there exists a fault detection filter $Q(\lambda)$ such that for each $j$, $j = 1, \ldots, m_f$, all control inputs $u$ and all disturbance inputs $d$, the residual $r \not=0$ if $f_j \not= 0$ and $f_k = 0$ for all $k\not = j$.
\end{definition}

We have the following results, proven in \cite{Varg17}, which characterize the fault detectability and the complete fault detectability properties.
\begin{proposition} \label{T-detectability} \index{fault detectability}
For the system (\ref{systemw}) the $j$-th fault is detectable if and only if
\be\label{fdetec} \rank \big[\, G_d(\lambda) \;\; G_{f_j}(\lambda) \,\big] > \rank G_d(\lambda) ,\ee
where $G_{f_j}(\lambda)$ is the $j$-th column of $G_{f}(\lambda)$ and $\rank \, (\cdot)$ is the normal rank (i.e., over rational functions) of a rational matrix.
\end{proposition}
\begin{theorem} \label{C-detectability} \index{fault detectability!complete}
The system (\ref{systemw}) is completely fault detectable if and only if
\be\label{cfdetec} \rank \big[\, G_d(\lambda) \;\; G_{f_j}(\lambda) \,\big] > \rank G_d(\lambda) , \; j = 1, \ldots, m_f \, .\ee
\end{theorem}


\index{fault detectability!strong}
{Strong fault detectability} is a concept related to the reliability and easiness of performing fault detection. The main idea behind this concept is the ability of the residual generators to produce persistent residual signals in the case of persistent fault excitation. For example, for reliable fault detection it is advantageous to have an asymptotically non-vanishing residual signal in the case of persistent faults as step or sinusoidal signals. On the contrary, the lack of strong fault detectability may make the detection of these type of faults more difficult, because their effects manifest in the residual only during possibly short transients, thus the effect disappears in the residual after an enough long time although the fault itself still persists.

The definitions of strong fault detectability and complete strong fault detectability cover several classes of persistent fault signals. Let $\partial\mathds{C}_s$ denote the boundary of the stability domain, which, in the case of a continuous-time system, is the extended imaginary axis (including also the infinity), while in the case of a discrete-time system, is the unit circle centered in the origin. Let $\Omega \subset \partial\mathds{C}_s$ be a set of complex frequencies, which characterize the classes of persistent fault signals in question. Common choices in a continuous-time setting are $\Omega = \{ 0 \}$ for a step signal or $\Omega = \{ \mathrm{i}\omega \}$ for a sinusoidal signal of frequency $\omega$. However, $\Omega$ may contain several such frequency values or even a whole interval of frequency values, such as $\Omega = \{\mathrm{i}\omega \mid \omega \in [\,\omega_1, \omega_2\,]\}$. We denote by $\mathcal F_\Omega$ the class of persistent fault signals characterized by $\Omega$.

\index{fault detectability!strong}
\begin{definition}\label{sfdetectability}
For the system (\ref{systemw}) and a given set of frequencies $\Omega \subset \partial\mathds{C}_s$, the $j$-th fault $f_j$ is \emph{strong fault detectable} with respect to $\Omega$ if there exists a stable fault detection filter $Q(\lambda)$ such that for all control inputs $u$ and all disturbance inputs $d$, the residual $r(t)  \not=0$ for $t \rightarrow\infty$ if $f_j \in \mathcal F_\Omega$ and $f_k = 0$ for all $k\not = j$.
\end{definition}

\index{fault detectability!complete, strong}
\begin{definition}\label{csfdetectability}
The system (\ref{systemw}) is \emph{completely strong fault detectable} with respect to a given set of frequencies $\Omega \subset \partial\mathds{C}_s$, if there exists a stable fault detection filter $Q(\lambda)$ such that for each $j = 1, \ldots, m_f$, all control inputs $u$ and all disturbance inputs $d$, the residual $r(t)  \not=0$ for $t \rightarrow\infty$ if $f_j \in \mathcal F_\Omega$ and $f_k = 0$ for all $k\not = j$.
\end{definition}

For a given stable filter $Q(\lambda)$ checking the strong detection property of the filter for the $j$-th fault $f_j$ involves to check that $R_{f_j}(\lambda)$ has no zeros in  $\Omega$.
A characterization of strong detectability as a system property  is given in what
follows.\index{fault detectability!strong}
\begin{theorem} \label{T-Sdetectability}
Let $\Omega \subset \partial\mathds{C}_s$ be a given set of frequencies.  For the system (\ref{systemw}), $f_j$ is strong fault detectable with respect to $\Omega$ if and only if $f_j$ is fault detectable and
the rational matrices $G_{e,j}(\lambda)$ and $\ba{c} G_{e,j}(\lambda)\\ F_{e}(\lambda)\ea$ have the same zero structure for each $\lambda_z \in \Omega$, where
\be\label{GejFe} G_{e,j}(\lambda) := \ba{ccc} G_{f_j}(\lambda) & G_u(\lambda) & G_d(\lambda)   \\
0 & I_{m_u}  & 0 \ea , \quad  F_{e}(\lambda) :=  [\, 1 \; \; 0_{1\times m_u} \;\; 0_{1\times m_d}\,] \, .\ee
\end{theorem}

\begin{remark}
Strong fault detectability implies fault detectability, which can be thus assimilated with a kind of \emph{weak} fault detectability property.
For the characterization of the strong fault detectability,  we can impose a weaker condition, involving only the existence of a filter $Q(\lambda)$ without poles in $\Omega$ (instead imposing stability). For such a filter $Q(\lambda)$, the stability can always be achieved by replacing $Q(\lambda)$ by $M(\lambda)Q(\lambda)$, where $M(\lambda)$  is a stable and invertible TFM without zeros in $\Omega$. Such an $M(\lambda)$ can be determined from a left coprime factorization with least order denominator of $[\, Q(\lambda) \; R_{f}\,(\lambda)\,]$.
\finr\end{remark}%

For complete strong fault detectability the strong fault detectability of each individual fault is necessary, however, it is not a sufficient condition.
The following theorem gives a general characterization of the complete strong fault detectability as a system property.\index{fault detectability!complete, strong}%

\begin{theorem} \label{T-sfdetecorig} Let $\Omega$ be the set of frequencies which characterize the persistent fault signals. The system (\ref{systemw}) with $w \equiv 0$  is completely strong fault detectable with respect to $\Omega$ if and only if each fault $f_j$, for $j = 1, \ldots, m_f$, is strong fault detectable with respect to $\Omega$ and all $G_{f_j}(\lambda)$, for $j = 1, \ldots, m_f$,  have the same pole structure in $\lambda_p$ for all  $\lambda_p \in \Omega$.
\end{theorem}

\subsection{Fault Isolability} \label{isolability}
\index{fault isolability}

While the detectability of a fault can be individually defined and checked, for the definition of fault isolability, we need to deal with the interactions among all fault inputs. Therefore for fault isolation, we assume a structuring of the residual vector $r$ into $n_b$ subvectors as in (\ref{rstruct}),  where each individual $q_i$-dimensional subvector $r^{(i)}$ is differently sensitive to faults. We assume that each fault $f_j$ is characterized by a distinct pattern of zeros and ones in a $n_b$-dimensional  vector $s_j$  called the \emph{signature} of the $j$-th fault. Then, fault isolation consists of recognizing which signature matches the resulting decision vector $\iota$ generated by the FDD system in Fig.~\ref{Fig_FDDSystem} according to the partitioning of $r$ in (\ref{rstruct}).

\index{fault isolability!structure matrix}
\index{fault isolability!structure matrix!fault signature}
\index{fault isolability!structure matrix!specification}
\index{structure matrix!fault signature}
\index{structure matrix!specification}
\index{structure matrix|ii}

For the discussion of fault isolability, we will assume that no noise input is present in the model (\ref{systemw}) ($w \equiv 0$). The structure of the residual vector in (\ref{rstruct}) corresponds to a $q\times m_f$ TFM $Q(\lambda)$ ($q = \sum_{i=1}^{n_b}q_i$) of the residual generation filter, built by stacking a bank of $n_b$ filters $Q^{(1)}(\lambda)$, $\ldots$, $Q^{(n_b)}(\lambda)$ as
\be\label{qbank} Q(\lambda) = \ba{c} Q^{(1)}(\lambda)\\ \vdots \\ Q^{(n_b)}(\lambda) \ea  \, .\ee
Thus, the $i$-th subvector $r^{(i)}$ is the output of the $i$-th filter  with the $q_i\times m_f$ TFM $Q^{(i)}(\lambda)$
\be\label{ri_fdip}
{\mathbf{r}}^{(i)}(\lambda) = Q^{(i)}(\lambda)\ba{c}
{\mathbf{y}}(\lambda)\\{\mathbf{u}}(\lambda)\ea \, .\ee
Let $R_f(\lambda)$  be the corresponding $q\times m_f$ fault-to-residual TFM
in (\ref{resys}) and we denote $R^{(i)}_{f_j}(\lambda) := Q^{(i)}(\lambda) \ba{c} G_{f_j}(\lambda) \\ 0 \ea$, the $q_i\times 1$ $(i,j)$-th block
of $R_f(\lambda)$ which describes how the $j$-th fault $f_j$ influences the $i$-th residual subvector $r^{(i)}$. Thus, $R_f(\lambda)$ is an $n_b\times m_f$ block-structured TFM of the form
\be\label{Rf_struct} R_f(\lambda) = \ba{ccc} R^{(1)}_{f_1}(\lambda)& \cdots &R^{(1)}_{f_{m_f}}(\lambda) \\
\vdots & \ddots & \vdots \\
 R^{(n_b)}_{f_1}(\lambda)& \cdots &R^{(n_b)}_{f_{m_f}}(\lambda) \ea \, .\ee
We associate to such a structured $R_f(\lambda)$ the $n_b\times m_f$
\emph{structure
matrix} $S_{R_f}$ whose $(i,j)$-th element is defined as
\be\label{structure_matrix}
{\arraycolsep=1mm\begin{array}{llrll} S_{R_f}(i,j) &=& 1 & \text{ if } & R^{(i)}_{f_j}(\lambda) \not=0 \; ,\\
S_{R_f}(i,j) &=& 0 & \text{ if } & R^{(i)}_{f_j}(\lambda) =0 \, .
\end{array}} \ee
If $S_{R_f}(i,j) = 1$
then we say that the residual component $r^{(i)}$ is sensitive to the $j$-th fault $f_j$, while if $S_{R_f}(i,j) = 0$ then the $j$-th fault $f_j$  is decoupled from $r^{(i)}$.

Fault isolability is a property which involves all faults and this is reflected in the following definition, which relates the fault isolability property to a certain structure matrix $S$. For a given structure matrix $S$, we refer to the
$i$-th row of $S$ as the \emph{specification} associated with the $i$-th residual component $r^{(i)}$,
while the
$j$-th column of $S$ is called the \emph{signature} (or \emph{code}) associated with the $j$-th fault
$f_j$.

\begin{definition}
For a given $n_b\times m_f$ structure matrix $S$, the model (\ref{systemw}) is $S$-\emph{fault isolable} if there exists a fault detection filter $Q(\lambda)$ such that $S_{R_f} = S$.
\end{definition}

When solving fault isolation problems, the choice of a suitable structure matrix $S$ is an important aspect. This choice is, in general, not unique and several  choices may lead to satisfactory synthesis results. In this context, the availability of the maximally achievable structure matrix is of paramount importance, because it allows to construct any $S$ by simply selecting a (minimal) number of achievable specifications (i.e., rows of this matrix). The M-function \texttt{\bfseries genspec}, allows to compute the maximally achievable structure matrix for a given system.

The choice of $S$ should usually reflect the fact that complete fault detectability must be a necessary condition for the $S$-fault isolability. This requirement is fulfilled if $S$ is chosen without zero columns. Also, for the unequivocal isolation of the $j$-th fault, the corresponding $j$-th column of $S$ must be different from all other columns. Structure matrices having all columns pairwise distinct are called \emph{weakly isolating}.\index{fault isolability!weak}
Fault signatures which results as (logical OR) combinations of two or more columns of the structure matrix, can be occasionally employed to isolate simultaneous faults, provided they are distinct from all columns of $S$. In this sense, a structure matrix $S$ which allows the isolation of an arbitrary number of simultaneously occurring faults is called \emph{strongly isolating}.\index{fault isolability!strong} It is important to mention in this context that a system which is not fault isolable for a given $S$ may still be fault isolable for another choice of the structure matrix.

To characterize the fault isolability property, we observe that each block row $Q^{(i)}(\lambda)$ of the  TFM $Q(\lambda)$ is itself a fault detection filter which must achieve the specification 
contained in the $i$-th row  of $S$. Thus, the isolability conditions will consist of a set of $n_b$ independent conditions, each of them characterizing the complete detectability of particular subsets of faults.
 We have the following straightforward characterization of fault isolability.

\index{fault isolability}
\begin{theorem}\label{T-isolability}
For a given $n_b\times m_f$ structure matrix $S$, the model (\ref{systemw}) is $S$-fault isolable  if and only if for $i = 1, \ldots, n_b$
\be\label{fdcondri} \rank \,[\, G_d(\lambda) \; \widehat G_d^{(i)}(\lambda)\;
G_{f_j}(\lambda)\, ] > \rank [\, G_d(\lambda) \; \widehat G_d^{(i)}(\lambda)\, ] , \quad \forall j, \;\; S_{ij} \not = 0\, ,\ee
where $\widehat G_d^{(i)}(\lambda)$ is formed from the columns $G_{f_j}(\lambda)$ of $G_f(\lambda)$ for which $S_{ij} = 0$.
\end{theorem}

The conditions  (\ref{fdcondri}) of Theorem~\ref{T-isolability} give a very general characterization of isolability of faults. An important particular case is \emph{strong fault isolability}, in which case  $S = I_{m_f}$, and thus diagonal. The following result characterizes the strong isolability.

\index{fault isolability!strong}
\begin{theorem}
\label{T-strong-isolability}
The model (\ref{systemw}) is strongly fault isolable  if and only if
\be\label{fdistrong} \rank \, [\, G_d(\lambda) \;
G_{f}(\lambda)\, ] = \rank  G_d(\lambda) + m_f  \, .\ee
\end{theorem}

\begin{remark} \label{left-invertibility}
In the case $m_d = 0$, the strong fault isolability  condition reduces to
the left invertibility condition
\be\label{VA:sufnecfdi}
\rank G_{f}(\lambda) = m_f \, .\ee
This condition is a necessary condition even in the case $m_d \not = 0$ (otherwise $R_f(\lambda)$ would not have full column rank).
\finr \end{remark}

\begin{remark} \label{rem:strong_structure}
The definition of the structure matrix $S_{R_f}$ associated with a given TFM $R_f(\lambda)$ can be extended to cover the strong fault detectability requirement defined by $\Omega \subset \partial\mathds{C}_s$, where $\Omega$ is the  set of relevant frequencies. For each $\lambda_z \in \Omega$, we can define the strong structure matrix at the complex frequency $\lambda_z$ as
\be\label{strong_structure_matrix}
{\arraycolsep=1mm\begin{array}{llrll} S_{R_f}(i,j) &=& 1 & \text{ if } & R^{(i)}_{f_j}(\lambda_z) \not=0 \; ,\\
S_{R_f}(i,j) &=& 0 & \text{ if } & R^{(i)}_{f_j}(\lambda_z) =0 \, .
\end{array}}  \ee
\finr\end{remark}

\subsection{Fault Detection and Isolation Problems} \label{fdp}
\index{a1@EFDP|see{fault detection problem}}
\index{a2@EFDIP|see{fault detection and isolation problem}}
\index{a5@EMMP|see{model-matching problem}}
\index{a6@EFEP|see{model-matching problem}}
\index{a3@AFDP|see{fault detection problem}}
\index{a4@AFDIP|see{fault detection and isolation problem}}
\index{a7@AMMP|see{model-matching problem}}
\index{a8@EMDP|see{model detection problem}}
\index{a9@AMDP|see{model detection problem}}


In this section we formulate several synthesis problems of fault detection and isolation filters for LTI systems. These problems can be considered as a minimal (canonical) set to cover the needs of most practical applications.
For the solution of these problems we seek linear residual generators  (or fault detection filters)
of the form (\ref{detec}), which
process the measurable system outputs $y(t)$ and known control
inputs $u(t)$ and generate the residual signals $r(t)$, which
serve for decision-making on the presence or absence of faults.
The standard requirements on all TFMs appearing in the implementation form (\ref{detec}) and internal form (\ref{resys}) of the fault detection filter are \emph{properness}  and \emph{stability}, to ensure physical realizability of the filter $Q(\lambda)$ and to guarantee a stable behaviour of the FDD system. The \textit{order} of the filter $Q(\lambda)$ is its
\textit{McMillan degree}, that is, the dimension of the state vector of a minimal state-space realization of $Q(\lambda)$. For practical purposes, lower order filters are preferable to larger order ones, and therefore, determining \emph{least order residual generators} is also a desirable synthesis goal. Finally, while the
dimension $q$ of the residual vector $r(t)$ depends on the fault detection problem to be solved, filters with the \emph{least number of outputs}, are always of interest for practical usage.

For the solution of fault detection and isolation problems it is always possible to completely decouple the control input $u(t)$ from the residual $r(t)$ by requiring $R_u(\lambda) = 0$. Regarding the disturbance input $d(t)$ and noise input $w(t)$ we aim to impose a similar condition on the disturbance input $d(t)$ by requiring $R_d(\lambda) = 0$, while minimizing simultaneously the effect of noise input $w(t)$ on the residual (e.g., by minimizing the norm of $R_w(\lambda)$). Thus, from a practical synthesis point of view, the  distinction between $d(t)$ and $w(t)$ lies solely in the way these signals are treated when solving the residual generator synthesis problem.

In all fault detection problems formulated in what follows, we require that by a suitable choice of a stable fault detection filter $Q(\lambda)$, we achieve that  the residual signal $r(t)$ is fully decoupled from the control input $u(t)$ and disturbance input $d(t)$. Thus, the following \emph{decoupling conditions} must be fulfilled for the filter synthesis
\be\label{ens} \begin{array}{ll}
  (i) & R_u(\lambda) = 0 \, ,\\
  (ii) & R_d(\lambda) = 0 \, .
\end{array}
\ee
In the case when condition $(ii)$ can not be fulfilled (e.g., due to lack of sufficient number of measurements), we can redefine some (or even all) components of $d(t)$ as noise inputs and include them in $w(t)$.

For each fault detection problem formulated in what follows, specific requirements have to be fulfilled, which are formulated as additional synthesis conditions. For all formulated problems  we also give the existence conditions of the solutions of these problems. For the proofs of the results consult \cite{Varg17}.

\subsubsection{EFDP -- Exact Fault Detection Problem} \label{sec:EFDP}
\index{fault detection and e@fault detection problem!a@exact (EFDP)|ii}
For the \emph{exact fault detection problem} (EFDP) the basic additional requirement is simply to achieve by a suitable choice of a stable and proper fault detection filter $Q(\lambda)$ that, in the absence of noise input (i.e., $w \equiv 0$), the residual  $r(t)$ is sensitive to all fault components $f_j(t)$, $j = 1, \ldots, m_f$. If a noise input $w(t)$ is present, then we assume the TFM $G_w(s)$ is stable (thus $R_w(\lambda)$ is stable too).
Thus, the following \emph{detection condition} has to be fulfilled:
\be\label{efdp}
  (iii) \;\; R_{f_j}(\lambda) \not = 0,\; j = 1, \ldots, m_f \;\; \textrm{with} \;\; R_f(\lambda) \;\; \textrm{stable.}
\ee
This is precisely the complete fault detectability requirement (without the stability condition) and leads to the following solvability condition:

\begin{theorem}\label{T-EFDP}
\index{fault detection and e@fault detection problem!a@exact (EFDP)!solvability|ii}
For the system (\ref{systemw}), the EFDP is solvable if and only if the system (\ref{systemw}) is completely fault detectable.
\end{theorem}

\index{fault detection and e@fault detection problem!a@exact (EFDP) with strong detectability|ii}
Let $\Omega \subset \partial\mathds{C}_s$ be a given set of frequencies which characterize the  relevant persistent faults. We can give a similar result in the case when the EFDP is solved with a  \emph{strong detection condition}:
\be\label{efdps}
  (iii)' \;\; R_{f_j}(\lambda_z) \not = 0, \;\;\forall \lambda_z \in \Omega, \; j = 1, \ldots, m_f \;\; \textrm{with} \;\; R_f(\lambda) \;\; \textrm{stable.}
\ee

\index{fault detection and e@fault detection problem!a@exact (EFDP) with strong detectability!solvability|ii}
The solvability condition of the EFDP with the strong detection condition above is precisely the complete strong fault detectability requirement as stated by the following theorem.
\begin{theorem}\label{T-EFDPS}
Let $\Omega$ be the set of frequencies which characterize the persistent fault signals.
For the system (\ref{systemw}), the EFDP with the strong detection condition (\ref{efdps}) is solvable if and only if the system (\ref{systemw}) is completely strong fault detectable with respect to $\Omega$.
\end{theorem}

\subsubsection{AFDP -- Approximate Fault Detection Problem}\label{sec:AFDP}
\index{fault detection and e@fault detection problem!approximate (AFDP)|ii}
The effects of the noise input $w(t)$ can usually not be fully decoupled from the residual $r(t)$. In this case, the basic requirements for the choice of $Q(\lambda)$  can be expressed to achieve that the residual  $r(t)$ is  influenced by all fault components $f_j(t)$ and the influence of the noise signal $w(t)$ is negligible. For the \emph{approximate fault detection problem} (AFDP) the following two additional conditions have to be fulfilled:
\be\label{afdp} \begin{array}{ll}
  (iii) & R_{f_j}(\lambda) \not = 0, \; j = 1, \ldots, m_f \;\; \textrm{with} \;\; R_f(\lambda) \;\; \textrm{stable;} \\
  (iv) & R_w(\lambda) \approx 0, \;\; \textrm{with} \;\; R_w(\lambda) \;\; \textrm{stable.}
\end{array}
\ee
Here, $(iii)$ is  the \emph{detection condition} of all faults employed also in the EFDP, while $(iv)$ is the  \emph{attenuation condition} for the noise input. The condition $R_w(\lambda) \approx 0$ expresses the requirement  that the transfer gain $\|R_w(\lambda)\|$ (measured by any suitable norm) can be made arbitrarily small.

The solvability conditions of the formulated AFDP can be easily established:
\begin{theorem}\label{T-AFDP}
For the system (\ref{systemw}) the AFDP is solvable if and only if the EFDP is solvable.
\end{theorem}
\index{fault detection and e@fault detection problem!approximate (AFDP)!solvability|ii}
\index{fault detection and e@fault detection problem!a@exact (EFDP)!solvability}

\begin{remark}
The above theorem is a pure mathematical result. The resulting filter $Q(\lambda)$, which makes $\|R_w(\lambda)\|$ ``small'', may simultaneously reduce $\|R_f(\lambda)\|$, such that while the fault detectability property is preserved, the filter has very limited practical use. In practice, the usefulness of  a solution $Q(\lambda)$ of the AFDP  must be judged by taking into account the maximum size of the noise signal and the desired minimum detectable sizes of faults.
\finr\end{remark}
\index{fault detection and e@fault detection problem!approximate (AFDP)}

\subsubsection{EFDIP -- Exact Fault Detection and Isolation Problem} \label{sec:EFDIP}
\index{fault detection and isolation problem!a@exact (EFDIP)|ii}
For a row-block structured fault detection filter $Q(\lambda)$ as in (\ref{qbank}), let $R_f(\lambda)$ be the corresponding block-structured fault-to-residual TFM  as defined in (\ref{Rf_struct}) with $n_b\times m_f$ blocks, and let $S_{R_f}$ be the corresponding $n_b\times m_f$ structure matrix  defined in (\ref{structure_matrix}) (see Section~\ref{isolability}).  Let $s_j$, $j = 1, \ldots, m_f$ be a set of $n_b$-dimensional binary signature vectors  associated to the faults $f_j$, $j = 1, \ldots, m_f$, which form the desired structure matrix $S := [\, s_1 \;\; \ldots  s_{m_f}\,]$. The \emph{exact fault detection and isolation problem} (EFDIP)   requires to determine for a given $n_b\times m_f$ structure matrix $S$,  a stable and proper filter $Q(\lambda)$ of the form (\ref{qbank}) such that the following condition is additionally fulfilled:
\be\label{efdip}
  (iii) \;\; S_{R_f} = S , \;\; \textrm{with} \; R_f(\lambda) \; \textrm{stable.}
\ee

We have the following straightforward solvability condition:
\index{fault detection and isolation problem!a@exact (EFDIP)!solvability|ii}
\begin{theorem}\label{T-EFDIP}
For the system (\ref{systemw}) with $w \equiv 0$   and a given structure matrix $S$, the EFDIP is solvable  if and only if the system (\ref{systemw}) is $S$-fault isolable.
\end{theorem}

\index{fault detection and isolation problem!a@exact (EFDIP) with strong isolability|ii}
A similar result can be established for the case when $S$ is the $m_f$-th order identity matrix $S = I_{m_f}$. We call the associated synthesis problem the \emph{strong}  EFDIP. The proof is similar to that of Theorem~\ref{T-EFDIP}.

\begin{theorem}\label{T-strongEFDIP}
\index{fault detection and isolation problem!a@exact (EFDIP) with strong isolability!solvability|ii}
For the system (\ref{systemw}) with $w \equiv 0$   and $S = I_{m_f}$, the EFDIP is solvable  if and only if the system (\ref{systemw}) is strongly fault isolable.
\end{theorem}

\subsubsection{AFDIP -- Approximate Fault Detection and Isolation Problem }\label{sec:AFDIP}
\index{fault detection and isolation problem!approximate (AFDIP)|ii}

Let $S$ be a desired $n_b\times m_f$ structure matrix targeted to be achieved by using a structured fault detection filter $Q(\lambda)$ with $n_b$ row blocks as in (\ref{qbank}). The $n_b\times m_f$ block structured fault-to-residual TFM $R_f(\lambda)$, corresponding to $Q(\lambda)$ is defined in (\ref{Rf_struct}),
can be additively decomposed as $R_f(\lambda) = \widetilde  R_f(\lambda) + \overline R_f(\lambda)$, where  $\widetilde  R_f(\lambda)$ and $\overline R_f(\lambda)$ have the same block structure as $R_f(\lambda)$ and have their $(i,j)$-th blocks defined as
\be\label{afdip-rtb} \widetilde  R^{(i)}_{f_j}(\lambda) = S_{ij}R^{(i)}_{f_j}(\lambda), \quad \overline R^{(i)}_{f_j}(\lambda) = (1-S_{ij})R^{(i)}_{f_j}(\lambda) \, .\ee
To address the approximate fault detection and isolation problem, we will target to enforce for
the part $\widetilde R_f(\lambda)$ of $R_f(\lambda)$ the desired structure matrix $S$, while the part $\overline R_f(\lambda)$ must be (ideally) negligible.
The \emph{soft approximate fault detection and isolation problem} (\emph{soft} AFDIP) can be formulated as follows. For a given $n_b\times m_f$ structure matrix $S$, determine a stable and proper filter $Q(\lambda)$  in the form (\ref{qbank}) such that the following conditions are additionally fulfilled:
\be\label{afdip-sr} \hspace*{-7mm}{\arraycolsep=1mm\begin{array}{ll}
  (iii) & S_{\widetilde R_f} = S, \; \overline R_f(\lambda) \approx 0 , \; \text{with} \; R_f(\lambda) \; \text{stable,} \;  \\
    (iv) & R_w(\lambda) \approx 0, \; \text{with} \; R_w(\lambda) \; \text{stable.}
\end{array}}
\ee

The  necessary and sufficient condition for the solvability of the \emph{soft} AFDIP is the solvability of the EFDP.

\begin{theorem}\label{T-AFDIP}
For the system (\ref{systemw}) and a given structure matrix $S$ without zero columns, the \emph{soft} AFDIP is solvable if and only if the EFDP is solvable.
\end{theorem}
\index{fault detection and isolation problem!approximate (AFDIP)!solvability|ii}
\index{fault detection and e@fault detection problem!a@exact (EFDP)!solvability}

\begin{remark} \label{rem:S_full}
If the given structure matrix $S$ has zero columns, then all faults corresponding to the zero columns of $S$ can be redefined as additional noise inputs. In this case, the Theorem~\ref{T-AFDIP} can be applied to a modified system with a reduced set of faults and increased set of noise inputs.
\finr\end{remark}

\index{fault detection and isolation problem!approximate (AFDIP)}

The solvability of the EFDIP is clearly a sufficient condition for the solvability of the \emph{soft} AFDIP, but is not, in general, also a necessary condition, unless we impose in the formulation of the AFDIP the stronger condition $\overline R_f(\lambda) = 0$ (instead  $\overline R_f(\lambda) \approx 0$). This is equivalent to require $S_{R_f} = S$. Therefore, we can alternatively formulate the \emph{strict} AFDIP  to fulfill the conditions:
\be\label{afdip-sr1} \hspace*{-7mm}{\arraycolsep=1mm\begin{array}{ll}
  (iii)' & S_{R_f} = S, \; \text{with} \; R_f(\lambda) \; \text{stable,} \;  \\
    (iv)' & R_w(\lambda) \approx 0, \; \text{with} \; R_w(\lambda) \; \text{stable.}
\end{array}}
\ee
In this case  we have the following result:
\begin{theorem}\label{T-AFDIPE}
\index{fault detection and isolation problem!approximate (AFDIP)!solvability|ii}
For the system (\ref{systemw}) and a given structure matrix $S$, the \emph{strict} AFDIP is solvable with $S_{R_f} = S$ if and only if the EFDIP is solvable.
\end{theorem}

\subsubsection{EMMP -- Exact Model-Matching Problem} \label{sec:EMMP}
\index{model-matching problem!a@exact (EMMP)|ii}

Let $M_{rf}(\lambda)$ be a given $q\times m_f$ TFM of a stable and proper reference model  specifying the desired input-output behaviour from the faults to residuals as ${\mathbf{r}}(\lambda) = M_{rf}(\lambda) {\mathbf{f}}(\lambda)$. Thus, we want to achieve by a suitable choice of a stable and proper $Q(\lambda)$  satisfying $(i)$ and $(ii)$ in (\ref{ens}), that we have additionally $R_f(\lambda) = M_{rf}(\lambda)$. For example, a typical choice for $M_{rf}(\lambda)$ is an $m_f \times m_f$  diagonal and invertible TFM, which ensures that each residual $r_i(t)$ is influenced only by the fault $f_i(t)$. The choice $M_{rf}(\lambda) = I_{m_f}$ targets the solution of an  \emph{exact fault estimation problem} (EFEP).

To determine $Q(\lambda)$, we have to solve the linear rational equation (\ref{resys1}), with the settings $R_u(\lambda) = 0$, $R_d(\lambda) = 0$, and $R_f(\lambda) = M_{rf}(\lambda)$ ($R_w(\lambda)$ and $G_w(\lambda)$ are assumed empty matrices). The choice of $M_{rf}(\lambda)$ may lead to a solution $Q(\lambda)$ which is not proper or is unstable or has both these undesirable properties. Therefore, besides determining $Q(\lambda)$, we also consider the determination of a suitable updating factor $M(\lambda)$ of $M_{rf}(\lambda)$ to ensure the stability and properness of the solution $Q(\lambda)$ for $R_f(\lambda) = M(\lambda) M_{rf}(\lambda)$. Obviously, $M(\lambda)$ must be chosen a proper, stable and invertible TFM. Additionally,  by choosing $M(\lambda)$ diagonal, the zero and nonzero entries of $M_{rf}(\lambda)$ can be also preserved in $R_f(\lambda)$ (see also Section \ref{sec:EFDIP}).
\index{model-matching problem!a@exact fault estimation (EFEP)|ii}

The \emph{exact model-matching problem} (EMMP) can be formulated as follows: given a stable and proper $M_{rf}(\lambda)$, it is required to determine a stable and proper filter $Q(\lambda)$ and a diagonal, proper, stable and invertible TFM $M(\lambda)$ such that, additionally to (\ref{ens}),  the following condition is fulfilled:
\be\label{emmp}
  (iii) \;\; R_f(\lambda) = M(\lambda)M_{rf}(\lambda)\, .\ee

The solvability condition of the EMMP  is the standard solvability condition of systems of linear equations:
\begin{theorem}\label{T-EMMP}
\index{model-matching problem!a@exact (EMMP)!solvability|ii}
For the system (\ref{systemw}) with $w \equiv 0$   and a given $M_{rf}(\lambda)$, the EMMP is solvable if and only if the following condition is fulfilled
\be\label{fdimmcond} \hspace*{-4mm}
\rank\, [\, G_d(\lambda)\; G_f(\lambda)\, ] =
\rank \, \ba{cc} G_d(\lambda) & G_f(\lambda)\\ 0 & M_{rf}(\lambda) \ea \, .
\ee
\end{theorem}

\begin{remark}
When $M_{rf}(\lambda)$ has full column rank $m_f$, the solvability condition (\ref{fdimmcond}) of the EMMP  reduces to the strong isolability condition (\ref{fdistrong}) (see also Theorem~\ref{T-strongEFDIP}). \finr \end{remark}

\begin{remark}\label{rem:gemm} It is possible to solve a slightly more general EMMP, to determine $Q(\lambda)$ and $M(\lambda)$ as before, such that, for given $M_r(\lambda) = [\, M_{ru}(\lambda)\; M_{rd}(\lambda)\; M_{rf}(\lambda)\; M_{rw}(\lambda\,]$, they satisfy
\be\label{emm:resys} {\arraycolsep=1mm Q(\lambda)  \ba{c|c|c|c} G_u(\lambda) & G_d(\lambda) & G_f(\lambda)  & G_w(\lambda) \\
         I_{m_u} & 0 & 0 & 0 \ea }  = M(\lambda)\ba{c|c|c|c} M_{ru}(\lambda) & M_{rd}(\lambda) & M_{rf}(\lambda) &  M_{rw}(\lambda)\ea
\, .
         \ee
This formulation may arise, for example, if $M_r(\lambda)$ is the internal form resulted from an approximate synthesis, for which $R_u(\lambda) \approx 0$, $R_d(\lambda) \approx 0$ and $R_w(\lambda) \approx 0$.

The solvability condition is simply that for solving the  linear system (\ref{emm:resys}) for $M(\lambda) = I$
\be\label{gemm:fdimmcond} \hspace*{-4mm}
\rank\, [\, G_d(\lambda)\; G_f(\lambda)\; G_w(\lambda)\, ] =
\rank \, \ba{ccc} G_d(\lambda) & G_f(\lambda) & G_w(\lambda)\\ M_{rd}(\lambda) & M_{rf}(\lambda) &  M_{rw}(\lambda) \ea \, .
\ee
\finr \end{remark}

The solvability conditions (see Theorem \ref{T-EMMP}) become more involved if we strive for a stable proper solution $Q(\lambda)$ for a given reference model $M_{rf}(\lambda)$ without allowing its updating. For example, this is the case when solving the EFEP for $M_{rf}(\lambda) = I_{m_f}$. For a slightly more general case, we have the following result.

\begin{theorem}\label{T-FEP}
\index{model-matching problem!a@exact (EMMP)}
\index{model-matching problem!a@exact fault estimation (EFEP)!solvability|ii}
For the system (\ref{systemw}) with $w \equiv 0$ and a given stable and minimum-phase $M_{rf}(\lambda)$ of full column rank, the EMMP is solvable with $M(\lambda)= I$ if and only if  the system is strongly fault isolable and $G_f(\lambda)$ is minimum phase.
\end{theorem}
\begin{remark}
If $G_f(\lambda)$ has unstable or infinite zeros, the solvability of the EMMP with $M(\lambda) = I$ is possible provided $M_{rf}(\lambda)$ is chosen such that
\be\label{EMMP-stable} \ba{cc} G_f(\lambda) & G_d(\lambda) \ea \text{    and    }  \ba{cc} G_f(\lambda) & G_d(\lambda) \\ M_{rf}(\lambda) & 0 \ea \ee
have the same unstable zero structure. For this it is necessary that  $M_{rf}(\lambda)$  has the same unstable and infinity zeros structure as $G_f(\lambda)$.
\finr\end{remark}

\subsubsection{AMMP -- Approximate Model-Matching Problem}\label{sec:AMMP}
\index{model-matching problem!approximate (AMMP)|ii}

\index{model-matching problem!a@exact (EMMP)}
Similarly to the formulation of the EMMP, we include the determination of an updating factor of the reference model in the standard formulation of the \emph{approximate model-matching problem} (AMMP). Specifically, for a given stable and proper TFM $M_{rf}(\lambda)$, it is required to determine a stable and proper filter $Q(\lambda)$ and a diagonal, proper, stable and invertible TFM $M(\lambda)$ such that the following conditions are additionally fulfilled:
\be\label{afdip-mm} \begin{array}{ll}
  (iii) & R_f(\lambda) \approx M(\lambda)M_{rf}(\lambda), \;\; \textrm{with} \;\; R_f(\lambda) \;\; \textrm{stable};\\
  (iv) & R_w(\lambda) \approx 0, \;\; \textrm{with} \;\; R_w(\lambda) \;\; \textrm{stable.}
\end{array}
\ee
The conditions $(iii)$ and $(iv)$ mean to simultaneously achieve that $\|R_f(\lambda) - M(\lambda)M_{rf}(\lambda)\| \approx 0$ and $\|R_w(\lambda)\| \approx 0$ (in some suitable norm).

A sufficient condition for the solvability of AMMP is the solvability of the EMMP.
\begin{proposition}\label{P-AMMP}
\index{model-matching problem!a@exact (EMMP)!solvability}
\index{model-matching problem!approximate (AMMP)!solvability|ii}
For the system (\ref{systemw}) and a given $M_{rf}(\lambda)$, the AMMP is solvable if the EMMP is solvable.
\end{proposition}

\begin{remark}\label{rem:gamm} It is possible to formulate a more general AMMP, to determine $Q(\lambda)$ and $M(\lambda)$ as before, such that, for given $M_r(\lambda) = [\, M_{ru}(\lambda)\; M_{rd}(\lambda)\; M_{rf}(\lambda)\; M_{rw}(\lambda)\,]$, they satisfy
\be\label{amm:resys} {\arraycolsep=1mm Q(\lambda)  \ba{c|c|c|c} G_u(\lambda) & G_d(\lambda) & G_f(\lambda)  & G_w(\lambda) \\
         I_{m_u} & 0 & 0 & 0 \ea }  \approx M(\lambda)\ba{c|c|c|c} M_{ru}(\lambda) & M_{rd}(\lambda) & M_{rf}(\lambda) &  M_{rw}(\lambda)\ea
\, .
         \ee
\finr\end{remark}

\subsection{Performance Evaluation of FDI Filters}\label{sec:FDIPerf}

Let $Q(\lambda)$ be a FDI filter of the form (\ref{detec}), which solves one of the six formulated FDI problems in Section \ref{fdp}. Accordingly, in the internal form (\ref{resys}) of the filter, the transfer function matrices $R_u(\lambda)$ and $R_d(\lambda)$ are zero to fulfill the decoupling conditions (\ref{ens}), $R_f(\lambda)$ is a stable transfer function matrix with $m_f$ columns, whose zero/nonzero structure characterizes the fault detection and  isolation properties, while $R_w(\lambda)$ will be generally assumed stable and nonzero. When solving fault detection and isolation problems with a targeted  $n_b\times m_f$ structure matrix $S$, the filters $Q(\lambda)$, $R_f(\lambda)$ and $R_w(\lambda)$ have a row partitioned structure, resulted by stacking banks of $n_b$ filters  as follows
\be\label{qbank-perfbl} Q(\lambda) = \ba{c} Q^{(1)}(\lambda)\\ \vdots \\ Q^{(n_b)}(\lambda) \ea , \quad
R_f(\lambda) = \ba{c} R_f^{(1)}(\lambda)\\ \vdots \\ R_f^{(n_b)}(\lambda) \ea, \quad R_w(\lambda) = \ba{c} R_w^{(1)}(\lambda)\\ \vdots \\ R_w^{(n_b)}(\lambda) \ea .\ee
The transfer function matrix $R_f(\lambda)$ has a block structure as in (\ref{Rf_struct}), which allows to define the associated binary structure matrix $S_{R_f}$, whose $(i,j)$-th element is 1 if $R^{(i)}_{f_j}(\lambda) \not=0$ and 0 if $R^{(i)}_{f_j}(\lambda) \not=0$.
If $S_{R_f}$ is the achieved structure matrix, then ideally $S_{R_f} = S$, but $S_{R_f}$ may also differ from $S$, as in the case of solving a \emph{soft} AFDIP (see Section \ref{sec:AFDIP}).
\index{residual generation!for fault detection and isolation}
\index{residual generator!internal form}

The performance of the fault diagnosis system can be assessed using specific performance criteria, which can also serve for optimization-based tuning of various free parameters which intervene in the synthesis of FDI filters. In what follows we discuss three categories of performance criteria of which, the first one can be used to assess the fault detectability properties of the diagnosis system, the second one characterizes the noise attenuation properties and the third one characterizes the model-matching performance. In the case of block structured filters as in (\ref{qbank-perfbl}), specific performance measures are defined, taking into account the assumed ``ideal'' structure matrix associated with the zero and nonzero columns of $R_f^{(i)}(\lambda)$, which is provided in the $i$-th row of the targeted structure matrix $S$.

\subsubsection{Fault Sensitivity Condition} \label{sec:fscond}
\index{performance evaluation!fault detection and isolation!fault sensitivity condition}
When solving fault detection problems, it is important to assess the sensitivity of the residual signal to individual fault components. The \emph{complete fault detectability} can be assessed by checking $R_{f_j}(\lambda) \neq 0$, for $j = 1, \ldots, m_f$. Alternatively, the assessment of complete fault detectability can be done by checking $\| R_{f}(\lambda) \|_{\infty -}  > 0$, where
\[ \| R_{f}(\lambda) \|_{\infty -} := \min_j \|R_{f_j}(\lambda)\|_\infty  \]
is the $\mathcal{H}_{\infty -}$-index defined in \cite{Varg17}, as a measure of the degree of complete fault detectability. If $\| R_{f}(\lambda) \|_{\infty -}  = 0$, then an least one fault component is not detectable in the residual signal $r$. The assessment of the \emph{strong complete fault detectability} with respect to a set of frequencies contained in a set $\Omega$ comes down to check  $R_{f_j}(\lambda_s) \neq 0$, for $\forall \lambda_s \in \Omega$ and for $j = 1, \ldots, m_f$. Alternatively, the assessment of strong complete fault detectability can be done by checking $\| R_{f}(\lambda) \|_{\Omega -}  > 0$, where
\[ \| R_{f}(\lambda) \|_{\Omega -} := \min_{j}   \{ \inf_{\lambda_s \in \Omega} \|R_{f_j}(\lambda_s)\|_2 \}  \]
is the (modified) $\mathcal{H}_{\infty -}$-index defined over the frequencies contained in $\Omega$ (see \cite{Varg17}).
Since nonzero values of $\| R_{f}(\lambda) \|_{\infty -}$ or $\| R_{f}(\lambda) \|_{\Omega -}$ are not invariant to scaling (e.g., when replacing $Q(\lambda)$ by $\alpha Q(\lambda)$), these quantities are less appropriate to quantitatively assess the degrees of complete detectability.

A scaling independent measure of complete fault detectability is the \emph{fault sensitivity condition} defined (over all frequencies) as
\[ J_{1} =   \| R_{f}(\lambda) \|_{\infty -} / \max_j \|R_{f_j}(\lambda)\|_\infty. \]
Similarly, scaling independent measure of the strong complete fault detectability is the \emph{fault sensitivity condition} defined (over the frequencies contained in $\Omega$) as
\[ \widetilde J_{1} =   \| R_{f}(\lambda) \|_{\Omega -} / \max_{j}\{ \sup_{\lambda_s \in \Omega} \|R_{f_j}(\lambda_s)\|_2 \}. \]
For a completely fault detectable system $J_1$ satisfies
\[ 0 < J_1 \leq 1  \]
and for a strong completely fault detectable system $\widetilde J_1$ satisfies
\[ 0 < \widetilde J_1 \leq 1 . \]
A value of $J_1$ (or of $\widetilde J_1$) near to 1, indicates nearly equal sensitivities of residual to all fault components, and makes easier the choice of suitable thresholds for fault detection. On contrary, a small value of $J_1$ (or of $\widetilde J_1$) indicates potential difficulties in detecting some components of the fault vector, due to a very low sensitivity of the residual to these fault components. In such cases, employing fault detection filters with several outputs ($q > 1$) could be advantageous.

When solving fault detection and isolation problems with a targeted structure matrix $S$, we obtain partitioned filters in the form (\ref{qbank-perfbl}) and we can define for each individual filter an associated fault condition number. Let $f^{(i)}$ be formed from the subset of faults corresponding to nonzero entries in the $i$-th row of $S$ and let $R_{f^{(i)}}^{(i)}(\lambda)$ be formed from the corresponding columns of $R_{f}^{(i)}(\lambda)$.  To characterize the  complete fault detectability of the subset of faults corresponding to nonzero entries in the $i$-th row of $S$ we can define the fault condition number of the $i$-th filter as
\[ J_{1}^{(i)} =   \big\| R_{f^{(i)}}^{(i)}(\lambda) \big\|_{\infty -} / \max_j \big\|R_{f_j}^{(i)}(\lambda)\big\|_\infty . \]
Similarly, to characterize the strong complete fault detectability of the subset of faults corresponding to nonzero entries in the $i$-th row of $S$,  we define the fault condition number of the $i$-th filter as
\[ \widetilde J_{1}^{(i)} =   \big\| R_{f^{(i)}}^{(i)}(\lambda) \big\|_{\Omega -} / \max_{j}\{ \sup_{\lambda_s \in \Omega} \big\|R_{f_j}^{(i)}(\lambda_s)\big\|_2 \} . \]

\subsubsection{Fault-to-Noise Gap} \label{sec:f2ngap}
\index{performance evaluation!fault detection and isolation!fault-to-noise gap}
A performance criterion relevant to solve approximate fault detection problems is the \emph{fault-to-noise gap } defined as
\[ J_2 = \| R_{f}(\lambda) \|_{\infty -} / \| R_{w}(\lambda) \|_{\infty} , \]
which represents a measure of the noise attenuation property of the designed filter. By convention, $J_2 = 0$ if $\| R_{f}(\lambda) \|_{\infty -} = 0$ and $J_2 = \infty$ if $\| R_{f}(\lambda) \|_{\infty -} > 0$ and $\| R_{w}(\lambda) \|_{\infty} = 0$ (e.g., when solving exact synthesis problems without noise inputs). A finite frequency variant of the above criterion, which allows to address strong fault detectability aspects for a given set $\Omega$ of relevant frequencies is
\[ \widetilde J_2 = \| R_{f}(\lambda) \|_{\Omega -} / \| R_{w}(\lambda) \|_{\infty} . \]
The higher the value of $J_2$ (or $\widetilde J_2$), the easier is to choose suitable thresholds to be used for fault detection purposes in the presence of noise. Therefore, the maximization of the above gaps is a valuable goal in improving the fault detection capabilities of the fault diagnosis system in the presence of exogenous noise.

For a partitioned filter in the form (\ref{qbank-perfbl}) and a targeted structure matrix $S$, we can define for the $i$-th filter component the associated value of the fault-to-noise gap, which  characterizes the noise attenuation properties of the $i$-th filter. Let $f^{(i)}$ be formed from the subset of faults corresponding to nonzero entries in the $i$-th row of $S$ and  let $\bar f^{(i)}$ be formed from the complementary subset of faults corresponding to zero entries in the $i$-th row of $S$. If $R_{f^{(i)}}^{(i)}(\lambda)$ and $R_{\bar f^{(i)}}^{(i)}(\lambda)$  are formed from the columns of $R_{f}^{(i)}(\lambda)$ corresponding to $f^{(i)}$ and $\bar f^{(i)}$, respectively, then the fault-to-noise gap of the $i$-th filter can be defined as
\[ J_{2}^{(i)} =   \big\| R_{f^{(i)}}^{(i)}(\lambda) \big\|_{\infty -} / \big\|\big[\, R_{\bar f^{(i)}}^{(i)}(\lambda) \;  R_{w}^{(i)}(\lambda)\,\big]\big\|_\infty . \]
This definition covers both the case of a \emph{soft} AFDIP as well as of a \emph{strict} AFDIP (see Section \ref{sec:AFDIP}).
For a similar characterization of the strong complete fault detectability of the subset of faults corresponding to nonzero entries in the $i$-th row of $S$,  we have
\[ \widetilde J_{2}^{(i)} =   \big\| R_{f^{(i)}}^{(i)}(\lambda) \big\|_{\Omega -} / \big\|\big[\, R_{\bar f^{(i)}}^{(i)}(\lambda) \;  R_{w}^{(i)}(\lambda)\,\big]\big\|_\infty . \]

\subsubsection{Model-matching performance} \label{sec:mmperf}
\index{performance evaluation!fault detection and isolation!model-matching performance}

A criterion suitable to characterize the solution of model-matching based syntheses is the  residual error norm
\[ J_3 = \big\| R(\lambda)- M(\lambda)M_r(\lambda)\big\|_{\infty/2}, \]
where $R(\lambda) = [\, R_u(\lambda)\; R_d(\lambda)\; R_f(\lambda)\; R_w(\lambda) \,\,]$  is the resulting internal form (\ref{resys1}), $M_r(\lambda)$ is a desired reference model $M_r(\lambda) = [\, M_{ru}(\lambda)\; M_{rd}(\lambda)\; M_{rf}(\lambda)\; M_{rw}(\lambda)\,]$ and $M(\lambda)$ is an updating factor. When applied to the results computed by other synthesis approaches (e.g., to solve the EFDP, AFDP, EFDIP, the \emph{strict} AFDIP or EMMP), this criterion can be formulated as
\[ \widetilde J_3 = \big\| R_w(\lambda)\big\|_{\infty/2}, \]
which corresponds to assume that $M(\lambda) = I$ and $M_r(\lambda) = [\, R_u(\lambda)\; R_d(\lambda)\; R_f(\lambda)\; 0 \,]$ (i.e., a perfect matching of control, disturbance and fault channels is always achieved).

In the case of solving an EFDIP or a \emph{strict} AFDIP, $R_w(\lambda)$ has the partitioned form in (\ref{qbank-perfbl}). For this case, we can define for the $i$-th filter component the associated model-matching performance $J_3^{(i)}$, characterizing the noise attenuation property of the $i$-th filter. $J_3^{(i)}$ is defined simply as
\[ J_3^{(i)} := \big\|R_w^{(i)}(\lambda)\big\|_{\infty/2} .\]
When solving a \emph{soft }AFDIP, we can use a  more general definition, which also accounts for possibly  no exact matching of a targeted structure matrix $S$ in the fault channel. Assuming the partitioned filter in the form (\ref{qbank-perfbl}) and a targeted structure matrix $S$, we build $\overline R_f^{(i)}(\lambda)$, with its $j$-th column defined as $\overline R_{f_j}^{(i)}(\lambda) := (1-S_{ij}) R_{f_j}^{(i)}(\lambda)$ (see also (\ref{afdip-rtb})).
We can define the model matching performance criterion of the $i$-th component filter as
\[ \widetilde J_3^{(i)} = \big\|  \big[\, \overline R_f^{(i)}(\lambda) \, R_w^{(i)}(\lambda) \,\big] \big\|_{\infty/2}.  \]
In the case of solving an EFDIP or a \emph{strict} AFDIP, $\overline R_{f_j}^{(i)}(\lambda) = 0$, and therefore $\widetilde J_3^{(i)} = J_3^{(i)}$.
\newpage
\section{Model Detection Basics}\label{sec:MDBasics}

In this section we describe first the basic model detection task and introduce and characterize the concept of model detectability. Two  model detection problems are formulated in the book \cite{Varg17} relying on LTI multiple models.  The formulated synthesis problems, involve the exact synthesis and the approximate synthesis of model detection filters. Jointly with the formulation of the model detection problems, general solvability conditions are given in terms of ranks of certain transfer function matrices. More details and the proofs of the  results are available in  Chapters 2 and 4 of \cite{Varg17}.

\subsection{Basic Model Detection Task}

\index{model detection}
Multiple models which describe various fault situations have been frequently used for fault detection purposes. In such applications, the detection of the occurrence of a fault comes down to identifying, using the available measurements from the measurable outputs and control inputs,  that model (from a collection of models) which best matches the dynamical behaviour of the faulty plant. The term \emph{model detection} describes the model identification task consisting of the selection of a model from a collection of $N$ models, which best matches the current dynamical behaviour of a plant.

\index{model detection}

\begin{figure}[thpb]
\begin{center}
\includegraphics[height=7.8cm]{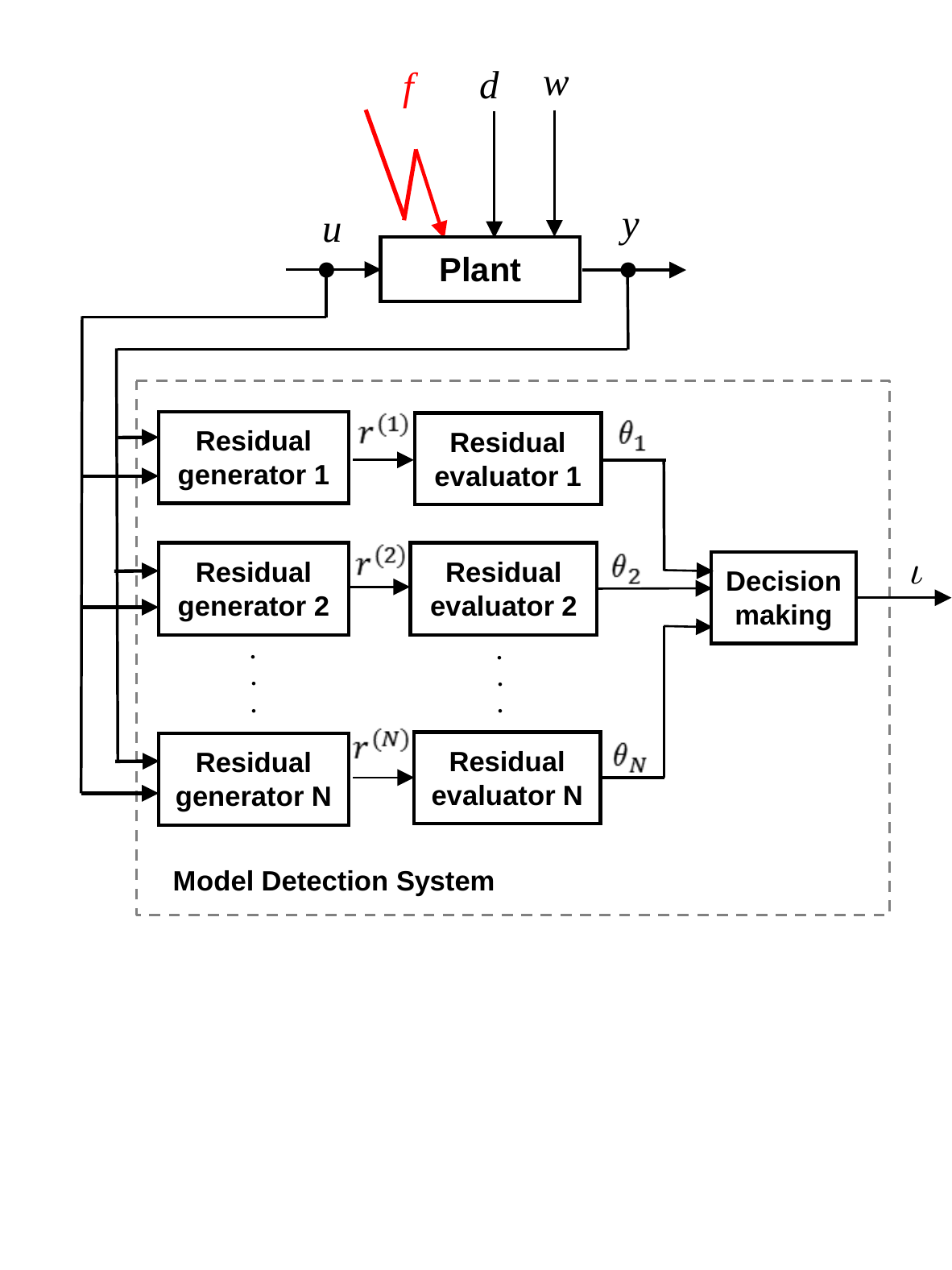}
\vspace*{-2mm}   
\caption{Basic model detection setup.}  
\label{Fig_MDSystem}                                 
\end{center}                                 
\end{figure}

A typical model detection setting is shown in Fig.~\ref{Fig_MDSystem}. A bank of $N$ residual generation filters (or residual generators) is used, with $r^{(i)}(t)$ being the output of the $i$-th residual generator. The $i$-th component $\theta_i$ of  the $N$-dimensional  evaluation vector $\theta$ usually represents an approximation of  $\|r^{(i)}\|_2$, the $\mathcal{L}_2$- or $\ell_2$-norm of $r^{(i)}$. The $i$-th component of the $N$-dimensional decision vector $\iota$ is set to 0 if $\theta_i \leq \tau_i$ and 1 otherwise, where $\tau_i$ is a suitable threshold. The $j$-th model is ``detected'' if $\iota_j =0$ and $\iota_i =1$ for all $i \not = j$. It follows that model detection can be interpreted as a particular type of
week fault isolation with $N$ signature vectors, where the $N$-dimensional $j$-th signature vector has all elements set to one, excepting the $j$-th entry which is set to zero. An alternative decision scheme can also be devised if $\theta_i$ can be associated with a distance function from the current model to the $i$-th model. In this case, $\iota$ is a scalar, set to $\iota = j$, where $j$ is the index for which $\theta_j = \min_{i=1:N} \theta_i$. Thus, the decision scheme selects that model $j$ which best fits with the current model characterized by the measured input and output data.

The underlying synthesis techniques of model detection systems rely on multiple-model
descriptions of physical fault cases.
Since different degrees of performance degradations can be easily described via multiple models, model detection techniques have potentially the capability to address certain fault identification aspects too.

\subsection{Multiple Physical Fault Models} \label{physfaultmodel}
For physically modelled faults, each fault mode leads
to a distinct model.
Assume that we have $N$ LTI models describing the fault-free and faulty systems,
 and for $j = 1, \ldots , N$ the $j$-th model  is specified in the input-output form
\be\label{systemi} {\mathbf{y}}^{(j)}(\lambda) =
G_u^{(j)}(\lambda){\mathbf{u}}(\lambda)
+ G_d^{(j)}(\lambda){\mathbf{d}}^{(j)}(\lambda)
+ G_w^{(j)}(\lambda){\mathbf{w}}^{(j)}(\lambda), \ee%
where $y^{(j)}(t) \in \mathds{R}^{p}$ is the output vector of the $i$-th system with control
input $u(t) \in \mathds{R}^{m_u}$, disturbance input $d^{(j)}(t) \in \mathds{R}^{m_d^{(j)}}$ and noise input $w^{(j)}(t) \in \mathds{R}^{m_w^{(j)}}$, and where
$G_u^{(j)}(\lambda)$, $G_d^{(j)}(\lambda)$ and $G_w^{(j)}(\lambda)$  are the TFMs from
the corresponding plant inputs to outputs. The significance of disturbance and noise inputs, and the basic difference between them, have already been discussed in Section~\ref{sec:additive_faults}.
The state-space realizations corresponding to the multiple model (\ref{systemi}) are for $j = 1, \ldots , N$ of the form
\be\label{ssystemi-dvar} \begin{array}{rcl}E^{(j)}\lambda x^{(j)}(t) &=& A^{(j)}x^{(j)}(t) + B^{(j)}_u u(t) + B^{(j)}_d d^{(j)}(t) + B^{(j)}_w w^{(j)}(t)  \, ,\\
y^{(j)}(t) &=& C^{(j)}x^{(j)}(t) + D^{(j)}_u u(t) + D^{(j)}_d d^{(j)}(t) + D^{(j)}_w w^{(j)}(t)   \, , \end{array} \ee
where $x^{(j)}(t) \in \mathds{R}^{n^{(j)}}$ is the state vector of the $j$-th system and,  generally, can have different dimensions for different systems.
\index{faulty system model!physical}
\index{faulty system model!multiple model}

The  multiple-model description represents a very general way to describe plant models with various faults. For example, {extreme}  variations of parameters representing the so-called {parametric faults}, can be easily described by multiple models.

\subsection{Residual Generation} \label{mdec_res}
\index{residual generation!for model detection}
Assume we have $N$ LTI models of the form (\ref{systemi}), for $j = 1, ..., N$, but the $N$ models originate from a common underlying system with $y(t) \in \mathds{R}^{p}$, the measurable output vector, and $u(t) \in \mathds{R}^{m_u}$, the known control input. Therefore,  $y^{(j)}(t) \in \mathds{R}^p$ is the output vector of the $j$-th system with the control input $u(t) \in \mathds{R}^{m_u}$, disturbance input $d^{(j)}(t) \in \mathds{R}^{m_d^{(j)}}$ and noise input $w^{(j)}(t) \in \mathds{R}^{m_w^{(j)}}$, respectively, and $G_u^{(j)}(\lambda)$, $G_d^{(j)}(\lambda)$ and $G_w^{(j)}(\lambda)$  are the TFMs from
the corresponding plant inputs to outputs.
We explicitly assumed that all models are controlled with the same control inputs $u(t)$, but the disturbance and noise inputs $d^{(j)}(t)$ and $w^{(j)}(t)$, respectively, may differ for each component model.
For complete generality of our problem formulations, we will allow that these TFMs are general rational matrices (proper or improper) for which
we will not \emph{a priori} assume any further properties.

\index{residual generator!implementation form}
Residual generation for model detection is performed using $N$ linear residual generators, which
process the measurable system outputs $y(t)$ and known control
inputs $u(t)$ and generate $N$ residual signals $r^{(i)}(t)$, $i = 1, \ldots, N$,  which
serve for decision making on which model best matches the current input-output measurement data.
As already mentioned, model detection can be interpreted as a week fault isolation problem with an $N\times N$ structure matrix $S$ having all its elements equal to one, excepting those on its diagonal which are zero. The task of model detection is thus to find out the model which best matches the measurements of outputs and inputs, by comparing the resulting decision vector $\iota$ with the set of signatures associated to each model and coded in the columns of $S$. The $N$ residual generation filters in their implementation form are described for $i=1,
\ldots, N$,  by the input-output relations
 \be\label{detecmi}
{\mathbf{r}}^{(i)}(\lambda) = Q^{(i)}(\lambda)\ba{c}
{\mathbf{y}}(\lambda)\\{\mathbf{u}}(\lambda)\ea  \, ,\ee
where $y$ and $u$ is the \emph{actual} measured system output and control input, respectively. The TFMs $Q^{(i)}(\lambda)$, for $i = 1, \ldots, N$, must be proper and stable.
The
dimension $q_i$ of the residual vector component $r^{(i)}(t)$ can be chosen always one, but occasionally values $q_i > 1$ may provide better sensitivity to model mismatches.

\index{residual generator!internal form}
Assuming $y(t) = y^{(j)}(t)$, the residual signal component $r^{(i)}(t)$ in (\ref{detecmi}) generally depends on all system inputs $u(t)$,
$d^{(j)}(t)$ and $w^{(j)}(t)$
via the system output $y^{(j)}(t)$.  The \emph{internal form} of the $i$-th filter driven by the $j$-th model is obtained by replacing in (\ref{detecmi}) ${\mathbf{y}}(\lambda)$ with
${\mathbf{y}}^{(j)}(\lambda)$  from (\ref{systemi}). To make explicit the dependence of $r^{(i)}$ on the $j$-th model, we will use $\widetilde r^{(i,j)}$, to denote the $i$-th residual output for the $j$-th model. After replacing in (\ref{detecmi}) ${\mathbf{y}}(\lambda)$ with ${\mathbf{y}}^{(j)}(\lambda)$ from (\ref{systemi}),  we obtain
\be\label{ri_internal}
\begin{array}{lrl}
{\widetilde{\mathbf{r}}}^{(i,j)}(\lambda) &:=& R^{(i,j)}(\lambda) \ba{c}{\mathbf{u}}(\lambda) \\ {\mathbf{d}}^{(j)}(\lambda)\\ {\mathbf{w}}^{(j)}(\lambda)\ea \\ \\[-2mm] &=&
R_u^{(i,j)}(\lambda){\mathbf{u}}(\lambda) +
R_d^{(i,j)}(\lambda){\mathbf{d}}^{(j)}(\lambda) +
R_w^{(i,j)}(\lambda){\mathbf{w}}^{(j)}(\lambda) \, ,
\end{array} \ee
with $R^{(i,j)}(\lambda) := \ba{c|c|c} R_u^{(i,j)}(\lambda) & R_d^{(i,j)}(\lambda)&R_w^{(i,j)}(\lambda)\ea$ defined as
\be\label{ri_internal1} \ba{c|c|c} R_u^{(i,j)}(\lambda) & R_d^{(i,j)}(\lambda)&R_w^{(i,j)}(\lambda)\ea :=
Q^{(i)}(\lambda)\ba{c|c|c} G_u^{(j)}(\lambda) & G_d^{(j)}(\lambda) & G_w^{(j)}(\lambda) \\
I_{m_u} & 0 & 0 \ea \, .\ee
For a successfully designed set of filters $Q^{(i)}(\lambda)$, $i = 1, \ldots, N$, the
corresponding internal representations $R^{(i,j)}(\lambda)$ in (\ref{ri_internal}) are also a proper and stable.

\subsection{Model Detectability} \label{sec:mdetectability}
\index{model detectability}

The concept of model detectability concerns with the sensitivity of the components of the residual vector to individual models from a given collection of models. Assume that we have
$N$ models, with  the $j$-th model specified in the input-output form (\ref{systemi}). For the discussion of the model detectability concept we will assume that no noise inputs are present in the models (\ref{systemi}) (i.e., $w^{(j)} \equiv 0$ for $j=1,\ldots,N$).
For model detection purposes, $N$ filters of the form (\ref{detecmi}) are employed. It follows from (\ref{ri_internal}) that the $i$-th component $r^{(i)}$ of the residual $r$ is sensitive to the $j$-th model provided
\be\label{mmdetec}   R^{(i,j)}(\lambda)  :=\ba{cc} R_u^{(i,j)}(\lambda) & R_d^{(i,j)}(\lambda) \ea \not = 0 \, . \ee
This condition involves the use of both control and disturbance inputs for model detection and can be useful even in the case of absence of control inputs.

For most of practical applications, it is however necessary to be able to perform model detection also in the (unlikely) case when the disturbance inputs are zero. Therefore, to achieve model detection independently of the presence or absence of disturbances, it is meaningful to impose instead (\ref{mmdetec}), the stronger condition
\be\label{mmdetecr}   R_u^{(i,j)}(\lambda)  \not = 0 \, . \ee
This condition involves the use of only control inputs for model detection purposes and is especially relevant to active methods for model detection based on employing special inputs to help the discrimination between models.

Depending on which of the condition (\ref{mmdetec}) or (\ref{mmdetecr}) are relevant for a particular model detection application,  we define the following two concepts of model detectability.


\index{model detectability!extended}
\begin{definition}\label{emdetectability}
 The multiple model defined by the $N$ component systems (\ref{systemi}) with $w^{(j)} \equiv 0$ for $j = 1, \ldots, N$,  is \emph{extended model detectable} if there exist $N$ filters of the form (\ref{detecmi}), such that $R^{(i,j)}(\lambda)$ defined in (\ref{mmdetec}) fulfills $R^{(i,i)}(\lambda) = 0$ for $i = 1, \ldots, N$ and $R^{(i,j)}(\lambda)\not= 0$ for all $i,j = 1,\ldots,N$ such that $i\not=j$.
 \end{definition}

\begin{definition}\label{mdetectability}
 The multiple model defined by the $N$ component systems (\ref{systemi}) with $w^{(j)} \equiv 0$ for $j = 1, \ldots, N$,  is \emph{model detectable} if there exist $N$ filters of the form (\ref{detecmi}), such that $R^{(i,j)}(\lambda)$ defined in (\ref{mmdetec}) fulfills $R^{(i,i)}(\lambda) = 0$ for $i = 1, \ldots, N$ and $R_u^{(i,j)}(\lambda)\not= 0$ for all $i,j = 1,\ldots,N$ such that $i\not=j$.
 \end{definition}
\noindent The Definition \ref{mdetectability} of model detectability involves the usage of only the control inputs for model detection purpose, and therefore implies the more general property of extended model detectability in Defintion \ref{emdetectability}. In the case of lack of disturbance inputs, the two definitions coincide.\\

The following result, proven in \cite{Varg17},  characterizes the  extended model detectability property.

%

\begin{theorem} \label{model_detectability}
The multiple model defined by the $N$ component systems (\ref{systemi}) with $w^{(j)} \equiv 0$ for $j = 1, \ldots, N$, is \emph{extended model detectable}
if and only if for $i =  1,
\ldots, N$
\be\label{mmdetectability}
\rank\, [\, G_d^{(i)}(\lambda)\;\;
G_d^{(j)}(\lambda)\;\;
G_u^{(i)}(\lambda)\!-\!G_u^{(j)}(\lambda)\, ] > \rank\,
G_d^{(i)}(\lambda)  \;\; \forall j \not = i  \, . \ee
\end{theorem}

The characterization of model detectability (using only control inputs) can be simply established as a  corollary of this theorem.

\begin{theorem} \label{model_detectabilityr}
The multiple model defined by the $N$ component systems (\ref{systemi}) with $w^{(j)} \equiv 0$ for $j = 1, \ldots, N$, is model detectable
if and only if for $i =  1,
\ldots, N$
\be\label{mmrdetectability}
\rank\, [\, G_d^{(i)}(\lambda)\;\;
G_u^{(i)}(\lambda)\!-\!G_u^{(j)}(\lambda)\, ] > \rank\,
G_d^{(i)}(\lambda)  \;\; \forall j \not = i  \, . \ee
\end{theorem}

We can also define the concepts of strong model detectability and strong extended model detectability with respect to classes of persistent control inputs characterized by a set of complex frequencies $\Omega \subset \partial\mathds{C}_s$. The following definitions formalize the aim that for each model $j$, there exists at least one excitation signal class characterized by a frequency  $\lambda_z \in \Omega$ for which all residual components $r^{(i)}(t)$ for $i\neq j$ are asymptotically nonzero and $r^{(j)}(t)$ asymptotically vanishes.

\index{model detectability!strong}
\begin{definition}\label{smdetectability}
 The multiple model defined by the $N$ component systems (\ref{systemi}) with $w^{(j)} \equiv 0$ for $j = 1, \ldots, N$,  is \emph{strong model detectable} with respect to a set of frequencies $\Omega \subset \partial\mathds{C}_s$ if there exist $N$ stable filters of the form (\ref{detecmi}), such that $R^{(i,j)}(\lambda)$ defined in (\ref{mmdetec}) fulfills $R^{(i,i)}(\lambda) = 0$ for $i = 1, \ldots, N$ and for all $i,j = 1,\ldots,N$ with $i\not=j$, $\exists$ $\lambda_z \in \Omega$ such that $R_u^{(i,j)}(\lambda_z)\not= 0$.
\end{definition}

\index{model detectability!strong}
\begin{definition}\label{smdetectability-ext}
 The multiple model defined by the $N$ component systems (\ref{systemi}) with $w^{(j)} \equiv 0$ for $j = 1, \ldots, N$,  is \emph{strong extended model detectable} with respect to a set of frequencies $\Omega \subset \partial\mathds{C}_s$ if there exist $N$ stable filters of the form (\ref{detecmi}), such that $R^{(i,j)}(\lambda)$ defined in (\ref{mmdetec}) fulfills $R^{(i,i)}(\lambda) = 0$ for $i = 1, \ldots, N$ and for all $i,j = 1,\ldots,N$ with $i\not=j$, $\exists$ $\lambda_z \in \Omega$ such that $R^{(i,j)}(\lambda_z)\not= 0$.
\end{definition}

The following results characterize the strong model detectability property and, respectively, the strong extended model detectability property.
\begin{theorem} \label{model_stmdetectability}
Let $\Omega \subset \partial\mathds{C}_s$ be a given set of frequencies, such that none of $\lambda_z \in \Omega$ is a pole of any of the component system (\ref{systemi}), for $j = 1, \ldots, N$. Then, the multiple model (\ref{systemi}) with $w^{(j)} \equiv 0$ for $j = 1, \ldots, N$, is strong model detectable with respect to $\Omega$
if and only if for $i =  1,
\ldots, N$
\be\label{mmsdetectability}
\forall j \not = i, \exists \lambda_z \in \Omega \text{ such that }
\rank\, [\, G_d^{(i)}(\lambda_z)\;\;
G_u^{(i)}(\lambda_z)\!-\!G_u^{(j)}(\lambda_z)\, ] > \rank\,
G_d^{(i)}(\lambda_z)   \, . \ee
\end{theorem}

\begin{theorem} \label{model_stmdetectability-ext}
Let $\Omega \subset \partial\mathds{C}_s$ be a given set of frequencies, such that none of $\lambda_z \in \Omega$ is a pole of any of the component system (\ref{systemi}), for $j = 1, \ldots, N$. Then, the multiple model (\ref{systemi}) with $w^{(j)} \equiv 0$ for $j = 1, \ldots, N$, is strong model detectable with respect to $\Omega$
if and only if for $i =  1,
\ldots, N$
\be\label{mmsdetectability-ext}
\forall j \not = i, \exists \lambda_z \in \Omega \text{ such that }
\rank\, \big[\, G_d^{(i)}(\lambda_z)\;\; G_d^{(j)}(\lambda_z)\;\;
G_u^{(i)}(\lambda_z)\!-\!G_u^{(j)}(\lambda_z)\, \big] > \rank\,
G_d^{(i)}(\lambda_z)   \, . \ee
\end{theorem}

\subsection{Model Detection Problems} \label{mdp}
In this section we formulate the exact and approximate synthesis problems of model detection filters for the collection of $N$ LTI systems (\ref{systemi}).
As in the case of the EFDIP or AFDIP, we seek $N$ linear residual generators  (or model
 detection filters)
of the form (\ref{detecmi}), which
process the measurable system outputs $y(t)$ and known control
inputs $u(t)$ and generate the $N$ residual signals $r^{(i)}(t)$  for $i = 1, \ldots, N$. These signals
serve for decision-making by comparing the pattern of fired and not fired residuals  with the signatures coded in the columns of the associated standard $N\times N$ structure matrix $S$ with zeros on the diagonal and ones elsewhere.
The standard requirements for the TFMs of the filters $Q^{(i)}(\lambda)$  in (\ref{detecmi}) are \emph{properness}  and \emph{stability}. For practical purposes, the orders of the filter $Q^{(i)}(\lambda)$ must be as small as possible. Least order filters $Q^{(i)}(\lambda)$ can be usually achieved by employing scalar output least order filters.

In analogy to the formulations of the EFDIP and AFDIP,
we use the internal form of the $i$-th residual generator  (\ref{ri_internal}) to formulate the basic model detection requirements. Independently of the presence of the noise inputs $w^{(j)}$, we will target that the $i$-th residual is exactly decoupled from the $i$-th model if $w^{(i)} \equiv 0$ and is sensitive to the $j$-th model, for all $j \not = i$.
These requirements can be easily translated into algebraic conditions using the internal form (\ref{ri_internal}) of the $i$-th residual generator. If both control and disturbance inputs are involved in the model detection then the following conditions have to be fulfilled
\be\label{EMDPe} \begin{array}{ll}
  (i) & [\, R_u^{(i,i)}(\lambda) \;\; R_d^{(i,i)}(\lambda)\,] = 0 , \;\; i = 1,\ldots, N \, ,\\
  (ii) & [\, R_u^{(i,j)}(\lambda) \;\; R_d^{(i,j)}(\lambda)\,] \not = 0, \;\; \forall j \not = i, \text{  with  } [\, R_u^{(i,j)}(\lambda) \;\; R_d^{(i,j)}(\lambda)\,] \text{  stable.}
\end{array}
\ee
while if only control inputs have to be employed for model detection, then the following conditions have to be fulfilled
\be\label{EMDP} \begin{array}{ll}
  (i) & [\, R_u^{(i,i)}(\lambda) \;\; R_d^{(i,i)}(\lambda)\,] = 0 , \;\; i = 1,\ldots, N \, ,\\
  (ii)' & R_u^{(i,j)}(\lambda) \not = 0, \;\; \forall j \not = i, \text{  with  } [\, R_u^{(i,j)}(\lambda) \;\; R_d^{(i,j)}(\lambda)\,] \text{  stable.}
\end{array}
\ee

Here, $(i)$ is the \emph{model decoupling condition} for the $i$-th model in the $i$-th residual component, while $(ii)$ and $(ii)'$ are the \emph{model sensitivity condition} of the $i$-th residual component to all models, excepting the $i$-th model.
In the case when condition $(i)$ cannot be fulfilled (e.g., due to lack of sufficient measurements), some (or even all) components of $d^{(i)}(t)$ can be redefined as noise inputs and included in $w^{(i)}(t)$.

In what follows, we formulate the exact and approximate model detection problems, for which we give  the existence conditions of the solutions. For the proof of the results consult \cite{Varg17}.

\subsubsection{EMDP -- Exact Model Detection Problem}\label{sec:EMDP}
\index{model detection problem!a@exact (EMDP)|ii}

The standard requirement for solving the \emph{exact model detection problem} (EMDP) is to determine  for the multiple model (\ref{systemi}), in the absence of noise input (i.e., $w^{(j)}\equiv 0$ for $j=1,\ldots, N$), a set of $N$ proper and stable filters $Q^{(i)}(\lambda)$ such that, for $i = 1, \ldots, N$, the conditions (\ref{EMDPe}) or (\ref{EMDP}) are fulfilled.
These conditions are similar to the model detectability requirement and lead to the following solvability condition:

\begin{theorem}\label{T-EMDPe}
\index{model detection problem!a@exact (EMDP)!solvability|ii}
For the multiple model (\ref{systemi}) with $w^{(j)}\equiv 0$ for $j=1,\ldots, N$, the EMDP is solvable with conditions (\ref{EMDPe}) if and only if the multiple model  (\ref{systemi}) is extended model detectable.
\end{theorem}

\begin{theorem}\label{T-EMDP}
\index{model detection problem!a@exact (EMDP)!solvability|ii}
For the multiple model (\ref{systemi}) with $w^{(j)}\equiv 0$ for $j=1,\ldots, N$, the EMDP is solvable with conditions (\ref{EMDP}) if and only if the multiple model  (\ref{systemi}) is model detectable.
\end{theorem}

\index{model detection problem!a@exact (EMDP) with strong model detectability|ii}
Let $\Omega \subset \partial\mathds{C}_s$ be a given set of frequencies which characterize the  relevant persistent input and disturbance signals. We can give a similar result in the case when the EMDP is solved, by replacing the condition $(ii)'$ in (\ref{EMDP}),  with the  \emph{strong model detection condition}:
\be\label{emdps}
  (ii)'' \;\;\forall j \not = i, \;\; \exists \lambda_z \in \Omega \text{ such that } R_u^{(i,j)}(\lambda_z) \not = 0, \;\; \text{  with  } [\, R_u^{(i,j)}(\lambda) \;\; R_d^{(i,j)}(\lambda)\,] \text{  stable.}
\ee

\index{model detection problem!a@exact (EMDP) with strong model detectability!solvability|ii}
The solvability condition of the EMDP with the strong model detection condition above is precisely the strong model detectability requirement as stated by the following theorem.
\begin{theorem}\label{T-EMDPS}
Let $\Omega$ be the set of frequencies which characterize the persistent control input signals.
For the multiple model (\ref{systemi}) with $w^{(j)}\equiv 0$ for $j=1,\ldots, N$, the EMDP with the strong model detectability condition (\ref{emdps}) is solvable if and only if the multiple model  (\ref{systemi}) is strong model detectable.
\end{theorem}

A similar result holds when targeting the strong extended model detectability property.
\subsubsection{AMMP -- Approximate Model Detection Problem}\label{sec:AMDP}
\index{model detection problem!approximate (AMDP)|ii}
The effects of the noise input $w^{(i)}(t)$ can usually not be fully decoupled from the residual $r^{(i)}(t)$. In this case, the basic requirements for the choice of $Q^{(i)}(\lambda)$  can be expressed as achieving that the residual  $r^{(i)}(t)$ is  influenced by all models in the multiple model (\ref{systemi}), while the influence of the $i$-th model is only due to the noise signal $w^{(i)}(t)$ and is negligible. Using the internal form (\ref{ri_internal}) of the $i$-th residual generator, for the \emph{approximate model detection problem} (AMDP) the following additional conditions to (\ref{EMDPe}) or (\ref{EMDP}) have to be fulfilled:
\be\label{amdp} \begin{array}{ll}
  (iii) & R_{w}^{(i,i)}(\lambda) \approx 0 ,\;\; \textrm{with} \;\; R_{w}^{(i,i)}(\lambda) \;\; \textrm{stable;} \\
  (iv) & R_{w}^{(i,j)}(\lambda) \;\; \textrm{stable} \; \forall j \not = i.
\end{array}
\ee
Here, $(iii)$ is the  \emph{attenuation condition} of the noise input.

The solvability conditions of the AMDP are precisely those of the EMDP:
\index{model detection problem!a@exact (EMDP)!solvability|ii}
\begin{theorem}\label{T-AMDP}\index{model detection problem!approximate (AMDP)!solvability|ii}
For the multiple model (\ref{systemi})   the AMDP is solvable if and only  the EMDP is solvable.
\end{theorem}

\subsection{Analysis and Performance Evaluation of Model Detection Filters} \label{sec:mdanalperf}
\subsubsection{Distances between Models} \label{sec:mdnugap}
\index{model detection!$\nu$-gap distances}
For the setup of model detection applications, an important first step is the selection of a representative set of component models to serve for the design of model detection filter. A practical requirement to set up multiple models as in (\ref{systemi}) or (\ref{ssystemi-dvar}) is to choose a set of component models, such that, each component model is sufficiently far away of the rest of models. A suitable tool to measure the distance between two models is the $\nu$-gap metric introduced in \cite{Vinn93}. For two transfer function matrices $G_1(\lambda)$ and $G_2(\lambda)$ of the same dimensions,  consider the normalized left coprime factorization $G_1(\lambda) = \widetilde M_1^{-1}(\lambda)\widetilde N_1(\lambda)$ (i.e., $\big [\, \widetilde N_1(\lambda) \; \widetilde M_1(\lambda)\,\big]$ is \emph{coinner}) and the normalized right coprime factorizations $G_1(\lambda) = N_1(\lambda)M_1^{-1}(\lambda)$ and $G_2(\lambda) = N_2(\lambda)M_2^{-1}(\lambda)$ (i.e., $\left[\begin{smallmatrix} N_i(\lambda) \\M_1(\lambda)\end{smallmatrix}\right]$ is \emph{inner} for $i = 1, 2$).  With
$\widetilde L_2(\lambda) := [\,-\widetilde M_2(\lambda) \; \widetilde N_2(\lambda)\,]$, $R_i(\lambda) := \left[ \begin{smallmatrix} N_i(\lambda) \\ M_i(\lambda) \end{smallmatrix}\right]$ for $i = 1, 2$, and $g(\lambda) := \det\big(R_2^\sim(\lambda) R_1(\lambda)\big)$, we have the following definition of the $\nu$-gap metric between the two transfer-function matrices:
\be\label{nugap} \delta_\nu(G_1(\lambda),G_2(\lambda)) := \left\{ \begin{array}{cl} \big\| \widetilde L_2(\lambda)R_1(\lambda) \big\|_\infty & \text{if } g(\lambda) \neq 0 \; \forall \lambda \in \partial\mathds{C}_s \text{ and } \wno(g) = 0 ,\\
1 & \text{otherwise} , \end{array}\right. \ee
where $\wno(g)$ denotes the \emph{winding number} of $g(\lambda)$ about the appropriate critical point for $\lambda$ following the corresponding standard Nyquist contour.  \index{transfer function!winding number}%
The winding number of $g(\lambda)$ can be determined as the difference between the number of unstable zeros of $g(\lambda)$ and the number of unstable poles of $g(\lambda)$ \cite{Vidy11}.
Generally, for any $G_1(\lambda)$ and $G_2(\lambda)$, we have  $0 \leq \delta_\nu(G_1(\lambda),G_2(\lambda))\leq 1$.
If $\delta_\nu\big(G_1(\lambda),G_2(\lambda)\big)$ is small, then we can say that
$G_1(\lambda)$ and $G_2(\lambda)$  are close and it is likely that a model detection filter suited for $G_1(\lambda)$ will also work with $G_2(\lambda)$, and therefore, one of the two models can be probably removed from the set of component models. On the other side, if $\delta_\nu\big(G_1(\lambda),G_2(\lambda)\big)$  is nearly equal to 1, then $G_1(\lambda)$ and $G_2(\lambda)$ are sufficiently distinct, such that an easy discrimination between the two models is possible. A common criticism of the $\nu$-gap metric is that there are many transfer function matrices $G_2(\lambda)$ at a distance $\delta_\nu\big(G_1(\lambda),G_2(\lambda)\big) = 1$ to a given $G_1(\lambda)$, but the metric fails to differentiate between them. However, this aspect should not rise difficulties in model detection applications.

In \cite{Vinn01}, the point-wise $\nu$-gap metric is also defined to evaluate the distance between two models in a single frequency point. If $\lambda_k$ is a fixed complex frequency, then the point-wise $\nu$-gap metric between two transfer-function matrices $G_1(\lambda)$ and $G_2(\lambda)$ at the frequency $\lambda_k$ is:
\be\label{nugap-pointwise} \delta_\nu(G_1(\lambda_k),G_2(\lambda_k)) := \left\{ \begin{array}{cl} \big\| \widetilde L_2(\lambda_k)R_1(\lambda_k) \big\|_2 & \text{if } g(\lambda) \neq 0 \; \forall \lambda \in \partial\mathds{C}_s \text{ and } \wno(g) = 0 ,\\
1 & \text{otherwise} , \end{array}\right. \ee

For a set of $N$ component models with input-output forms as in (\ref{systemi}), it is useful to determine the pairwise $\nu$-gap distances between the control input channels of the component models by defining the symmetric matrix $\Delta$, whose $(i,j)$-th entry is the $\nu$-gap distance between the transfer-function matrices of the $i$-th and $j$-th model
\be\label{Delta_nu-gaps} \Delta_{ij} := \delta_\nu\big(G_u^{(i)}(\lambda),G_u^{(j)}(\lambda)\big) . \ee
It follows that $\Delta$ has all its diagonal elements zero. For model detection applications all  off-diagonal elements of $\Delta$ must be nonzero, otherwise there are models which can not be potentially discriminated.
The definition (\ref{Delta_nu-gaps}) of the distances between the $i$-th and $j$-th models focuses only on the control input channels. In most of practical applications of the model detection, this is perfectly justified by the fact that, a certain control activity is always necessary, to ensure reliable discrimination among models, independently of the presence or absence of disturbances. However, if the disturbance inputs are relevant to perform model detection (e.g., there are no control inputs), and all component models share the same disturbance inputs (i.e., $d^{(j)}(t) = d(t)$ for $j = 1, \ldots, N$), then the definition of $\Delta$ in (\ref{Delta_nu-gaps}) can be modified to include the disturbance inputs as well
\be\label{Delta_nu-gaps-ext}  \Delta_{ij} := \delta_\nu\big(\big[\,G_u^{(i)}(\lambda)\;G_d^{(i)}(\lambda)\,\big],
\big[\,G_u^{(j)}(\lambda)\;G_d^{(j)}(\lambda)\,\big]\big) . \ee
If $\lambda_k$, $k = 1, \ldots, n_f$, is a set of $n_f$ frequency values, then, instead (\ref{Delta_nu-gaps}), we can use the maximum of the point-wise distances
\[ \Delta_{ij} := \max_k\delta_\nu\big(G_u^{(i)}(\lambda_k),G_u^{(j)}(\lambda_k)\big) ,\]
and similarly, instead (\ref{Delta_nu-gaps-ext}), we can use the maximum of the point-wise distances
\[ \Delta_{ij} := \max_k\delta_\nu\big(\big[\,G_u^{(i)}(\lambda_k)\;G_d^{(i)}(\lambda_k)\,\big],
\big[\,G_u^{(j)}(\lambda_k)\;G_d^{(j)}(\lambda_k)\,\big]\big) .\]

\index{model detection!$\mathcal{H}_\infty$-norm distances}
\index{model detection!$\mathcal{H}_2$-norm distances}
Besides the $\nu$-gap distance between two transfer function matrices, it is possible to use distances defined in terms of the $\mathcal{H}_\infty$ norm or the $\mathcal{H}_2$ norm of the difference between them. Thus we can  use instead (\ref{Delta_nu-gaps})
 \[ \Delta_{ij} := \big\|G_u^{(i)}(\lambda)-G_u^{(j)}(\lambda)\big\|_\infty \]
 or
\[ \Delta_{ij} := \big\|G_u^{(i)}(\lambda)-G_u^{(j)}(\lambda)\big\|_2. \]
If $\lambda_k$, $k = 1, \ldots, n_f$, is a set of $n_f$ frequency values, then, instead of the above norm-based distances, we can use the maximum of the point-wise distances
\[ \Delta_{ij} := \max_k\big\|G_u^{(i)}(\lambda_k)-G_u^{(j)}(\lambda_k)\big\|_2. \]

\subsubsection{Distances to a Current Model} \label{sec:mddist2c}
\index{model detection!$\nu$-gap distances}

An important aspect which arises in model detection applications, where the use of $\nu$-gap metric could be instrumental, is to assess the nearness of a current model, with the input-output form
\be\label{systemref} {\mathbf{y}}(\lambda) =
\widetilde G_u(\lambda){\mathbf{u}}(\lambda)
+ \widetilde G_d(\lambda){\mathbf{d}}(\lambda)
+ \widetilde G_w(\lambda){\mathbf{w}}(\lambda), \ee%
to the component models in (\ref{systemi}). This involves evaluating, for $j = 1, \ldots, N$, the distances between the control input channels of the models (\ref{systemi}) and  (\ref{systemref}) as
\be\label{dist_nu_gap} \eta_j := \delta_\nu\big(G_u^{(j)}(\lambda),\widetilde G_u(\lambda)\big) . \ee
It is also of interest to determine the index $\ell$ of that component model for  which $\eta_\ell$ is the least distance. This allows to assign the model (\ref{systemref}) to the (open) set of nearby models to the $\ell$-th component model and can serve for checking the preservation of this property by the mapping achieved by the model detection filters via the norms of internal forms $R_u^{(i,j)}(\lambda)$  in (\ref{ri_internal}).

If the disturbance inputs are also relevant to the model detection application, then a similar extension as above is possible to assess the distances between a current model and a set of component models by redefining $\eta_j$ in (\ref{dist_nu_gap}) as
\be\label{dist_nu_gap_ext} \eta_j := \delta_\nu\big(\big[\,G_u^{(j)}(\lambda)\;G_d^{(j)}(\lambda)\,\big],
\big[\,\widetilde G_u(\lambda)\;\widetilde G_d(\lambda)\,\big]\big) . \ee

As before, we can alternatively use distances defined in terms of the $\mathcal{H}_\infty$ norm or the $\mathcal{H}_2$ norm. Thus, instead (\ref{dist_nu_gap}), we can use
\[ \eta_j := \big\|G_u^{(j)}(\lambda)-\widetilde G_u(\lambda)\big\|_\infty \]
or
\[ \eta_j := \big\|G_u^{(j)}(\lambda)-\widetilde G_u(\lambda)\big\|_2 . \]
If $\lambda_k$, $k = 1, \ldots, n_f$, is a given set of $n_f$ frequency values, then, instead of the above peak distances, we can use the maximum of the point-wise distances over the finite set of frequency values.

\subsubsection{Distance Mapping Performance } \label{sec:mdperf}
\index{model detection!distance mapping}
\index{performance evaluation!model detection!distance mapping}
One of the goals of the model detection is to achieve a special mapping of the distances between component models using $N$ model detection filters of the form (\ref{detecmi}) such that the norms of the transfer-function matrices $R_u^{(i,j)}(\lambda)$  or of $\big[\, R_u^{(i,j)}(\lambda)\; R_d^{(i,j)}(\lambda)\,\big]$ in the internal forms of the filters (\ref{ri_internal}) qualitatively reproduce the $\nu$-gap distances expressed by the $\Delta$ matrix, whose $(i,j)$-th entries are defined in (\ref{Delta_nu-gaps}) or (\ref{Delta_nu-gaps-ext}, respectively. The preservation of this distance  mapping property is highly desirable, and the choice of model detection filters must be able to ensure this feature (at least partially for the nearest models).  For example, the choice of the $i$-th filter $Q^{(i)}(\lambda)$ as a left annihilator  of $\left[\begin{smallmatrix} G_u^{(i)}(\lambda) & G_d^{(i)}(\lambda) \\ I_{m_u} & 0 \end{smallmatrix}\right]$ ensures (see \cite[Remark 6.1]{Varg17}) that norm of $\big[\, R_u^{(i,j)}(\lambda)\; R_u^{(i,j)}(\lambda)\,\big]$ can be interpreted as a weighted distance between the $i$-th and $j$-th component models. It follows that the distance mapping performance of a set of model detection filters $Q^{(i)}(\lambda)$, $i = 1, \ldots, N$ can be assessed by computing mapped distance matrix $\Gamma$, whose $(i,j)$-th entry is
\be\label{mapdist} \Gamma_{ij} = \big\| R_u^{(i,j)}(\lambda) \big\|_\infty \ee
or, if the disturbance inputs are relevant,
\be\label{mapdist-ext} \Gamma_{ij} = \big\| \big[\, R_u^{(i,j)}(\lambda)\; R_d^{(i,j)}(\lambda)\,\big]\big\|_\infty .\ee
Using the above choice of the filter $Q^{(i)}(\lambda)$, we have that all diagonal elements of $\Gamma$ are zero. Additionally, to guarantee model detectability or extended model detectability (see Section \ref{sec:mdetectability}), any valid design of the model detection filters must guarantee that all off-diagonal elements of $\Gamma$ are nonzero. These two properties of $\Gamma$ allows to unequivocally identify the exact matching of the current model with one (and only one) of the $N$ component models.

Two other properties of $\Gamma$ are desirable, when solving model detection applications. The first property is the symmetry of $\Gamma$. In contrast to $\Delta$,   $\Gamma$ is generally not symmetric, excepting for some particular classes of component models and for special choices of model detection filters. For example, this property can be ensured if all component models are stable and have no disturbance inputs, by choosing $Q^{(i)}(\lambda) = \big[\, I \; -G_u^{(i)}(\lambda)\,\big]$, in which case $R_u^{(i,j)}(\lambda) = -R_u^{(j,i)}(\lambda)$. Ensuring the symmetry of $\Gamma$, although very desirable,  is in general difficult to be achieved. In practice, it is often sufficient to ensure via suitable scaling that the gains of first row and first column are equal.

The second desirable property of the mapping $\Delta_{ij} \rightarrow \Gamma_{ij}$ is the \emph{monotonic mapping property} of distances, which is the requirement that for all $i$ and $k$ ($i, k = 1, \ldots, N$), if $\Delta_{ij} < \Delta_{ik}$, then $\Gamma_{ij} < \Gamma_{ik}$. Ensuring this property, make easier to address model identification problems for which no exact matching between the current model and any one of the component models can be assumed.

If $\lambda_k$, $k = 1, \ldots, n_f$, is a given set of $n_f$ frequency values, then, instead of the peak distances in (\ref{mapdist}) or in (\ref{mapdist-ext}), we can use the maximum of the point-wise distances over the finite set of frequency values, to assess the strong model detectability or the extended strong model detectability, respectively (see Section \ref{sec:mdetectability}).

\subsubsection{Distance Matching Performance } \label{sec:mdmatch}
\index{model detection!distance matching}
\index{performance evaluation!model detection!distance matching}
To evaluate the distance matching property of the model detection filters in the case when no exact matching between the current model (\ref{systemref}) and any one of the component models (\ref{systemi}) can be assumed, we can define the corresponding current internal forms as
\be\label{ri_internal1-cur} \ba{c|c|c} \widetilde R_u^{(i)}(\lambda) & \widetilde R_d^{(i)}(\lambda)&\widetilde R_w^{(i)}(\lambda)\ea :=
Q^{(i)}(\lambda)\ba{c|c|c} \widetilde G_u(\lambda) & \widetilde G_d(\lambda) & \widetilde G_w(\lambda) \\
I_{m_u} & 0 & 0 \ea \ee
and evaluate the mapped distances $\gamma_i$, for $i = 1, \ldots, N$, defined as
\be\label{mapdistcur}  \gamma_i := \big\| \widetilde R_u^{(i)}(\lambda) \big\|_\infty \ee
or, if the disturbance inputs are relevant,
\be\label{mapdistcur-ext} \gamma_i := \big\| \big[\,  \widetilde R_u^{(i)}(\lambda)\; \widetilde R_d^{(i)}(\lambda)\,\big]\big\|_\infty .\ee
The index $\ell$ of the smallest value $\gamma_\ell$ provides (for a well designed set of model detection filters) the index of the best matching component model of the current model.

If $\lambda_k$, $k = 1, \ldots, n_f$, is a given set of $n_f$ frequency values, then, instead of the above peak distances, we can use the maximum of the point-wise distances over the finite set of frequency values.

\subsubsection{Model Detection Noise Gaps } \label{sec:mdgap}
\index{model detection!noise gap}
\index{performance evaluation!model detection!noise gap}
The noise attenuation performance of model detection filters can be characterized via the noise gaps achieved by individual filters. The noise gap for the $i$-th filter can be defined in terms of the resulting internal forms (\ref{ri_internal}) as the ratio $\eta_i := \beta_i/\gamma_i$, where
\be\label{mdbetai} \beta_i := \min_{j\neq i} \big\|R_u^{(i,j)}(\lambda)\big\|_\infty \ee
and
\be\label{mdgammai}  \gamma_i := \big\|R_w^{(i,i)}(\lambda)\big\|_\infty .\ee
The values of $\beta_i > 0$, for $i = 1, \ldots, N$ characterize the model detectability property of the collection of the $N$ component models  (\ref{systemi}), while $\gamma_i$ characterizes the worst-case influence of noise inputs on the $i$-th residual component. If $\gamma_i = 0$ (no noise), then $\eta_i = \infty$.

If the disturbance inputs are relevant for the model detection, then instead (\ref{mdbetai}) we can use the following definition of $\beta_i$
\be\label{mdbetaid} \beta_i := \min_{j\neq i} \big\|\big[\,R_u^{(i,j)}(\lambda)\; R_d^{(i,j)}(\lambda)\,\big]\big\|_\infty .\ee
In this case, $\beta_i > 0$, for $i = 1, \ldots, N$ characterize the extended model detectability property of the collection of the $N$ component models  (\ref{systemi}).

If $\lambda_k$, $k = 1, \ldots, n_f$, is a given set of $n_f$ frequency values, then, instead of
(\ref{mdbetai}) we use the maximum of the point-wise distances over the finite set of frequency values
\be\label{mdbetai-pw} \beta_i := \min_{j\neq i} \max_k\big\|R_u^{(i,j)}(\lambda_k)\big\|_\infty \ee
and instead of
(\ref{mdbetaid}) we use
\be\label{mdbetaid-pw} \beta_i := \min_{j\neq i} \max_k\big\|\big[\,R_u^{(i,j)}(\lambda_k)\; R_d^{(i,j)}(\lambda_k)\,\big]\big\|_\infty .\ee

\section{Description of FDITOOLS} \label{sec:userguide}
This user's guide is intended to provide users  basic information on the \textbf{FDITOOLS} collection to solve the fault detection and isolation problems formulated in Section~\ref{fdp} and the model detection problem formulated in Section~\ref{mdp}.
The notations and terminology used throughout this guide have been introduced and extensively discussed in the accompanying book \cite{Varg17}, which also represents the main reference for the implemented  computational methods underlying the analysis and synthesis functions of \textbf{FDITOOLS}. Information on the requirements for installing \textbf{FDITOOLS} are given in Appendix \ref{appA}.

In this section, we present first a short overview of the existing functions of \textbf{FDITOOLS} and then, we illustrate a typical work flow by solving an EFDIP. In-depth information on the command syntax of the functions of the \textbf{FDITOOLS} collection is given is Sections \ref{fditools:analysis} and \ref{fditools:synthesis}.
To execute the examples presented in this guide, simply paste the presented code sequences into the MATLAB command window. More involved examples are given in several case studies presented in \cite{Varg17}.\footnote{Use \url{https://sites.google.com/site/andreasvargacontact/home/book/matlab} to download the case study examples presented in \cite{Varg17}.}

\subsection{Quick Reference Tables}
The current release of \textbf{FDITOOLS} is version V1.0, dated November 30, 2018. The corresponding \texttt{Contents.m} file is listed in Appendix \ref{app:contents}.
This section contains quick reference tables for the functions of the \textbf{FDITOOLS} collection.
The M-files available in the current version of \textbf{FDITOOLS}, which are documented in this user's guide, are listed below by category, with short descriptions.

\begin{longtable}{|p{2.5cm}|p{12cm}|} \hline
\multicolumn{2}{|c|}{\textbf{Demonstration}} \\ \hline
  \texttt{\bfseries FDIToolsdemo}& Demonstration of Fault Detection and Isolation Tools\\
  \hline
\multicolumn{2}{c}{} \\ \hline
\multicolumn{2}{|c|}{\textbf{Setup of synthesis models}} \\ \hline
\texttt{\bfseries fdimodset}&  Setup of models for solving FDI synthesis problems.\\ \hline
\texttt{\bfseries mdmodset} &Setup of models for solving model detection synthesis problems.\\ \hline
\multicolumn{2}{c}{} \\ \hline
\multicolumn{2}{|c|}{\textbf{FDI Related Analysis}} \\ \hline
\texttt{\bfseries fdigenspec}& Generation of achievable FDI specifications.\\
  \hline
\texttt{\bfseries fdichkspec}& Feasibility analysis of a set of FDI specifications.\\
  \hline
\multicolumn{2}{c}{} \\ \hline
\multicolumn{2}{|c|}{\textbf{Model Detection Related Analysis}} \\ \hline
\texttt{\bfseries mddist}& Computation of distances between component models.\\
  \hline
\texttt{\bfseries mddist2c}& Computation of distances to a set of component models.\\
  \hline
\end{longtable}
\newpage
\begin{longtable}{|p{2.5cm}|p{12cm}|} 
\multicolumn{2}{c}{} \\ \hline
\multicolumn{2}{|c|}{\textbf{Performance evaluation of FDI filters}} \\ \hline
\texttt{\bfseries fditspec}& Computation of the weak or strong structure matrix.\\ \hline
\texttt{\bfseries fdisspec} &Computation of the strong structure matrix.\\ \hline
  \texttt{\bfseries fdifscond}&  Fault sensitivity condition of FDI filters.\\ \hline
  \texttt{\bfseries fdif2ngap}& Fault-to-noise gap of FDI filters.\\ \hline \samepage
  \texttt{\bfseries fdimmperf}& Model-matching performance of FDI filters.\\ \hline
\multicolumn{2}{c}{} \\ \hline
\multicolumn{2}{|c|}{\textbf{Performance evaluation of model detection filters}} \\ \hline
  \texttt{\bfseries mdperf}&  Model detection distance mapping performance.\\ \hline
  \texttt{\bfseries mdmatch}& Model detection distance matching performance.\\ \hline
  \texttt{\bfseries mdgap}& Noise gaps of model detection filters.\\ \hline
\multicolumn{2}{c}{} \\ \hline
\multicolumn{2}{|c|}{\textbf{Synthesis of fault detection filters}} \\ \hline
  \texttt{\bfseries efdsyn}& Exact synthesis of fault detection filters.\\ \hline
  \texttt{\bfseries afdsyn}& Approximate synthesis of fault detection filters.\\ \hline
  \texttt{\bfseries efdisyn}& Exact synthesis of fault detection and isolation filters.\\
  \hline
  \texttt{\bfseries afdisyn}& Approximate synthesis of fault detection and isolation filters.\\
  \hline
  \texttt{\bfseries emmsyn}& Exact model matching based synthesis of FDI filters.\\
  \hline
  \texttt{\bfseries ammsyn}& Approximate model matching based synthesis of FDI filters.\\
  \hline
\multicolumn{2}{c}{} \\ \hline
\multicolumn{2}{|c|}{\textbf{Synthesis of model detection filters}} \\ \hline
  \texttt{\bfseries emdsyn}& Exact synthesis of model detection filters.\\ \hline
  \texttt{\bfseries amdsyn}& Approximate synthesis of model detection filters.\\ \hline
\end{longtable}


\subsection{Getting Started}

In this section  we shortly illustrate the typical steps of solving a fault detection and isolation problem, starting with
building an adequate fault model, performing preliminary analysis, selecting the suitable synthesis approach, and evaluating the computed results.

\subsubsection{Building Models with Additive Faults}

In-depth information on how to create and manipulate LTI system models and arrays of LTI system models are available in the online documentation of the Control System Toolbox and in its User' Guide \cite{MLCO15}. These types of models are the basis of the data objects used in the \textbf{FDITOOLS} collection.

The input plant models with additive faults used by all synthesis functions of the \textbf{FDITOOLS} collection are LTI models of the form (\ref{systemw}), given via their equivalent descriptor system state-space realizations of the form (\ref{ssystemw}). The object-oriented framework employed in the Control System Toolbox has been used to define the LTI plant models, by defining several input groups corresponding to various input signal. The employed standard definitions of input groups are: \texttt{\bfseries 'controls'} for the control inputs $u(t)$, \texttt{\bfseries 'disturbances'} for the disturbance inputs $d(t)$, \texttt{\bfseries 'faults'} for the fault inputs $f(t)$, and \texttt{\bfseries 'noise'} for the noise inputs $w(t)$. For convenience, occasionally an input group \texttt{\bfseries 'aux'} can be also defined for additional inputs.  For different ways to define input groups, see the documentation of the Control System Toolbox \cite{MLCO15}.

Once you have a plant model for a system without faults, you can construct models with faults using simple commands in the Control System Toolbox or using the model setup function  \texttt{fdimodset}.
For example, consider a plant model \texttt{sys} with 3 inputs, 3 outputs and 3 state components. Assume that the first two inputs are control inputs which are susceptible to actuator faults and the third input is a disturbance input, which is not measurable and therefore is considered as an unknown input. All outputs are measurable, and assume that the first output is susceptible to sensor fault. The following commands generate a plant model with additive faults as described above:
\begin{verbatim}
rng(50); sys = rss(3,3,3);
inputs = struct('c',1:2,'d',3,'f',1:2,'fs',1);
sysf = fdimodset(sys,inputs);
\end{verbatim}

\subsubsection{Determining the Achievable FDI Specifications}
To determine the achievable strong FDI specifications with respect to constant faults, the function \texttt{fdigenspec} can be used as follows:
\begin{verbatim}
S = fdigenspec(sysf,struct('FDFreq',0));
\end{verbatim}
For the above example, the possible fault signatures are contained in the generic
structure matrix
\[ S = \ba{ccc}  0 & 1 & 1\\ 1 & 0 & 1 \\ 1 & 1 & 0 \\ 1 & 1 & 1\ea ,\]
which indicates that the EFDP is solvable (last row of $S$) and the EFDIP is also solvable
using the specifications contained in the rows of
\[ S_{FDI} = \ba{ccc} 0 & 1 & 1\\ 1 & 0 & 1 \\ 1 & 1 & 0 \ea .\]

\subsubsection{Designing an FDI Filter Using \texttt{efdisyn}}

To solve the EFDIP, an option structure is used to specify various user options for the synthesis function \texttt{efdisyn}. Frequently used options are the desired stability degree for the poles of the fault detection filter, the requirement for performing least order synthesis, or values for the tolerances used for rank computations or fault detectability tests.
The user options can be specified by setting appropriately the respective fields in a MATLAB structure \texttt{options}.
For example, the desired stability degree of $-2$ of the filter, the targeted fault signature specification $S_{FDI}$, and the frequency $0$ for strong synthesis (for constant faults), can be set using
\begin{verbatim}
options = struct('sdeg',-2,'SFDI',S(1:3,:),'FDFreq',0);
\end{verbatim}

\noindent A solution of the EFDIP, for the selected structure matrix $S_{FDI}$, can be computed using the function \texttt{efdisyn} as given below
\begin{verbatim}
[Q,R,info] = efdisyn(sysf,options);
\end{verbatim}
The resulting bank of scalar output fault detection filters, in implementation form, is contained in \texttt{Q} (stored as an one-dimensional cell array of systems), while the corresponding internal forms of the bank of filters are contained in the cell array \texttt{R}.
The information structure \texttt{info} contains further information on the resulting designs.

\subsubsection{Assessing the Residual Generator}

\begin{figure}[h]
\begin{center}
\vspace*{-4mm}
\includegraphics[height=8cm]{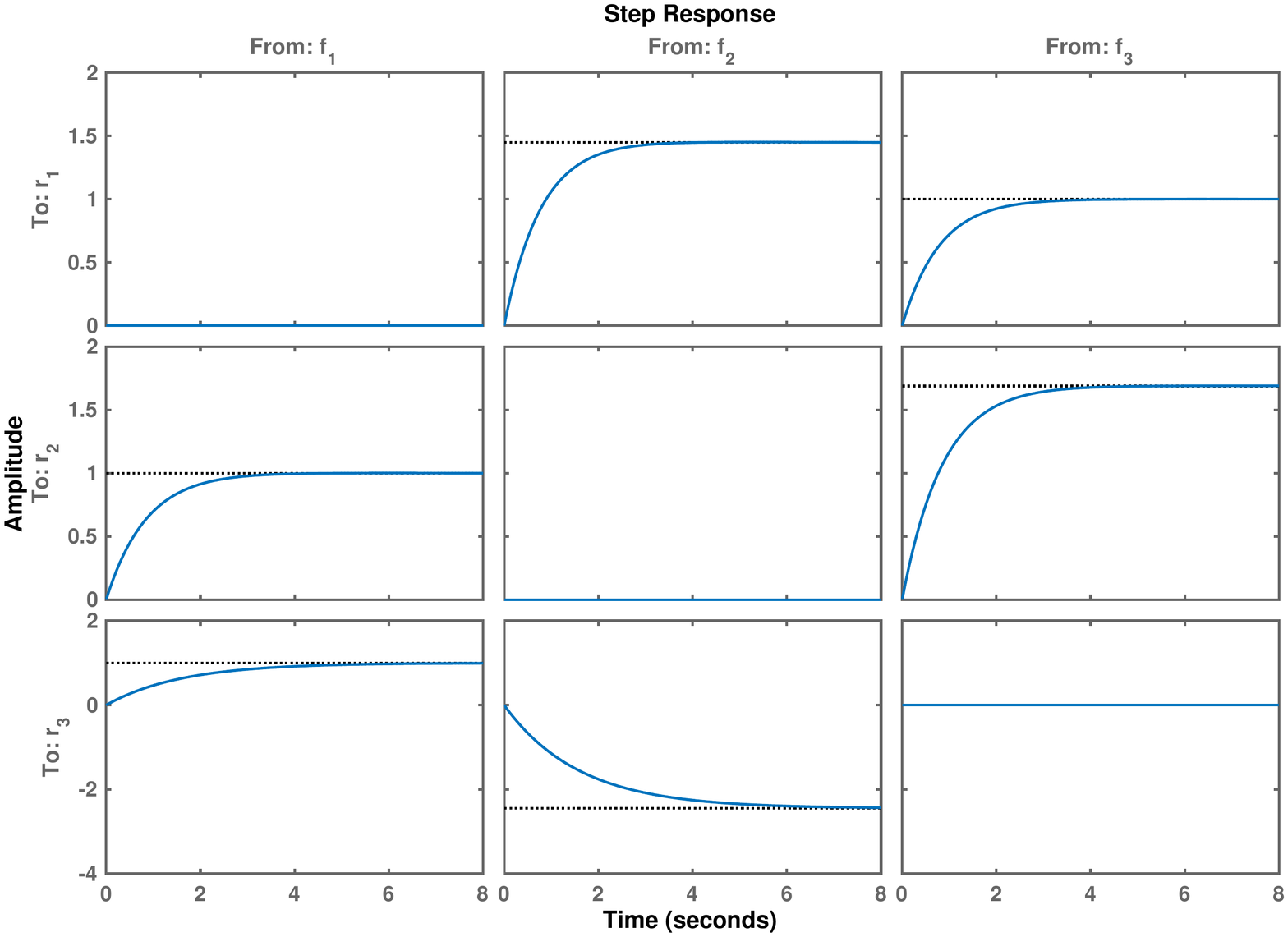}
\vspace*{-5mm}
\caption{Step responses from the fault inputs.}
\label{efdipstep}
 \end{center}
\end{figure}

For assessment purposes, often simulations performed using the resulting internal form provide sufficient qualitative information to verify the obtained results.
The example below illustrates how to simulate step inputs from faults using the computed cell array \texttt{R} containing the internal form of the filter.
\begin{verbatim}
Rf = vertcat(R{:});  % build the global internal form of the filter
Rf.OutputName = strcat(strseq('r_{',1:3),'}');
Rf.InputName  = strcat(strseq('f_{',1:3),'}');
step(Rf,8);
\end{verbatim}
A typical output of this computation can be used to assess the achieved fault signatures as shown in
Fig.\ \ref{efdipstep}. As it can be observed, the diagonal entries of the overall transfer-function matrix of \texttt{Rf} are zero, while all off-diagonal entries are nonzero.

\newpage

The resulting strong structure matrix and the fault sensitivity conditions the resulting bank of internal filters can be obtained directly from the computed internal form \texttt{R}:
\begin{verbatim}
S_strong = fdisspec(R)
fscond   = fdifscond(R,0,S_strong)
\end{verbatim}
The resulting values of the fault sensitivity conditions $\{0.6905, 0.5918, 0.4087 \}$ indicate an acceptable sensitivity of all residual components to individual faults.

%
%

\subsection{Functions for the Setup of Synthesis Models} \label{fditools:setup}

The functions for the setup of synthesis models allow to easily define models in forms suitable for the use of analysis and synthesis functions.\\[-7mm]

\subsubsection{\texttt{\bfseries fdimodset}}

\subsubsection*{Syntax}
\index{M-functions!\texttt{\bfseries fdimodset}}
\begin{verbatim}
SYSF = fdimodset(SYS,INPUTS)
\end{verbatim}

\subsubsection*{Description}

\noindent \texttt{fdimodset} builds synthesis models with additive faults to be used in conjunction with the analysis and synthesis functions of FDI filters.

\subsubsection*{Input data}
\begin{description}
\item
\texttt{SYS} is a LTI system in a descriptor system state-space form
\be\label{fdimodset:sysss}
\begin{aligned}
E\lambda x(t)  &=   Ax(t)+ B\widetilde u(t)  ,\\
y(t) &=  C x(t)+ D \widetilde u(t)  ,
\end{aligned}
\ee
where $y(t) \in \mathds{R}^p$ is the system output and $\widetilde u(t) \in \mathds{R}^{m}$ is the system global input, which usually includes the control and disturbance inputs, but may also explicitly include fault inputs, noise inputs and auxiliary inputs.
\item
 \texttt{INPUTS} is a MATLAB structure used to specify the indices of the columns of the matrices $B$ and, respectively $D$, which correspond to the control, disturbance, fault, noise and auxiliary inputs, using the following fields:
{\setlength\LTleft{30pt}\begin{longtable}{|l|p{11.6cm}|} \hline \textbf{\texttt{INPUTS} fields} & \textbf{Description} \\ \hline
       \texttt{controls}       & vector of indices of the control inputs (Default: void)\\ \hline
       \texttt{c}       & alternative short form to specify the indices of
                        the control inputs \newline (Default: void)\\ \hline
       \texttt{disturbances}       & vector of indices of the disturbance inputs (Default: void)\\ \hline
       \texttt{d}       & alternative short form to specify the indices of
                        the disturbance inputs (Default: void)\\ \hline
       \texttt{faults}       & vector of indices of the fault inputs (Default: void)\\ \hline
       \texttt{f}       & alternative short form to specify the indices of
                        the fault inputs \newline (Default: void)\\ \hline
         \url{faults\_sen}       & vector of indices of the  outputs subject to sensor faults (Default: void)\\ \hline
       \texttt{fs}       & alternative short form to specify the indices of the  outputs subject to sensor faults  (Default: void)\\ \hline
     \texttt{noise}       & vector of indices of the noise inputs (Default: void)\\ \hline
       \texttt{n}       & alternative short form to specify the indices of
                        the noise inputs \newline (Default: void)\\ \hline
       \texttt{aux}       & vector of indices of the auxiliary inputs (Default: void)\\ \hline
\end{longtable}     }
\end{description}

\subsubsection*{Output data}
\begin{description}
\item
\texttt{SYSF} is a  LTI system  in the state-space form
\be\label{fdimodset:sysss1}
{\begin{aligned}
E\lambda x(t)  & =  Ax(t)+ B_u u(t)+ B_d d(t) + B_f f(t)+ B_w w(t) + B_{v} v(t)  , \\
y(t) & =  C x(t) + D_u u(t)+ D_d d(t) + D_f f(t) + D_w w(t)+ B_{v} v(t) .
\end{aligned}}
\ee
where $u(t)\in \mathds{R}^{m_u}$ are the control inputs, $d(t)\in \mathds{R}^{m_d}$ are the disturbance inputs, $f(t)\in \mathds{R}^{m_f}$ are the fault inputs,  $w(t)\in \mathds{R}^{m_w}$ are the noise inputs and $v(t)\in \mathds{R}^{m_v}$ are auxiliary inputs.
Any of the inputs components $u(t)$, $d(t)$, $f(t)$, $w(t)$  or $v(t)$ can be void. 
The input groups for $u(t)$, $d(t)$, $f(t)$, $w(t)$ and $v(t)$ have the standard names \texttt{\bfseries 'controls'}, \texttt{\bfseries 'disturbances'}, \texttt{\bfseries 'faults'}, \texttt{\bfseries 'noise'} and \texttt{'aux'}, respectively. The resulting model \texttt{SYSF} inherits the sampling time of the original model \texttt{SYS}.
\end{description}

\subsubsection*{Remark on input and output data} The function \texttt{\bfseries fdimodset} can also be employed if \texttt{SYS} is an $N \times 1$ or $1 \times N$ array of LTI  models, in which case the resulting \texttt{SYSF} is also an $N \times 1$ or $1 \times N$  array  of LTI models, respectively.

\subsubsection*{Method}

The system matrices $B_u$, $B_d$, $B_f$, $B_w$, $B_v$, and $D_u$, $D_d$, $D_f$, $D_w$, $D_v$ in the resulting model (\ref{fdimodset:sysss1}) are defined by specifying the indices of the columns of the matrices $B$ and, respectively $D$, in the model (\ref{fdimodset:sysss}) which correspond to the control, disturbance, fault, noise and auxiliary inputs.
If the system \texttt{SYS} in (\ref{fdimodset:sysss}) has the equivalent input-output form
\be\label{fdimodset:iosys} {\mathbf{y}}(\lambda) =
G(\lambda)\widetilde{\mathbf{u}}(\lambda)  \ee
and the resulting system \texttt{SYSF} in (\ref{fdimodset:sysss1}) has the equivalent input-output form
\be\label{fdimodset:iosys1} {\mathbf{y}}(\lambda) =
G_u(\lambda){\mathbf{u}}(\lambda) +
G_d(\lambda){\mathbf{d}}(\lambda) +
G_f(\lambda){\mathbf{f}}(\lambda) +
G_w(\lambda){\mathbf{w}}(\lambda) +
G_v(\lambda){\mathbf{v}}(\lambda) ,
 \ee
then the following relations exist among the above transfer function matrices
\[ \begin{array}{lcl}G_u(\lambda) &=& G(\lambda)S_u, \\ G_d(\lambda) &=& G(\lambda)S_d, \\ G_f(\lambda) &=& [\, G(\lambda)S_f \; S_s \,], \\ G_w(\lambda) &=& G(\lambda)S_w, \\
G_v(\lambda) &=& G(\lambda)S_v, \end{array}\]
where $S_u$, $S_d$, $S_f$, $S_w$, and $S_v$ are columns of the identity matrix $I_m$ and
$S_s$ is formed from the columns of the indentity matrix $I_p$. These (selection) matrices are used to select the corresponding columns of $G(\lambda)$, and thus to obtain the input and feedthrough matrices of the model (\ref{fdimodset:sysss1}) from those of the model (\ref{fdimodset:sysss}). The indices of the selected columns are specified by the vectors of indices contained in the \texttt{INPUTS} structure.

\subsubsection*{Example}

\begin{example}\label{ex:fdimodset}
For the setup of a LTI synthesis model with actuator and sensor faults consider the continuous-time input-output model defined with the transfer function matrices
\[ G_u(s) = \ba{c}  \frac{s+1}{s+2} \\ \\[-2mm] \frac{s+2}{s+3} \ea, \quad G_d(s) = \ba{c}  \frac{s-1}{s+2} \\ \\[-2mm]0 \ea, \quad G_f(s) = [\, G_u(s)\; I\,] .\]
As it can be observed, the fault inputs correspond to an actuator fault and two sensor faults for both output measurements. To setup the state-space synthesis model, the following MATLAB commands can be employed:
\begin{verbatim}
% setup the system with additive faults
% [Gu(s) Gd(s) Gf(s)], where Gf(s) = [ Gu(s) I]
s = tf('s'); % define the Laplace variable s
Gu = [(s+1)/(s+2); (s+2)/(s+3)];  % enter Gu(s)
Gd = [(s-1)/(s+2); 0];            % enter Gd(s)
% build state space model of [Gu(s) Gd(s) Gf(s)] and set input groups
sysf = fdimodset(ss([Gu Gd]),struct('c',1,'d',2,'f',1,'fs',1:2));
\end{verbatim}
\end{example}
\subsubsection{\texttt{\bfseries mdmodset}}

\subsubsection*{Syntax}
\index{M-functions!\texttt{\bfseries mdmodset}}
\begin{verbatim}
SYSM = mdmodset(SYS,INPUTS)
\end{verbatim}

\subsubsection*{Description}

\noindent \texttt{mdmodset} builds synthesis models to be used in conjunction with the  synthesis functions of model detection filters.

\subsubsection*{Input data}
\begin{description}

\item
\texttt{SYS} is a multiple model which contains $N$ LTI systems, with the $j$-th model having the state-space form
\be\label{mdmodset:sysssj}
\begin{array}{rcl}E^{(j)}\lambda x^{(j)}(t) &=& A^{(j)}x^{(j)}(t) + B^{(j)} \widetilde u^{(j)}(t)  \, ,\\
y^{(j)}(t) &=& C^{(j)}x^{(j)}(t) + D^{(j)} \widetilde u^{(j)}(t)  \, , \end{array}
\ee
where $y^{(j)}(t) \in \mathds{R}^p$, $x^{(j)}(t) \in \mathds{R}^{n^{(j)}}$, and  $\widetilde u^{(j)}(t) \in \mathds{R}^{m^{(j)}}$ are the output, state and input vectors  of the $j$-th model. The (global) input $\widetilde u^{(j)}(t)$ usually includes the control inputs  and may also include disturbance and noise inputs. The multiple model \texttt{SYS} is either an one-dimensional array of $N$ LTI systems of the form (\ref{mdmodset:sysssj}), in which case $m^{(j)} = m$ $\forall j$, or is a $1\times N$ cell array, with \texttt{SYSM\{$j$\}} containing the $j$-th component system in the form (\ref{mdmodset:sysssj}).
\item
 \texttt{INPUTS} is a MATLAB structure used to specify the indices of the columns of the matrices $B^{(j)}$ and, respectively $D^{(j)}$, which correspond to the control, disturbance, and noise inputs, using the following fields:
{\setlength\LTleft{30pt}\begin{longtable}{|l|p{11.6cm}|} \hline \textbf{\texttt{INPUTS} fields} & \textbf{Description} \\ \hline
       \texttt{controls}       & vector of indices of the control inputs (Default: void)\\ \hline
       \texttt{c}       & alternative short form to specify the indices of
                        the control inputs \newline (Default: void)\\ \hline
       \texttt{disturbances}       & vector of indices of the disturbance inputs or an $N$-dimensional cell array, with \texttt{INPUTS.disturbances\{$j$\}} containing the vector of indices of the disturbance inputs of the $j$-th component
                        model  (Default: void)\\ \hline
       \texttt{d}       & alternative short form to specify the indices of
                        the disturbance inputs or an $N$-dimensional cell array, with \texttt{INPUTS.d\{$j$\}} containing the vector of indices of the disturbance inputs of the $j$-th component
                        model  (Default: void)\\ \hline
     \texttt{noise}       & vector of indices of the noise inputs, or an $N$-dimensional cell array, with \texttt{INPUTS.noise\{$j$\}} containing the vector of indices of the noise inputs of the $j$-th component
                        model  (Default: void)\\ \hline
       \texttt{n}       & alternative short form to specify the indices of
                        the noise inputs, or an $N$-dimensional cell array, with \texttt{INPUTS.n\{$j$\}} containing the vector of indices of the noise inputs of the $j$-th component model \newline (Default: void)\\ \hline
    \end{longtable}     }
\end{description}

\subsubsection*{Output data}
\begin{description}
\item
\texttt{SYSM} is a multiple LTI system, with the $j$-th model  in the state-space form
\be\label{mdmodset:sysss1j}
\begin{array}{rcl}E^{(j)}\lambda x^{(j)}(t) &=& A^{(j)}x^{(j)}(t) + B^{(j)}_{u} u^{(j)}(t) + B^{(j)}_d d^{(j)}(t) + B^{(j)}_w w^{(j)}(t)  \, ,\\
y^{(j)}(t) &=& C^{(j)}x^{(j)}(t) + D^{(j)}_u u^{(j)}(t) + D^{(j)}_d d^{(j)}(t) + D^{(j)}_w w^{(j)}(t)   \, , \end{array} \ee
where $u^{(j)}(t)\in \mathds{R}^{m_u}$ are the control inputs, $d^{(j)}(t)\in \mathds{R}^{m_d^{(j)}}$ are the disturbance inputs, and  $w^{(j)}(t)\in \mathds{R}^{m_w^{(j)}}$ are the noise inputs.
Any of the inputs components $u^{(j)}(t)$, $d^{(j)}(t)$, or $w^{(j)}(t)$  can be void.
The input groups for $u^{(j)}(t)$, $d^{(j)}(t)$, and $w^{(j)}(t)$  have the standard names \texttt{\bfseries 'controls'}, \texttt{\bfseries 'disturbances'}, and \texttt{\bfseries 'noise'}, respectively. The resulting multiple model \texttt{SYSM} has the same representation as the original model \texttt{SYS} (i.e., either an one-dimensional array of $N$ LTI systems or a $1\times N$ cell array) and
inherits the sampling time of \texttt{SYS}.
\end{description}

\subsubsection*{Method}

The system matrices $B_u^{(j)}$, $B_d^{(j)}$, $B_w^{(j)}$, and $D_u^{(j)}$, $D_d^{(j)}$, $D_w^{(j)}$, are defined by specifying the indices of the columns of the matrices $B^{(j)}$ and, respectively $D^{(j)}$, which correspond to the control, disturbance, and noise inputs.
If the $j$-th component system of \texttt{SYS} in (\ref{mdmodset:sysssj}) has the equivalent input-output form
\be\label{mdmodset:iosysj} {\mathbf{y}}^{(j)}(\lambda) =
G^{(j)}(\lambda)\widetilde{\mathbf{u}}^{(j)}(\lambda)  \ee
and the resulting $j$-th component system of \texttt{SYSM} in (\ref{mdmodset:sysss1j}) has the equivalent input-output form
\be\label{mdmodset:iosys1j} {\mathbf{y}}^{(j)}(\lambda) =
G_u^{(j)}(\lambda){\mathbf{u}}(\lambda) +
G_d^{(j)}(\lambda){\mathbf{d}}^{(j)}(\lambda) +
G_w^{(j)}(\lambda){\mathbf{w}}^{(j)}(\lambda) ,
 \ee
then the following relations exist among the above transfer function matrices
\[ \begin{array}{lcl}G_u^{(j)}(\lambda) &=& G^{(j)}(\lambda)S_u, \\ G_d^{(j)}(\lambda) &=& G^{(j)}(\lambda)S_d^{(j)}, \\  G_w^{(j)}(\lambda) &=& G^{(j)}(\lambda)S_w^{(j)}, \end{array}\]
where $S_u$, $S_d^{(j)}$, and $S_w^{(j)}$, are columns of the identity matrix $I_{m^{(j)}}$ and are used to select the corresponding columns of $G^{(j)}(\lambda)$. The indices of the selected columns are specified by the vectors of indices contained in the \texttt{INPUTS} structure.

\subsubsection*{Examples}

\begin{example}\label{ex:act_faults}
Consider the first-order input-output flight actuator model
\[ G(s,k) = \frac{k}{s+k}  \; , \]
where $k$ is the actuator gain. An input-output multiple  model of the form (\ref{systemi}), defined as
\[ G_u^{(j)}(s) := G(s,k^{(j)}) , \quad j = 1, ..., 4, \]
covers, via suitable choices of the values of the gain $k$,  the normal case for $k = k^{(1)}$ as well as three main classes of parametric actuator faults $k = k^{(j)}$, for $j = 2, 3, 4$, as follows:
\begin{center}\begin{tabular}{lll}
$k^{(1)} = 14$ & --& normal (nominal) case \\
$k^{(2)} = 0.5k^{(1)}$ & --& loss of efficiency (LOE) fault  \\
$k^{(3)} =  10k^{(1)}$ & --& surface disconnection fault  \\
$k^{(4)} =  0.01k^{(1)}$ & --& stall load fault
\end{tabular}
\end{center}

For the setup of the multiple synthesis model to be used for model detection purposes, the following MATLAB commands can be used:
\begin{verbatim}
% Generation of a multiple model for actuator faults
s = tf('s');
k1 = 14;  % nominal gain
sysact(:,:,1) = k1/(s+k1);           % nominal case
sysact(:,:,2) = 0.5*k1/(s+0.5*k1);   % 50% LOE
sysact(:,:,3) = 10*k1/(s+10*k1);     % disconnection
sysact(:,:,4) = 0.01*k1/(s+0.01*k1); % stall load
% setup the multiple synthesis model
sysmact = mdmodset(ss(sysact),struct('c',1))
\end{verbatim}
\end{example}

\begin{example}\label{ex:cell_md}
This example illustrates the setup of  a multiple synthesis model with variable numbers of disturbance and noise inputs. Let $N = 2$ be the number of models of the form (\ref{mdmodset:sysss1j}), with $p = 3$ outputs, $m_u = 2$ control inputs,  $m_d^{(1)} = 1$ and  $m_d^{(2)} = 2$ disturbance inputs, $m_w^{(1)} = 2$ and  $m_w^{(2)} = 1$ noise inputs.

For the setup of the multiple synthesis model to be used for model detection purposes, the following MATLAB commands can be used:
\begin{verbatim}
% Generation of a multiple model with two component models
p = 3; mu = 2; md1 = 1; md2 = 2; mw1 = 2; mw2 = 1;
sysm{1} = rss(4,p,mu+md1+mw1);
sysm{2} = rss(2,p,mu+md2+mw2);
% setup the multiple synthesis model
% note the compulsory use of double braces here
sysm = mdmodset(sysm,struct('c',1,'d',{{2,2:3}},'n',{{3:4,4}}))
\end{verbatim}
\end{example}


\subsection{Functions for FDI Related Analysis} \label{fditools:analysis}

These functions cover the generation of achievable weak and strong FDI specifications to be used for solving synthesis problems of FDI filters and the analysis of the feasibility of a set of FDI specifications. For the definitions of isolability related concepts see Section \ref{isolability}.

\subsubsection{\texttt{\bfseries fdigenspec}}

\subsubsection*{Syntax}
\index{M-functions!\texttt{\bfseries fdigenspec}}
\begin{verbatim}
S = fdigenspec(SYSF,OPTIONS)
\end{verbatim}

\subsubsection*{Description}
\index{structure matrix}

\texttt{\bfseries fdigenspec} determines all achievable fault detection specifications for the
  LTI state-space system \texttt{SYSF} with additive faults.

\subsubsection*{Input data}
\begin{description}
\item
\texttt{SYSF} is a  LTI system  in the state-space form
\be\label{fdigenspec:sysss1}
\begin{aligned}
E\lambda x(t)  & =  Ax(t)+ B_u u(t)+ B_d d(t) + B_f f(t)   , \\
y(t) & =  C x(t) + D_u u(t)+ D_d d(t) + D_f f(t)  ,
\end{aligned}
\ee
where any of the inputs components $u(t)$, $d(t)$, and $f(t)$  can be void.  For the system \texttt{SYSF}, the input groups for $u(t)$, $d(t)$, and $f(t)$, have the standard names \texttt{\bfseries 'controls'}, \texttt{\bfseries 'disturbances'}, and \texttt{\bfseries 'faults'},  respectively. Any additionally defined input groups are ignored.

If no standard input groups are explicitly defined, then \texttt{SYSF} is assumed to be a partitioned  LTI system
\texttt{SYSF = [SYS1 SYS2]}  in a state-space form
\be\label{fdigenspec:sysss2}
\begin{aligned}
E\lambda x(t)  &=   Ax(t)+ B_d d(t) + B_f f(t) ,\\
y(t) &=  C x(t)+ D_d d(t) + D_f f(t) ,
\end{aligned}
\ee
where the inputs components $d(t) \in \mathds{R}^{m_1}$, and  $f(t) \in \mathds{R}^{m_2}$ (both input components can be void). \texttt{SYS1} has $d(t)$ as input vector and the corresponding state-space realization is $(A-\lambda E,B_d,C,D_d)$, while \texttt{SYS2} has $f(t)$ as input vector and the corresponding realization is $(A-\lambda E,B_f,C,D_f)$. The dimension $m_1$ of the input vector $d(t)$ is specified by the \texttt{OPTIONS} field \texttt{OPTIONS.m1} (see below). For compatibility with a previous version, if \texttt{OPTIONS.m1} is specified, then the form (\ref{fdigenspec:sysss2}) is assumed, even if the standard input
groups have been explicitly defined.
\item
 \texttt{OPTIONS} is a MATLAB structure used to specify various  options and has the following fields:
\begin{center}
\begin{longtable}{|l|p{11.7cm}|} \hline \textbf{Option fields} & \textbf{Description} \\ \hline
       \texttt{tol}       & tolerance for rank determinations\newline
                 (Default: internally computed)\\ \hline
       \texttt{FDTol}     & threshold for fault detectability checks \newline (Default: 0.0001)\\ \hline
       \texttt{FDGainTol} & threshold for strong fault detectability checks  \newline (Default: 0.01)\\ \hline
       \texttt{m1}  &  the number $m_1$ of the inputs of \texttt{SYS1} (Default: 0); if \texttt{OPTIONS.m1} is explicitly specified, then \texttt{SYSF} is assumed to be partitioned as \texttt{SYSF = [SYS1 SYS2]} with a state-space realization of the form (\ref{fdigenspec:sysss2}) and the
                      definitions of input groups are ignored. \\ \hline
       \texttt{FDFreq}  &  vector of $n_f$ real frequency values $\omega_k$, $k = 1, \ldots, n_f$, for strong fault detectability checks. To each real frequency $\omega_k$, corresponds a complex frequency $\lambda_k = \mathrm{i}\omega_k$, in the continuous-time case, and $\lambda_k = \exp (\mathrm{i}\omega_k T)$, in the discrete-time case, where $T$ is the sampling time of the system. \newline (Default: \texttt{[ ]}) \\ \hline
\pagebreak[4] \hline
       \texttt{sdeg}   & prescribed stability degree for the poles of the internally generated filters (see \textbf{Method}): in the continuous-time case, the real parts of filters poles must be less than or equal to  \texttt{OPTIONS.sdeg}, while in discrete-time case, the magnitudes of filter poles must be less than or equal to \texttt{OPTIONS.sdeg}; \\
       & (Default: if \texttt{OPTIONS.FDFreq} is empty, then \texttt{OPTIONS.sdeg = [ ]}, i.e., no stabilization is performed; if  \texttt{OPTIONS.FDFreq} is nonempty, then \texttt{OPTIONS.sdeg} = $-0.05$ in the continuous-time case and \texttt{OPTIONS.sdeg} = 0.9 in the  discrete-time case).\\ \hline
  \end{longtable}
\end{center}
\end{description}

\subsubsection*{Output data}
\begin{description}
\item
\texttt{S} is a logical array, whose rows contains the achievable fault detection specifications. Specifically, the $i$-th row of \texttt{S} contains the $i$-th achievable specification,
obtainable by using a certain (e.g., scalar output) fault detection filter $Q^{(i)}(\lambda)$, whose internal form  is $R_f^{(i)}(\lambda)$, with  $R_f^{(i)}(\lambda) \neq  0$   (see \textbf{Method}). Thus,   the row \texttt{S}($i$,:) is the  block-structure based structure matrix of $R_f^{(i)}(\lambda)$, such that
  \texttt{S}($i,j$) = \texttt{true} if $R_{f_j}^{(i)}(\lambda) \neq 0$  and \texttt{S}($i,j$) = \texttt{false}  if $R_{f_j}^{(i)}(\lambda) = 0$.
  If the real frequency values $\omega_k$, $k = 1, \ldots, n_f$,  are provided in \texttt{OPTIONS.FDFreq} for determining strong specifications, then  \texttt{S}($i,j$) = \texttt{true} if
  $\big\|R_{f_j}^{(i)}(\lambda_k)\big\| \geq$ \texttt{OPTIONS.FDGainTol} for all $\lambda_k$, $k = 1, \ldots, n_f$, where  $\lambda_k$ is the complex frequency corresponding to $\omega_k$, and \texttt{S}($i,j$) = \texttt{false} if there exists $\lambda_k$ such that $\big\|R_{f_j}^{(i)}(\lambda_k)\big\| <$ \texttt{OPTIONS.FDGainTol}.
\end{description}

\subsubsection*{Method}

The implementation of \texttt{\bfseries fdigenspec} is based on the \textbf{Procedure GENSPEC} from \cite[Sect.\ 5.4]{Varg17}. The nullspace
method of \cite{Varg08a} is recursively employed to generate the complete set of
achievable specifications, obtainable using suitable fault detection filters. The method is also described in \cite{Varg09g}. In what follows we give some details of this approach.

Assume the system \texttt{SYSF} in (\ref{fdigenspec:sysss1}) has the input-output form
\be\label{fdigenspec:sysio} {\mathbf{y}}(\lambda) =
G_u(\lambda){\mathbf{u}}(\lambda) +
G_d(\lambda){\mathbf{d}}(\lambda) +
G_f(\lambda){\mathbf{f}}(\lambda) .
 \ee
If \texttt{SYSF} has the form (\ref{fdigenspec:sysss2}), then we simply assume that $u(t)$ is void in (\ref{fdigenspec:sysio}). To determine the $i$-th row of \texttt{S}, which contains the $i$-th achievable specification, a certain (e.g., scalar output) fault detection filter is employed,
with the input-output implementation form
\be\label{fdigenspec:detio}
{\mathbf{r}}^{(i)}(\lambda) = Q^{(i)}(\lambda)\ba{c}
{\mathbf{y}}(\lambda)\\{\mathbf{u}}(\lambda)\ea  \,
\ee
and its internal form
\be\label{fdigenspec:detinio} {\mathbf{r}}^{(i)}(\lambda) =
R_f^{(i)}(\lambda){\mathbf{f}}(\lambda) ,\ee
with $R^{(i)}_f(\lambda)$ defined as
\be\label{fdigenspec:detintfm}  R_f^{(i)}(\lambda) :=
{\arraycolsep=1mm Q^{(i)}(\lambda)  \ba{c} G_f(\lambda) \\
          0 \ea } \, .
         \ee
The resulting $i$-th row of $S$ is the structure matrix (weak or strong) of $R_f^{(i)}(\lambda)$.               Recursive filter updating based on nullspace techniques is employed to systematically generate particular filters which are sensitive to certain fault inputs and insensitive to the rest of inputs.

The check for nonzero elements of $R_f^{(i)}(\lambda)$ is performed by using the function
  \texttt{\bfseries fditspec} to evaluate the corresponding weak specifications. The corresponding threshold is specified via \texttt{OPTIONS.FDTol}.
  If frequency values for strong detectability tests are provided in \texttt{OPTIONS.FDFreq}, then the magnitudes of the elements of $R_f^{(i)}(\lambda)$ must be above a certain threshold for all complex frequencies corresponding to the specified real frequency
  values in \texttt{OPTIONS.FDFreq}. For this purpose, the function \texttt{\bfseries fdisspec}  is used to
  evaluate the corresponding strong specifications. The corresponding threshold is specified via \texttt{OPTIONS.FDGainTol}. The call of \texttt{\bfseries fdisspec} requires that none of the complex frequencies $\lambda_k$, $k = 1, \ldots, n_f$, corresponding to the real frequencies $\omega_k$, $k = 1, \ldots, n_f$,  specified in \texttt{OPTIONS.FDFreq}, belongs to the set of poles of $R_f^{(i)}(\lambda)$. This condition is fulfilled by ensuring a certain stability degree for the poles of $R_f^{(i)}(\lambda)$, specified  via \texttt{OPTIONS.sdeg}.

\subsubsection*{Example}
\begin{example}\label{ex:Yuan}
This is the  example of \cite{Yuan97} of a continuous-time state-space model of the form (\ref{fdigenspec:sysss1}) with $E = I_4$,
\[ A = {\arraycolsep=1mm\ba{rrrr} -1 & 1 &0 &0 \\
1 &-2 &1& 0 \\
0 &1& -2& 1 \\
0 &0 &1 &-2\ea }, \quad B_u = \ba{r} 1\\ 0\\ 0\\ 0\ea, \quad B_d = 0, \quad
B_f = {\arraycolsep=1.5mm\ba{rrrrrrrr}
1 &0 &0 &0 & 1&  0&  0& 0 \\
0 &1 &0 &0 &-1&  1&  0& 0 \\
0 &0 &1 &0 & 0& -1&  1& 0 \\
0 &0 &0 &1 & 0&  0& -1& 1\ea} ,\]
\[ C = \ba{cccc}
1 &0 &0 &0 \\
0 &0 &1 &0 \\
0 &0 &0 &1 \ea, \quad D_u = 0 , \quad  D_d = 0 , \quad  D_f = 0. \]
The achievable 18 weak fault specifications and 12 strong fault specifications, computed with the following script, are:
\[ S_{weak} = \ba{cccccccc}
     0&     0 &    0 &    1 &    0 &    0  &   1 &    1\\
     0&     1 &    1 &    0 &    1 &    1  &   1 &    0\\
     0&     1 &    1 &    1 &    1 &    1  &   0 &    1\\
     0&     1 &    1 &    1 &    1 &    1  &   1 &    1\\
     1&     0 &    1 &    0 &    1 &    1  &   1 &    0\\
     1&     0 &    1 &    1 &    1 &    1  &   0 &    1\\
     1&     0 &    1 &    1 &    1 &    1  &   1 &    1\\
     1&     1 &    0 &    0 &    1 &    1  &   0 &    0\\
     1&     1 &    0 &    1 &    1 &    1  &   1 &    1\\
     1&     1 &    1 &    0 &    0 &    1  &   1 &    0 \\
     1&     1 &    1 &    0 &    1 &    0  &   1 &    0 \\
     1&     1 &    1 &    0 &    1 &    1  &   1 &    0 \\
     1&     1 &    1 &    1 &    0 &    1  &   0 &    1 \\
     1&     1 &    1 &    1 &    0 &    1  &   1 &    1 \\
     1&     1 &    1 &    1 &    1 &    0  &   0 &    1 \\
     1&     1 &    1 &    1 &    1 &    0  &   1 &    1\\
     1&     1 &    1 &    1 &    1 &    1  &   0 &    1\\
     1&     1 &    1 &    1 &    1 &    1  &   1 &    1 \ea , \qquad
S_{strong} = \ba{cccccccc}
     0&     0&     0&     1&     0&     0&     1&     1 \\
     0&     1&     1&     0&     1&     1&     1&     0 \\
     0&     1&     1&     1&     1&     1&     0&     1 \\
     0&     1&     1&     1&     1&     1&     1&     1 \\
     1&     0&     1&     0&     1&     1&     1&     0 \\
     1&     0&     1&     1&     1&     1&     0&     1 \\
     1&     0&     1&     1&     1&     1&     1&     1 \\
     1&     1&     0&     0&     1&     1&     0&     0 \\
     1&     1&     0&     1&     1&     1&     1&     1 \\
     1&     1&     1&     0&     1&     1&     1&     0 \\
     1&     1&     1&     1&     1&     1&     0&     1 \\
     1&     1&     1&     1&     1&     1&     1&     1 \ea
\]
Observe that there are 6 weak specifications, which are not strong specifications.

\begin{verbatim}
% Example of Yuan et al. IJC (1997)
p = 3; mu = 1; mf = 8;
A = [ -1 1 0 0; 1 -2 1 0; 0 1 -2 1; 0 0 1 -2  ]; Bu = [1 0 0 0]';
Bf = [ 1 0 0 0 1 0 0 0; 0 1 0 0 -1 1 0 0; 0 0 1 0 0 -1 1 0; 0 0 0 1 0 0 -1 1];
C = [ 1 0 0 0; 0 0 1 0; 0 0 0 1];
Du = zeros(p,mu); Df = zeros(p,mf);
% setup the model with additive faults
sysf = ss(A,[Bu Bf],C,[Du Df]);
% set input groups
set(sysf,'InputGroup',struct('controls',1:mu,'faults',mu+(1:mf)));

% compute the achievable weak specifications
opt = struct('tol',1.e-7,'FDTol',1.e-5);
S_weak = fdigenspec(sysf,opt), size(S_weak)

% compute the achievable strong specifications for constant faults
opt = struct('tol',1.e-7,'FDTol',0.0001,'FDGainTol',.001,...
    'FDFreq',0,'sdeg',-0.05);
S_strong = fdigenspec(sysf,opt), size(S_strong)
\end{verbatim}
\end{example}

\subsubsection{\texttt{\bfseries fdichkspec}}

\subsubsection*{Syntax}
\index{M-functions!\texttt{\bfseries fdichkspec}}
\begin{verbatim}
[RDIMS,ORDERS,LEASTORDERS] = fdichkspec(SYSF)
[RDIMS,ORDERS,LEASTORDERS] = fdichkspec(SYSF,SFDI,OPTIONS)
\end{verbatim}

\subsubsection*{Description}
\index{structure matrix}

\texttt{\bfseries fdichkspec} checks for the
  LTI state-space system \texttt{SYSF} with additive faults, the feasibility of a given set of FDI specifications \texttt{SFDI} and determines information related to the synthesis of FDI filters to achieve the feasible specifications.

\subsubsection*{Input data}
\begin{description}
\item
\texttt{SYSF} is a  LTI system  in the state-space form
\be\label{fdichkspec:sysss1}
\begin{aligned}
E\lambda x(t)  & =  Ax(t)+ B_u u(t)+ B_d d(t) + B_f f(t)   , \\
y(t) & =  C x(t) + D_u u(t)+ D_d d(t) + D_f f(t)  ,
\end{aligned}
\ee
where any of the inputs components $u(t)$, $d(t)$, and $f(t)$  can be void.  For the system \texttt{SYSF}, the input groups for $u(t)$, $d(t)$, and $f(t)$, have the standard names \texttt{\bfseries 'controls'}, \texttt{\bfseries 'disturbances'}, and \texttt{\bfseries 'faults'},  respectively. Any additionally defined input groups are ignored.
\item
\texttt{SFDI} is an $N\times m_f$ logical array whose rows contain the set of FDI specifications, whose feasibility has to be checked. If \texttt{SFDI} is empty or not specified, then the fault inputs are considered void and therefore ignored.
\item
 \texttt{OPTIONS} is a MATLAB structure used to specify various synthesis  options and has the following fields:
\begin{center}
\begin{tabular}{|l|p{11.7cm}|} \hline \textbf{Option fields} & \textbf{Description} \\ \hline
       \texttt{tol}       & tolerance for rank determinations
                 (Default: internally computed)\\ \hline
       \texttt{tolmin}       & absolute tolerance for observability tests \newline
                 (Default: internally computed)\\ \hline
       \texttt{FDTol}     & threshold for fault detectability checks (Default: 0.0001)\\ \hline
       \texttt{FDGainTol} & threshold for strong fault detectability checks (Default: 0.01)\\ \hline
       \texttt{FDFreq}  &  vector of $n_f$ real frequency values $\omega_k$, $k = 1, \ldots, n_f$, for strong fault detectability checks. To each real frequency $\omega_k$, corresponds a complex frequency $\lambda_k = \mathrm{i}\omega_k$, in the continuous-time case, and $\lambda_k = \exp (\mathrm{i}\omega_k T)$, in the discrete-time case, where $T$ is the sampling time of the system. \newline (Default: \texttt{[ ]}) \\ \hline
  \end{tabular}
\end{center}
\end{description}

\subsubsection*{Output data}

\begin{description}
\item
\texttt{RDIMS} is an $N$-dimensional integer vector, whose $i$-th component \texttt{RDIMS$(i)$}, if nonzero,  contains the number of residual
  outputs of a FDI filter based on a minimal nullspace basis, which can be used
  to achieve the $i$-th specification contained in \texttt{SFDI($i$,:)}.
  If \texttt{RDIMS$(i)$} = 0, then the $i$-th specification is not feasible.
\item
\texttt{ORDERS} is an $N$-dimensional integer vector, whose $i$-th component \texttt{ORDERS$(i)$} contains, for a feasible specification \texttt{SFDI($i$,:)}, the order of the minimal
  nullspace basis based FDI filter (see above). If the $i$-th specification
  is not feasible, then \texttt{ORDERS$(i)$} is set to $-1$.
\item
\texttt{LEASTORDERS} is an $N$-dimensional integer vector, whose $i$-th component \texttt{LEASTORDERS$(i)$} contains, for a feasible specification \texttt{SFDI($i$,:)}, the
  least achievable order for a scalar output FDI filter which can be used
  to achieve the $i$-th specification. If the $i$-th specification
  is not feasible, then \texttt{LEASTORDERS$(i)$} is set to $-1$.

\end{description}

\subsubsection*{Method}

The nullspace
method of \cite{Varg08a} is successively employed to determine FDI filters as minimal left nullspace bases which solve suitably formulated fault detection problems.  In what follows we give some details of this approach.

Assume the system \texttt{SYSF} in (\ref{fdichkspec:sysss1}) has the input-output form
\be\label{fdichkspec:sysio} {\mathbf{y}}(\lambda) =
G_u(\lambda){\mathbf{u}}(\lambda) +
G_d(\lambda){\mathbf{d}}(\lambda) +
G_f(\lambda){\mathbf{f}}(\lambda) .
 \ee
To determine a FDI filter which achieves the $i$-th specification contained in the $i$-th row of \texttt{SFDI}, we can reformulate this FDI problem as a fault detection problem for modified sets of disturbance and fault inputs.
Let $f^{(i)}$ be formed from the subset of faults corresponding to nonzero entries in the $i$-th row of \texttt{SFDI} and let $G_f^{(i)}(\lambda)$ be formed from the corresponding columns of $G_{f}(\lambda)$. Similarly, let $d^{(i)}$ be formed from the subset of faults corresponding to zero entries in the $i$-th row of \texttt{SFDI} and let $G_d^{(i)}(\lambda)$ be formed from the corresponding columns of $G_{f}(\lambda)$. The solution of the EFDIP for the $i$-th row of \texttt{SFDI} is thus equivalent to solve the EFDP for the modified system
\be\label{fdichkspec:sysiom} {\mathbf{y}}(\lambda) =
G_u(\lambda){\mathbf{u}}(\lambda) +
\big[\,G_d(\lambda) \; G_d^{(i)}(\lambda)\,\big] \ba{c}{\mathbf{d}}(\lambda)\\ {\mathbf{d}}^{(i)}(\lambda)\ea +
G_f^{(i)}(\lambda){\mathbf{f}}^{(i)}(\lambda) .
 \ee
A candidate fault detector filter $Q^{(i)}(\lambda)$ can be determined as a left proper nullspace basis of the transfer function matrix
\[ G^{(i)}(\lambda) := \ba{ccc} G_u(\lambda) & G_d(\lambda) & G_d^{(i)}(\lambda) \\
I_{m_u} & 0 & 0 \ea \]
satisfying $Q^{(i)}(\lambda)G^{(i)}(\lambda) = 0$. The corresponding internal form is \[ {\mathbf{r}}^{(i)}(\lambda) = R_f^{(i)}(\lambda){\mathbf{f}}^{(i)}(\lambda) , \]
where
\[ R_f^{(i)}(\lambda) := Q^{(i)}(\lambda) \ba{c}G_f^{(i)}(\lambda)\\ 0 \ea \]
can be determined proper as well.
If the nullspace basis is nonempty (i.e., $Q^{(i)}(\lambda)$  has at least one row) and all columns of the resulting $R_f^{(i)}(\lambda)$ are nonzero, then the $i$-th specification is feasible.            In this case, the number of basis vectors (i.e., the number of rows of $Q^{(i)}(\lambda)$) is returned in \texttt{RDIMS$(i)$} and the order of the realization of $Q^{(i)}(\lambda)$ (and also of $R_f^{(i)}(\lambda)$) is returned in \texttt{ORDERS$(i)$}.

The check for nonzero elements of $R_f^{(i)}(\lambda)$ is performed by using the function
  \texttt{\bfseries fditspec} to evaluate the corresponding weak specifications. The corresponding threshold is specified via \texttt{OPTIONS.FDTol}.
  If $n_f$ frequency values $\omega_k$, $k = 1, \ldots, n_f$, are provided in the vector \texttt{OPTIONS.FDFreq} for strong detectability tests, then the magnitudes of the elements of $R_f^{(i)}(\lambda_k)$ must be above a certain threshold for the complex frequencies $\lambda_k$, $k = 1, \ldots, n_f$, corresponding to the specified real frequency
  values in \texttt{OPTIONS.FDFreq}. For this purpose, the function \texttt{\bfseries fdisspec}  is used to
  evaluate the corresponding strong specifications. The corresponding threshold is specified via \texttt{OPTIONS.FDGainTol}. The call of \texttt{\bfseries fdisspec} requires that the set of poles of $R_f^{(i)}(\lambda)$  and the complex frequencies $\lambda_k$, $k = 1, \ldots, n_f$, corresponding to the real frequencies specified in \texttt{OPTIONS.FDFreq}, are disjoint. This condition is fulfilled by ensuring a certain stability degree for the poles of $R_f^{(i)}(\lambda)$.

A least order scalar output filter, which fulfills the above fault detectability conditions, can be determined using minimum dynamic cover techniques \cite{Varg17}. This computation essentially involves the determination of a linear combination of the basis vectors using a rational vector $h(\lambda)$ such that $h(\lambda)R_f^{(i)}(\lambda)$ has all columns nonzero and $h(\lambda)Q^{(i)}(\lambda)$ has the least McMillan degree. The resulting least order of the scalar output FDI filter $h(\lambda)Q^{(i)}(\lambda)$ is returned in \texttt{LEASTORDERS$(i)$}.

\subsubsection*{Example}
\begin{example}\label{ex:Yuan2}
This is the  example of \cite{Yuan97} already considered in Example \ref{ex:Yuan}.
Of the 18 weak achievable fault specifications 12 are strong fault specifications for constant faults. This can be also checked using the following MATLAB script, where
the strong fault detectability checks are performed on the set of 18 weak specifications. The resulting least orders of the scalar output FDI filters to achieve the 12 feasible specifications are:
 1,     2,     2,     2,     1,     1,     1,     2,     2,     2,     2,     2.

\begin{verbatim}
% Example of Yuan et al. IJC (1997)
p = 3; mu = 1; mf = 8;
A = [ -1 1 0 0; 1 -2 1 0; 0 1 -2 1; 0 0 1 -2  ]; Bu = [1 0 0 0]';
Bf = [ 1 0 0 0 1 0 0 0; 0 1 0 0 -1 1 0 0; 0 0 1 0 0 -1 1 0; 0 0 0 1 0 0 -1 1];
C = [ 1 0 0 0; 0 0 1 0; 0 0 0 1];
Du = zeros(p,mu); Df = zeros(p,mf);
% setup the model with additive faults
sysf = ss(A,[Bu Bf],C,[Du Df]);
% set input groups
set(sysf,'InputGroup',struct('controls',1:mu,'faults',mu+(1:mf)));

% compute the achievable weak specifications
opt = struct('tol',1.e-7,'FDTol',1.e-5);
S_weak = fdigenspec(sysf,opt), size(S_weak)

% check for the achievable strong specifications for constant faults
opt = struct('tol',1.e-7,'FDTol',0.0001,'FDGainTol',.001,...
    'FDFreq',0);
[rdims,orders,leastorders] = fdichkspec(sysf,S_weak,opt);

% select strong specifications and display the least achievable orders
S_strong = S_weak(rdims > 0,:), size(S_strong)
leastord = leastorders(rdims>0)'
\end{verbatim}
\end{example}

\subsection{Functions for Model Detection Related Analysis} \label{fditools:analysismd}

These functions cover the evaluation of the pairwise distances between the component models of a multiple model or between the component models and a given model as defined in Section \ref{sec:mdnugap}.

\subsubsection{\texttt{\bfseries mddist}}
\index{M-functions!\texttt{\bfseries mddist}}
\index{model detection!$\nu$-gap distances}
\subsubsection*{Syntax}
\begin{verbatim}
[DIST,FPEAK,PERM,RELDIST] = mddist(SYSM,OPTIONS)
\end{verbatim}

\subsubsection*{Description}

\texttt{\bfseries mddist} determines the pairwise distances between the component models of a given LTI multiple model \texttt{SYSM} containing $N$ models.

\subsubsection*{Input data}
\begin{description}
\item
\texttt{SYSM} is a multiple model which contains $N$  LTI systems in the state-space form
\be\label{mddist:sysiss}
\begin{array}{rcl}E^{(j)}\lambda x^{(j)}(t) &=& A^{(j)}x^{(j)}(t) + B^{(j)}_u u(t) + B^{(j)}_d d^{(j)}(t) + B^{(j)}_w w^{(j)}(t)  \, ,\\
y^{(j)}(t) &=& C^{(j)}x^{(j)}(t) + D^{(j)}_u u(t) + D^{(j)}_d d^{(j)}(t) + D^{(j)}_w w^{(j)}(t)   \, , \end{array} \ee
where $x^{(j)}(t) \in \mathds{R}^{n^{(j)}}$ is the state vector of the $j$-th system with control input $u(t) \in \mathds{R}^{m_u}$, disturbance input $d^{(j)}(t) \in \mathds{R}^{m_d^{(j)}}$ and noise input $w^{(j)}(t) \in \mathds{R}^{m_w^{(j)}}$, and
\index{faulty system model!physical}%
\index{faulty system model!multiple model}%
where any of the inputs components $u(t)$, $d^{(j)}(t)$, or $w^{(j)}(t)$  can be void. The multiple model \texttt{SYSM} is either an array of $N$ LTI systems of the form (\ref{mddist:sysiss}), in which case $m_d^{(j)} = m_d$ and $m_w^{(j)} = m_w$ for $j = 1, \ldots, N$,  or is an$1\times N$ cell array, with \texttt{SYSM\{$j$\}} containing the $j$-th component system in the form (\ref{mddist:sysiss}). The input groups for $u(t)$, $d^{(j)}(t)$, and $w^{(j)}(t)$  have the standard names \texttt{\bfseries 'controls'}, \texttt{\bfseries 'disturbances'}, and \texttt{\bfseries 'noise'}, respectively. If \texttt{OPTIONS.cdinp = true} (see below), then
the same disturbance input $d$ is assumed for all component models (i.e., $d^{(j)} = d$ and $m_d^{(j)} = m_d$).
The state-space form (\ref{mddist:sysiss}) corresponds to the input-output form
\be\label{mddist:systemi} {\mathbf{y}}^{(j)}(\lambda) =
G_u^{(j)}(\lambda){\mathbf{u}}^{(j)}(\lambda)
+ G_d^{(j)}(\lambda){\mathbf{d}}^{(j)}(\lambda)
+ G_w^{(j)}(\lambda){\mathbf{w}}^{(j)}(\lambda) , \ee
where $G_u^{(j)}(\lambda)$, $G_d^{(j)}(\lambda)$ and $G_w^{(j)}(\lambda)$  are the TFMs from
the corresponding inputs to outputs.
\item
 \texttt{OPTIONS} is a MATLAB structure used to specify various  options and has the following fields:
{\tabcolsep=1mm
\setlength\LTleft{30pt}\begin{longtable}{|l|lcp{10cm}|} \hline
\textbf{\texttt{OPTIONS} fields} & \multicolumn{3}{l|}{\textbf{Description}} \\ \hline
 \texttt{MDSelect}   & \multicolumn{3}{p{12cm}|}{$M$-dimensional integer vector $\sigma$ with increasing elements containing the indices of the selected component models to which the distances have to be evaluated
                     (Default: $[\,1, \ldots, N\,]$)}\\
                                        \hline
 \texttt{tol}   & \multicolumn{3}{l|}{relative tolerance for rank computations (Default: internally computed)} \\ \hline
  \texttt{distance}   & \multicolumn{3}{p{11.5cm}|}{option for the selection of the distance function $\dist(G_1,G_2)$ between two transfer function matrices $G_1(\lambda)$ and $G_2(\lambda)$:}\\
                 &  \texttt{'nugap'}&--& $\dist(G_1,G_2) = \delta_\nu(G_1,G_2)$, the $\nu$-gap distance (default)   \\
                 &  \texttt{'Inf'}&--& $\dist(G_1,G_2) = \|G_1-G_2\|_\infty$, the $\mathcal{H}_\infty$-norm based distance \\
                 &  \texttt{'2'}&--& $\dist(G_1,G_2) = \|G_1-G_2\|_2$, the $\mathcal{H}_2$-norm based distance \\
                                        \hline
\texttt{MDFreq}  &  \multicolumn{3}{p{12cm}|}{real vector, which contains the frequency values $\omega_k$, $k = 1, \ldots, n_f$, for which the point-wise distances have to be computed. For each real frequency  $\omega_k$, there corresponds a complex frequency $\lambda_k$ which is used to evaluate the point-wise distance. Depending on the system type, $\lambda_k = \mathrm{i}\omega_k$, in the continuous-time case, and $\lambda_k = \exp (\mathrm{i}\omega_k T)$, in the discrete-time case, where $T$ is the common sampling time of the component models.  (Default: \texttt{[ ]}) } \\ \hline
 \texttt{offset}   & \multicolumn{3}{p{12cm}|}{stability
 boundary offset $\beta$, to be used  to assess the finite zeros which belong to $\partial\mathds{C}_s$ (the boundary of the stability domain) as follows: in the
 continuous-time case these are the finite
  zeros having real parts in the interval $[-\beta, \beta]$, while in the
 discrete-time case these are the finite zeros having moduli in the
 interval $[1-\beta, 1+\beta]$.  (Default: $\beta = 1.4901\cdot 10^{-08}$). }\\
                                        \hline
\texttt{cdinp}   & \multicolumn{3}{p{12cm}|}{option to use both control and
                     disturbance input channels to evaluate the $\nu$-gap
                     distances, as follows:}\\
                 &  \texttt{true} &--& use both control and disturbance input channels; \\
                 &  \texttt{false}&--& use only the control input channels (default)  \\
                                        \hline
 \texttt{MDIndex}   & \multicolumn{3}{p{12cm}|}{index $\ell$ of the $\ell$-th smallest                      distances to be used to evaluate the relative
                     distances to the second smallest distances
                     (Default: $\ell = 3$)} \\ \hline
\end{longtable}}
\end{description}

\subsubsection*{Output data}
\begin{description}
\item
\texttt{DIST} is an $M\times N$ nonnegative matrix, whose $(i,j)$-th element
\texttt{DIST$(i,j)$}  contains the computed distance (see \texttt{OPTIONS.distance}) between the selected input channels of the $\sigma_i$-th and $j$-th component models as follows:\\
-- if \texttt{OPTIONS.MDFreq = []} and \texttt{OPTIONS.cdinp = false} then
\[ \texttt{DIST}(i,j) = \dist\big(G_u^{(\sigma_i)}(\lambda),G_u^{(j)}(\lambda)\big) \]
-- if \texttt{OPTIONS.MDFreq} is nonempty and \texttt{OPTIONS.cdinp = false} then
\[ \texttt{DIST}(i,j) = \max_{k} \dist\big(G_u^{(\sigma_i)}(\lambda_k),G_u^{(j)}(\lambda_k)\big) \]
-- if \texttt{OPTIONS.MDFreq = []} and \texttt{OPTIONS.cdinp = true} then
\[ \texttt{DIST}(i,j) = \dist\big(\big[\,G_u^{(\sigma_i)}(\lambda)\;G_d^{(\sigma_i)}(\lambda)\,\big],
\big[\,G_u^{(j)}(\lambda)\;G_d^{(j)}(\lambda)\,\big]\big) \]
-- if \texttt{OPTIONS.MDFreq} is nonempty and \texttt{OPTIONS.cdinp = true} then
\[ \texttt{DIST}(i,j) = \max_{k} \dist\big(\big[\,G_u^{(\sigma_i)}(\lambda_k)\;G_d^{(\sigma_i)}(\lambda_k)\,\big],
\big[\,G_u^{(j)}(\lambda_k)\;G_d^{(j)}(\lambda_k)\,\big]\big) \]
\item
\texttt{FPEAK} is an $M\times N$ nonnegative matrix, whose $(i,j)$-th element
\texttt{FPEAK$(i,j)$}  contains the peak
  frequency (in rad/TimeUnit), where \texttt{DIST$(i,j)$} is achieved. \\
\item
\texttt{PERM} is an $M\times N$ integer matrix, whose $i$-th row
contains the permutation to be applied to  increasingly reorder the i-th row of \texttt{DIST}. \\
\item
\texttt{RELDIST} is an $M$-dimensional vector, whose $i$-th element \texttt{RELDIST$(i)$} contains the ratio of the second and $\ell$-th smallest distances in the $i$-th row of \texttt{DIST}, where $\ell = $ \texttt{OPTIONS.MDIndex}.
\\
\end{description}

\subsubsection*{Method}
The definition of the distances between component models is given in Section \ref{sec:mdnugap}.
The evaluation of the $\nu$-gap distances relies on the definition
   proposed in \cite{Vinn93}. For efficiency purposes, the intervening normalized
   factorizations of the components systems are performed only once and
   all existing symmetries are exploited. The point-wise distances $\dist\big(G_u^{(i)}(\lambda_k),G_u^{(j)}(\lambda_k)\big)$ for the $\mathcal{H}_\infty$- and $\mathcal{H}_2$ norms are simply the 2-norm of the difference of the frequency responses
   \( \dist\big(G_u^{(j)}(\lambda_k),G_u^{(j)}(\lambda_k)\big) = \big\|G_u^{(j)}(\lambda_k)-G_u^{(j)}(\lambda_k)\big\|_2. \) Similar formulas apply if the disturbance inputs are also selected.

\subsubsection{\texttt{\bfseries mddist2c}}
\index{M-functions!\texttt{\bfseries mddist2c}}
\index{model detection!$\nu$-gap distances}
\index{model detection!$\mathcal{H}_\infty$-norm distances}
\index{model detection!$\mathcal{H}_2$-norm distances}
\subsubsection*{Syntax}
\begin{verbatim}
[DIST,FPEAK,MIND] = mddist2c(SYSM,SYS,OPTIONS)
\end{verbatim}

\subsubsection*{Description}

\texttt{\bfseries mddist2c} determines the distances of the component models of a  LTI multiple model \texttt{SYSM} to the current model \texttt{SYS}.

\subsubsection*{Input data}
\begin{description}
\item
\texttt{SYSM} is a multiple model which contains $N$  LTI systems in the state-space form
\be\label{mddist2c:sysiss}
\begin{array}{rcl}E^{(j)}\lambda x^{(j)}(t) &=& A^{(j)}x^{(j)}(t) + B^{(j)}_u u(t) + B^{(j)}_d d^{(j)}(t) + B^{(j)}_w w^{(j)}(t)  \, ,\\
y^{(j)}(t) &=& C^{(j)}x^{(j)}(t) + D^{(j)}_u u(t) + D^{(j)}_d d^{(j)}(t) + D^{(j)}_w w^{(j)}(t)   \, , \end{array} \ee
where $x^{(j)}(t) \in \mathds{R}^{n^{(j)}}$ is the state vector of the $j$-th system with control input $u(t) \in \mathds{R}^{m_u}$, disturbance input $d^{(j)}(t) \in \mathds{R}^{m_d^{(j)}}$ and noise input $w^{(j)}(t) \in \mathds{R}^{m_w^{(j)}}$, and
\index{faulty system model!physical}%
\index{faulty system model!multiple model}%
where any of the inputs components $u(t)$, $d^{(j)}(t)$, or $w^{(j)}(t)$  can be void. The multiple model \texttt{SYSM} is either an array of $N$ LTI systems of the form (\ref{mddist2c:sysiss}), in which case $m_d^{(j)} = m_d$ and $m_w^{(j)} = m_w$ for $j = 1, \ldots, N$,  or is an $1\times N$ cell array, with \texttt{SYSM\{$j$\}} containing the $j$-th component system in the form (\ref{mddist2c:sysiss}). The input groups for $u(t)$, $d^{(j)}(t)$, and $w^{(j)}(t)$  have the standard names \texttt{\bfseries 'controls'}, \texttt{\bfseries 'disturbances'}, and \texttt{\bfseries 'noise'}, respectively. If \texttt{OPTIONS.cdinp = true} (see below), then
the same disturbance input $d$ is assumed for all component models (i.e., $d^{(j)} = d$ and $m_d^{(j)} = m_d$).
The state-space form (\ref{mddist2c:sysiss}) corresponds to the input-output form
\be\label{mddist2c:systemi} {\mathbf{y}}^{(j)}(\lambda) =
G_u^{(j)}(\lambda){\mathbf{u}}^{(j)}(\lambda)
+ G_d^{(j)}(\lambda){\mathbf{d}}^{(j)}(\lambda)
+ G_w^{(j)}(\lambda){\mathbf{w}}^{(j)}(\lambda) , \ee
where $G_u^{(j)}(\lambda)$, $G_d^{(j)}(\lambda)$ and $G_w^{(j)}(\lambda)$  are the TFMs from
the corresponding inputs to outputs.
\item
\texttt{SYS} is a LTI model  in the state-space form
\be\label{mddist2c:sysissref}
\begin{array}{rcl}E\lambda x(t) &=& Ax(t) + B_u u(t) + B_d d(t) + B_w w(t)  \, ,\\
y(t) &=& Cx(t) + D_u u(t) + D_d d(t) + D_w w(t)   \, , \end{array} \ee
where $x(t) \in \mathds{R}^{n}$ is the state vector of the system with control input $u(t) \in \mathds{R}^{m_u}$, disturbance input $d(t) \in \mathds{R}^{m_d}$ and noise input $w(t) \in \mathds{R}^{m_w}$, and
where any of the inputs components $u(t)$, $d(t)$, or $w(t)$  can be void. The input groups for $u(t)$, $d(t)$, and $w(t)$  have the standard names \texttt{\bfseries 'controls'}, \texttt{\bfseries 'disturbances'}, and \texttt{\bfseries 'noise'}, respectively.
The state-space form (\ref{mddist2c:sysissref}) corresponds to the input-output form
\be\label{mddist2c:system} {\mathbf{y}}(\lambda) =
G_u(\lambda){\mathbf{u}}(\lambda)
+ G_d(\lambda){\mathbf{d}}(\lambda)
+ G_w(\lambda){\mathbf{w}}(\lambda) , \ee
where $G_u(\lambda)$, $G_d(\lambda)$ and $G_w(\lambda)$  are the TFMs from
the corresponding inputs to output.
\item
 \texttt{OPTIONS} is a MATLAB structure used to specify various  options and has the following fields:
{\tabcolsep=1mm
\setlength\LTleft{30pt}\begin{longtable}{|l|lcp{10cm}|} \hline
\textbf{\texttt{OPTIONS} fields} & \multicolumn{3}{l|}{\textbf{Description}} \\ \hline
 \texttt{tol}   & \multicolumn{3}{l|}{relative tolerance for rank computations (Default: internally computed)} \\ \hline
 \texttt{distance}   & \multicolumn{3}{p{11.5cm}|}{option for the selection of the distance function $\dist(G_1,G_2)$ between two transfer function matrices $G_1(\lambda)$ and $G_2(\lambda)$:}\\
                 &  \texttt{'nugap'}&--& $\dist(G_1,G_2) = \delta_\nu(G_1,G_2)$, the $\nu$-gap distance (default)   \\
                 &  \texttt{'Inf'}&--& $\dist(G_1,G_2) = \|G_1-G_2\|_\infty$, the $\mathcal{H}_\infty$-norm based distance \\
                 &  \texttt{'2'}&--& $\dist(G_1,G_2) = \|G_1-G_2\|_2$, the $\mathcal{H}_2$-norm based distance \\
                                        \hline
 \texttt{MDFreq}  &  \multicolumn{3}{p{12cm}|}{real vector, which contains the frequency values $\omega_k$, $k = 1, \ldots, n_f$, for which the point-wise distances have to be computed. For each real frequency  $\omega_k$, there corresponds a complex frequency $\lambda_k$ which is used to evaluate the point-wise distance. Depending on the system type, $\lambda_k = \mathrm{i}\omega_k$, in the continuous-time case, and $\lambda_k = \exp (\mathrm{i}\omega_k T)$, in the discrete-time case, where $T$ is the common sampling time of the component models.  (Default: \texttt{[ ]}) } \\ \hline
 \texttt{offset}   & \multicolumn{3}{p{12cm}|}{stability
 boundary offset $\beta$, to be used  to assess the finite zeros which belong to $\partial\mathds{C}_s$ (the boundary of the stability domain) as follows: in the
 continuous-time case these are the finite
  zeros having real parts in the interval $[-\beta, \beta]$, while in the
 discrete-time case these are the finite zeros having moduli in the
 interval $[1-\beta, 1+\beta]$.  (Default: $\beta = 1.4901\cdot 10^{-08}$). }\\
                                        \hline
\texttt{cdinp}   & \multicolumn{3}{p{12cm}|}{option to use both control and
                     disturbance input channels to evaluate the
                     distances, as follows:}\\
                 &  \texttt{true} &--& use both control and disturbance input channels; \\
                 &  \texttt{false}&--& use only the control input channels (default)  \\
                                        \hline
\end{longtable}}
\end{description}

\subsubsection*{Output data}
\begin{description}
\item
\texttt{DIST} is an $N$-dimensional row vector with nonnegative elements whose $j$-th element
\texttt{DIST$(j)$}  contains the computed distance (see \texttt{OPTIONS.distance}) between the selected input channels of \texttt{SYS} and the $j$-th component model \texttt{SYSM$(j)$} as follows:\\
-- if \texttt{OPTIONS.MDFreq = []} and \texttt{OPTIONS.cdinp = false} then
\[ \texttt{DIST}(j) = \dist\big(G_u(\lambda),G_u^{(j)}(\lambda)\big) \]
-- if \texttt{OPTIONS.MDFreq} is nonempty and \texttt{OPTIONS.cdinp = false} then
\[ \texttt{DIST}(j) = \max_{k} \dist\big(G_u(\lambda_k),G_u^{(j)}(\lambda_k)\big) \]
-- if \texttt{OPTIONS.MDFreq = []} and \texttt{OPTIONS.cdinp = true} then
\[ \texttt{DIST}(j) = \dist\big(\big[\,G_u(\lambda)\;G_d(\lambda)\,\big],
\big[\,G_u^{(j)}(\lambda)\;G_d^{(j)}(\lambda)\,\big]\big) \]
-- if \texttt{OPTIONS.MDFreq} is nonempty and \texttt{OPTIONS.cdinp = true} then
\[ \texttt{DIST}(j) = \max_{k} \dist\big(\big[\,G_u(\lambda_k)\;G_d(\lambda_k)\,\big],
\big[\,G_u^{(j)}(\lambda_k)\;G_d^{(j)}(\lambda_k)\,\big]\big) \]
\item
\texttt{FPEAK} is an $N$-dimensional row vector, whose $j$-th element
\texttt{FPEAK$(j)$}  contains the peak
  frequency (in rad/TimeUnit), where \texttt{DIST$(j)$} is achieved.
\item
\texttt{MIND} is the index $\ell$ of the component model for which the minimum value of the distances in \texttt{DIST}
 is achieved.
\end{description}

\subsubsection*{Method}
The definition of the distances of a set of component models to a current model is given in Section \ref{sec:mddist2c}.
The evaluation of the $\nu$-gap distances relies on the definition
   proposed in \cite{Vinn93}. The point-wise distances $\dist\big(G_u(\lambda_k),G_u^{(j)}(\lambda_k)\big)$ for the $\mathcal{H}_\infty$- and $\mathcal{H}_2$ norms are simply the 2-norm of the difference of the frequency responses
   \( \dist\big(G_u(\lambda_k),G_u^{(j)}(\lambda_k)\big) = \big\|G_u(\lambda_k)-G_u^{(j)}(\lambda_k)\big\|_2. \) Similar formulas apply if the disturbance inputs are also selected.

\subsubsection*{Example}
\begin{example} \label{ex:Ex6.1bis}
This is \emph{Example} 6.1 from the book \cite{Varg17}, which deals with a continuous-time state-space model, describing, in the fault-free case,  the lateral dynamics
of an F-16 aircraft with the matrices
\[ A^{(1)} = \left[\begin{array}{rrrr}
   -0.4492&    0.046&    0.0053&   -0.9926\\
         0&         0&    1.0000&    0.0067\\
   -50.8436&         0&   -5.2184&    0.7220\\
   ~16.4148&         0&    0.0026&   -0.6627
\end{array}\right]
 , \quad
 B_u^{(1)} = \left[\begin{array}{rr}
    0.0004&    0.0011\\
     0&        0\\
   -1.4161&    0.2621\\
   -0.0633&   -0.1205
\end{array}\right] ,\] \[ \;\, C^{(1)} = I_4, \;\qquad D_u^{(1)} = 0_{4\times 2} \, .\]
The four state variables are the sideslip angle, roll angle, roll rate and yaw rate,
and the two input variables are the aileron deflection and rudder
deflection. The model detection problem addresses the synthesis of model detection filters for the detection and identification of loss of efficiency of the two flight actuators, which control the deflections of the aileron and rudder.
The individual fault models correspond to different degrees of surface
efficiency degradation. A multiple model with  $N = 9$ component models is used, which correspond to a two-dimensional
 parameter grid for $N$ values of the parameter vector
 $\rho := [\rho_1,\rho_2]^T$.  For each component of $\rho$, we employ the three grid points
 $\{0,0.5,1\}$.
The component system matrices in (\ref{emdsyn:sysiss}) are defined
for $i = 1, 2, \ldots, N$ as: $E^{(i)} = I_4$,
$A^{(i)} = A^{(1)}$, $C^{(i)} = C^{(1)}$, and $B_u^{(i)} = B^{(1)}_u \Gamma^{(i)}$, where $\Gamma^{(i)} = \diag \big(1-\rho_1^{(i)},1-\rho_2^{(i)}\big)$
and $\big(\rho_1^{(i)},\rho_2^{(i)}\big)$ are the values of parameters $(\rho_1,\rho_2)$
on the chosen grid: \begin{center}
{\tabcolsep=2mm\begin{tabular}{|r|rrrrrrrrr|} \hline
 $\rho_1:$ &  0  &       0 &   0  &  0.5  &  0.5 &   0.5 &   1 &     1 &   1 \\
 $\rho_2:$ &  0  &     0.5 &   1  &    0  &  0.5 &     1 &   0 &   0.5 &   1 \\\hline
\end{tabular}  . }
\end{center} For example, $\big(\rho_1^{(1)},\rho_2^{(1)}\big) = (0,0)$ corresponds to
the fault-free situation, while $\big(\rho_1^{(9)},\rho_2^{(9)}\big) =
(1,1)$ corresponds to complete failure of both control surfaces. It follows, that the TFM  $G_u^{(i)}(s)$ of the $i$-th system can be expressed as
\be\label{mdnnugapdist:Gui_ex} G_u^{(i)}(s) = G_u^{(1)}(s)\Gamma^{(i)}, \ee
where
\[ G_u^{(1)}(s) = C^{(1)}\big(sI-A^{(1)}\big)^{-1}B^{(1)}_u \]
is the TFM of the fault-free system. Note that $G_u^{(N)}(s) = 0$ describes the case of complete failure.

We evaluate the distances between a potential current model to the set of component models using a finer grid of the damage parameters with a length of 0.05, leading to a 441 parameter combinations. For each pair of values $(\rho_1,\rho_2)$ on this grid we determine the current transfer function matrix
\[  G_u(s) := C^{(1)}\big(sI-A^{(1)}\big)^{-1}B^{(1)}\left[\begin{smallmatrix} 1-\rho_1 & 0\\ 0 & 1-\rho_2\end{smallmatrix}\right] \]
and compute the distances to the component models $G_u^{(i)}(s)$ defined in (\ref{mdnnugapdist:Gui_ex}). For the evaluation of least distances, we employ four methods. The first is simply to determine the least distance between the current values of $(\rho_1,\rho_2)$ to the values defined in the above coarse grid. The second and third evaluations of the least distance are based on evaluating the minimum $\mathcal{H}_\infty$- or $\mathcal{H}_2$-norm of the difference $G_u^{(i)}(s)-G_u(s)$, respectively. The fourth evaluation computes the minimum of $\nu$-gap distances  $\delta_\nu\big(G_u^{(i)}(s),G_u(s)\big)$. The resulting numbers of matches of the models defined on the finer grid with those on the original coarse grid result by counting the model indices at the least distances, which are plotted in the histogram in Fig.\ \ref{fig:hist-dist}. As it can be observed, there are differences between the least distances determined with different methods. While the direct comparison of parameters and the  $\mathcal{H}_\infty$- and $\mathcal{H}_2$-norm based values agree reasonably well, the information provided by the computation of $\nu$-gap distances significantly differ, showing strong preference for the 4-th model and significantly less preferences for the 7-th, 8-th, and 9-th models. The largest number of matches occurs for the 5-th model, which corresponds to $\rho_1 = 0.5$ and  $\rho_2 = 0.5$.

\begin{figure}[thpb]
  \begin{center}
  \vspace*{-5mm}
    \includegraphics[width=15cm]{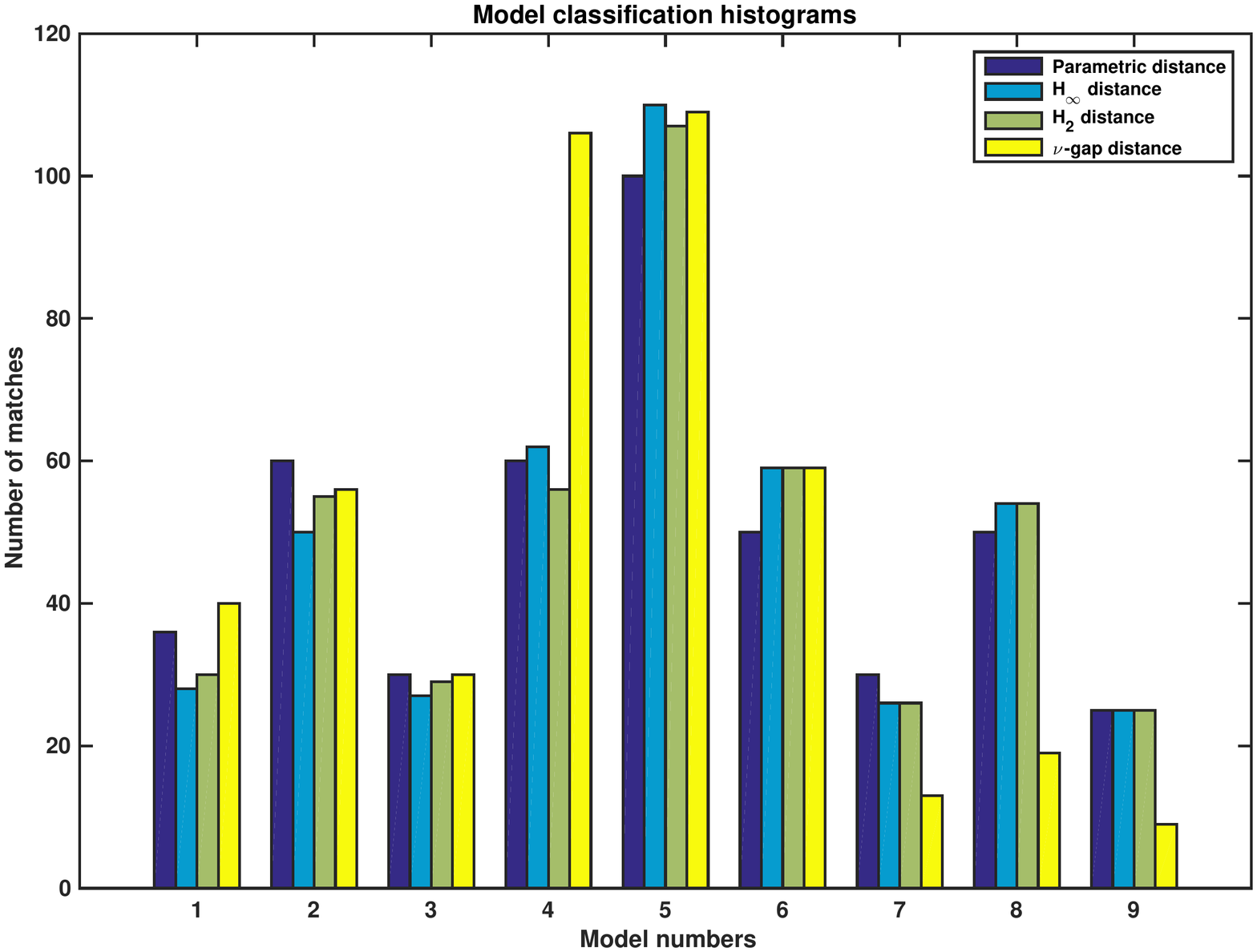} \vspace*{-10mm}
    \caption{Classification of 441 models using different distances: parametric distance (dark blue), $\mathcal{H}_\infty$-norm (blue),  $\mathcal{H}_2$-norm (green), $\nu$-gap distance (yellow) }
    \label{fig:hist-dist}
  \end{center}
\end{figure}

The following MATLAB script produces the histogram in Fig.\ \ref{fig:hist-dist}.

\begin{verbatim}
% Example - Comparison of estimations of minimum distances

% Define a lateral aircraft dynamics model with
% n  = 4 states, mu = 2 control inputs, p  = 4 measurable outputs
A = [-.4492 0.046 .0053 -.9926;
       0    0     1     0.0067;
   -50.8436 0   -5.2184  .722;
    16.4148 0     .0026 -.6627];
Bu = [0.0004 0.0011; 0 0; -1.4161 .2621; -0.0633 -0.1205];
C = eye(4); p = size(C,1); mu = size(Bu,2);

% define the loss of efficiency (LOE) faults as input scaling gains
% Gamma(i,:) = [ 1-rho1(i) 1-rho2(i) ]
Gamma = 1 - [ 0  0 0 .5 .5 .5 1  1 1;
              0 .5 1  0 .5  1 0 .5 1 ]';
N = size(Gamma,1);  % number of LOE cases

% define a multiple physical fault model Gui = Gu*diag(Gamma(i,:))
sysu = ss(zeros(p,mu,N,1));
for i=1:N
    sysu(:,:,i,1) = ss(A,Bu*diag(Gamma(i,:)),C,0);
end
% setup the multiple model
sysu = mdmodset(sysu,struct('controls',1:mu));

% define fine grid and number of samples
rhogrid = 0:0.05:1; K = length(rhogrid)^2;
% perform least distance based model classification
ind = zeros(K,4);
i = 0;
for rho1 = rhogrid
    for rho2 = rhogrid
        i = i+1;
        % define actual model
        rho = [ rho1 rho2];
        sys = ss(A,Bu*diag(1-rho),C,0);
        set(sys,'InputGroup',struct('controls',1:mu));
        temp = Gamma - repmat(1-rho,N,1);
        [~,ind(i,1)]   = min(sqrt(temp(:,1).^2+temp(:,2).^2));
        [~,~,ind(i,2)] = mddist2c(sysu,sys,struct('distance','Inf'));
        [~,~,ind(i,3)] = mddist2c(sysu,sys,struct('distance','2'));
        [~,~,ind(i,4)] = mddist2c(sysu,sys);
    end
end

% plot classification histograms
hist(ind,1:N)
title('Model classification histograms')
xlabel('\bf Model numbers')
ylabel('\bf Number of matches')
legend('\bf Parametric distance','\bf H_\infty distance',...
       '\bf H_2 distance','\bf \nu-gap distance')
\end{verbatim}
\end{example}

\subsection{Functions for Performance Evaluation of FDI Filters} \label{fditools:performance}

These functions address the determination of the structure matrices defined in Section \ref{isolability} and the computation of the performance criteria of FDI filters defined in Section \ref{sec:FDIPerf}.  All functions are fully compatible with the results computed by the synthesis functions of FDI filters described in Section \ref{fditools:synthesis}.
\subsubsection{\texttt{\bfseries fditspec}}

\subsubsection*{Syntax}
\index{M-functions!\texttt{\bfseries fditspec}}
\begin{verbatim}
SMAT = fditspec(R)
SMAT = fditspec(R,TOL)
SMAT = fditspec(R,TOL,FDTOL)
SMAT = fditspec(R,TOL,FDTOL,FREQ)
SMAT = fditspec(R,TOL,FDTOL,[],BLKOPT)
SMAT = fditspec(R,TOL,FDTOL,FREQ,BLKOPT)

\end{verbatim}

\subsubsection*{Description}
\index{fault isolability!structure matrix}
\index{structure matrix}

\noindent \texttt{fditspec} determines the weak or strong structure
matrix corresponding to the fault inputs of the internal form of a FDI filter or
of a collection of internal forms of FDI filters.

\subsubsection*{Input data}
\begin{description}
\item
\texttt{R} is a LTI system or a cell array of LTI systems.

If \texttt{R} is a LTI system representing the internal form of a FDI filter, then it is in a descriptor system state-space form
\be\label{fditspec:sysss}
\begin{aligned}
E_R\lambda x_R(t)  &=   A_Rx_R(t)+ B_{R_f} f(t) + B_{R_v} v(t)  ,\\
r(t) &=  C_R x_R(t)+ D_{R_f} f(t) + D_{R_v} v(t) ,
\end{aligned}
\ee
where $r(t) \in \mathds{R}^q$ is the residual output, $f(t) \in \mathds{R}^{m_f}$ is the fault input and $v(t)$ contains all additional inputs. Any of the input components $f(t)$ and $v(t)$ can be void. For the fault input $f(t)$  the input group \texttt{'faults'} has to be defined. If there is no input group \texttt{'faults'} defined, then, by default, all system inputs are considered faults and the auxiliary input is assumed void. The input-output form of \texttt{R} corresponding to (\ref{fditspec:sysss}) is
\be\label{fditspec:sysio} {\mathbf{r}}(\lambda) =
R_f(\lambda){\mathbf{f}}(\lambda)+R_v(\lambda){\mathbf{v}}(\lambda),
\ee
where $R_f(\lambda)$ and $R_v(\lambda)$ are  the transfer function matrices from the corresponding inputs, respectively. If \texttt{R} is specified in the input-output representation (\ref{fditspec:sysio}), then it is automatically converted to an equivalent  minimal order state-space form as in (\ref{fditspec:sysss}).

If \texttt{R} is a $N\times 1$ cell array of LTI systems representing the internal forms of $N$ FDI filters, then the $i$-th component system \texttt{R\{$i$\}} is in the state-space form
\be\label{fditspec:sysiss}
{\begin{aligned}
E_R^{(i)}\lambda x_R^{(i)}(t)  & =   A_R^{(i)}x_R^{(i)}(t)+ B_{R_f}^{(i)}f(t)+ B_{R_v}^{(i)}v(t), \\
r^{(i)}(t) & =  C_R^{(i)} x_R^{(i)}(t) + D_{R_f}^{(i)}f(t)+ D_{R_v}^{(i)}v(t) ,
\end{aligned}}
\ee
where $r^{(i)}(t) \in \mathds{R}^{q^{(i)}}$ is the $i$-th residual component, $f(t) \in \mathds{R}^{m_f}$ is the fault input and $v(t)$ contains the rest of inputs.
Any of the input components $f(t)$ and $v(t)$ can be void. For the fault input $f(t)$  the input group \texttt{'faults'} has to be defined, while $v(t)$ includes any other (not relevant) system inputs. If there is no input group \texttt{'faults'} defined, then, by default, all system inputs are considered faults and the auxiliary input is assumed void. All component systems \texttt{R\{$i$\}}, $i = 1, \ldots, N$, must have the same number of fault inputs and the same sampling time.
The input-output form of  \texttt{R\{$i$\}} corresponding to (\ref{fditspec:sysiss}) is
\be\label{fditspec:sysiio} {\mathbf{r}}^{(i)}(\lambda) =
R_f^{(i)}(\lambda){\mathbf{f}}(\lambda)+R_v^{(i)}(\lambda){\mathbf{v}}(\lambda),
\ee
where $R_f^{(i)}(\lambda)$ and $R_v^{(i)}(\lambda)$ are the transfer function matrices from the corresponding inputs, respectively.
If \texttt{R\{$i$\}}  is specified in the input-output representation (\ref{fditspec:sysiio}), then it is automatically converted to an equivalent  minimal order state-space form as in (\ref{fditspec:sysiss}).

\item
\texttt{TOL} is a relative tolerance used for controllability
tests. A default value is internally computed if \texttt{TOL} $\leq 0$ or is not specified at input.
\item
\texttt{FDTOL} is an absolute threshold for the magnitudes of the zero elements
in the system matrices $B_{R_f}$, $C_R$ and $D_{R_f}$, in the case of model (\ref{fditspec:sysss}), or in the matrices $B_{R_f}^{(i)}$, $C_R^{(i)}$ and $D_{R_f}^{(i)}$, in the case of models of the form  (\ref{fditspec:sysiss}).
Any element of these matrices whose magnitude does not exceed \texttt{FDTOL} is considered zero. Additionally, if \texttt{FREQ} is nonempty, \texttt{FDTOL} is also used for the singular-value-based rank tests performed on the system matrix $\Big[\begin{smallmatrix} A_R-\lambda E_R & B_{R_f}\\ C_R & D_{R_f} \end{smallmatrix}\Big]$, in the case of model (\ref{fditspec:sysss}), or on the system matrices $\bigg[\begin{smallmatrix} A_R^{(i)}-\lambda E_R^{(i)} & B_{R_f}^{(i)}\\ C_R^{(i)} & D_{R_f}^{(i)} \end{smallmatrix}\bigg]$ for $i = 1, \ldots, N$, in the case of models of the form (\ref{fditspec:sysiss}).
If \texttt{FDTOL} $\leq 0$ or not specified at input, the default value \texttt{FDTOL} = $10^{-4}\max\big(1,\big\|B_{R_f}\big\|_1,\big\|C_R\big\|_\infty,\big\|D_{R_f}\big\|_1\big)$ is used in the case of model (\ref{fditspec:sysss}). In the case of models of the form (\ref{fditspec:sysiss}), the default value \texttt{FDTOL} = $10^{-4}\max\big(1,\big\|B_{R_f}^{(i)}\big\|_1,\big\|C_R^{(i)}\big\|_\infty,\big\|D_{R_f}^{(i)}\big\|_1\big)$ is  used for handling the $i$-th model (\ref{fditspec:sysiss}).
If \texttt{FDTOL} $\leq 0$  and if \texttt{FREQ} is nonempty,  the default value \texttt{FDTOL} = $10^{-4} \max\Big( 1, \Big\|\left[\begin{smallmatrix}  A_R & B_{R_f}\\C_R& D_{R_f}\end{smallmatrix}\right] \Big\|_1,\big\|E_R\big\|_1\Big)$  is used for the rank tests on the
system matrix in the case of model (\ref{fditspec:sysss}). In the case of models of the form (\ref{fditspec:sysiss}), the default value \texttt{FDTOL} = $10^{-4} \max\Big( 1, \Bigg\|\left[\begin{smallmatrix}  A_R^{(i)} & B_{R_f}^{(i)}\\C_R^{(i)}& D_{R_f}^{(i)}\end{smallmatrix}\right] \Bigg\|_1,\big\|E_R^{(i)}\big\|_1\Big)$ is  used for handling the system matrix of the $i$-th model (\ref{fditspec:sysiss}).
\item
\texttt{FREQ} is a real vector, which contains the frequency values $\omega_k$, $k = 1, \ldots, n_f$, to be used to check for zeros of the individual (rational) elements or individual columns of the transfer function matrix $R_f(\lambda)$ in the case of model (\ref{fditspec:sysss}), or of the transfer function matrices $R_f^{(i)}(\lambda)$ in the case of models (\ref{fditspec:sysiss}). By default, \texttt{FREQ} is empty if it is not specified. For each real frequency  $\omega_k$ contained in \texttt{FREQ}, there corresponds a complex frequency $\lambda_k$, which is used to check the elements or columns of the respective transfer function matrices to have $\lambda_k$ as zero. Depending on the system type, $\lambda_k = \mathrm{i}\omega_k$, in the continuous-time case, and $\lambda_k = \exp (\mathrm{i}\omega_k T)$, in the discrete-time case, where $T$ is the sampling time of the system.
\item
\texttt{BLKOPT} is a character variable to be set to \texttt{'block'} to specify the block-structure based evaluation option of the structure matrix.
\end{description}

\subsubsection*{Output data}

\begin{description}
\item
\texttt{SMAT} is a logical array which contains the resulting structure matrix.

In the case of a LTI system \texttt{R},  \texttt{SMAT} is determined depending on the selected option \texttt{BLKOPT} and the frequency values specified in \texttt{FREQ}, as follows:
 \begin{itemize}
 \item If \texttt{BLKOPT} is not specified (or empty), then \texttt{SMAT} is the structure matrix  corresponding to the zero and nonzero elements of the transfer function matrix $R_f(\lambda)$:
     \begin{itemize}
     \item
If \texttt{FREQ} is empty or not specified at input, then \texttt{SMAT} is a $q\times m_f$  logical array, which contains the \emph{weak} structure matrix corresponding to the zero and nonzero elements of $R_f(\lambda)$  (see (\ref{structure_matrix}) for the definition of the weak structure matrix). Accordingly,  \texttt{SMAT}$(i,j)$ = \texttt{true}, if the $(i,j)$-th
element of $R_f(\lambda)$ is nonzero. Otherwise, \texttt{SMAT}$(i,j)$ = \texttt{false}.
\item
If \texttt{FREQ} is nonempty, then \texttt{SMAT} is a $q\times m_f \times n_f$ logical array which contains in the $k$-th page $\texttt{SMAT}(:,:,k)$, the \emph{strong} structure matrix corresponding to the presence or absence of zeros of the elements of $R_f(\lambda)$  in the complex frequency $\lambda_k$ corresponding to the $k$-th real frequency $\omega_k$ contained in \texttt{FREQ} (see description of \texttt{FREQ})  (see also (\ref{strong_structure_matrix}) for the definition of the strong structure matrix at a complex frequency $\lambda_k$).
Accordingly, $\texttt{SMAT}(i,j,k)$ = \texttt{true}, if the $(i,j)$-th
element of $R_f(\lambda)$ has no zero in $\lambda_k$. Otherwise, $\texttt{SMAT}(i,j,k)$ = \texttt{false}.
\end{itemize}
 \item If \texttt{BLKOPT = 'block'} is specified, then \texttt{SMAT} is the structure (row) vector corresponding to the zero and nonzero columns of the transfer function matrix $R_f(\lambda)$:
     \begin{itemize}
     \item
If \texttt{FREQ} is empty or not specified at input, then \texttt{SMAT} is a $1\times m_f$  logical (row) vector, which contains the \emph{weak} structure matrix corresponding to the zero and nonzero columns of  $R_f(\lambda)$ (see (\ref{structure_matrix}) for the block-structured definition of the weak structure matrix). Accordingly,  \texttt{SMAT}$(1,j)$ = \texttt{true}, if the $j$-th
column of $R_f(\lambda)$ is nonzero. Otherwise, \texttt{SMAT}$(1,j)$ = \texttt{false}.
\item
If \texttt{FREQ} is nonempty, then \texttt{SMAT} is a $1\times m_f \times n_f$ logical array which contains in the $k$-th page $\texttt{SMAT}(:,:,k)$, the \emph{strong} structure vector corresponding to the presence or absence of a zero in the columns of  $R_f(\lambda)$ in the complex frequency $\lambda_k$ corresponding to the $k$-th real frequency $\omega_k$ contained in \texttt{FREQ} (see description of \texttt{FREQ}).
Accordingly, $\texttt{SMAT}(1,j,k)$ = \texttt{true}, if the $j$-th
column of $R_f(\lambda)$ has no zero in the complex frequency $\lambda_k$. Otherwise, $\texttt{SMAT}(1,j,k)$ = \texttt{false}.
\end{itemize}
\end{itemize}
In the case of a cell array of  LTI systems \texttt{R\{$i$\}}, $i = 1, \ldots, N$,   \texttt{SMAT} is determined depending on the frequency values specified in \texttt{FREQ}, as follows:
\begin{itemize}
\item[$-$]
If \texttt{FREQ} is empty or not specified at input, then \texttt{SMAT} is an $N\times m_f$  logical matrix, whose $i$-th row contains the \emph{weak} structure vector corresponding to the zero and nonzero columns of the transfer function matrix $R_f^{(i)}(\lambda)$ (see (\ref{structure_matrix}) for the block-structured definition of the weak structure matrix). Accordingly,  \texttt{SMAT}$(i,j)$ = \texttt{true}, if the $j$-th
column of $R_f^{(i)}(\lambda)$ is nonzero. Otherwise, \texttt{SMAT}$(i,j)$ = \texttt{false}. All entries of the $i$-th row of \texttt{SMAT} are set to \texttt{false} if \texttt{R\{$i$\}} is empty.
\item[$-$]
If \texttt{FREQ} is nonempty, then \texttt{SMAT} is a $N\times m_f \times n_f$ logical array which contains in the $i$-th row of the $k$-th page $\texttt{SMAT}(:,:,k)$, the \emph{strong} structure (row) vector corresponding to the presence or absence of a zero in the  columns of $R_f^{(i)}(\lambda)$  in the complex frequency $\lambda_k$ corresponding to the $k$-th real frequency $\omega_k$ contained in \texttt{FREQ} (see description of \texttt{FREQ})  (see also (\ref{strong_structure_matrix}) for the definition of the strong structure matrix at a complex frequency $\lambda_k$).
Accordingly, $\texttt{SMAT}(i,j,k)$ = \texttt{true}, if the $j$-th
column of $R_f^{(i)}(\lambda)$ has no zero in $\lambda_k$. Otherwise, $\texttt{SMAT}(i,j,k)$ = \texttt{false}. All entries of the $i$-th row of \texttt{SMAT} are set to \texttt{false} if \texttt{R\{$i$\}} is empty.
\end{itemize}

\end{description}

\subsubsection*{Method}
We first describe the implemented analysis method for the case when \texttt{R} is a LTI system in a state-space form as in (\ref{fditspec:sysss}) and  $R_f(\lambda)$ is the transfer function matrix from the fault inputs to the residual output as in the input-output model (\ref{fditspec:sysio}).
For the definition of the weak and strong structure matrices, see Section \ref{isolability}. For the
  determination of the weak structure matrix, controllable realizations
  are determined for each column of $R_f(\lambda)$ and tests
  are performed to identify the nonzero elements in the respective column of $R_f(\lambda)$ by using \cite[Corollary 7.1]{Varg17} in a controllability related dual formulation. The block-structure based evaluation is based on the input observability tests of \cite[Corollary 7.1]{Varg17}, performed for the controllable realizations of each column of $R_f(\lambda)$.

  For the determination of the strong structure matrix, minimal
  realizations are determined for each element of $R_f(\lambda)$ and
  the absence of zeros is assessed by checking the full rank of the corresponding
  system matrix for all complex frequencies corresponding to the real frequencies specified in \texttt{FREQ} (see \cite[Corollary 7.2]{Varg17}). For the block-structure based evaluation, controllable realizations
  are determined for each column of $R_f(\lambda)$ and the test used in \cite[Corollary 7.2]{Varg17} is employed.

For the case when \texttt{R} is a cell array containing $N$  LTI systems in state-space forms as in (\ref{fditspec:sysiss}) and  $R_f^{(i)}(\lambda)$ is the transfer function matrix of $i$-th LTI system \texttt{R\{$i$\}} from the fault inputs to the $i$-th residual component as in the input-output models (\ref{fditspec:sysiio}), the block-structure based evaluations, described above, are employed for each $R_f^{(i)}(\lambda)$ to determine the corresponding $i$-th row of the structure matrix \texttt{SMAT}.

\subsubsection{\texttt{\bfseries fdisspec}}

\subsubsection*{Syntax}
\index{M-functions!\texttt{\bfseries fdisspec}}
\begin{verbatim}
[SMAT,GAINS] = fdisspec(R)
[SMAT,GAINS] = fdisspec(R,FDGAINTOL)
[SMAT,GAINS] = fdisspec(R,FDGAINTOL,FREQ)
[SMAT,GAINS] = fdisspec(R,FDGAINTOL,FREQ,BLKOPT)
\end{verbatim}

\subsubsection*{Description}
\index{fault isolability!structure matrix}
\index{structure matrix}

\noindent \texttt{fdisspec} determines the strong structure
matrix corresponding to the fault inputs of the internal form of a FDI filter or
of a collection of internal forms of FDI filters.

\subsubsection*{Input data}

\begin{description}
\item
\texttt{R} is a LTI system or a cell array of LTI systems.

If \texttt{R} is a LTI system representing the internal form of a FDI filter, then it is in a descriptor system state-space form
\be\label{fdisspec:sysss}
\begin{aligned}
E_R\lambda x_R(t)  &=   A_Rx_R(t)+ B_{R_f} f(t) + B_{R_v} v(t)  ,\\
r(t) &=  C_R x_R(t)+ D_{R_f} f(t) + D_{R_v} v(t) ,
\end{aligned}
\ee
where $r(t) \in \mathds{R}^q$ is the residual output, $f(t) \in \mathds{R}^{m_f}$ is the fault input and $v(t)$ contains all additional inputs. Any of the input components $f(t)$ and $v(t)$ can be void. For the fault input $f(t)$  the input group \texttt{'faults'} has to be defined. If there is no input group \texttt{'faults'} defined, then, by default, all system inputs are considered faults and the auxiliary input is assumed void. The input-output form of \texttt{R} corresponding to (\ref{fdisspec:sysss}) is
\be\label{fdisspec:sysio} {\mathbf{r}}(\lambda) =
R_f(\lambda){\mathbf{f}}(\lambda)+R_v(\lambda){\mathbf{v}}(\lambda),
\ee
where $R_f(\lambda)$ and $R_v(\lambda)$ are  the transfer function matrices from the corresponding inputs. If \texttt{R} is specified in the input-output representation (\ref{fdisspec:sysio}), then it is automatically converted to an equivalent  minimal order state-space form as in (\ref{fdisspec:sysss}).

If \texttt{R} is an $N\times 1$ cell array of LTI systems representing the internal forms of $N$ FDI filters, then the $i$-th component system \texttt{R\{$i$\}} is in the state-space form
\be\label{fdisspec:sysiss}
{\begin{aligned}
E_R^{(i)}\lambda x_R^{(i)}(t)  & =   A_R^{(i)}x_R^{(i)}(t)+ B_{R_f}^{(i)}f(t)+ B_{R_v}^{(i)}v(t), \\
r^{(i)}(t) & =  C_R^{(i)} x_R^{(i)}(t) + D_{R_f}^{(i)}f(t)+ D_{R_v}^{(i)}v(t) ,
\end{aligned}}
\ee
where $r^{(i)}(t) \in \mathds{R}^{q^{(i)}}$ is the $i$-th residual component, $f(t) \in \mathds{R}^{m_f}$ is the fault input and $v(t)$ contains the rest of inputs.
Any of the input components $f(t)$ and $v(t)$ can be void. For the fault input $f(t)$  the input group \texttt{'faults'} has to be defined, while $v(t)$ includes any other (not-relevant) system inputs. If there is no input group \texttt{'faults'} defined, then, by default, all system inputs are considered faults and the auxiliary input is assumed void. All component systems \texttt{R\{$i$\}}, $i = 1, \ldots, N$, must have the same number of fault inputs and the same sampling time.
The input-output form of  \texttt{R\{$i$\}} corresponding to (\ref{fdisspec:sysiss}) is
\be\label{fdisspec:sysiio} {\mathbf{r}}^{(i)}(\lambda) =
R_f^{(i)}(\lambda){\mathbf{f}}(\lambda)+R_v^{(i)}(\lambda){\mathbf{v}}(\lambda),
\ee
where $R_f^{(i)}(\lambda)$ and $R_v^{(i)}(\lambda)$ are the transfer function matrices from the corresponding inputs.
If \texttt{R\{$i$\}}  is specified in the input-output representation (\ref{fdisspec:sysiio}), then it is automatically converted to an equivalent  minimal order state-space form as in (\ref{fdisspec:sysiss}).

\item
\texttt{FDGAINTOL} is a threshold for the magnitudes of the frequency-response gains of the transfer function matrix $R_f(\lambda)$ in the case of model (\ref{fdisspec:sysss}), or of the transfer function matrices $R_f^{(i)}(\lambda)$ in the case of models (\ref{fdisspec:sysiss}).    If \texttt{FDGAINTOL} = 0 or not specified at input, the default value \texttt{FDGAINTOL} = 0.01 is used.
\item
\texttt{FREQ} is a real vector, which contains the real frequency values $\omega_k$, $k = 1, \ldots, n_f$, to be used to check for the existence of zeros of the (rational) elements or columns of the transfer function matrix $R_f(\lambda)$ in the case of model (\ref{fditspec:sysss}), or of the transfer function matrices $R_f^{(i)}(\lambda)$ in the case of models (\ref{fditspec:sysiss}). By default, \texttt{FREQ} = 0, if it is empty  or not specified.  For each real frequency  $\omega_k$ contained in \texttt{FREQ}, there corresponds a complex frequency $\lambda_k$ which is used to evaluate $R_f(\lambda_k)$, to check that the elements or columns of the respective transfer function matrices have $\lambda_k$ as a zero. Depending on the system type, $\lambda_k = \mathrm{i}\omega_k$, in the continuous-time case, and $\lambda_k = \exp (\mathrm{i}\omega_k T)$, in the discrete-time case, where $T$ is the sampling time of the system. The complex frequencies corresponding to the real frequencies specified in \texttt{FREQ}
must be disjoint from the set of poles of $R_f(\lambda)$.
\item
\texttt{BLKOPT} is a character variable to be set to \texttt{'block'} to specify the block-structure based evaluation option of the structure matrix.
\end{description}

\subsubsection*{Output data}

\begin{description}
\item
\texttt{SMAT} is a logical array which contains the resulting structure matrix.

In the case \texttt{R} is a LTI system,  \texttt{SMAT} is determined depending on the selected option \texttt{BLKOPT}, as follows:
 \begin{itemize}
 \item If \texttt{BLKOPT} is not specified (or empty), then \texttt{SMAT} is a $q\times m_f \times n_f$ logical array which contains in the $k$-th page $\texttt{SMAT}(:,:,k)$, the \emph{strong} structure matrix corresponding to the presence or absence of zeros of the elements of $R_f(\lambda)$  in the complex frequency $\lambda_k$ corresponding to the $k$-th real frequency $\omega_k$ contained in \texttt{FREQ} (see description of \texttt{FREQ})  (see also (\ref{strong_structure_matrix}) for the definition of the strong structure matrix at a complex frequency $\lambda_k$).
Accordingly, $\texttt{SMAT}(i,j,k)$ = \texttt{true}, if the magnitude of the $(i,j)$-th
element of $R_f(\lambda_k)$ is greater than or equal to \texttt{FDGAINTOL}. Otherwise, $\texttt{SMAT}(i,j,k)$ = \texttt{false}.
 \item If \texttt{BLKOPT = 'block'} is specified, then \texttt{SMAT} is a $1\times m_f \times n_f$ logical array which contains in the $k$-th page $\texttt{SMAT}(:,:,k)$, the \emph{strong} structure (row) vector corresponding to the presence or absence of a zero in the columns of  $R_f(\lambda)$ in the complex frequency $\lambda_k$ corresponding to the $k$-th real frequency $\omega_k$ contained in \texttt{FREQ} (see description of \texttt{FREQ}).
Accordingly, $\texttt{SMAT}(1,j,k)$ = \texttt{true}, if the norm of the $j$-th
column of $R_f(\lambda_k)$ is greater than or equal to \texttt{FDGAINTOL} (i.e., $\big\|R_{f_j}(\lambda_k)\big\|_2 \geq$ \texttt{FDGAINTOL}). Otherwise, $\texttt{SMAT}(1,j,k)$ = \texttt{false}.
\end{itemize}

In the case \texttt{R} is an $N\times 1$  cell array of  LTI systems,  \texttt{SMAT} is an $N\times m_f \times n_f$ logical array which contains in the $i$-th row of the $k$-th page $\texttt{SMAT}(:,:,k)$, the \emph{strong} structure (row) vector corresponding to the presence or absence of a zero in the  columns of $R_f^{(i)}(\lambda)$  in the complex frequency $\lambda_k$ corresponding to the $k$-th real frequency $\omega_k$ contained in \texttt{FREQ} (see description of \texttt{FREQ})  (see also (\ref{strong_structure_matrix}) for the definition of the strong structure matrix at a complex frequency $\lambda_k$).
Accordingly, $\texttt{SMAT}(i,j,k)$ = \texttt{true}, if the norm of the $j$-th
column of $R_f^{(i)}(\lambda_k)$ is greater than or equal to \texttt{FDGAINTOL} (i.e., $\big\|R_{f_j}^{(i)}(\lambda_k)\big\|_2 \geq$ \texttt{FDGAINTOL}). Otherwise, $\texttt{SMAT}(i,j,k)$ = \texttt{false}. If \texttt{R\{$i$\}} is empty, then all entries of the $i$-th row of \texttt{SMAT} are set to \texttt{false}.

\item
 \texttt{GAINS} is a real nonnegative array.

 In the case \texttt{R} is a LTI system,  \texttt{GAINS} is determined depending on the selected option \texttt{BLKOPT}, as follows:
 \begin{itemize}
 \item If \texttt{BLKOPT} is not specified (or empty), then \texttt{GAINS} is $q\times m_f$ matrix, whose $(i,j)$-th element contains the minimum value of the frequency-response gains of the $(i,j)$-th element of $R_f(\lambda)$ evaluated
  over all complex frequencies corresponding to \texttt{FREQ} (see description of \texttt{FREQ}). This value is a particular instance of the $\mathcal{H}_-$-index of the $(i,j)$-th element of $R_f(\lambda)$.
 \item If \texttt{BLKOPT = 'block'} is specified, then \texttt{GAINS} is a $m_f$-dimensional row vector, whose $j$-th element contains the minimum of the norms of the frequency responses of the $j$-th column of $R_f(\lambda)$ evaluated
  over all complex frequencies corresponding to \texttt{FREQ} (see description of \texttt{FREQ}). This value is a particular instance of the $\mathcal{H}_-$-index of the $j$-th column of $R_f(\lambda)$.
\end{itemize}

In the case \texttt{R} is an $N\times 1$ cell array of  LTI systems, \texttt{GAINS} is an $N\times m_f$ matrix, whose $(i,j)$-th element contains the minimum of the norms of the frequency responses of the $j$-th column of $R_f^{(i)}(\lambda)$ evaluated
  over all complex frequencies corresponding to \texttt{FREQ} (see description of \texttt{FREQ}). If \texttt{R\{$i$\}} is empty, then all entries of the $i$-th row of \texttt{GAINS} are set to zero.
\end{description}
\subsubsection*{Method}
For the definition of the strong structure matrix at a given frequency, see Remark \ref{rem:strong_structure} of Section \ref{isolability}.
For the case when \texttt{R} is a LTI system in a state-space form as in (\ref{fdisspec:sysss}) and  $R_f(\lambda)$ is the transfer function matrix from the fault inputs to the residual output as in the input-output model (\ref{fdisspec:sysio}), the element-wise or column-wise (if \texttt{BLKOPT = 'block'}) evaluations of the $\mathcal{H}_-$-index on a discrete set of frequency values (see \cite[Section 5.3]{Varg17}) are employed for each element or, respectively, each column of $R_f(\lambda)$, to determine the corresponding element of the structure matrix \texttt{SMAT} and associated \texttt{GAINS}.
The resulting entries of \texttt{GAINS}  correspond to an element-wise or column-wise evaluation of the $\mathcal{H}_-$-index on a discrete set of frequency values (see \cite[Section 5.3]{Varg17}).

For the case when \texttt{R} is a cell array containing $N$  LTI systems in state-space forms as in (\ref{fdisspec:sysiss}) and  $R_f^{(i)}(\lambda)$ is the transfer function matrix of $i$-th LTI system \texttt{R\{$i$\}} from the fault inputs to the $i$-th residual component as in the input-output models (\ref{fdisspec:sysiio}), the column-wise evaluations of the $\mathcal{H}_-$-index on a discrete set of frequency values (see \cite[Section 5.3]{Varg17}) are employed for each $R_f^{(i)}(\lambda)$ to determine the corresponding $i$-row of the structure matrix \texttt{SMAT} and the associated $i$-th row of \texttt{GAINS}.

\subsubsection{\texttt{\bfseries fdifscond}}
\index{M-functions!\texttt{\bfseries fdifscond}}
\subsubsection*{Syntax}
\begin{verbatim}
FSCOND = fdifscond(R)
FSCOND = fdifscond(R,FREQ)
FSCOND = fdifscond(R,[],S)
FSCOND = fdifscond(R,FREQ,S)
[BETA,GAMMA] = fdifscond(...)
\end{verbatim}

\subsubsection*{Description}
\index{performance evaluation!fault detection and isolation!fault sensitivity condition}

\noindent \texttt{fdifscond} evaluates the fault sensitivity condition of the internal form of a FDI filter or the fault sensitivity conditions of the internal forms of a collection of FDI filters.

\subsubsection*{Input data}
\begin{description}
\item
\texttt{R} is a LTI system or a cell array of LTI systems.

If \texttt{R} is a LTI system representing the internal form of a FDI filter, then it is in a descriptor system state-space form
\be\label{fdifscon:sysss}
{\begin{aligned}
E_R\lambda x_R(t)  & =   A_Rx_R(t)+ B_{R_f}f(t)+ B_{R_{v}}v(t) ,\\
r(t) & =  C_R x_R(t) + D_{R_f}f(t)+ D_{R_{v}}v(t) ,
\end{aligned}}
\ee
where $r(t) \in \mathds{R}^q$ is the residual output, $f(t) \in \mathds{R}^{m_f}$ is the fault input, and $v(t)$ contains all
additional inputs. For the fault input $f(t)$  the input group \texttt{'faults'} has to be defined. The input-output form of  \texttt{R} corresponding to (\ref{fdifscon:sysss}) is
\be\label{fdifscon:sysio} {\mathbf{r}}(\lambda) =
R_f(\lambda){\mathbf{f}}(\lambda)+R_v(\lambda){\mathbf{v}}(\lambda),
\ee
where $R_f(\lambda)$ and $R_v(\lambda)$ are the transfer function matrices from the corresponding inputs.

If \texttt{R} is an $N\times 1$ cell array of LTI systems representing the internal forms of $N$ FDI filters, then the $i$-th component system \texttt{R\{$i$\}} is in the state-space form
\be\label{fdifscon:sysiss}
{\begin{aligned}
E_R^{(i)}\lambda x_R^{(i)}(t)  & =   A_R^{(i)}x_R^{(i)}(t)+ B_{R_f}^{(i)}f(t)+ B_{R_v}^{(i)}v(t), \\
r^{(i)}(t) & =  C_R^{(i)} x_R^{(i)}(t) + D_{R_f}^{(i)}f(t)+ D_{R_v}^{(i)}v(t) ,
\end{aligned}}
\ee
where $r^{(i)}(t) \in \mathds{R}^{q^{(i)}}$ is the $i$-th residual component, $f(t) \in \mathds{R}^{m_f}$ is the fault input and $v(t)$ contains the rest of inputs. For the fault input $f(t)$  the input group \texttt{'faults'} has to be defined.
The input-output form of  \texttt{R\{$i$\}} corresponding to (\ref{fdifscon:sysiss}) is
\be\label{fdifscon:sysiio} {\mathbf{r}}^{(i)}(\lambda) =
R_f^{(i)}(\lambda){\mathbf{f}}(\lambda)+R_v^{(i)}(\lambda){\mathbf{v}}(\lambda),
\ee
where $R_f^{(i)}(\lambda)$ and $R_v^{(i)}(\lambda)$ are the transfer function matrices from the corresponding inputs.

\item
\texttt{FREQ} is a real vector, which contains the frequency values $\omega_k$, $k = 1, \ldots, n_f$, to be used to evaluate the fault condition number of the transfer function matrix $R_f(\lambda)$ in (\ref{fdifscon:sysio}), or of the transfer function matrices $R_f^{(i)}(\lambda)$ in  (\ref{fdifscon:sysiio}). By default, \texttt{FREQ} is empty if it is not specified. For each real frequency  $\omega_k$ contained in \texttt{FREQ}, there corresponds a complex frequency $\lambda_k$, which is used to define the complex set $\Omega = \{ \lambda_1, \ldots, \lambda_{n_f}\}$. Depending on the system type, $\lambda_k = \mathrm{i}\omega_k$, in the continuous-time case, and $\lambda_k = \exp (\mathrm{i}\omega_k T)$, in the discrete-time case, where $T$ is the sampling time of the system.
\item
\texttt{S} is a $q\times m_f$ logical structure matrix if \texttt{R} is the internal form of a FDI filter with $q$ residual outputs or is an $N\times m_f$ logical structure matrix if \texttt{R} is a collection of $N$ internal forms of fault detection and isolation filters, where \texttt{R$\{i\}$} is the internal form of the $i$-th filter.
\end{description}

\subsubsection*{Output data}

\begin{description}
\item
\texttt{FSCOND} is the computed fault sensitivity condition, which is either a scalar or a vector, depending on the  input variables.

In the case of calling \texttt{fdifscond} as
\begin{verbatim}
   FSCOND = fdifscond(R)
\end{verbatim}
then:
 \begin{itemize}
\item
If \texttt{R} is a LTI system, then \texttt{FSCOND} is the fault sensitivity condition computed as $\beta / \gamma$, where $\beta = \| R_{\!f}(\lambda) \|_{\infty -}$ and $\gamma = \displaystyle\max_j \|R_{\!f_j}(\lambda)\|_\infty$ ;
 \item
 If \texttt{R} is a collection of $N$ LTI systems, then \texttt{FSCOND} is an $N$-dimensional vector of fault sensitivity conditions, with \texttt{FSCOND\{$i$\}}, the $i$-th fault sensitivity condition, computed as $\beta_i / \gamma_i$, where $\beta_i = \| R_{\!f}^{(i)}(\lambda) \|_{\infty -}$ and $\gamma_i = \displaystyle\max_j \|R_{\!f_j}^{(i)}(\lambda)\|_\infty$. If \texttt{R\{$i$\}} is empty, then \texttt{FSCOND\{$i$\}} is set to \texttt{NaN}.
\end{itemize}

In the case of calling \texttt{fdifscond} as
\begin{verbatim}
   FSCOND = fdifscond(R,FREQ)
\end{verbatim}
where \texttt{FREQ} is a nonempty vector of real frequencies, which define the set of complex frequencies $\Omega$ (see description of \texttt{FREQ}), then:
 \begin{itemize}
\item
If \texttt{R} is a LTI system, then \texttt{FSCOND} is the fault sensitivity condition computed as $\beta / \gamma$, where $\beta = \big\| R_{\!f}(\lambda) \big\|_{\Omega -}$ and $\gamma = \displaystyle\max_{j}\big\{ \sup_{\lambda_s \in \Omega} \big\|R_{\!f_j}(\lambda_s)\big\|_2 \big\}$ ;
 \item
 If \texttt{R} is a collection of $N$ LTI systems, then \texttt{FSCOND} is an $N$-dimensional vector of fault sensitivity conditions, with \texttt{FSCOND\{$i$\}}, the $i$-th fault sensitivity condition, computed as $\beta_i / \gamma_i$, where $\beta_i = \big\| R_{\!f}^{(i)}(\lambda) \big\|_{\Omega -}$ and $\gamma_i = \displaystyle\max_{j}\big\{ \sup_{\lambda_s \in \Omega} \big\|R_{\!f_j}^{(i)}(\lambda_s)\big\|_2 \big\}$. If \texttt{R\{$i$\}} is empty, then \texttt{FSCOND\{$i$\}} is set to  \texttt{NaN}.
\end{itemize}

In the case of calling \texttt{fdifscond} as
\begin{verbatim}
   FSCOND = fdifscond(R,[],S)
\end{verbatim}
then:
 \begin{itemize}
\item
If \texttt{R} is a LTI system with $q$ residual outputs and \texttt{S} is a $q\times m_f$ structure matrix,  then \texttt{FSCOND} is a $q$-dimensional vector of fault sensitivity conditions. \texttt{FSCOND\{$i$\}} is the fault sensitivity condition of the $i$-th row $R_{\!f}^{(i)}(\lambda)$ of $R_{\!f}(\lambda)$, computed as $\beta_i / \gamma_i$, with $\beta_i = \big\| R_{\!f^{(i)}}^{(i)}(\lambda) \big\|_{\infty -}$ and $\gamma_i = \displaystyle\max_j \big\|R_{\!f_j}^{(i)}(\lambda)\big\|_\infty$, where $R_{\!f^{(i)}}^{(i)}(\lambda)$ is formed from the elements of $R_{\!f}^{(i)}(\lambda)$ which correspond to \texttt{true} values in the $i$-th row of \texttt{S} and  $R_{\!f_j}^{(i)}(\lambda)$ is the $(i,j)$-th element of $R_{\!f}(\lambda)$;
 \item
 If \texttt{R} is a collection of $N$ LTI systems and \texttt{S} is an $N\times m_f$ structure matrix,  then \texttt{FSCOND} is an $N$-dimensional vector of fault sensitivity conditions, with \texttt{FSCOND\{$i$\}}, the $i$-th fault sensitivity condition, computed as $\beta_i / \gamma_i$, with $\beta_i = \big\| R_{\!f^{(i)}}^{(i)}(\lambda) \big\|_{\infty -}$ and $\gamma_i = \displaystyle\max_j \big\|R_{\!f_j}^{(i)}(\lambda)\big\|_\infty$, where $R_{\!f^{(i)}}^{(i)}(\lambda)$ is formed from the columns of $R_{\!f}^{(i)}(\lambda)$ which correspond to \texttt{true} values in the $i$-th row of \texttt{S} and  $R_{\!f_j}^{(i)}(\lambda)$ is the $j$-th column of $R_{\!f}^{(i)}(\lambda)$. If \texttt{R\{$i$\}} is empty, then \texttt{FSCOND\{$i$\}} is set to  \texttt{NaN}.
\end{itemize}

In the case of calling \texttt{fdifscond} as
\begin{verbatim}
   FSCOND = fdifscond(R,FREQ,S)
\end{verbatim}
where both \texttt{FREQ} and \texttt{S} are nonempty then:
 \begin{itemize}
\item
If \texttt{R} is a LTI system with $q$ residual outputs and \texttt{S} is a $q\times m_f$ structure matrix,  then \texttt{FSCOND} is a $q$-dimensional vector of fault sensitivity conditions. \texttt{FSCOND\{$i$\}} is the fault sensitivity condition of the $i$-th row $R_{\!f}^{(i)}(\lambda)$ of $R_{\!f}(\lambda)$, computed as $\beta_i / \gamma_i$, with $\beta_i = \big\| R_{\!f^{(i)}}^{(i)}(\lambda) \big\|_{\Omega -}$ and $\gamma_i = \displaystyle\max_{j}\big\{ \sup_{\lambda_s \in \Omega} \big\|R_{\!f_j}^{(i)}(\lambda_s)\big\|_2\big\}$, where $R_{\!f^{(i)}}^{(i)}(\lambda)$ is formed from the elements of $R_{\!f}^{(i)}(\lambda)$ which correspond to \texttt{true} values in the $i$-th row of \texttt{S} and  $R_{\!f_j}^{(i)}(\lambda)$ is the $(i,j)$-th element of $R_{\!f}(\lambda)$;
\item
If \texttt{R} is a collection of $N$ LTI systems and \texttt{S} is an $N\times m_f$ structure matrix,  then \texttt{FSCOND} is an $N$-dimensional vector of fault sensitivity conditions. \texttt{FSCOND\{$i$\}} is the fault sensitivity condition of the $i$-th system \texttt{R\{$i$\}}, computed as $\beta_i / \gamma_i$, with $\beta_i = \big\| R_{\!f^{(i)}}^{(i)}(\lambda) \big\|_{\Omega -}$ and $\gamma_i = \displaystyle\max_{j}\big\{ \sup_{\lambda_s \in \Omega} \big\|R_{\!f_j}^{(i)}(\lambda_s)\big\|_2 \big\}$, where $R_{\!f^{(i)}}^{(i)}(\lambda)$ is formed from the columns of $R_{\!f}^{(i)}(\lambda)$ which correspond to \texttt{true} values in the $i$-th row of \texttt{S} and  $R_{\!f_j}^{(i)}(\lambda)$ is the $j$-th column of $R_{\!f}^{(i)}(\lambda)$. If \texttt{R\{$i$\}} is empty, then \texttt{FSCOND\{$i$\}} is set to  \texttt{NaN}.
\end{itemize}

In the case of calling \texttt{fdifscond} as
\begin{verbatim}
   [BETA,GAMMA] = fdifscond(...)
\end{verbatim}
then:

If \texttt{BETA} and \texttt{GAMMA} are scalar values, then they contain the values of $\beta$ and $\gamma$, respectively, whose ratio represents the fault sensitivity condition.
If \texttt{BETA} and \texttt{GAMMA} are vectors, then \texttt{BETA\{$i$\}} and \texttt{GAMMA\{$i$\}} contain the values of $\beta_i$ and $\gamma_i$, respectively, whose ratio represents the $i$-th fault sensitivity condition. If \texttt{R\{$i$\}} is empty, then \texttt{BETA\{$i$\}} and \texttt{GAMMA\{$i$\}} are set to zero.
\end{description}

\subsubsection*{Method}
The definitions of the fault sensitivity condition in terms of the $\mathcal{H}_{\infty -}$-index and its finite frequency counterpart, the $\mathcal{H}_{\Omega -}$-index, are given in Section \ref{sec:fscond}.

\subsubsection{\texttt{\bfseries fdif2ngap}}
\index{M-functions!\texttt{\bfseries fdif2ngap}}
\subsubsection*{Syntax}
\begin{verbatim}
GAP = fdif2ngap(R)
GAP = fdif2ngap(R,FREQ)
GAP = fdif2ngap(R,[],S)
GAP = fdif2ngap(R,FREQ,S)
[BETA,GAMMA] = fdif2ngap(...)
\end{verbatim}

\subsubsection*{Description}
\index{performance evaluation!fault detection and isolation!fault-to-noise gap}

\noindent \texttt{fdif2ngap} evaluates the fault-to-noise gap of the internal form of a FDI filter or the fault-to-noise gaps of the internal forms of a collection of FDI filters.

\subsubsection*{Input data}
\begin{description}
\item
\texttt{R} is a LTI system or a cell array of LTI systems.

If \texttt{R} is a LTI system representing the internal form of a FDI filter, then it is in a descriptor system state-space form
\be\label{fdif2ngap:sysss}
\begin{aligned}
E_R\lambda x_R(t)  & =   A_Rx_R(t)+ B_{R_f}f(t)+ B_{R_w}w(t) + B_{R_{v}}v(t) ,\\
r(t) & =  C_R x_R(t) + D_{R_f}f(t)+ D_{R_w}w(t) + D_{R_{v}}v(t) ,
\end{aligned}
\ee
where $r(t) \in \mathds{R}^q$ is the residual output, $f(t) \in \mathds{R}^{m_f}$ is the fault input, $w(t)\in \mathds{R}^{m_w}$ is the noise input  and $v(t)$ is the auxiliary input. For the fault input $f(t)$ and noise input $w(t)$ the input groups \texttt{'faults'} and \texttt{'noise'}, respectively,  have to be defined. The input-output form of  \texttt{R} corresponding to (\ref{fdif2ngap:sysss}) is
\be\label{fdif2ngap:sysio} {\mathbf{y}}(\lambda) =
G_f(\lambda){\mathbf{f}}(\lambda)+G_w(\lambda){\mathbf{w}}(\lambda)+G_v(\lambda){\mathbf{v}}(\lambda),
\ee
where $R_f(\lambda)$, $R_w(\lambda)$ and $R_v(\lambda)$ are the transfer function matrices from the corresponding inputs.

If \texttt{R} is an $N\times 1$ cell array of LTI systems representing the internal forms of $N$ FDI filters, then the $i$-th component system \texttt{R\{$i$\}} is in the state-space form
\be\label{fdif2ngap:sysiss}
{\begin{aligned}
E_R^{(i)}\lambda x_R^{(i)}(t)  & =   A_R^{(i)}x_R^{(i)}(t)+ B_{R_f}^{(i)}f(t)+ B_{R_w}^{(i)}w(t), \\
r^{(i)}(t) & =  C_R^{(i)} x_R^{(i)}(t) + D_{R_f}^{(i)}f(t)+ D_{R_w}^{(i)}w(t),
\end{aligned}}
\ee
where $r^{(i)}(t) \in \mathds{R}^{q^{(i)}}$ is the $i$-th residual component, $f(t) \in \mathds{R}^{m_f}$ is the fault input and $w(t)$ is the noise input. For the fault input $f(t)$ and noise input $w(t)$ the input groups \texttt{'faults'} and \texttt{'noise'}, respectively,  have to be defined.
The input-output form of  \texttt{R\{$i$\}} corresponding to (\ref{fdif2ngap:sysiss}) is
\be\label{fdif2ngap:sysiio} {\mathbf{r}}^{(i)}(\lambda) =
R_f^{(i)}(\lambda){\mathbf{f}}(\lambda)+R_w^{(i)}(\lambda){\mathbf{w}}(\lambda),
\ee
where $R_f^{(i)}(\lambda)$ and $R_w^{(i)}(\lambda)$ are the transfer function matrices from the corresponding
inputs.

%
%

\item
\texttt{FREQ} is a real vector, which contains the frequency values $\omega_k$, $k = 1, \ldots, n_f$, to be used to evaluate the gap between  the transfer function matrices $R_f(\lambda)$ and $R_w(\lambda)$ in  (\ref{fdif2ngap:sysio}), or between the transfer function matrices $R_f^{(i)}(\lambda)$ and $R_w^{(i)}(\lambda)$ in  (\ref{fdif2ngap:sysiio}). By default, \texttt{FREQ} is empty if it is not specified. For each real frequency  $\omega_k$ contained in \texttt{FREQ}, there corresponds a complex frequency $\lambda_k$, which is used to define the complex set $\Omega = \{ \lambda_1, \ldots, \lambda_{n_f}\}$. Depending on the system type, $\lambda_k = \mathrm{i}\omega_k$, in the continuous-time case, and $\lambda_k = \exp (\mathrm{i}\omega_k T)$, in the discrete-time case, where $T$ is the sampling time of the system.
\item
\texttt{S} is a $q\times m_f$ logical structure matrix if \texttt{R} is the internal form of fault detection filter with $q$ residual outputs or is an $N\times m_f$ logical structure matrix if \texttt{R} is a collection of $N$ internal forms of fault detection and isolation filters, where \texttt{R$\{i\}$} is the internal form of the $i$-th filter.
\end{description}

\subsubsection*{Output data}

\begin{description}
\item
\texttt{GAP} is the computed fault-to-noise gap, which is either a scalar or a vector, depending on the input variables.

In the case of calling \texttt{fdif2ngap} as
\begin{verbatim}
   GAP = fdif2ngap(R)
\end{verbatim}
then:
 \begin{itemize}
\item
If \texttt{R} is a LTI system, then \texttt{GAP} is the fault-to-noise gap computed as $\beta / \gamma$, where $\beta = \big\| R_{\!f}(\lambda) \big\|_{\infty -}$ and $\gamma = \big\|R_w(\lambda)\big\|_\infty$ ;
 \item
 If \texttt{R} is a collection of $N$ LTI systems, then \texttt{GAP} is an $N$-dimensional vector, with \texttt{GAP\{$i$\}}, the $i$-th fault-to-noise gap, computed as $\beta_i / \gamma_i$, where $\beta_i = \big\| R_{\!f}^{(i)}(\lambda) \big\|_{\infty -}$ and $\gamma_i = \big\|R_{w}^{(i)}(\lambda)\big\|_\infty$. If \texttt{R\{$i$\}} is empty, then \texttt{GAP\{$i$\}} is set to  \texttt{NaN}.
\end{itemize}

In the case of calling \texttt{fdif2ngap} as
\begin{verbatim}
   GAP = fdif2ngap(R,FREQ)
\end{verbatim}
where \texttt{FREQ} is a nonempty vector of real frequencies, which defines the set of complex frequencies $\Omega$ (see description of \texttt{FREQ}), then:
 \begin{itemize}
\item
If \texttt{R} is a LTI system, then \texttt{GAP} is the fault-to-noise gap computed as $\beta / \gamma$, where $\beta = \big\| R_{\!f}(\lambda) \big\|_{\Omega -}$ and $\gamma = \big\|R_w(\lambda)\big\|_\infty$ ;
 \item
 If \texttt{R} is a collection of $N$ LTI systems, then \texttt{GAP} is an $N$-dimensional vector, with \texttt{GAP\{$i$\}}, the $i$-th fault-to-noise gap, computed as $\beta_i / \gamma_i$, where $\beta_i = \big\| R_{\!f}^{(i)}(\lambda) \big\|_{\Omega -}$ and $\gamma_i = \big\|R_{w}^{(i)}(\lambda)\big\|_\infty$. If \texttt{R\{$i$\}} is empty, then \texttt{GAP\{$i$\}} is set to  \texttt{NaN}.
\end{itemize}

In the case of calling \texttt{fdif2ngap} as
\begin{verbatim}
   GAP = fdif2ngap(R,[],S)
\end{verbatim}
then:
 \begin{itemize}
\item
If \texttt{R} is a LTI system with $q$ residual outputs and \texttt{S} is a $q\times m_f$ structure matrix,  then \texttt{GAP} is a $q$-dimensional vector, whose $i$-th component \texttt{GAP\{$i$\}} is the fault-to-noise gap between the $i$-th rows $R_{\!f}^{(i)}(\lambda)$ and $R_{w}^{(i)}(\lambda)$ of $R_{\!f}(\lambda)$ and  $R_{w}(\lambda)$, respectively, computed as $\beta_i / \gamma_i$, with $\beta_i = \big\| R_{\!f^{(i)}}^{(i)}(\lambda) \big\|_{\infty -}$ and $\gamma_i =  \big\|\big[\, R_{\!\bar f^{(i)}}^{(i)}(\lambda) \; R_{w}^{(i)}(\lambda)\,\big]\big\|_\infty$, where $R_{\!f^{(i)}}^{(i)}(\lambda)$ and $R_{\!\bar f^{(i)}}^{(i)}(\lambda)$ are formed from the elements of $R_{\!f}^{(i)}(\lambda)$ which correspond to the \texttt{true} and, respectively,  \texttt{false} values, in the $i$-th row of \texttt{S};
 \item
 If \texttt{R} is a collection of $N$ LTI systems and \texttt{S} is an $N\times m_f$ structure matrix,  then \texttt{GAP} is an $N$-dimensional vector, whose $i$-th component  \texttt{GAP\{$i$\}} is the $i$-th fault-to-noise gap, computed as $\beta_i / \gamma_i$, with $\beta_i = \big\| R_{\!f^{(i)}}^{(i)}(\lambda) \big\|_{\infty -}$ and $\gamma_i = \big\|\big[\, R_{\!\bar f^{(i)}}^{(i)}(\lambda) \; R_{w}^{(i)}(\lambda)\,\big]\big\|_\infty$, where $R_{\!f^{(i)}}^{(i)}(\lambda)$ and $R_{\!\bar f^{(i)}}^{(i)}(\lambda)$ are formed from the columns of $R_{\!f}^{(i)}(\lambda)$ which correspond to the \texttt{true} and, respectively,  \texttt{false} values, in the $i$-th row of \texttt{S}. If \texttt{R\{$i$\}} is empty, then \texttt{GAP\{$i$\}} is set to  \texttt{NaN}.
\end{itemize}

In the case of calling \texttt{fdif2ngap} as
\begin{verbatim}
   GAP = fdif2ngap(R,FREQ,S)
\end{verbatim}
where both \texttt{FREQ} and \texttt{S} are nonempty then:
 \begin{itemize}
\item
If \texttt{R} is a LTI system with $q$ residual outputs and \texttt{S} is a $q\times m_f$ structure matrix,  then \texttt{GAP} is a $q$-dimensional vector, whose $i$-th component \texttt{GAP\{$i$\}} is the fault-to-noise gap between the $i$-th rows $R_{\!f}^{(i)}(\lambda)$ and $R_{w}^{(i)}(\lambda)$ of $R_{\!f}(\lambda)$ and  $R_{w}(\lambda)$, respectively, computed as $\beta_i / \gamma_i$, with $\beta_i = \big\| R_{\!f^{(i)}}^{(i)}(\lambda) \big\|_{\Omega -}$ and $\gamma_i =  \big\|\big[\, R_{\!\bar f^{(i)}}^{(i)}(\lambda) \; R_{w}^{(i)}(\lambda)\,\big]\big\|_\infty$, where $R_{\!f^{(i)}}^{(i)}(\lambda)$ and $R_{\!\bar f^{(i)}}^{(i)}(\lambda)$ are formed from the elements of $R_{\!f}^{(i)}(\lambda)$ which correspond to the \texttt{true} and, respectively,  \texttt{false} values, in the $i$-th row of \texttt{S};
 \item
 If \texttt{R} is a collection of $N$ LTI systems and \texttt{S} is an $N\times m_f$ structure matrix,  then \texttt{GAP} is an $N$-dimensional vector, whose $i$-th component  \texttt{GAP\{$i$\}} is the $i$-th fault-to-noise gap, computed as $\beta_i / \gamma_i$, with $\beta_i = \big\| R_{\!f^{(i)}}^{(i)}(\lambda) \big\|_{\Omega -}$ and $\gamma_i = \big\|\big[\,  R_{\!\bar f^{(i)}}^{(i)}(\lambda) \; R_{w}^{(i)}(\lambda)\,]\big\|_\infty$, where $R_{\!f^{(i)}}^{(i)}(\lambda)$ and $R_{\!\bar f^{(i)}}^{(i)}(\lambda)$ are formed from the columns of $R_{\!f}^{(i)}(\lambda)$ which correspond to the \texttt{true} and, respectively,  \texttt{false} values, in the $i$-th row of \texttt{S}. If \texttt{R\{$i$\}} is empty, then \texttt{GAP\{$i$\}} is set to  \texttt{NaN}.
\end{itemize}

In the case of calling \texttt{fdif2ngap} as
\begin{verbatim}
   [BETA,GAMMA] = fdif2ngap(...)
\end{verbatim}
then:

If \texttt{BETA} and \texttt{GAMMA} are scalar values, then they contain the values of $\beta$ and $\gamma$, respectively,  whose ratio represents the fault-to-noise gap.
If \texttt{BETA} and \texttt{GAMMA} are vectors, then \texttt{BETA\{$i$\}} and \texttt{GAMMA\{$i$\}} contain the values of $\beta_i$ and $\gamma_i$, respectively, whose ratio represents the $i$-th fault-to-noise gap. If \texttt{R\{$i$\}} is empty, then \texttt{BETA\{$i$\}} and \texttt{GAMMA\{$i$\}} are set to  zero.
\end{description}

\subsubsection*{Method}
The definitions of the fault-to-noise gap in terms of the $\mathcal{H}_{\infty -}$-index and its finite frequency counterpart, the $\mathcal{H}_{\Omega -}$-index, are given in Section \ref{sec:f2ngap}.

\subsubsection{\texttt{\bfseries fdimmperf}}
\index{M-functions!\texttt{\bfseries fdimmperf}}
\subsubsection*{Syntax}
\begin{verbatim}
GAMMA = fdimmperf(R)
GAMMA = fdimmperf(R,SYSR)
GAMMA = fdimmperf(R,SYSR,NRMFLAG)
GAMMA = fdimmperf(R,[],NRMFLAG)
GAMMA = fdimmperf(R,[],NRMFLAG,S)
\end{verbatim}

\subsubsection*{Description}
\index{performance evaluation!fault detection and isolation!model-matching performance}

\noindent \texttt{fdimmperf} evaluates the model-matching performance of the internal form of a FDI filter or the model-matching performance of the internal forms of a collection of FDI filters.

\subsubsection*{Input data}
\begin{description}
\item
\texttt{R} is a stable LTI system or a cell array of stable LTI systems.

If \texttt{R} is a LTI system representing the internal form of a FDI filter, then it is in a descriptor system state-space form
\be\label{fdimmperf:sysss}
{\begin{aligned}
E_R\lambda x_R(t)  & =   A_Rx_R(t)+ B_{R_u} u(t)+ B_{R_d} d(t) + B_{R_f} f(t)+ B_{R_w} w(t)   , \\
r(t) & =  C_R x_R(t) + D_{R_u} u(t)+ D_{R_d} d(t) + D_{R_f} f(t) + D_{R_w} w(t) ,
\end{aligned}}
\ee
where $r(t)$ is a $q$-dimensional residual output vector and any of the inputs components $u(t)$, $d(t)$,  $f(t)$ or $w(t)$  can be void.  For the system \texttt{R}, the input groups for $u(t)$, $d(t)$, $f(t)$, and $w(t)$ must have the standard names \texttt{\bfseries 'controls'}, \texttt{\bfseries 'disturbances'}, \texttt{\bfseries 'faults'}, and \texttt{\bfseries 'noise'}, respectively. The input-output form of  \texttt{R} corresponding to (\ref{fdimmperf:sysss}) is
\be\label{fdimmperf:sysio} {\mathbf{r}}(\lambda) = R_u(\lambda){\mathbf{u}}(\lambda)+R_d(\lambda){\mathbf{d}}(\lambda)+
R_f(\lambda){\mathbf{f}}(\lambda)+R_w(\lambda){\mathbf{w}}(\lambda),
\ee
where $R_u(\lambda)$, $R_d(\lambda)$, $R_f(\lambda)$, and $R_w(\lambda)$ are the transfer function matrices from the control, disturbance, fault, and noise inputs, respectively. In the case of void inputs, the corresponding transfer function matrices are assumed to be zero.

If \texttt{R} is an $N\times 1$ cell array of LTI systems representing the internal forms of $N$ FDI filters, then the $i$-th component system \texttt{R\{$i$\}} is in the state-space form
\be\label{fdimmperf:sysiss}
\begin{array}{rcl}E_R^{(i)}\lambda x_R^{(i)}(t) &=& A_R^{(i)}x_R^{(i)}(t) + B_{R_u}^{(i)}u(t) + B_{R_d}^{(i)} d(t) + B_{R_f}^{(i)} f(t) + B_{R_w}^{(i)} w(t)   \, ,\\
r^{(i)}(t) &=& C_R^{(i)}x_R^{(i)}(t) + D_{R_u}^{(i)} u(t) + D_{R_d}^{(i)} d(t) + D_{R_f}^{(i)} f(t) +  D_{R_w}^{(i)} w(t)   \, , \end{array} \ee
where $r^{(i)}(t)$ is a $q_i$-dimensional output vector representing the $i$-th residual component and any of the inputs components $u(t)$, $d(t)$, $f(t)$ or $w(t)$  can be void.  For each component system \texttt{R\{$i$\}}, the input groups for $u(t)$, $d(t)$, $f(t)$, and $w(t)$ must have the standard names \texttt{\bfseries 'controls'}, \texttt{\bfseries 'disturbances'}, \texttt{\bfseries 'faults'}, and \texttt{\bfseries 'noise'}, respectively. The input-output form of  \texttt{R\{$i$\}} corresponding to (\ref{fdimmperf:sysiss}) is
\be\label{fdimmperf:sysiio} {\mathbf{r}}^{(i)}(\lambda) = R_u^{(i)}(\lambda){\mathbf{u}}(\lambda)+R_d^{(i)}(\lambda){\mathbf{d}}(\lambda)+
R_f^{(i)}(\lambda){\mathbf{f}}(\lambda)+R_w^{(i)}(\lambda){\mathbf{w}}(\lambda),
\ee
where $R_u^{(i)}(\lambda)$, $R_d^{(i)}(\lambda)$, $R_f^{(i)}(\lambda)$, and $R_w^{(i)}(\lambda)$ are the transfer function matrices from the control, disturbance, fault, and noise inputs, respectively, for the $i$-th component system.

\item
\texttt{SYSR}, if nonempty, is a stable LTI system or a cell array of stable LTI systems.

If \texttt{SYSR} is a LTI system representing a reference model for the internal form of the FDI filter \texttt{R}, then it is in the state-space form
\be\label{fdimmperf:sysrss}
{\begin{aligned}
\lambda x_r(t)  & =   A_rx_r(t)+ B_{ru} u(t)+ B_{rd} d(t) + B_{rf} f(t)+ B_{rw} w(t)   , \\
y_r(t) & =  C_r x_r(t) + D_{ru} u(t)+ D_{rd} d(t) + D_{rf} f(t) + D_{rw} w(t) ,
\end{aligned}}
\ee
where the $y_r(t)$ is a $q$-dimensional vector and
any of the inputs components $u(t)$, $d(t)$, $f(t)$, or $w(t)$  can be void.  For the system \texttt{SYSR}, the input groups for $u(t)$, $d(t)$, $f(t)$, and $w(t)$ must have the standard names \texttt{\bfseries 'controls'}, \texttt{\bfseries 'disturbances'}, \texttt{\bfseries 'faults'}, and \texttt{\bfseries 'noise'}, respectively. The input-output form corresponding to (\ref{fdimmperf:sysrss}) is
\be\label{fdimmperf:sysrio} {\mathbf{y}}_r(\lambda) = M_{ru}(\lambda){\mathbf{u}}(\lambda)+M_{rd}(\lambda){\mathbf{d}}(\lambda)+
M_{rf}(\lambda){\mathbf{f}}(\lambda)+M_{rw}(\lambda){\mathbf{w}}(\lambda),
\ee
where $M_{ru}(\lambda)$, $M_{rd}(\lambda)$, $M_{rf}(\lambda)$, and $M_{rw}(\lambda)$ are the transfer function matrices from the control, disturbance, fault, and noise inputs, respectively. In the case of void inputs, the corresponding transfer function matrices are assumed to be zero.

If \texttt{SYSR} is an $N\times 1$ cell array of LTI systems  representing a  collection of $N$ reference models for the internal forms of $N$ FDI filters in the $N\times 1$ cell array \texttt{R}, then the $i$-th component system \texttt{SYSR\{$i$\}} is in the state-space form
\be\label{fdimmperf:sysriss}
\begin{array}{rcl}
\lambda x_r^{(i)}(t)  & =   A_r^{(i)}x_r^{(i)}(t)+ B_{ru}^{(i)} u(t)+ B_{rd}^{(i)} d(t) + B_{rf}^{(i)} f(t)+ B_{rw}^{(i)} w(t)   , \\
y_r^{(i)}(t) & =  C_r^{(i)} x_r^{(i)}(t) + D_{ru}^{(i)} u(t)+ D_{rd}^{(i)} d(t) + D_{rf}^{(i)} f(t) + D_{rw}^{(i)} w(t)  \, , \end{array} \ee
where $y_r^{(i)}(t)$ is a $q_i$-dimensional vector and any of the inputs components $u(t)$, $d(t)$, $f(t)$ or $w(t)$  can be void.  For each component system \texttt{SYSR\{$i$\}}, the input groups for $u(t)$, $d(t)$, $f(t)$, and $w(t)$ must have the standard names \texttt{\bfseries 'controls'}, \texttt{\bfseries 'disturbances'}, \texttt{\bfseries 'faults'}, and \texttt{\bfseries 'noise'}, respectively. The input-output form of \texttt{SYSR\{$i$\}} corresponding to (\ref{fdimmperf:sysriss}) is
\be\label{fdimmperf:sysriio} {\mathbf{y}}_r^{(i)}(\lambda) = M_{ru}^{(i)}(\lambda){\mathbf{u}}(\lambda)+M_{rd}^{(i)}(\lambda){\mathbf{d}}(\lambda)+
M_{rf}^{(i)}(\lambda){\mathbf{f}}(\lambda)+M_{rw}^{(i)}(\lambda){\mathbf{w}}(\lambda),
\ee
where $M_{ru}^{(i)}(\lambda)$, $M_{rd}^{(i)}(\lambda)$, $M_{rf}^{(i)}(\lambda)$, and $M_{rw}^{(i)}(\lambda)$ are the transfer function matrices from the control, disturbance, fault, and noise inputs, respectively, for the $i$-th component system.

\item
\texttt{NRMFLAG} specifies the used system norm and, if specified, must be either 2 for using the $\mathcal{H}_2$-norm or \texttt{Inf} for using the $\mathcal{H}_\infty$-norm. By default, \texttt{NRMFLAG} = \texttt{Inf} (if not specified).
\item
\texttt{S} is a $q\times m_f$ logical structure matrix if \texttt{R} is a LTI system with $q$ outputs or is an $N\times m_f$ logical structure matrix if \texttt{R} is a collection of $N$ LTI systems.
\end{description}

\subsubsection*{Output data}

\begin{description}
\item
\texttt{GAMMA} is the computed model-matching performance, depending on the input variables.

In the case of calling \texttt{fdimmperf} as
\begin{verbatim}
   GAMMA = fdimmperf(R,SYSR,NRMFLAG)
\end{verbatim}
with \texttt{NRMFLAG} = $\alpha$ ($\alpha = 2$ or $\alpha = \infty$), then:
\begin{itemize}
\item
If \texttt{R} and \texttt{SYSR} are LTI systems having input-output descriptions of the form (\ref{fdimmperf:sysio}) and (\ref{fdimmperf:sysrio}), respectively, then the model-matching performance $\texttt{GAMMA} =\gamma$ is computed as
\[ \gamma = \big\|\big[\ R_u(\lambda)\!-\!M_{ru}(\lambda) \;\; R_d(\lambda)\!-\!M_{rd}(\lambda) \;\; R_f(\lambda)\!-\!M_{rf}(\lambda) \;\; R_w(\lambda)\!-\!M_{rw}(\lambda) \, \big] \big\|_{\alpha}\, .\]
\item
If \texttt{R} and \texttt{SYSR} are cell arrays of $N$ LTI systems having input-output descriptions of the form (\ref{fdimmperf:sysiio}) and (\ref{fdimmperf:sysriio}), respectively, then $\texttt{GAMMA}$ is an $N$-dimensional vector, whose $i$-th component  $\texttt{GAMMA}(i) = \gamma_i$ is computed as
\[ \gamma_i = \big\|\big[\ R_u^{(i)}(\lambda)\!-\!M_{ru}^{(i)}(\lambda) \;\; R_d^{(i)}(\lambda)\!-\!M_{rd}^{(i)}(\lambda) \;\; R_f^{(i)}(\lambda)\!-\!M_{rf}^{(i)}(\lambda) \;\; R_w^{(i)}(\lambda)\!-\!M_{rw}^{(i)}(\lambda) \, \big] \big\|_{\alpha}\, .\]
\end{itemize}

The call of \texttt{fdimmperf} as
\begin{verbatim}
   GAMMA = fdimmperf(R,SYSR)
\end{verbatim}
is equivalent to
\begin{verbatim}
   GAMMA = fdimmperf(R,SYSR,Inf) .
\end{verbatim}

In the case of calling \texttt{fdimmperf} as
\begin{verbatim}
   GAMMA = fdimmperf(R,[],NRMFLAG)
\end{verbatim}
then, for $\texttt{NRMFLAG} = \alpha$:
 \begin{itemize}
\item
If \texttt{R} is a LTI system having the input-output description of the form (\ref{fdimmperf:sysio}), then the model-matching performance $\texttt{GAMMA} =\gamma$  is computed as $\gamma = \big\|R_w(\lambda)\big\|_\alpha$;
 \item
 If \texttt{R} is a collection of $N$ LTI systems having the input-output descriptions (\ref{fdimmperf:sysiio}), then $\texttt{GAMMA}$ is an $N$-dimensional vector, whose $i$-th component  $\texttt{GAMMA}(i) = \gamma_i$ is computed as
 $\gamma_i = \big\| R_w^{(i)}(\lambda) \big\|_\alpha$.
\end{itemize}

The call of \texttt{fdimmperf} as
\begin{verbatim}
   GAMMA = fdimmperf(R)
\end{verbatim}
is equivalent to
\begin{verbatim}
   GAMMA = fdimmperf(R,[],Inf) .
\end{verbatim}

In the case of calling \texttt{fdimmperf} as
\begin{verbatim}
   GAMMA = fdimmperf(R,[],NRMFLAG,S)
\end{verbatim}
with $\texttt{NRMFLAG} = \alpha$, then:
 \begin{itemize}
\item
If \texttt{R} is a LTI system having the input-output description of the form (\ref{fdimmperf:sysio}) and \texttt{S} is a $q\times m_f$ structure matrix,  then the model-matching performance $\texttt{GAMMA} =\gamma$  is computed as $\gamma = \big\| \big[\, \overline R_f(\lambda) \; R_w(\lambda) \, \big]\big\|_\alpha$, where $\overline R_f(\lambda)$ is a $q\times m_f$ transfer function matrix whose $(i,j)$-th element is 0 if $S_{ij} = \texttt{true}$ and is equal to the $(i,j)$-th element of $R_f(\lambda)$ if $S_{ij} = \texttt{false}$.
 \item
 If \texttt{R} is a collection of $N$ LTI systems having the input-output descriptions (\ref{fdimmperf:sysiio}), then $\texttt{GAMMA}$ is an $N$-dimensional vector, whose $i$-th component  $\texttt{GAMMA}(i) = \gamma_i$ is computed as
 \[ \gamma_i = \big\| \big[\, \overline R_f^{(i)}(\lambda) \; R_w^{(i)}(\lambda)\,\big] \big\|_\alpha ,\]
where $\overline R_f^{(i)}(\lambda)$ is a transfer function matrix whose $j$-th column is 0 if $S_{ij} = \texttt{true}$ and is equal to the $j$-th column of $R_f^{(i)}(\lambda)$ if $S_{ij} = \texttt{false}$.
\end{itemize}

\end{description}

\subsubsection*{Method}
The definitions of the model-matching performance  are given in Section \ref{sec:mmperf}.

\subsection{Functions for Performance Evaluation of Model Detection Filters} \label{fditools:performancemd}
The functions for performance evaluation address the computation of the performance criteria of model detection filters defined in Section \ref{sec:mdanalperf}.  All functions are fully compatible with the results computed by the synthesis functions of model detection filters described in Section \ref{fditools:mdsynthesis}.

\subsubsection{\texttt{\bfseries mdperf}}
\index{M-functions!\texttt{\bfseries mdperf}}
\index{model detection!distance mapping}
\index{performance evaluation!model detection!distance mapping}
\subsubsection*{Syntax}
\begin{verbatim}
[MDGAIN,FPEAK,P,RELGAIN] = mdperf(R,OPTIONS)
\end{verbatim}

\subsubsection*{Description}

\texttt{\bfseries mdperf} evaluates the distance mapping performance of a collection of model detection filters.

\subsubsection*{Input data}
\begin{description}
\item \texttt{R} is an $N\times N$ cell array of filters, where the $(i,j)$-th filter \texttt{R\{$i,j$\}}, is the internal form of the $i$-th model detection filter
  acting on the $j$-th model.
Each nonempty \texttt{R\{$i,j$\}}  has a standard state-space representation
\be\label{mdperf:ssystemiR}
{\begin{aligned}
\lambda x_R^{(i,j)}(t)  & =   A_R^{(i,j)}x_R^{(i,j)}(t)+ B_{R_u}^{(i,j)}u(t)+  B_{R_d}^{(i,j)}d^{(j)}(t)+ B_{R_w}^{(i,j)}w^{(j)}(t), \\
r^{(i,j)}(t) & =  C_R^{(i,j)} x_R^{(i,j)}(t) + D_{R_u}^{(i,j)}u(t)+ D_{R_d}^{(i,j)}d^{(j)}(t)+ D_{R_w}^{(i,j)}w^{(j)}(t) ,
\end{aligned}}
\ee
where $x_R^{(i,j)}(t)$ is the state vector of the $(i,j)$-th filter with the residual output $r^{(i,j)}(t)$,   control input $u(t) \in \mathds{R}^{m_u}$, disturbance input $d^{(j)}(t) \in \mathds{R}^{m_d^{(j)}}$ and noise input $w^{(j)}(t) \in \mathds{R}^{m_w^{(j)}}$, and
where any of the inputs components $u(t)$, $d^{(j)}(t)$, or $w^{(j)}(t)$  can be void. The input groups for $u(t)$, $d^{(j)}(t)$, and $w^{(j)}(t)$  have the standard names \texttt{\bfseries 'controls'}, \texttt{\bfseries 'disturbances'}, and \texttt{\bfseries 'noise'}, respectively. If \texttt{OPTIONS.cdinp = true} (see below), then
the same disturbance input $d$ is assumed for all  filters (i.e., $d^{(j)} = d$).
The state-space form (\ref{mdperf:ssystemiR}) corresponds to the input-output form
\be\label{mdperf:systemiR}
{{\mathbf{r}}}^{(i,j)}(\lambda) =
R_u^{(i,j)}(\lambda){\mathbf{u}}(\lambda) +
R_d^{(i,j)}(\lambda){\mathbf{d}}^{(j)}(\lambda) +
R_w^{(i,j)}(\lambda){\mathbf{w}}^{(j)}(\lambda) \, ,
\ee
where $R_u^{(i,j)}(\lambda)$, $R_d^{(i,j)}(\lambda)$ and $R_w^{(i,j)}(\lambda)$  are the TFMs from
the corresponding inputs to the residual output.
\item
 \texttt{OPTIONS} is a MATLAB structure used to specify various options and has the following fields:
{\tabcolsep=1mm
\setlength\LTleft{30pt}\begin{longtable}{|l|lcp{9cm}|} \hline
\textbf{\texttt{OPTIONS} fields} & \multicolumn{3}{l|}{\textbf{Description}} \\ \hline
 \texttt{MDSelect}   & \multicolumn{3}{p{12cm}|}{$M$-dimensional integer vector with increasing elements $\sigma_i$, $i = 1, \ldots, M$, containing the indices of the selected model detection filters for which the gains of the corresponding internal forms have to be evaluated \newline
                     (Default: $[\,1, \ldots, N\,]$)}\\
                                        \hline
 \texttt{MDFreq}  &  \multicolumn{3}{p{12cm}|}{real vector, which contains the frequency values $\omega_k$, $k = 1, \ldots, n_f$, for which the point-wise gains have to be computed. For each real frequency  $\omega_k$, there corresponds a complex frequency $\lambda_k$ which is used to evaluate the point-wise gain. Depending on the system type, $\lambda_k = \mathrm{i}\omega_k$, in the continuous-time case, and $\lambda_k = \exp (\mathrm{i}\omega_k T)$, in the discrete-time case, where $T$ is the common sampling time of the component systems.  (Default: \texttt{[ ]}) } \\ \hline
\texttt{cdinp}   & \multicolumn{3}{p{12cm}|}{option to use both control and
                     disturbance input channels to evaluate the distance mapping performance, as follows:}\\
                 &  \texttt{true} &--& use both control and disturbance input channels; \\
                 &  \texttt{false}&--& use only the control input channels (default)  \\
                                        \hline
 \texttt{MDIndex}   & \multicolumn{3}{p{12cm}|}{index $\ell$ of the $\ell$-th smallest gains to be used to evaluate the relative
                     gains to the second smallest gains
                     (Default: $\ell = 3$)} \\ \hline
\end{longtable}}
\end{description}

\subsubsection*{Output data}
\begin{description}
\item
\texttt{MDGAIN} is an $M\times N$ nonnegative matrix, whose $(i,j)$-th element
\texttt{MDGAIN$(i,j)$}  contains the computed peak gain for the selected input channels of the $(\sigma_i,j)$-th filter as follows:\\
-- if \texttt{OPTIONS.MDFreq} is empty and \texttt{OPTIONS.cdinp = false} then
\[ \texttt{MDGAIN}(i,j) = \big\|R_u^{(\sigma_i,j)}(\lambda)\big\|_\infty ;\]
-- if \texttt{OPTIONS.MDFreq} is nonempty and \texttt{OPTIONS.cdinp = false} then
\[ \texttt{MDGAIN}(i,j) = \max_{k} \big\|R_u^{(\sigma_i,j)}(\lambda_k)\big\|_2 ; \]
-- if \texttt{OPTIONS.MDFreq} is empty  and \texttt{OPTIONS.cdinp = true} then
\[ \texttt{MDGAIN}(i,j) = \big\| \big[\, R_u^{(\sigma_i,j)}(\lambda)\; R_d^{(\sigma_i,j)}(\lambda)\,\big]\big\|_\infty ;\]
-- if \texttt{OPTIONS.MDFreq} is nonempty and \texttt{OPTIONS.cdinp = true} then
\[ \texttt{MDGAIN}(i,j) = \max_{k} \big\| \big[\, R_u^{(\sigma_i,j)}(\lambda_k)\; R_d^{(\sigma_i,j)}(\lambda_k)\,\big]\big\|_2 .\]
\item
\texttt{FPEAK} is an $M\times N$ nonnegative matrix, whose $(i,j)$-th element
\texttt{FPEAK$(i,j)$}  contains the peak
  frequency (in rad/TimeUnit), where \texttt{MDGAIN$(i,j)$} is achieved. \\
\item
\texttt{P} is an $M\times N$ integer matrix, whose $i$-th row
contains the permutation to be applied to  increasingly reorder the i-th row of \texttt{MDGAIN}. \\
\item
\texttt{RELGAIN} is an $M$-dimensional vector, whose $i$-th element \texttt{RELGAIN$(i)$} contains the ratio of the second and $\ell$-th smallest gains in the $i$-th row of \texttt{MDGAIN}, where $\ell = $ \texttt{OPTIONS.MDIndex}.
\\
\end{description}

\subsubsection*{Method}
The definition of the distance mapping performance of a set of model detection filters is given in Section \ref{sec:mdperf}.

\subsubsection{\texttt{\bfseries mdmatch}}
\index{M-functions!\texttt{\bfseries mdmatch}}
\index{model detection!distance matching}
\index{performance evaluation!model detection!distance matching}
\subsubsection*{Syntax}
\begin{verbatim}
[MDGAIN,FPEAK,MIND] = mdmatch(Q,SYS,OPTIONS)
\end{verbatim}

\subsubsection*{Description}

\texttt{\bfseries mdmatch}  evaluates the distance matching performance of a collection of model detection filters acting on a given model and determines the index of the best matching component model.

\subsubsection*{Input data}
\begin{description}
\item
\texttt{Q} is an $N\times 1$ cell array of stable model detection filters, where \texttt{Q\{$i$\}} contains the $i$-th filter  in a standard state-space representation
\be\label{mdmatch:ssystemiQ}
{\begin{aligned}
\lambda x_Q^{(i)}(t)  & =   A_Q^{(i)}x_Q^{(i)}(t)+ B_{Q_y}^{(i)}y(t)+ B_{Q_u}^{(i)}u(t) ,\\
r^{(i)}(t) & =  C_Q^{(i)} x_Q^{(i)}(t) + D_{Q_y}^{(i)}y(t)+ D_{Q_u}^{(i)}u(t) ,
\end{aligned}}
\ee
where $x_Q^{(i)}(t)$ is the state vector of the $i$-th filter with the residual signal $r^{(i)}(t)$ as output and the  measured outputs  $y(t)$ and control inputs $u(t)$ as inputs. The input groups for $y(t)$ and $u(t)$  have the standard names \texttt{\bfseries 'outputs'} and \texttt{\bfseries 'controls'}, respectively.
The state-space form (\ref{mdmatch:ssystemiQ}) corresponds to the input-output form
\be\label{mdmatch:systemiQ}
{{\mathbf{r}}}^{(i)}(\lambda) =
Q^{(i)}(\lambda) \ba{c}{\mathbf{y}}(\lambda) \\ {\mathbf{u}}(\lambda) \ea \, .
\ee
\texttt{Q\{$i$\}} may be empty.
\item
\texttt{SYS} is a stable LTI model  in the state-space form
\be\label{mdmatch:sysssref}
\begin{array}{rcl}E\lambda x(t) &=& Ax(t) + B_u u(t) + B_d d(t) + B_w w(t)  \, ,\\
y(t) &=& Cx(t) + D_u u(t) + D_d d(t) + D_w w(t)   \, , \end{array} \ee
where $x(t) \in \mathds{R}^{n}$ is the state vector of the system with control input $u(t) \in \mathds{R}^{m_u}$, disturbance input $d(t) \in \mathds{R}^{m_d}$ and noise input $w(t) \in \mathds{R}^{m_w}$, and
where any of the inputs components $u(t)$, $d(t)$, or $w(t)$  can be void. The input groups for $u(t)$, $d(t)$, and $w(t)$  have the standard names \texttt{\bfseries 'controls'}, \texttt{\bfseries 'disturbances'}, and \texttt{\bfseries 'noise'}, respectively.
The state-space form (\ref{mdmatch:sysssref}) corresponds to the input-output form
\be\label{mdmatch:system} {\mathbf{y}}(\lambda) =
G_u(\lambda){\mathbf{u}}(\lambda)
+ G_d(\lambda){\mathbf{d}}(\lambda)
+ G_w(\lambda){\mathbf{w}}(\lambda) , \ee
where $G_u(\lambda)$, $G_d(\lambda)$ and $G_w(\lambda)$  are the TFMs from
the corresponding inputs to the output.
\item
 \texttt{OPTIONS} is a MATLAB structure used to specify various options and has the following fields:
{\tabcolsep=1mm
\setlength\LTleft{30pt}\begin{longtable}{|l|lcp{9cm}|} \hline
\textbf{\texttt{OPTIONS} fields} & \multicolumn{3}{l|}{\textbf{Description}} \\ \hline
 \texttt{MDFreq}  &  \multicolumn{3}{p{12cm}|}{real vector, which contains the frequency values $\omega_k$, $k = 1, \ldots, n_f$, for which the point-wise gains have to be computed. For each real frequency  $\omega_k$, there corresponds a complex frequency $\lambda_k$ which is used to evaluate the point-wise gain. Depending on the system type, $\lambda_k = \mathrm{i}\omega_k$, in the continuous-time case, and $\lambda_k = \exp (\mathrm{i}\omega_k T)$, in the discrete-time case, where $T$ is the common sampling time of the component systems.  (Default: \texttt{[ ]}) } \\ \hline
\texttt{cdinp}   & \multicolumn{3}{p{12cm}|}{option to use both control and
                     disturbance input channels to evaluate the distance matching performance, as follows:}\\
                 &  \texttt{true} &--& use both control and disturbance input channels; \\
                 &  \texttt{false}&--& use only the control input channels (default)  \\
                                        \hline
\end{longtable}}
\end{description}

\subsubsection*{Output data}
\begin{description}
\item
\texttt{MDGAIN} is an $N$-dimensional column vector, whose $i$-th element
\texttt{MDGAIN$(i)$}  contains, for a nonempty filter \texttt{Q\{$i$\}}, the computed peak gain for the selected input channels of the $i$-th internal form as follows:\\
-- if \texttt{OPTIONS.MDFreq} is empty and \texttt{OPTIONS.cdinp = false} then
\[ \texttt{MDGAIN}(i) = \left\|Q^{(i)}(\lambda)\ba{c}G_u(\lambda)\\ I_{m_u} \ea\right\|_\infty ;\]
-- if \texttt{OPTIONS.MDFreq} is nonempty and \texttt{OPTIONS.cdinp = false} then
\[ \texttt{MDGAIN}(i) = \max_{k} \left\|Q^{(i)}(\lambda_k)\ba{c}G_u(\lambda_k)\\ I_{m_u} \ea\right\|_\infty ; \]
-- if \texttt{OPTIONS.MDFreq} is empty and \texttt{OPTIONS.cdinp = true} then
\[ \texttt{MDGAIN}(i) = \left\|Q^{(i)}(\lambda)\ba{cc}G_u(\lambda) & G_d(\lambda) \\ I_{m_u} & 0 \ea\right\|_\infty ;\]
-- if \texttt{OPTIONS.MDFreq} is nonempty and \texttt{OPTIONS.cdinp = true} then
\[ \texttt{MDGAIN}(i) = \max_{k} \left\|Q^{(i)}(\lambda_k)\ba{cc}G_u(\lambda_k) & G_d(\lambda_k) \\ I_{m_u} & 0 \ea\right\|_\infty .\]
\texttt{MDGAIN$(i) = 0$} if \texttt{Q\{$i$\}} is empty.
\item
\texttt{FPEAK} is an $N$-dimensional vector, whose $i$-th element
\texttt{FPEAK$(i)$}  contains the peak
  frequency (in rad/TimeUnit), where \texttt{MDGAIN$(i)$} is achieved. \\
\item
\texttt{MIND} is the index $\ell$ of the component of \texttt{MDGAIN} for which the minimum value
  of the peak gains is achieved. For a properly designed filter \texttt{Q}, this is
  also the index of the best matching model to the current model \texttt{SYS}.
\end{description}

\subsubsection*{Method}
The definitions related to the model matching performance of a set of model detection filters are given in Section \ref{sec:mdmatch}.

\subsubsection{\texttt{\bfseries mdgap}}
\index{M-functions!\texttt{\bfseries mdgap}}
\index{model detection!noise gap}
\index{performance evaluation!model detection!noise gap}
\subsubsection*{Syntax}
\begin{verbatim}
GAP = mdgap(R,OPTIONS)
[BETA,GAMMA] = mdgap(R,OPTIONS)
\end{verbatim}

\subsubsection*{Description}

\texttt{\bfseries mdgap} computes the  noise gaps of model detection filters.

\subsubsection*{Input data}
\begin{description}
\item \texttt{R} is an $N\times N$ cell array of filters, where the $(i,j)$-th filter \texttt{R\{$i,j$\}}, is the internal form of the $i$-th model detection filter
  acting on the $j$-th model.
Each nonempty \texttt{R\{$i,j$\}}  has a standard state-space representation
\be\label{mdgap:ssystemiR}
{\begin{aligned}
\lambda x_R^{(i,j)}(t)  & =   A_R^{(i,j)}x_R^{(i,j)}(t)+ B_{R_u}^{(i,j)}u(t)+  B_{R_d}^{(i,j)}d^{(j)}(t)+ B_{R_w}^{(i,j)}w^{(j)}(t), \\
r^{(i,j)}(t) & =  C_R^{(i,j)} x_R^{(i,j)}(t) + D_{R_u}^{(i,j)}u(t)+ D_{R_d}^{(i,j)}d^{(j)}(t)+ D_{R_w}^{(i,j)}w^{(j)}(t)
\end{aligned}}
\ee
where $x_R^{(i,j)}(t)$ is the state vector of the $(i,j)$-th filter with the residual output $r^{(i,j)}(t)$, control input $u(t) \in \mathds{R}^{m_u}$, disturbance input $d^{(j)}(t) \in \mathds{R}^{m_d^{(j)}}$ and noise input $w^{(j)}(t) \in \mathds{R}^{m_w^{(j)}}$, and
where any of the inputs components $u(t)$, $d^{(j)}(t)$, or $w^{(j)}(t)$  can be void. The input groups for $u(t)$, $d^{(j)}(t)$, and $w^{(j)}(t)$  have the standard names \texttt{\bfseries 'controls'}, \texttt{\bfseries 'disturbances'}, and \texttt{\bfseries 'noise'}, respectively. If \texttt{OPTIONS.cdinp = true} (see below), then
the same disturbance input $d$ is assumed for all  filters (i.e., $d^{(j)} = d$).
The state-space form (\ref{mdgap:ssystemiR}) corresponds to the input-output form
\be\label{mdgap:systemiR}
{{\mathbf{r}}}^{(i,j)}(\lambda) =
R_u^{(i,j)}(\lambda){\mathbf{u}}(\lambda) +
R_d^{(i,j)}(\lambda){\mathbf{d}}^{(j)}(\lambda) +
R_w^{(i,j)}(\lambda){\mathbf{w}}^{(j)}(\lambda) \, ,
\ee
where $R_u^{(i,j)}(\lambda)$, $R_d^{(i,j)}(\lambda)$ and $R_w^{(i,j)}(\lambda)$  are the TFMs from
the corresponding inputs to the residual output.
\item
 \texttt{OPTIONS} is a MATLAB structure used to specify various options and has the following fields:
{\tabcolsep=1mm
\setlength\LTleft{30pt}\begin{longtable}{|l|lcp{9cm}|} \hline
\textbf{\texttt{OPTIONS} fields} & \multicolumn{3}{l|}{\textbf{Description}} \\ \hline
 \texttt{MDSelect}   & \multicolumn{3}{p{12cm}|}{$M$-dimensional integer vector with increasing elements $\sigma_i$, $i = 1, \ldots, M$,  containing the indices of the selected model detection filters for which the gaps of the corresponding internal forms have to be evaluated\newline
                     (Default: $[\,1, \ldots, N\,]$)}\\
                                        \hline
 \texttt{MDFreq}  &  \multicolumn{3}{p{12cm}|}{real vector, which contains the frequency values $\omega_k$, $k = 1, \ldots, n_f$, for which the point-wise gains have to be computed. For each real frequency  $\omega_k$, there corresponds a complex frequency $\lambda_k$ which is used to evaluate the point-wise gain. Depending on the system type, $\lambda_k = \mathrm{i}\omega_k$, in the continuous-time case, and $\lambda_k = \exp (\mathrm{i}\omega_k T)$, in the discrete-time case, where $T$ is the common sampling time of the component systems.  (Default: \texttt{[ ]}) } \\ \hline
\texttt{cdinp}   & \multicolumn{3}{p{12cm}|}{option to use both control and
                     disturbance input channels to evaluate the noise gaps, as follows:}\\
                 &  \texttt{true} &--& use both control and disturbance input channels; \\
                 &  \texttt{false}&--& use only the control input channels (default)  \\
                                        \hline
\end{longtable}}
\end{description}

\pagebreak[3]
\subsubsection*{Output data}
In the case of calling \texttt{mdgap} as
\begin{verbatim}
   GAP = mdgap(R,OPTIONS)
\end{verbatim}
then
\texttt{GAP} is an $M$-dimensional vector, whose $i$-th element
\texttt{GAP$(i)$}  contains the computed noise gap for the selected input channels of the $(\sigma_i,j)$-th filter as follows:\\
-- if \texttt{OPTIONS.MDFreq} is empty and \texttt{OPTIONS.cdinp = false} then
\[ \texttt{GAP$(i)$}= \min_{j\neq \sigma_i} \big\|R_u^{(\sigma_i,j)}(\lambda)\big\|_\infty /\big\|R_w^{(\sigma_i,\sigma_i)}(\lambda)\big\|_\infty ; \]
-- if \texttt{OPTIONS.MDFreq} is nonempty and \texttt{OPTIONS.cdinp = false} then
\[ \texttt{GAP$(i)$} = \min_{j\neq \sigma_i}\max_{k} \big\|R_u^{(\sigma_i,j)}(\lambda_k)\big\|_\infty /\big\|R_w^{(\sigma_i,\sigma_i)}(\lambda)\big\|_\infty ; \]
-- if \texttt{OPTIONS.MDFreq} is empty and \texttt{OPTIONS.cdinp = true} then
\[ \texttt{GAP$(i)$} = \min_{j\neq \sigma_i}\big\| \big[\, R_u^{(\sigma_i,j)}(\lambda)\; R_d^{(\sigma_i,j)}(\lambda)\,\big]\big\|_\infty /\big\|R_w^{(\sigma_i,\sigma_i)}(\lambda)\big\|_\infty ;\]
-- if \texttt{OPTIONS.MDFreq} is nonempty and \texttt{OPTIONS.cdinp = true} then
\[ \texttt{GAP$(i)$} = \min_{j\neq \sigma_i}\max_{k} \big\| \big[\, R_u^{(\sigma_i,j)}(\lambda_k)\; R_d^{(\sigma_i,j)}(\lambda_k)\,\big]\big\|_\infty
/\big\|R_w^{(\sigma_i,\sigma_i)}(\lambda)\big\|_\infty . \]

In the case of calling \texttt{mdgap} as
\begin{verbatim}
   [BETA,GAMMA] = mdgap(R,OPTIONS)
\end{verbatim}
then
\texttt{BETA} and \texttt{GAMM}A are $M$-dimensional vector, whose $i$-th elements
\texttt{BETA$(i)$}  and \texttt{GAMMA$(i)$} contain the values whose ratio represents the noise gaps for the selected input channels of the $(\sigma_i,j)$-th filter as follows:\\
-- if \texttt{OPTIONS.MDFreq} is empty and \texttt{OPTIONS.cdinp = false} then
\[ \texttt{BETA$(i)$}= \min_{j\neq \sigma_i} \big\|R_u^{(\sigma_i,j)}(\lambda)\big\|_\infty, \quad \texttt{GAMMA$(i)$} =  \big\|R_w^{(\sigma_i,\sigma_i)}(\lambda)\big\|_\infty ; \]
-- if \texttt{OPTIONS.MDFreq} is nonempty and \texttt{OPTIONS.cdinp = false} then
\[ \texttt{BETA$(i)$}  = \min_{j\neq \sigma_i}\max_{k} \big\|R_u^{(\sigma_i,j)}(\lambda_k)\big\|_2, \quad \texttt{GAMMA$(i)$} = \big\|R_w^{(\sigma_i,\sigma_i)}(\lambda)\big\|_\infty ; \]
-- if \texttt{OPTIONS.MDFreq} is empty and \texttt{OPTIONS.cdinp = true} then
\[ \texttt{BETA$(i)$}  = \min_{j\neq \sigma_i}\big\| \big[\, R_u^{(\sigma_i,j)}(\lambda)\; R_d^{(\sigma_i,j)}(\lambda)\,\big]\big\|_\infty, \quad \texttt{GAMMA$(i)$} = \big\|R_w^{(\sigma_i,\sigma_i)}(\lambda)\big\|_\infty ;\]
-- if \texttt{OPTIONS.MDFreq} is nonempty and \texttt{OPTIONS.cdinp = true} then
\[ \texttt{BETA$(i)$}=  \min_{j\neq \sigma_i}\max_{k} \big\| \big[\, R_u^{(\sigma_i,j)}(\lambda_k)\; R_d^{(\sigma_i,j)}(\lambda_k)\,\big]\big\|_2, \quad \texttt{GAMMA$(i)$} = \big\|R_w^{(\sigma_i,\sigma_i)}(\lambda)\big\|_\infty . \]

\subsubsection*{Method}
The definition of the noise gap of a set of model detection filters is given in Section \ref{sec:mdgap}.

\newpage
\subsection{Functions for the Synthesis of FDI Filters }\label{fditools:synthesis}

\subsubsection{\texttt{\bfseries efdsyn}}
\index{M-functions!\texttt{\bfseries efdsyn}}
\subsubsection*{Syntax}
\begin{verbatim}
[Q,R,INFO] = efdsyn(SYSF,OPTIONS)
\end{verbatim}

\subsubsection*{Description}
\index{fault detection and e@fault detection problem!a@exact (EFDP)}

\texttt{\bfseries efdsyn} solves   the \emph{exact fault detection problem} (EFDP) (see Section \ref{sec:EFDP}), for a given LTI system \texttt{SYSF} with additive faults. Two stable and proper filters, \texttt{Q} and \texttt{R}, are computed, where \texttt{Q} contains the fault detection filter representing the solution of the EFDP,  and \texttt{R} contains its internal form.

\subsubsection*{Input data}
\begin{description}
\item
\texttt{SYSF} is a  LTI system  in the state-space form
\be\label{efdsyn:sysss}
{\begin{aligned}
E\lambda x(t)  & =  Ax(t)+ B_u u(t)+ B_d d(t) + B_f f(t)+ B_w w(t) + B_{v} v(t)  , \\
y(t) & =  C x(t) + D_u u(t)+ D_d d(t) + D_f f(t) + D_w w(t)+ B_{v} v(t) ,
\end{aligned}}
\ee
where any of the inputs components $u(t)$, $d(t)$, $f(t)$, $w(t)$  or $v(t)$ can be void. The auxiliary input signal $v(t)$ can be used for convenience. For the system \texttt{SYSF}, the input groups for $u(t)$, $d(t)$, $f(t)$, and $w(t)$ have the standard names \texttt{\bfseries 'controls'}, \texttt{\bfseries 'disturbances'}, \texttt{\bfseries 'faults'}, and \texttt{\bfseries 'noise'}, respectively. For the auxiliary input $v(t)$ the standard input group \texttt{'aux'} can be used.
\item
 \texttt{OPTIONS} is a MATLAB structure used to specify various synthesis  options and has the following fields:
{\setlength\LTleft{30pt}\begin{longtable}{|l|p{11.6cm}|} \hline \textbf{\texttt{OPTIONS} fields} & \textbf{Description} \\ \hline
       \texttt{tol}       & relative tolerance for rank computations \newline
                 (Default: internally computed)\\ \hline
       \texttt{tolmin}       & absolute tolerance for observability tests \newline
                 (Default: internally computed)\\ \hline
       \texttt{FDTol}     & threshold for fault detectability checks
                 (Default: $10^{-4}$)\\ \hline
       \texttt{FDGainTol} & threshold for strong fault detectability checks (Default: $10^{-2}$)\\ \hline
       \texttt{rdim}       & desired number $q$ of residual outputs for \texttt{Q} and \texttt{R}\\
                 & (Default: \hspace*{-2.5mm}\begin{tabular}[t]{p{10cm}} \texttt{[ ]}, in which case:  if \texttt{OPTIONS.HDesign} is empty, then \newline $q = 1$, if \texttt{OPTIONS.minimal} = \texttt{true}, or \newline $q = p-r_d$, if \texttt{OPTIONS.minimal} = \texttt{false} (see \textbf{Method}); \\if \texttt{OPTIONS.HDesign} is nonempty, then $q$ is the row dimension of the design matrix $H$ contained in \texttt{OPTIONS.HDesign})\end{tabular}\\ \hline
       \texttt{FDFreq}  &  vector of real frequency values for strong detectability checks \newline (Default: \texttt{[ ]})  \\ \hline
       \texttt{smarg}   & stability margin for the poles of the filters \texttt{Q} and \texttt{R}\\
                 &  (Default: \texttt{-sqrt(eps)} for a continuous-time system \texttt{SYSF}; \\
                 &  \hspace*{4.5em}\texttt{1-sqrt(eps)} for a discrete-time system \texttt{SYSF}).\\ \hline
      \texttt{sdeg}   & prescribed stability degree for the poles of the filters \texttt{Q} and \texttt{R}\\
                 &  (Default: $-0.05$ for a continuous-time system \texttt{SYSF}; \\
                 &  \hspace*{4.8em} $0.95$ for a discrete-time system \texttt{SYSF}).\\ \hline
       \texttt{poles}   & complex vector containing a complex conjugate set of desired poles (within the stability domain) to be assigned for the filters \texttt{Q} and \texttt{R} \newline (Default: \texttt{[ ]})
                   \\\hline
   \texttt{nullspace}   & option to use a specific proper nullspace basis\\
                 & \hspace*{-.9mm}{\tabcolsep=0.7mm\begin{tabular}[t]{lcp{10cm}} \texttt{true } &--& use a minimal proper basis (default); \\
                 \texttt{false} &--& use a full-order observer based basis (see \textbf{Method}). \newline \emph{Note:} This option can  only be used if no disturbance inputs are present in (\ref{efdsyn:sysss}) and $E$ is invertible.
                 \end{tabular}} \\  \hline
     \texttt{simple} & option to employ a simple proper basis for filter synthesis\\
                 &  \texttt{true }  -- use a simple basis; \\
                 &  \texttt{false}  -- use a non-simple basis (default)\\\hline
       \texttt{minimal} & option to perform a least order filter synthesis\\
                 &  \texttt{true }  -- perform least order synthesis (default); \\
                 &  \texttt{false}  -- perform full order synthesis. \\\hline
       \texttt{tcond} & maximum alowed condition number of the employed non-orthogonal transformations (Default: $10^4$).\\ \hline
 \texttt{HDesign}   & full row rank design matrix $H$ to build \texttt{OPTIONS.rdim} linear
                      combinations of the left nullspace basis vectors (see \textbf{Method}) \newline
                      (Default: \texttt{[ ]})\\
                                        \hline
\end{longtable}}
\end{description}
\subsubsection*{Output data}
\begin{description}
\item
\texttt{Q} is the resulting fault detection filter in a standard state-space form
\be\label{efdsyn:detss}
{\begin{aligned}
\lambda x_Q(t)  & =   A_Qx_Q(t)+ B_{Q_y}y(t)+ B_{Q_u}u(t) ,\\
r(t) & =  C_Q x_Q(t) + D_{Q_y}y(t)+ D_{Q_u}u(t) ,
\end{aligned}}
\ee
where the residual signal $r(t)$ is a $q$-dimensional vector. The resulting value of $q$ depends on the selected options \texttt{OPTIONS.rdim} and \texttt{OPTIONS.minimal} (see \textbf{Method}). For the system object \texttt{Q}, two input groups \texttt{\bfseries 'outputs'} and \texttt{\bfseries 'controls'} are defined for $y(t)$ and $u(t)$, respectively, and the output group \texttt{\bfseries 'residuals'} is defined for the residual signal $r(t)$.

\item
\texttt{R} is the resulting internal form of the fault detection filter and has a standard state-space representation of the form
\be\label{efdsyn:detinss}
{\begin{aligned}
\lambda x_R(t)  & =   A_Qx_R(t)+ B_{R_f}f(t)+ B_{R_w}w(t) + B_{R_{v}}v(t) ,\\
r(t) & =  C_Q x_R(t) + D_{R_f}f(t)+ D_{R_w}w(t) + D_{R_{v}}v(t) .
\end{aligned}}
\ee
The input groups \texttt{\bfseries 'faults'}, \texttt{\bfseries 'noise'} and \texttt{'aux'} are defined for $f(t)$, $w(t)$, and $v(t)$, respectively, and the output group \texttt{\bfseries 'residuals'} is defined for the residual signal $r(t)$. Notice that the realizations of \texttt{Q} and \texttt{R} share the matrices $A_Q$ and $C_Q$.

\item
\texttt{INFO} is a MATLAB structure containing additional information as follows:
\begin{center}
\begin{tabular}{|l|p{12cm}|} \hline \textbf{\texttt{INFO} fields} & \textbf{Description} \\ \hline
\texttt{tcond} & maximum of the condition numbers of the employed
                      non-orthogonal transformation matrices; a warning is
                      issued if \texttt{INFO.tcond $\geq$ OPTIONS.tcond}.\\ \hline
\texttt{degs}     & if \texttt{OPTIONS.simple} = \texttt{true}, the orders of the basis vectors of the employed simple nullspace basis;
if \texttt{OPTIONS.simple} = \texttt{false}, the degrees of the basis vectors of an equivalent polynomial nullspace basis. \texttt{INFO.degs = [ ]} if \texttt{OPTIONS.nullspace = false} is used.  \\ \hline
\texttt{S}     & binary structure matrix corresponding to $H\overline G_f(\lambda)$  (see \textbf{Method})  \\ \hline
\texttt{HDesign}     & design matrix $H$ employed for the synthesis of the fault detection filter (see \textbf{Method}) \\ \hline
\end{tabular}
\end{center}

\end{description}
\subsubsection*{Method}
The function \texttt{efdsyn} implements an extension of the \textbf{Procedure EFD} from \cite[Sect.\ 5.2]{Varg17}, which
relies on the nullspace-based synthesis method proposed in \cite{Varg03b}. In what follows, we succinctly present this extended procedure, in terms of the input-output descriptions. Full details of the employed state-space based computational algorithms are given in \cite{Varg17}[Chapter 7].

Let assume the system \texttt{SYSF} in (\ref{efdsyn:sysss}) has the equivalent input-output form
\be\label{efdsyn:sysio} {\mathbf{y}}(\lambda) =
G_u(\lambda){\mathbf{u}}(\lambda) +
G_d(\lambda){\mathbf{d}}(\lambda) +
G_f(\lambda){\mathbf{f}}(\lambda) +
G_w(\lambda){\mathbf{w}}(\lambda) +
G_v(\lambda){\mathbf{v}}(\lambda),
 \ee
where the vectors $y$, $u$, $d$, $f$, $w$ and $v$ have dimensions $p$, $m_u$, $m_d$, $m_f$, $m_w$ and $m_v$, respectively. The resulting fault detection filter in (\ref{efdsyn:detss}) has the input-output form
 \be\label{efdsyn:detio}
{\mathbf{r}}(\lambda) = Q(\lambda)\ba{c}
{\mathbf{y}}(\lambda)\\{\mathbf{u}}(\lambda)\ea ,  \ee
where the residual vector $r(t)$ has $q$ components.

The synthesis method which underlies \textbf{Procedure EFD}, essentially determines the  filter $Q(\lambda)$ as a stable rational left annihilator of
\be\label{efdsyn:gtfm}  G(\lambda) := \ba{cc} G_u(\lambda) & G_d(\lambda) \\
 I_{m_u} & 0 \ea ,\ee
such that $R_{f_j}(\lambda) \neq 0$, for $j = 1, \ldots, m_f$, where
\be\label{efdsyn:detintfm} R(\lambda) := \ba{c|c|c} R_f(\lambda) & R_w(\lambda) & R_v(\lambda) \ea :=
{\arraycolsep=1mm Q(\lambda)  \ba{c|c|c} G_f(\lambda) & G_w(\lambda) & G_v(\lambda) \\
          0 & 0 & 0 \ea } \, .
         \ee
The resulting internal form of the fault detection filter (\ref{efdsyn:detio}) is
\be\label{efdsyn:detinio} \hspace*{-5mm}{\mathbf{r}}(\lambda) = R(\lambda)\!{\arraycolsep=0mm \ba{c}{\mathbf{f}}(\lambda)\\
{\mathbf{w}}(\lambda) \\ {\mathbf{v}}(\lambda) \ea }\! =
R_f(\lambda){\mathbf{f}}(\lambda) \!+\!
R_w(\lambda){\mathbf{w}}(\lambda) \!+\!
R_v(\lambda){\mathbf{v}}(\lambda)  \, , \hspace*{-4mm}\ee
with $R(\lambda)$, defined in (\ref{efdsyn:detintfm}),
 stable.

The filter $Q(\lambda)$ is determined in the product form
\be\label{efdsyn:Qprod} Q(\lambda) = Q_3(\lambda)Q_2(\lambda)Q_1(\lambda) , \ee
where the factors are  determined as follows:
 \begin{itemize}
 \item[(a)] $Q_1(\lambda) = N_l(\lambda)$, with $N_l(\lambda)$ a $\big(p-r_d\big) \times (p+m_u)$ proper rational left nullspace basis satisfying $N_l(\lambda)G(\lambda) = 0$, with $r_d := \text{rank}\, G_d(\lambda)$;
     \item[(b)] $Q_2(\lambda)$ is an admissible factor (i.e., guaranteeing complete fault detectability) to perform least order synthesis;
     \item[(c)] $Q_3(\lambda)$ is a stable invertible factor determined  such that $Q(\lambda)$ and the associated $R(\lambda)$ in (\ref{efdsyn:detintfm}) have a desired dynamics.
 \end{itemize}
The computations of  individual factors depend on the user's options and specific choices are discussed in what follows.

\subsubsection*{Computation of $Q_1(\lambda)$}
If \texttt{OPTIONS.nullspace = true} or $m_d > 0$ or $E$ is singular, then $N_l(\lambda)$ is determined as a minimal proper nullspace basis. In this case, if \texttt{OPTIONS.simple = true}, then $N_l(\lambda)$ is determined as a simple rational basis and the orders of the basis vectors are provided in \texttt{INFO.degs}. If \texttt{OPTIONS.simple = false}, then $N_l(\lambda)$ is determined as a proper rational basis and \texttt{INFO.degs} contains the degrees of the basis vectors of an equivalent polynomial nullspace basis (see \cite[Section 9.1.3]{Varg17} for  definitions). A stable basis is determined if \texttt{OPTIONS.FDfreq} is not empty.

If \texttt{OPTIONS.nullspace = false}, $m_d = 0$ and $E$ is invertible, then $N_l(\lambda) = [ \, I_{p} \; -G_u(\lambda)\,]$ is used, which corresponds to a full-order Luenberger observer. If \texttt{OPTIONS.FDfreq} is not empty and the system (\ref{efdsyn:sysss}) is unstable, then $\widetilde N_l(\lambda) = M(\lambda)N_l(\lambda)$ is used instead $N_l(\lambda)$, where $M(\lambda)$ and $\widetilde N_l(\lambda)$ are the stable factors of a left coprime factorization $N_l(\lambda) = M^{-1}(\lambda)\widetilde N_l(\lambda)$.

To check the solvability of the EFDP, the transfer function matrix  $\overline G_f(\lambda) :=  Q_1(\lambda)
\left[\begin{smallmatrix}  G_f(\lambda) \\ 0 \end{smallmatrix}\right]$ and the structure matrix
$S_{H\overline G_f}$ of $H\overline G_f(\lambda)$ are determined, where $H$ is the design  matrix specified in  a nonempty \texttt{OPTIONS.HDesign} (otherwise $H = I_{p-r_d}$ is used). The EFDP is solvable provided $S_{H\overline G_f}$ has all its columns nonzero. The resulted $S_{H\overline G_f}$ is provided in \texttt{INFO.S}.

\subsubsection*{Computation of $Q_2(\lambda)$}

For a nonempty \texttt{OPTIONS.rdim}, the resulting dimension $q$ of the residual vector $r(t)$ is  $q = \min(\text{\texttt{OPTIONS.rdim}},p\!-\!r_d)$, where $r_d = \rank G_d(\lambda)$. If \texttt{OPTIONS.rdim} is empty, then a default value of $q$ is used (see the description of  \texttt{OPTIONS.rdim}).

If \texttt{OPTIONS.minimal = false}, then $Q_2(\lambda) = H$, where $H$ is a suitable $q\times \big(p-r_d\big)$ full row rank design matrix. $H$ is set as follows. If \texttt{OPTIONS.HDesign} is nonempty, then $H = \texttt{OPTIONS.HDesign}$.
If \texttt{OPTIONS.HDesign} is empty, then the matrix $H$  is chosen to build $q$ linear combinations of the $p-r_d$ left nullspace basis vectors, such that
$HQ_1(\lambda)$ has full row rank. If $q = p-r_d$ then the choice $H = I_{p-r_d}$ is used, otherwise $H$ is chosen a randomly generated $q \times \big(p-r_d\big)$ real matrix.

If \texttt{OPTIONS.minimal = true}, then $Q_2(\lambda)$ is a $q\times \big(p-r_d\big)$ transfer function matrix, with $q$ chosen as above.  $Q_2(\lambda)$ is determined in the form
\[  Q_2(\lambda) = H+Y_2(\lambda) \, , \]
such that $Q_2(\lambda)Q_1(\lambda)$ $\big(\! = HN_l(\lambda)+Y_2(\lambda)N_l(\lambda)\big)$ and $Y_2(\lambda)$ are  the least order solution of a left minimal cover problem \cite{Varg17g}.  If \texttt{OPTIONS.HDesign} is nonempty, then $H = \texttt{OPTIONS.HDesign}$, and if \texttt{OPTIONS.HDesign} is empty, then a suitable randomly generated $H$ is employed (see above).

The structure field \texttt{INFO.HDesign} contains the employed value of the design matrix $H$.

\subsubsection*{Computation of $Q_3(\lambda)$}
 $Q_3(\lambda)$ is a stable invertible transfer function matrix determined such that $Q(\lambda)$ in (\ref{efdsyn:Qprod}) and the associated $R(\lambda)$ in (\ref{efdsyn:detintfm}) have a desired dynamics (specified via \texttt{OPTIONS.sdeg} and \texttt{OPTIONS.poles}).

\subsubsection*{Example}

\begin{example}
This is Example 5.4 from the book \cite{Varg17} of an unstable continuous-time system with the TFMs
\[ G_u(s) = \ba{c}  \displaystyle\frac{s+1}{s-2} \\ \\[-2mm]\displaystyle \frac{s+2}{s-3} \ea, \;\; G_d(s) = \ba{c}  \displaystyle\frac{s-1}{s+2} \\ \\[-2mm] 0 \ea , \;\; G_f(s) = \ba{cc} \displaystyle\frac{s+1}{s-2} & 0\\ \\[-2mm] \displaystyle\frac{s+2}{s-3} & 1 \ea, \;\; G_w(s) = 0, \;\; G_v(s) = 0, \]
where the fault input $f_1$ corresponds to an additive actuator fault, while the fault input $f_2$ describes an additive sensor fault in the second output $y_2$.  The TFM $G_d(s)$ is non-minimum phase, having an unstable zero at 1.

We want to design a least order fault detection filter $Q(s)$ with scalar output, and a stability degree of $-3$ for the poles,  which fulfills:
\begin{itemize}
\item[--] the decoupling condition: $Q(s)\left[\begin{smallmatrix} G_u(s) & G_d(s) \\ I_{m_u} & 0  \end{smallmatrix}\right] = 0$;
\item[--] the fault detectability condition: $R_{f_j}(\lambda) \neq 0, \quad j = 1, \ldots, m_f $.
\end{itemize}

The results computed with the following script are
\[ Q(s) = \ba{ccc} 0 & \displaystyle\frac{s-3}{s+3} &  -\displaystyle\frac{s+2}{s+3} \ea , \quad R_f(s) = \ba{cc}  \displaystyle\frac{s+2}{s+3} & \displaystyle\frac{s-3}{s+3} \ea \, . \]

\begin{verbatim}

% Example - Solution of an exact fault detection problem (EFDP)

s = tf('s'); % define the Laplace variable s
% define Gu(s) and Gd(s)
Gu = [(s+1)/(s-2); (s+2)/(s-3)];     % enter Gu(s)
Gd = [(s-1)/(s+2); 0];               % enter Gd(s)
p = 2; mu = 1; md = 1; mf = 2;       % set dimensions

% setup the synthesis model with faults
sysf = fdimodset(ss([Gu Gd]),struct('c',1,'d',2,'f',1,'fs',2));

% call of EFDSYN with the options for stability degree -3 and the synthesis
% of a scalar output filter
[Q,R] = efdsyn(sysf,struct('sdeg',-3,'rdim',1));

% display the implementation form Q and the internal form Rf of the
% resulting fault detection filter
tf(Q), tf(R)

% check synthesis conditions: Q[Gu Gd;I 0] = 0 and Q[Gf; 0] = R
syse = [sysf;eye(mu,mu+md+mf)];  % form Ge = [Gu Gd Gf;I 0 0];
norm_Ru_Rd = norm(Q*syse(:,{'controls','disturbances'}),inf)
norm_rez = norm(Q*syse(:,'faults')-R,inf)

% check weak and strong fault detectability
S_weak = fditspec(R)
[S_strong,abs_dcgains] = fdisspec(R)

% determine the fault sensitivity condition
FSCOND = fdifscond(R,0)

% evaluate step responses
set(R,'InputName',{'f_1','f_2'},'OutputName','r');
step(R);
title('Step responses from the fault inputs'), ylabel('')
\end{verbatim}

\end{example}

\newpage
\subsubsection{\texttt{\bfseries afdsyn}}
\index{M-functions!\texttt{\bfseries afdsyn}}
\subsubsection*{Syntax}
\begin{verbatim}
[Q,R,INFO] = afdsyn(SYSF,OPTIONS)
\end{verbatim}

\subsubsection*{Description}
\index{fault detection and e@fault detection problem!approximate (AFDP)}

\texttt{\bfseries afdsyn} solves   the \emph{approximate fault detection problem} (AFDP) (see Section \ref{sec:AFDP}), for a given LTI system \texttt{SYSF} with additive faults. Two stable and proper filters, \texttt{Q} and \texttt{R}, are computed, where \texttt{Q} contains the fault detection filter representing the solution of the AFDP,  and \texttt{R} contains its internal form.

\subsubsection*{Input data}
\begin{description}
\item
\texttt{SYSF} is a  LTI system  in the state-space form
\be\label{afdsyn:sysss}
{\begin{aligned}
E\lambda x(t)  & =  Ax(t)+ B_u u(t)+ B_d d(t) + B_f f(t)+ B_w w(t) + B_{v} v(t)  , \\
y(t) & =  C x(t) + D_u u(t)+ D_d d(t) + D_f f(t) + D_w w(t)+ B_{v} v(t) ,
\end{aligned}}
\ee
where any of the inputs components $u(t)$, $d(t)$, $f(t)$, $w(t)$  or $v(t)$ can be void. The auxiliary input signal $v(t)$ can be used for convenience. For the system \texttt{SYSF}, the input groups for $u(t)$, $d(t)$, $f(t)$, and $w(t)$ have the standard names \texttt{\bfseries 'controls'}, \texttt{\bfseries 'disturbances'}, \texttt{\bfseries 'faults'}, and \texttt{\bfseries 'noise'}, respectively. For the auxiliary input $v(t)$ the standard input group \texttt{'aux'} can be used.
\item
 \texttt{OPTIONS} is a MATLAB structure used to specify various synthesis  options and has the following fields:
{\setlength\LTleft{30pt}\begin{longtable}{|l|p{11.6cm}|} \hline \textbf{\texttt{OPTIONS} fields} & \textbf{Description} \\ \hline
       \texttt{tol}       & relative tolerance for rank computations \newline
                 (Default: internally computed)\\ \hline
       \texttt{tolmin}       & absolute tolerance for observability tests \newline
                 (Default: internally computed)\\ \hline
       \texttt{FDTol}     & threshold for fault detectability checks
                 (Default: $10^{-4}$)\\ \hline
       \texttt{FDGainTol} & threshold for strong fault detectability checks (Default: $10^{-2}$)\\ \hline
       \pagebreak[4] \hline
       \texttt{rdim}       & desired number $q$ of residual outputs for \texttt{Q} and \texttt{R}\\
                 & (Default: \hspace*{-2.5mm}\begin{tabular}[t]{p{10cm}} \texttt{[ ]}, in which case $q = q_1+q_2$, with $q_1$ and $q_2$ selected as follows: \newline if \texttt{OPTIONS.HDesign} is empty, then \newline $q_1 = \min(1,r_w)$ if \texttt{OPTIONS.minimal} = \texttt{true}, or \newline $q_1 = r_w$, if \texttt{OPTIONS.minimal} = \texttt{false} (see \textbf{Method}); \\if \texttt{OPTIONS.HDesign} is nonempty, then $q_1$ is the row dimension of the design matrix $H_1$ contained in \texttt{OPTIONS.HDesign} \\
                 if \texttt{OPTIONS.HDesign2} is empty, then \newline $q_2 = 1-\min(1,r_w)$ if \texttt{OPTIONS.minimal} = \texttt{true}, or \newline $q_2 = p-r_d-r_w$, if \texttt{OPTIONS.minimal} = \texttt{false} (see \textbf{Method}); \\if \texttt{OPTIONS.HDesign2} is nonempty, then $q_2$ is the row dimension of the design matrix $H_2$ contained in \texttt{OPTIONS.HDesign2})
                 \end{tabular}\\ \hline
       \texttt{FDFreq}  &  vector of real frequency values for strong detectability checks \newline (Default: \texttt{[ ]})  \\ \hline
       \texttt{smarg}   & stability margin for the poles of the filters \texttt{Q} and \texttt{R}\\
                 &  (Default: \texttt{-sqrt(eps)} for a continuous-time system \texttt{SYSF}; \\
                 &  \hspace*{4.5em}\texttt{1-sqrt(eps)} for a discrete-time system \texttt{SYSF}).\\ \hline
      \texttt{sdeg}   & prescribed stability degree for the poles of the filters \texttt{Q} and \texttt{R}\\
                 &  (Default: $-0.05$ for a continuous-time system \texttt{SYSF}; \\
                 &  \hspace*{4.8em} $0.95$ for a discrete-time system \texttt{SYSF}).\\ \hline
       \texttt{poles}   & complex vector containing a complex conjugate set of desired poles (within the stability domain) to be assigned for the filters \texttt{Q} and \texttt{R} \newline (Default: \texttt{[ ]})
                   \\\hline
   \texttt{nullspace}   & option to use a specific proper nullspace basis\\
                 & \hspace*{-.9mm}{\tabcolsep=0.7mm\begin{tabular}[t]{lcp{10cm}} \texttt{true } &--& use a minimal proper basis (default); \\
                 \texttt{false} &--& use a full-order observer based basis (see \textbf{Method}). \newline \emph{Note:} This option can be only used  if no disturbance inputs are present in (\ref{afdsyn:sysss}) and $E$ is invertible.
                 \end{tabular}} \\  \hline
       \texttt{simple} & option to employ a simple proper basis for filter synthesis\\
                 &  \texttt{true }  -- use a simple basis; \\
                 &  \texttt{false}  -- use a non-simple basis (default)\\\hline
       \texttt{minimal} & option to perform a least order filter synthesis\\
                 &  \texttt{true }  -- perform least order synthesis (default); \\
                 &  \texttt{false}  -- perform full order synthesis. \\\hline
       \texttt{exact} & option to perform exact filter synthesis \\
                 &  \texttt{true }  -- perform exact synthesis (i.e., no optimization performed); \\
                 &  \texttt{false}  -- perform approximate synthesis (default). \\\hline
       \texttt{tcond} & maximum alowed condition number of the employed non-orthogonal transformations (Default: $10^4$).\\ \hline
\texttt{freq}   & complex frequency value to be employed to check
                      the full row rank admissibility condition (see \textbf{Method}) \newline
                      (Default:\texttt{[ ]}, i.e., a randomly generated frequency). \\
                                        \hline
 \texttt{HDesign}   & design matrix $H_1$, with full row rank $q_1$, to build $q_1$ linear
                      combinations of the left nullspace basis vectors of $G_1(\lambda) := \left[\begin{smallmatrix} G_u(\lambda) & G_d(\lambda) \\ I & 0 \end{smallmatrix} \right]$; \newline $H_1$ is used for the synthesis of the filter components $Q^{(1)}(\lambda)$ and  $R^{(1)}(\lambda)$ (see \textbf{Method}) 
                      (Default: \texttt{[ ]})\\
                                        \hline
 \texttt{HDesign2}   & design matrix $H_2$, with full row rank $q_2$, to build $q_2$ linear
                      combinations of the left nullspace basis vectors of $G_2(\lambda) := \left[\begin{smallmatrix} G_u(\lambda) & G_d(\lambda) & G_w(\lambda) \\ I & 0 & 0\end{smallmatrix} \right]$;  $H_2$ is used for the synthesis of the filter components $Q^{(2)}(\lambda)$ and  $R^{(2)}(\lambda)$ (see \textbf{Method}) 
                      (Default: \texttt{[ ]})\\
                                        \hline
\texttt{gamma}   & upper bound on the resulting $\|R_w(\lambda)\|_\infty$ (see \textbf{Method}) \newline (Default: 1) \\ \hline
\texttt{epsreg}   & regularization parameter (Default: 0.1) \\ \hline
\texttt{sdegzer}   & prescribed stability degree for zeros shifting \\ &  (Default: $-0.05$ for a continuous-time system \texttt{SYSF}; \\
                 &  \hspace*{4.8em} $0.95$ for a discrete-time system \texttt{SYSF}).\\ \hline
\texttt{nonstd}   & option to handle nonstandard optimization problems (see \textbf{Method})\\
                 &  1 -- use the quasi-co-outer--co-inner factorization (default); \\
                 &  2 -- use the modified co-outer--co-inner factorization
                          with the \\
                 & \hspace{1.7em}regularization parameter \texttt{OPTIONS.epsreg};  \\
                 &  3 -- use the Wiener-Hopf type co-outer--co-inner
                          factorization.  \\
                 &     4 -- use the Wiener-Hopf type co-outer-co-inner factorization with\\
                 & \hspace{1.7em}zero shifting of the  non-minimum phase factor using the\\
                 & \hspace{1.7em}stabilization parameter \texttt{OPTIONS.sdegzer} \\
                 &     5 -- use the Wiener-Hopf type co-outer-co-inner factorization with \\
                 & \hspace{1.7em}the regularization of the non-minimum phase factor using the \\
                 & \hspace{1.7em}regularization parameter \texttt{OPTIONS.epsreg}  \\                                       \hline
\end{longtable}}
\end{description}
\subsubsection*{Output data}
\begin{description}
\item
\texttt{Q} is the resulting fault detection filter in a standard state-space form
\be\label{afdsyn:detss}
{\begin{aligned}
\lambda x_Q(t)  & =   A_Qx_Q(t)+ B_{Q_y}y(t)+ B_{Q_u}u(t) ,\\
r(t) & =  C_Q x_Q(t) + D_{Q_y}y(t)+ D_{Q_u}u(t) ,
\end{aligned}}
\ee
where the residual signal $r(t)$ is a $q$-dimensional vector. The resulting value of $q$ depends on the selected options \texttt{OPTIONS.rdim} and \texttt{OPTIONS.minimal} (see \textbf{Method}). For the system object \texttt{Q}, two input groups \texttt{\bfseries 'outputs'} and \texttt{\bfseries 'controls'} are defined for $y(t)$ and $u(t)$, respectively, and the output group \texttt{\bfseries 'residuals'} is defined for the residual signal $r(t)$.

\item
\texttt{R} is the resulting internal form of the fault detection filter and has a standard state-space representation of the form
\be\label{afdsyn:detinss}
{\begin{aligned}
\lambda x_R(t)  & =   A_Qx_R(t)+ B_{R_f}f(t)+ B_{R_w}w(t) + B_{R_{v}}v(t) ,\\
r(t) & =  C_Q x_R(t) + D_{R_f}f(t)+ D_{R_w}w(t) + D_{R_{v}}v(t) .
\end{aligned}}
\ee
The input groups \texttt{\bfseries 'faults'}, \texttt{\bfseries 'noise'} and \texttt{'aux'} are defined for $f(t)$, $w(t)$, and $v(t)$, respectively, and the output group \texttt{\bfseries 'residuals'} is defined for the residual signal $r(t)$. Notice that the realizations of \texttt{Q} and \texttt{R} share the matrices $A_Q$ and $C_Q$.

\item
\texttt{INFO} is a MATLAB structure containing additional information as follows:
\begin{center}
\begin{tabular}{|l|p{12cm}|} \hline \textbf{\texttt{INFO} fields} & \textbf{Description} \\ \hline
\texttt{tcond} & maximum of the condition numbers of the employed
                      non-orthogonal transformation matrices; a warning is
                      issued if \texttt{INFO.tcond $\geq$ OPTIONS.tcond}.\\ \hline
\texttt{degs}     & If \texttt{OPTIONS.nullspace} = \texttt{true}, then: \newline  if \texttt{OPTIONS.simple} = \texttt{true}, the orders of the basis vectors of the employed simple left nullspace basis of $G_1(\lambda)$; \newline
if \texttt{OPTIONS.simple} = \texttt{false}, the degrees of the basis vectors of an equivalent polynomial nullspace basis  (see \textbf{Method}). \newline If \texttt{OPTIONS.nullspace = false} then \texttt{INFO.degs = [ ]}. \\ \hline
\texttt{degs2}     & if \texttt{OPTIONS.simple} = \texttt{true}, the orders of the basis vectors of the employed simple left nullspace basis of $\overline G_w(\lambda)$; \newline
if \texttt{OPTIONS.simple} = \texttt{false}, the degrees of the basis vectors of an equivalent polynomial nullspace basis  (see \textbf{Method}) \\ \hline
\texttt{S}     & binary structure matrix $S_1$ corresponding to $H_1\overline G_f(\lambda)$ (see \textbf{Method})  \\ \hline
\texttt{S2}     & binary structure matrix $S_2$ corresponding to $H_2\overline G_f^{(2)}(\lambda)$ (see \textbf{Method})  \\ \hline
\texttt{HDesign}     & design matrix $H_1$ employed for the synthesis of the fault detection filter (see \textbf{Method}) \\ \hline
\texttt{HDesign2}     & design matrix $H_2$ employed for the synthesis of the fault detection filter (see \textbf{Method}) \\ \hline
\texttt{freq}   & complex frequency value employed to check the full row rank
                      admissibility condition (see \textbf{Method}) \\
                                        \hline
\texttt{gap}     & achieved gap $\|R_{f}(\lambda)\|_{\infty -}/\|R_w(\lambda)\|_\infty$, where the $\mathcal{H}_-$-index is computed over
                      the whole frequency range, if \texttt{OPTIONS.FDFreq} is
                      empty, or over the frequency values contained in
                      \texttt{OPTIONS.FDFreq}.  (see \textbf{Method}) \\ \hline
\end{tabular}
\end{center}

\end{description}

\subsubsection*{Method}
The function \texttt{afdsyn} implements an extension of the \textbf{Procedure AFD} from \cite[Sect.\ 5.3]{Varg17} as proposed in \emph{Remark 5.10} in \cite{Varg17}, which
relies on the nullspace-based synthesis method proposed in  \cite{Varg09b}, with extensions discussed in \cite{Glov11}. In what follows, we discuss succinctly the main steps of this extended procedure, in terms of the input-output descriptions. Full details of the employed state-space based computational algorithms are given in \cite{Varg17}[Chapter 7].

Let assume the system \texttt{SYSF} in (\ref{afdsyn:sysss}) has the equivalent input-output form
\be\label{afdsyn:sysio} {\mathbf{y}}(\lambda) =
G_u(\lambda){\mathbf{u}}(\lambda) +
G_d(\lambda){\mathbf{d}}(\lambda) +
G_f(\lambda){\mathbf{f}}(\lambda) +
G_w(\lambda){\mathbf{w}}(\lambda) +
G_v(\lambda){\mathbf{v}}(\lambda),
 \ee
where the vectors $y$, $u$, $d$, $f$, $w$ and $v$ have dimensions $p$, $m_u$, $m_d$, $m_f$, $m_w$ and $m_v$, respectively. The resulting fault detection filter in (\ref{afdsyn:detss}) has the input-output form
 \be\label{afdsyn:detio}
{\mathbf{r}}(\lambda) = Q(\lambda)\ba{c}
{\mathbf{y}}(\lambda)\\{\mathbf{u}}(\lambda)\ea ,  \ee
where the residual vector $r(t)$ has $q$ components.


The implemented synthesis method  essentially determines the  filter $Q(\lambda)$ as a stable rational left annihilator of
\be\label{afdsyn:gtfm}  G_1(\lambda) := \ba{cc} G_u(\lambda) & G_d(\lambda) \\
 I_{m_u} & 0 \ea ,\ee
such that $R_{f_j}(\lambda) \neq 0$, for $j = 1, \ldots, m_f$, where
\be\label{afdsyn:detintfm} R(\lambda) := \ba{c|c|c} R_f(\lambda) & R_w(\lambda) & R_v(\lambda) \ea :=
{\arraycolsep=1mm Q(\lambda)  \ba{c|c|c} G_f(\lambda) & G_w(\lambda) & G_v(\lambda) \\
          0 & 0 & 0 \ea } \, .
         \ee
The resulting internal form of the fault detection filter (\ref{afdsyn:detio}) is
\be\label{afdsyn:detinio} \hspace*{-5mm}{\mathbf{r}}(\lambda) = R(\lambda)\!{\arraycolsep=0mm \ba{c}{\mathbf{f}}(\lambda)\\
{\mathbf{w}}(\lambda) \\ {\mathbf{v}}(\lambda) \ea }\! =
R_f(\lambda){\mathbf{f}}(\lambda) \!+\!
R_w(\lambda){\mathbf{w}}(\lambda) \!+\!
R_v(\lambda){\mathbf{v}}(\lambda)  \, , \hspace*{-4mm}\ee
with $R(\lambda)$, defined in (\ref{afdsyn:detintfm}),
 stable.

An initial synthesis step implements the nullspace based synthesis approach and determines $Q(\lambda)$ in the product form   $Q(\lambda) = \overline Q_1(\lambda)Q_1(\lambda)$, where $Q_1(\lambda)$ is a proper left nullspace basis of $G_1(\lambda)$ in (\ref{afdsyn:gtfm}). This step ensures the decoupling of control and disturbance inputs in the residual (as is apparent in (\ref{afdsyn:detinio})) and allows to determine of $\overline Q_1(\lambda)$ by solving an AFDP for the reduced system
\be\label{afdsyn:redsys} \overline  {\mathbf{y}}(\lambda) := \overline G_f(\lambda){\mathbf{f}}(\lambda) \!+\!
\overline G_w(\lambda){\mathbf{w}}(\lambda) \!+\!
\overline G_v(\lambda){\mathbf{v}}(\lambda)  \, , \ee
where
\be\label{afdsyn:redsysmat} \ba{c|c|c} \overline G_f(\lambda) & \overline G_w(\lambda) & \overline G_v(\lambda) \ea = Q_1(\lambda)  \ba{c|c|c} G_f(\lambda) & G_w(\lambda) & G_v(\lambda) \\
          0 & 0 & 0 \ea ,\ee
to obtain
 \be\label{afdsyn:detio1}
{\mathbf{r}}(\lambda) = \overline Q_1(\lambda)\overline  {\mathbf{y}}(\lambda) .  \ee

If $r_d := \rank G_d(\lambda)$, then $Q_1(\lambda)$ has $p-r_d$ rows, which are the basis vectors of the left nullspace of $G_1(\lambda)$. Let $r_w = \rank \overline G_w(\lambda)$, which satisfies $0 \leq r_w \leq p-r_d$. If $r_w = 0$, then we have to solve an EFDP for the reduced system (\ref{afdsyn:redsys}) with $\overline G_w(\lambda) = 0$ (e.g., by using \textbf{Procedure EFD} in \cite{Varg17}), while for $r_w > 0$ and $q \leq r_w$ we have to solve an AFDP, for which \textbf{Procedure AFD} in \cite{Varg17} can be applied. In the case $0 < r_w < q \leq p-r_d$, we can apply the approach suggested in \emph{Remark 5.10} in  \cite{Varg17}), to determine $\overline Q_1(\lambda)$ in the row partitioned form
\[ \overline Q_1(\lambda) = \ba{cc} \overline Q_1^{(1)}(\lambda) \\ \overline Q_1^{(2)}(\lambda) \ea , \]
where $\overline Q_1^{(1)}(\lambda)$ has $q_1 = r_w$ rows and can be determined by solving an AFDP for the reduced system (\ref{afdsyn:redsys}), while $\overline Q_1^{(2)}(\lambda)$ has $q_2 = q-r_w$ rows and can be determined by solving an EFDP for the same reduced system (\ref{afdsyn:redsys}), but with the noise inputs redefined as disturbances and a part of the fault inputs (those which become undetectable due to the decoupling of noise inputs) redefined as auxiliary inputs. More precisely, let
$Q_2^{(2)}(\lambda)$ be a left nullspace basis of $\overline G_w(\lambda)$ and let $\overline Q_1^{(2)}(\lambda) = \overline Q_2^{(2)}(\lambda) Q_2^{(2)}(\lambda)$. Then, a second reduced system is obtained as
\be\label{afdsyn:redsys2} \overline  {\mathbf{y}}^{(2)}(\lambda) := \overline G_f^{(2)}(\lambda){\mathbf{f}}(\lambda) +
\overline G_v^{(2)}(\lambda){\mathbf{v}}(\lambda)  \, , \ee
where
\[ \ba{c|c} \overline G_f^{(2)}(\lambda) & \overline G_v^{(2)}(\lambda) \ea = Q_2^{(2)}(\lambda)  \ba{c|c} \overline G_f(\lambda) & \overline G_v(\lambda) \ea .\]
To determine $\overline Q_2^{(2)}(\lambda)$, we first determine $S_2$, the structure matrix of $\overline G_f^{(2)}(\lambda)$ and separate the fault components in two parts: $f^{(1)}$, which correspond to nonzero columns in $S_2$ and $f^{(2)}$, which correspond to zero columns in $S_2$. If we denote $\overline G_{f^{(1)}}^{(2)}(\lambda)$ and $\overline G_{f^{(2)}}^{(2)}(\lambda)$ the columns of $\overline G_{f}^{(2)}(\lambda)$ corresponding to $f^{(1)}$ and $f^{(2)}$, respectively, we can rewrite the reduced system (\ref{afdsyn:redsys2}) as
\be\label{afdsyn:redsys3} \overline  {\mathbf{y}}^{(2)}(\lambda) :=
\overline G_{f^{(1)}}^{(2)}(\lambda){\mathbf{f}}^{(1)}(\lambda) +
\overline G_{f^{(2)}}^{(2)}(\lambda){\mathbf{f}}^{(2)}(\lambda) +
\overline G_v^{(2)}(\lambda){\mathbf{v}}(\lambda)  \, .  \ee
The filter component $\overline Q_2^{(2)}(\lambda)$ can be determined by solving an EFDP for the reduced system (\ref{afdsyn:redsys3}), with $f^{(1)}$ as fault inputs and $f^{(2)}$ and $v(t)$ as auxiliary inputs.

To check the solvability of the AFDP, let $S_1$ be the structure matrix of $\overline G_f(\lambda)$. According to Corollary 5.4 of \cite{Varg17}, the AFDP is solvable if and only if $\overline G_{f_j}(\lambda) \not = 0$ for $j = 1, \ldots, m_f$, or equivalently all columns of $S_1$ are nonzero. More generally, if $H_1$ is the design  matrix specified in  a nonempty \texttt{OPTIONS.HDesign} (otherwise $H_1 = I_{p-r_d}$ is used) and $H_2$ is the design  matrix specified in  a nonempty \texttt{OPTIONS.HDesign2} (otherwise $H_2 = I_{p-r_d-r_w}$ is used), then let $S_1$ be the structure matrix of $H_1\overline G_f(\lambda)$ and let $S_2$ be the structure matrix of $H_2\overline G_f^{(2)}(\lambda)$. It follows that the AFDP is solvable if $S = \left[\begin{smallmatrix} S_1\\S_2\end{smallmatrix} \right]$ has all its columns nonzero.

 The solution of the AFDP for the reduced system  (\ref{afdsyn:redsys}) to determine $\overline Q_1^{(1)}(\lambda)$ involves the  solution of a $\mathcal{H}_{\infty -}/\mathcal{H}_\infty$ optimization problem
 \be\label{afdpopt} \beta = \max_{\overline Q_1^{(1)}(\lambda)} \Big\{ \, \left\|R_f^{(1)}(\lambda)\right\|_{\infty -} \,\, \Big| \,\, \left\|R_w^{(1)}(\lambda)\right\|_{\infty} \leq \gamma \,\Big\} ,\ee
where $\big[\, R_f^{(1)}(\lambda)\; R_w^{(1)}(\lambda) \,\big] := \overline Q_1^{(1)}(\lambda)\big[\, \overline G_f(\lambda) \; \overline G_w(\lambda)\,\big]$ and $\gamma$ is a given upper bound on the resulting $\big\|R_w^{(1)}(\lambda)\big\|_\infty$ (specified via \texttt{OPTIONS.gamma}). The $\mathcal{H}_{\infty -}$-index $\|\cdot\|_{\infty -}$ characterizes the complete fault detectability property of the system (\ref{afdsyn:sysio}) and is evaluated, for \texttt{OPTIONS.FDFreq} empty,  as
 \be\label{minh} \|R_{f}^{(1)}(\lambda)\|_{\infty -} :=
\min_{1\leq j \leq m_f} \|R_{f_j}^{(1)}(\lambda)\|_{\infty}, \ee%
while for \texttt{OPTIONS.FDFreq} containing a nonempty set of real frequencies, which define a complex frequency domain $\Omega$, the $\mathcal{H}_{\Omega -}$-index
\be\label{minusindex1} \|R_{f}^{(1)}(\lambda)\|_{\Omega -} := \min_{1\leq j \leq m_f} \big\{\inf_{\lambda_s \in \Omega} \big\|R_{f_j}^{(1)}(\lambda_s)\big\|_2 \big\} \, \ee
is used instead.
The value of the achieved fault-to-noise gap $\eta := \beta/\gamma$ is provided in \texttt{INFO.gap} and represents a measure of the noise attenuation quality of the fault detection filter. For $\gamma = 0$,  the exact solution of an EFDP with $w \equiv 0$ is targeted and the corresponding gap $\eta = \infty$. \index{performance evaluation!fault detection and isolation!fault-to-noise gap}

In general, the filter $Q(\lambda)$ and its corresponding internal form $R(\lambda)$ are determined in the partitioned forms
\be\label{QRpart} Q(\lambda) = \ba{c}Q^{(1)}(\lambda) \\ Q^{(2)}(\lambda) \ea = \ba{cc} \overline Q_1^{(1)}(\lambda) \\ \overline Q_2^{(2)}(\lambda) Q_2^{(2)}(\lambda) \ea Q_1(\lambda), \quad
R(\lambda) = \ba{c}R^{(1)}(\lambda) \\ R^{(2)}(\lambda) \ea, \ee
where the filters $Q^{(1)}(\lambda)$ and $R^{(1)}(\lambda)$ with $q_1$ residual outputs are a solution of an AFDP, while $Q^{(2)}(\lambda)$ and $R^{(2)}(\lambda)$ with $q_2$ residual outputs are a solution of an EFDP. In what follows, we first describe the determination of $Q_1(\lambda)$ and $Q_2^{(2)}(\lambda)$, and discuss the verification of the solvability conditions. Then we discuss the determination of the remaining factors $\overline Q_1^{(1)}(\lambda)$ and $\overline Q_2^{(2)}(\lambda)$.

\subsubsection*{Computation of $Q_1(\lambda)$}
$Q_1(\lambda) = N_l(\lambda)$, with $N_l(\lambda)$ a $\big(p-r_d\big) \times (p+m_u)$ proper rational left nullspace basis satisfying $N_l(\lambda)G_1(\lambda) = 0$, where $r_d := \text{rank}\, G_d(\lambda)$.

If \texttt{OPTIONS.nullspace = true} or $m_d > 0$ or $E$ is singular, then $N_l(\lambda)$ is determined as a minimal proper nullspace basis. In this case, if \texttt{OPTIONS.simple = true}, then $N_l(\lambda)$ is determined as a simple rational basis and the orders of the basis vectors are provided in \texttt{INFO.degs}. If \texttt{OPTIONS.simple = false}, then $N_l(\lambda)$ is determined as a proper rational basis and \texttt{INFO.degs} contains the degrees of the basis vectors of an equivalent polynomial nullspace basis. A stable basis is determined if \texttt{OPTIONS.FDfreq} is not empty.

If \texttt{OPTIONS.nullspace = false}, $m_d = 0$ and $E$ is invertible, then $N_l(\lambda) = [ \, I_{p} \; -G_u(\lambda)\,]$ is used, which corresponds to a full-order Luenberger observer. If \texttt{OPTIONS.FDfreq} is not empty and the system (\ref{afdsyn:sysss}) is unstable, then $\widetilde N_l(\lambda) = M(\lambda)N_l(\lambda)$ is used instead $N_l(\lambda)$, where $M(\lambda)$ and $\widetilde N_l(\lambda)$ are the stable factors of a left coprime factorization $N_l(\lambda) = M^{-1}(\lambda)\widetilde N_l(\lambda)$. In this case \texttt{INFO.degs = [ ]}.

\subsubsection*{Computation of $Q_2^{(2)}(\lambda)$}
In the case when $r_w < p-r_d$, $Q_2^{(2)}(\lambda) = \overline N_{l,w}(\lambda)$, with $\overline N_{l,w}(\lambda)$ a $\big(p-r_d-r_w)\times (p-r_d)$  proper left nullspace basis satisfying $\overline N_{l,w}(\lambda)\overline G_w(\lambda) = 0$, where $r_w := \rank \overline G_w(\lambda)$. It follows, that $\overline N_{l,w}(\lambda)N_l(\lambda)$ is a proper left nullspace basis of
\be\label{afdsyn:gtfm2}  G_2(\lambda) := \ba{ccc} G_u(\lambda) & G_d(\lambda) & G_w(\lambda)\\
 I_{m_u} & 0 & 0 \ea .\ee
 If \texttt{OPTIONS.simple = true}, then $\overline N_{l,w}(\lambda)N_l(\lambda)$ is determined as a simple rational basis and the orders of the basis vectors are provided in \texttt{INFO.degs2}. If \texttt{OPTIONS.simple = false}, then $\overline N_{l,w}(\lambda)N_l(\lambda)$ is determined as a proper rational basis and \texttt{INFO.degs2} contains the degrees of the basis vectors of an equivalent polynomial nullspace basis. A stable basis is determined if \texttt{OPTIONS.FDfreq} is not empty.

\subsubsection*{Checking the solvability conditions of the AFDP}

Let $\overline G_f(\lambda) :=  Q_1(\lambda)
\left[\begin{smallmatrix}  G_f(\lambda) \\ 0 \end{smallmatrix}\right]$ and $\overline G_w(\lambda) :=  Q_1(\lambda)
\left[\begin{smallmatrix}  G_w(\lambda) \\ 0 \end{smallmatrix}\right]$ be the TFMs of the reduced model (\ref{afdsyn:redsys}) and let $\overline G_f^{(2)}(\lambda) = Q_2^{(2)}(\lambda)\overline G_f(\lambda)$ be the TFM from the fault inputs in the reduced model (\ref{afdsyn:redsys2}).
To check the solvability of the AFDP, the structure matrices
$S_1$ of $H_1\overline G_f(\lambda)$ and $S_2$ of $H_2\overline G_f^{(2)}(\lambda)$ are determined, where $H_1$ is a full row rank design matrix with $q_1 \leq p-r_d$ rows specified in  a nonempty \texttt{OPTIONS.HDesign} (otherwise $H_1 = I_{p-r_d}$ is used) and $H_2$ is a full row rank design matrix specified with $q_2 \leq p-r_d-r_w$ rows in  a nonempty \texttt{OPTIONS.HDesign2} (otherwise $H_2 = I_{p-r_d-r_w}$ is used). The AFDP is solvable if $S = \left[\begin{smallmatrix} S_1\\S_2\end{smallmatrix} \right]$ has all its columns nonzero. The resulted $S_1$ and $S_2$ are provided in \texttt{INFO.S} and \texttt{INFO.S2}, respectively.

\subsubsection*{Computation of  $\overline Q_1^{(1)}(\lambda)$}

If $r_w > 0$, the filter  $\overline Q_1^{(1)}(\lambda)$ is determined in the product form
\be\label{afdsym:Qprod1}  \overline Q_1^{(1)}(\lambda) = Q_4^{(1)}(\lambda)Q_3^{(1)}(\lambda)Q_2^{(1)}(\lambda) , \ee
where the factors are  determined as follows:
 \begin{itemize}
 \item[(a)]  $Q_2^{(1)}(\lambda)$ is an admissible regularization factor;
     \item[(b)] $Q_3^{(1)}(\lambda)$ represents an optimal choice which maximizes the  gap $\eta$;
     \item[(c)] $Q_4^{(1)}(\lambda)$ is a stable invertible factor determined  such that $Q^{(1)}(\lambda)$ and $R^{(1)}(\lambda)$  have a desired dynamics.
 \end{itemize}
The computations of  individual factors depend on the user's options and specific choices are discussed in what follows.

\subsubsection*{Computation of $Q_2^{(1)}(\lambda)$}
Let
$q =\texttt{OPTIONS.rdim}$ if \texttt{OPTIONS.rdim} is nonempty. If \texttt{OPTIONS.rdim} is empty, $q_1$ is set to a default value as follows:  if \texttt{OPTIONS.HDesign} is empty, then $q_1 = 1$ if \texttt{OPTIONS.minimal} = \texttt{true}, or $q_1 = r_w$, if \texttt{OPTIONS.minimal} = \texttt{false}. If \texttt{OPTIONS.HDesign} is nonempty, then $q_1$ is the row dimension of the full row rank design matrix $H_1$ contained in \texttt{OPTIONS.HDesign}. To be admissible, $H_1$ must also fulfill  $\rank H_1\overline G_w(\lambda) = q_1$.

If \texttt{OPTIONS.minimal = false}, then $Q_2^{(1)}(\lambda) = H_1$, where $H_1$ is a suitable $q_1\times \big(p-r_d\big)$ full row rank design matrix.
If both \texttt{OPTIONS.HDesign} and \texttt{OPTIONS.HDesign2} are empty, then $H_1$  is chosen to build $q_1 = \min(q,r_w)$ linear combinations of the $p-r_d$ left nullspace basis vectors, such that
$\rank H_1\overline G_w(\lambda) = q_1$, and, additionally, $H_1\overline G_f(\lambda)$ has the same nonzero columns as $S_1$.
If \texttt{OPTIONS.HDesign} is empty but \texttt{OPTIONS.HDesign2} is nonempty, then $q_1 = q-q_2$.  If $q_1 = p-r_d$ then the choice $H_1 = I_{p-r_d}$ is used, otherwise $H_1$ is chosen a randomly generated $q_1 \times \big(p-r_d\big)$ real matrix.

If \texttt{OPTIONS.minimal = true}, then $Q_2^{(1)}(\lambda)$ is a $q_1\times \big(p-r_d\big)$ transfer function matrix, with $q_1$ chosen as above.  $Q_2^{(1)}(\lambda)$ is determined as
\[  Q_2^{(1)}(\lambda) = \widetilde Q_2(\lambda) \, , \]
where $\widetilde Q_2(\lambda) := H_1+Y_2(\lambda)$, $\widetilde Q_2(\lambda)Q_1(\lambda)$ $\big(\! = H_1N_l(\lambda)+Y_2(\lambda)N_l(\lambda)\big)$ and $Y_2(\lambda)$ are  the least order solution of a left minimal cover problem \cite{Varg17g}. If \texttt{OPTIONS.HDesign} is nonempty, then $H_1 = \texttt{OPTIONS.HDesign}$, and if \texttt{OPTIONS.HDesign} is empty, then $q_1$ is chosen as above and  a suitable randomly generated $H_1$ is employed (see above). To be admissible, $\widetilde Q_2(\lambda)$ must fulfill $\rank \widetilde Q_2(\lambda)\overline G_w(\lambda) = q_1$ and, additionally, $\widetilde Q_2(\lambda)\overline G_f(\lambda)$ has the same nonzero columns as $S_1$.
The above rank condition is checked as
\[ \rank \widetilde G_w(\lambda_s) = q_1 , \]
where $\widetilde G_w(\lambda) := \widetilde Q_2(\lambda)\overline G_w(\lambda)$ and  $\lambda_s$ is a suitable frequency value, which can be specified via the \texttt{OPTIONS.freq}. In the case when \texttt{OPTIONS.freq} is empty, the employed frequency value $\lambda_s$ is provided in \texttt{INFO.freq}.

\subsubsection*{Computation of $Q_3^{(1)}(\lambda)$}
Let redefine $\widetilde G_w(\lambda) :=  Q_2^{(1)}(\lambda)\overline G_w(\lambda)$ and compute the  quasi-co-outer--co-inner factorization of $\widetilde G_w(\lambda)$  as
\[ \widetilde G_w(\lambda) = R_{wo}(\lambda)R_{wi}(\lambda) ,\]
where $R_{wo}(\lambda)$ is an invertible quasi-co-outer factor and  $R_{wi}(\lambda)$ is a (full row rank) co-inner factor. In the \emph{standard case} $R_{wo}(\lambda)$ is outer (i.e., has no zeros on the boundary of the stability domain $\partial\mathds{C}_s$) and we choose $Q_3^{(1)}(\lambda) = R_{wo}^{-1}(\lambda)$.
This is an optimal choice which ensures that the optimal gap $\eta$ is achieved.

In the \emph{non-standard case} $R_{wo}(\lambda)$ is only quasi-outer and thus,  has zeros on the boundary of the stability domain $\partial\mathds{C}_s$. Depending on the selected option to  handle nonstandard optimization problems \texttt{OPTIONS.nonstd}, several choices are possible for $Q_3^{(1)}(\lambda)$ in this case:
 \begin{itemize}
 \item If \texttt{OPTIONS.nonstd} = 1, then $Q_3^{(1)}(\lambda) =R_{wo}^{-1}(\lambda)$ is used.
 \item If \texttt{OPTIONS.nonstd} = 2, then a modified co-outer--co-inner factorization of $[\,R_{wo}(\lambda)\;\epsilon I\,]$ is computed, whose co-outer factor $R_{wo,\epsilon}(\lambda)$ satisfies
\[ R_{wo,\epsilon}(\lambda)\big(R_{wo,\epsilon}(\lambda)\big)^\sim = \epsilon^2 I + R_{wo}(\lambda)\big(R_{wo}(\lambda)\big)^\sim . \]
Then $Q_3^{(1)}(\lambda) = R^{-1}_{wo,\epsilon}(\lambda)$ is used. The value of the regularization parameter $\epsilon$ is specified via \texttt{OPTIONS.epsreg}.
\item If \texttt{OPTIONS.nonstd} = 3, then a Wiener-Hopf type co-outer--co-inner factorization is computed in the form
\be\label{WHfact} \widetilde G_w(\lambda) = R_{wo}(\lambda)R_{wb}(\lambda)R_{wi}(\lambda) ,\ee
where $R_{wo}(\lambda)$ is co-outer, $R_{wi}(\lambda)$ is co-inner,  and $R_{wb}(\lambda)$ is a square stable factor whose zeros are precisely the zeros of $\widetilde G_w(\lambda)$ in $\partial\mathds{C}_s$.
$Q_3^{(1)}(\lambda)$ is determined as before  $Q_3^{(1)}(\lambda) = R_{wo}^{-1}(\lambda)$.
\item If \texttt{OPTIONS.nonstd} = 4, then the Wiener-Hopf type co-outer--co-inner factorization (\ref{WHfact}) is computed and
$Q_3^{(1)}(\lambda)$ is determined as  $Q_3^{(1)}(\lambda) = R_{wb}^{-1}(\tilde{\lambda})R_{wo}^{-1}(\lambda)$, where $\tilde{\lambda}$ is a small perturbation of $\lambda$ to move all zeros of $R_{wb}(\lambda)$ into the stable domain. In the continuous-time case $\tilde s = \frac{s-\beta_z}{1-\beta_zs}$, while in the discrete-time case $\tilde z = z/\beta_z$, where the zero shifting parameter $\beta_z$ is the prescribed stability degree for the zeros specified in  \texttt{OPTIONS.sdegzer}. For the evaluation of $R_{wb}(\tilde\lambda)$, a suitable bilinear transformation is performed.
\item If \texttt{OPTIONS.nonstd} = 5, then the Wiener-Hopf type co-outer--co-inner factorization (\ref{WHfact}) is computed and
$Q_3^{(1)}(\lambda)$ is determined as  $Q_3^{(1)}(\lambda) = R_{wb,\epsilon}^{-1}(\lambda)R_{wo}^{-1}(\lambda)$, where $R_{wb,\epsilon}(\lambda)$ is the co-outer factor of the co-outer--co-inner factorization of $\big[\, R_{wb}(\lambda) \; \epsilon I\,\big]$ and satisfies
\[ R_{wb,\epsilon}(\lambda)\big(R_{wb,\epsilon}(\lambda)\big)^\sim = \epsilon^2 I + R_{wb}(\lambda)\big(R_{wb}(\lambda)\big)^\sim . \]
The value of the regularization parameter $\epsilon$ is specified via \texttt{OPTIONS.epsreg}.
\end{itemize}

A typical feature of the non-standard case is that, with the exception of using the option \texttt{OPTIONS.nonstd} = 3, all other choices of \texttt{OPTIONS.nonstd} lead to a poor dynamical performance of the resulting filter, albeit an arbitrary large gap $\eta$ can be occasionally achieved.

\subsubsection*{Computation of $Q_4^{(1)}(\lambda)$}
In the standard case, $Q_4^{(1)}(\lambda) = I$. In the non-standard case,
 $Q_4^{(1)}(\lambda)$ is a stable invertible transfer function matrix determined such that $Q^{(1)}(\lambda)$ and $R^{(1)}(\lambda)$ in (\ref{QRpart}) have a desired dynamics (specified via \texttt{OPTIONS.sdeg} and \texttt{OPTIONS.poles}).

\subsubsection*{Computation of $\overline Q^{(2)}(\lambda)$}

If $0 \leq r_w < p-r_d$ or \texttt{OPTIONS.exact = true}, then the filter $Q^{(2)}(\lambda)$  in (\ref{QRpart}) is determined with $\overline Q^{(2)}(\lambda)$ in the product form
\be\label{afdsym:Qprod2} \overline Q^{(2)}(\lambda) = Q_4^{(2)}(\lambda)Q_3^{(2)}(\lambda) , \ee
where the factors are  determined as follows:
 \begin{itemize}
 \item[(a)] $Q_3^{(2)}(\lambda)$ is an  admissible regularization factor;
     \item[(b)] $Q_4^{(2)}(\lambda)$ is a stable invertible factor determined  such that $Q^{(2)}(\lambda)$ and $R^{(2)}(\lambda)$ in (\ref{QRpart}) have a desired dynamics.
 \end{itemize}
The computations of  individual factors depend on the user's options and specific choices are discussed in what follows.

 \subsubsection*{Computation of $Q_3^{(2)}(\lambda)$}

 Let
$q =\texttt{OPTIONS.rdim}$ if \texttt{OPTIONS.rdim} is nonempty. If \texttt{OPTIONS.rdim} is empty, $q_2$ is set to a default value as follows:  if \texttt{OPTIONS.HDesign2} is empty, then $q_2 = 1-\min(1,r_w)$ if \texttt{OPTIONS.minimal} = \texttt{true}, or $q_2 = p-r_d-r_w$, if \texttt{OPTIONS.minimal} = \texttt{false}. In the case when \texttt{OPTIONS.HDesign2} is nonempty, then $q_2$ is the row dimension of the full row rank design matrix $H_2$ contained in \texttt{OPTIONS.HDesign2}.

If \texttt{OPTIONS.minimal = false}, then $Q_3^{(2)}(\lambda) = H_2$, where $H_2$ is a suitable $q_2\times \big(p-r_d-r_w\big)$ full row rank design matrix.
If both \texttt{OPTIONS.HDesign} and \texttt{OPTIONS.HDesign2} are empty, then $H_2$  is chosen to build $q_2 = \min(q,r_w)$ linear combinations of the $p-r_d-r_w$ left nullspace basis vectors, such that
$H_2\overline G_f^{(2)}(\lambda)$ has the same nonzero columns as $S_2$.
If \texttt{OPTIONS.HDesign2} is empty but \texttt{OPTIONS.HDesign} is nonempty, then $q_2 = q-q_1$.  If $q_2 = p-r_d-r_w$ then the choice $H_2 = I_{p-r_d-r_w}$ is used, otherwise $H_2$ is chosen a randomly generated $q_2 \times \big(p-r_d-r_w\big)$ real matrix.

If \texttt{OPTIONS.minimal = true}, then $Q_3^{(2)}(\lambda)$ is a $q_2\times \big(p-r_d-r_w\big)$ transfer function matrix, with $q_2$ chosen as above.  $Q_3^{(2)}(\lambda)$ is determined as
\[  Q_3^{(2)}(\lambda) = \widetilde Q_3(\lambda) \, , \]
where $\widetilde Q_3(\lambda) := H_2+Y_3(\lambda)$, $\widetilde Q_3(\lambda)Q_2^{(2)}(\lambda)$ $\big(\! = H_2\overline N_{l,w}(\lambda)+Y_3(\lambda)\overline N_{l,w}(\lambda)\big)$ and $Y_3(\lambda)$ are  the least order solution of a left minimal cover problem \cite{Varg17g}. If \texttt{OPTIONS.HDesign2} is nonempty, then $H_2 = \texttt{OPTIONS.HDesign}$, and if \texttt{OPTIONS.HDesign2} is empty, then $q_2$ is chosen as above and  a suitable randomly generated $H_2$ is employed (see above). To be admissible, $\widetilde Q_2(\lambda)$ must ensure that $\widetilde Q_3(\lambda)\overline G_f^{(2)}(\lambda)$ has the same nonzero columns as $S_2$.

\subsubsection*{Computation of $Q_4^{(2)}(\lambda)$}
 $Q_4^{(2)}(\lambda)$ is a stable invertible transfer function matrix determined such that $Q^{(2)}(\lambda)$ and $R^{(2)}(\lambda)$ in (\ref{QRpart}) have a desired dynamics (specified via \texttt{OPTIONS.sdeg} and \texttt{OPTIONS.poles}).

 \subsubsection*{Example}

\begin{example}
This is Example 5.3 from the book \cite{Varg17}, with the disturbance redefined as a noise input and by adding a sensor fault for the first output measurement. The TFMs of the system are:
\[ G_u(s) = {\arraycolsep=1mm\ba{c}  \displaystyle\frac{s+1}{s+2} \\ \\[-2mm] \displaystyle\frac{s+2}{s+3} \ea, \quad G_d(s) = 0, \quad G_f(s) = [\, G_u(s)\; I\,], \quad G_w(s) = \ba{c}  \displaystyle\frac{s-1}{s+2} \\ \\[-2mm]0 \ea} ,\]
where the fault input $f_1$ corresponds to an additive actuator fault, while the fault inputs $f_2$  and $f_3$ describe additive sensor faults in the outputs $y_1$ and $y_2$, respectively.
The transfer function matrix $G_w(s)$ is non-minimum phase, having an unstable zero at 1. Interestingly, the EFDP formulated with $G_d(s) = G_w(s)$ is not solvable.

We want to design a least order fault detection filter $Q(s)$, which fulfills:
\begin{itemize}
\item[--] the decoupling condition: $Q(s)\left[\begin{smallmatrix} G_u(s) & G_d(s) \\ I_{m_u} & 0  \end{smallmatrix}\right] = 0$;
\item[--] the fault detectability condition: $\|R_f(s)\|_{\infty -} > 0$;
\item[--]  the maximization of the noise attenuation gap: $\eta := \|R_f(s)\|_{\infty -}/\|R_w(s)\|_\infty = \max$.
\end{itemize}

The results computed with the following script are
\[ Q(s) = {\arraycolsep=.5mm\ba{ccc} \displaystyle\frac{s+2}{s+1}  & \displaystyle\frac{s+3}{s+1} &  -\displaystyle\frac{2s+3}{s+1} \ea , \quad\; R_f(s) = \ba{ccc} \displaystyle\frac{2s+3}{s+1} &  \displaystyle\frac{s+2}{s+1} & \displaystyle\frac{s+3}{s+1} \ea }, \quad R_w(s) = \displaystyle\frac{s-1}{s+1} . \]

\begin{verbatim}
% Example - Solution of an approximate fault detection problem (AFDP)

% Example 5.3 of (V,2017) (modified)
s = tf('s'); % define the Laplace variable s
Gu = [(s+1)/(s+2); (s+2)/(s+3)];  % enter Gu(s)
Gw = [(s-1)/(s+2); 0];            % enter Gw(s)

% build model with additive faults with Gf = [Gu eye(p)];
sysf = fdimodset(ss([Gu Gw]),struct('c',1,'n',2,'f',1,'fs',1:2));
mu = 1; mw = 1; p = 2; mf = mu+p; % set dimensions


% perform synthesis with  AFDSYN, using default options
[Q,R,info] = afdsyn(sysf);  % R(s) = [Rf(s) Rw(s)]

% display the implementation form Q(s) and the internal forms Rf(s) and Rw(s)
% of resulting fault detection filter
tf(Q), tf(R(:,'faults')),   tf(R(:,'noise'))

% display the resulting gap and fault condition number
info.gap
FSCOND = fdifscond(R)

% check results: R(s) := Q(s)*Ge(s) = [0 Rf(s) Rw(s)],
% with Ge(s) = [Gu(s) Gf(s) Gw(s); I 0 0]
syse = [sysf;eye(mu,mu+mf+mw)]; % form Ge(s)
norm_Ru = norm(Q*syse(:,'controls'),inf)
norm_rez = norm(Q*syse(:,{'faults','noise'})-R,inf)
gap = fdif2ngap(R,0)            % check gap

% check strong fault detectability
S_strong = fdisspec(R(:,'faults'))

% simulate step responses from fault and noise inputs
inpnames = {'f_1','f_2','f_3','w'};
set(R,'InputName',inpnames,'OutputName','r');
step(R); ylabel('')
title('Step responses from fault and noise inputs')
\end{verbatim}

\end{example}

\subsubsection{\texttt{\bfseries efdisyn}}
\index{M-functions!\texttt{\bfseries efdisyn}}
\index{fault detection and isolation problem!a@exact (EFDIP)}
\subsubsection*{Syntax}
\begin{verbatim}
[Q,R,INFO] = efdisyn(SYSF,OPTIONS)
\end{verbatim}
\subsubsection*{Description}
\texttt{\bfseries efdisyn} solves   the \emph{exact fault detection and isolation problem} (EFDIP) (see Section \ref{sec:EFDIP}), for a given LTI system \texttt{SYSF} with additive faults and a given structure matrix $S_{FDI}$ (specified via the \texttt{OPTIONS} structure). Two banks of stable and proper filters are computed in the $n_b$-dimensional cell arrays \texttt{Q} and \texttt{R}, where $n_b$ is the number of specifications contained in $S_{FDI}$ (i.e., the number of rows of the structure matrix $S_{FDI}$).  \texttt{Q\{i\}} contains the $i$-th fault detection filter (\ref{ri_fdip}) in the overall solution (\ref{qbank}) of the EFDIP  and \texttt{R\{i\}} contains its internal form.\\[-7mm]

\subsubsection*{Input data}
\begin{description}
\item
\texttt{SYSF} is a  LTI system  in the state-space form
\be\label{efdisyn:sysss}
{\begin{aligned}
E\lambda x(t)  & =   Ax(t)+ B_u u(t)+ B_d d(t) + B_f f(t)+ B_w w(t)   , \\
y(t) & =  C x(t) + D_u u(t)+ D_d d(t) + D_f f(t) + D_w w(t) ,
\end{aligned}}
\ee
where any of the inputs components $u(t)$, $d(t)$, $f(t)$, or $w(t)$  can be void.  For the system \texttt{SYSF}, the input groups for $u(t)$, $d(t)$, $f(t)$, and $w(t)$ have the standard names \texttt{\bfseries 'controls'}, \texttt{\bfseries 'disturbances'}, \texttt{\bfseries 'faults'}, and \texttt{\bfseries 'noise'}, respectively.\\[-6mm]
\item
 \texttt{OPTIONS} is a MATLAB structure used to specify various synthesis  options and has the following fields:\\[-6mm]
{\setlength\LTleft{30pt}\begin{longtable}{|l|p{11.6cm}|} \hline \textbf{\texttt{OPTIONS} fields} & \textbf{Description} \\ \hline
       \texttt{SFDI}       & the desired structure matrix $S_{FDI}$ to solve the EFDIP\newline
                 (Default: \texttt{[1 ... 1]}, i.e., solve an exact fault
                                detection problem)\\ \hline
       \texttt{tol}       & relative tolerance for rank computations \newline
                 (Default: internally computed)\\ \hline
       \texttt{tolmin}       & absolute tolerance for observability tests \newline
                 (Default: internally computed)\\ \hline
       \texttt{FDTol}     & threshold for fault detectability checks
                 (Default: 0.0001)\\ \hline
       \texttt{FDGainTol} & threshold for strong fault detectability checks (Default: 0.01)\\ \hline
       \pagebreak[4]  \hline
       \texttt{rdim}       & vector, whose $i$-th component $q_i$, specifies the desired number of residual outputs for the $i$-th component filters \texttt{Q\{$i$\}} and \texttt{R\{$i$\}}; if \texttt{OPTIONS.rdim} is a scalar $q$, then  a vector with all components $q_i = q$ is assumed.     \\
                 & (Default: \hspace*{-2.5mm}\begin{tabular}[t]{l} \texttt{[ ]}, in which case: \\ \hspace*{-2em} --  if \texttt{OPTIONS.HDesign\{$i$\}} is empty, then \\ $q_i = 1$, if \texttt{OPTIONS.minimal} = \texttt{true}, or \\ $q_i$ is the dimension of the left nullspace which
                                underlies  the \\ synthesis of \texttt{Q\{$i$\}} and \texttt{R\{$i$\}}, if \texttt{OPTIONS.minimal} = \texttt{false}; \\ \hspace*{-2em} -- if \texttt{OPTIONS.HDesign\{$i$\}} is nonempty, then $q_i$ is the \\ row dimension of the design matrix contained in \\ \texttt{OPTIONS.HDesign\{$i$\}}.)
                                \end{tabular}\\ \hline
         \texttt{FDFreq}  &  vector of real frequency values for strong detectability checks \newline (Default: \texttt{[ ]})  \\ \hline
       \texttt{smarg}   & stability margin for the poles of the component filters \texttt{Q\{$i$\}} and \texttt{R\{$i$\}}\\
                 &  (Default: \texttt{-sqrt(eps)} for a continuous-time system \texttt{SYSF}; \\
                 &  \hspace*{4.5em}\texttt{1-sqrt(eps)} for a discrete-time system \texttt{SYSF}).\\ \hline
      \texttt{sdeg}   & prescribed stability degree for the poles of the component filters \texttt{Q\{$i$\}} and \texttt{R\{$i$\}}\\
                 &  (Default: $-0.05$ for a continuous-time system \texttt{SYSF}; \\
                 &  \hspace*{4.8em} $0.95$ for a discrete-time system \texttt{SYSF}).\\ \hline
       \texttt{poles}   & complex vector containing a complex conjugate set of desired poles (within the stability domain) to be assigned for the component filters \texttt{Q\{$i$\}} and \texttt{R\{$i$\}} (Default: \texttt{[ ]})
                   \\\hline
 \texttt{nullspace}   & option to use a specific proper nullspace basis to be employed at the initial reduction step\\
                 & \hspace*{-.9mm}{\tabcolsep=0.7mm\begin{tabular}[t]{lcp{10cm}} \texttt{true } &--& use a minimal proper basis (default); \\
                 \texttt{false} &--& use a full-order observer based basis (see \textbf{Method}). \newline \emph{Note:} This option can  only be used if no disturbance inputs are present in (\ref{efdisyn:sysss}) and $E$ is invertible.
                 \end{tabular}} \\  \hline
       \texttt{simple} & option to employ simple proper bases for the synthesis of the component filters \texttt{Q\{$i$\}} and \texttt{R\{$i$\}} \\
                 &  \texttt{true }  -- use simple bases; \\
                 &  \texttt{false}  -- use non-simple bases (default)\\\hline
       \texttt{minimal} & option to perform least order synthesis of the component filters \texttt{Q\{$i$\}} and \texttt{R\{$i$\}} \\
                 &  \texttt{true }  -- perform least order synthesis (default); \\
                 &  \texttt{false}  -- perform full order synthesis \\\hline
       \texttt{tcond} & maximum alowed condition number of the employed non-orthogonal transformations (Default: $10^4$).\\ \hline
 \texttt{FDSelect}   & integer vector with increasing elements containing the indices of the desired filters to be designed
                     (Default: $[\,1, \ldots, n_b\,]$)\\
                                        \hline
 \texttt{HDesign}   & $n_b$-dimensional cell array; \texttt{OPTIONS.HDesign\{$i$\}}, if not empty,  is a
                      full row rank design matrix employed for the
                      synthesis of the $i$-th fault detection filter
                      (Default: \texttt{[ ]})\\
                                        \hline
\end{longtable}}
\end{description}
\subsubsection*{Output data}
\begin{description}
\item
\texttt{Q} an $n_b$-dimensional cell array, with \texttt{Q\{$i$\}} containing  the resulting $i$-th filter in a standard state-space representation
\[
{\begin{aligned}
\lambda x_Q^{(i)}(t)  & =   A_Q^{(i)}x_Q^{(i)}(t)+ B_{Q_y}^{(i)}y(t)+ B_{Q_u}^{(i)}u(t) ,\\
r^{(i)}(t) & =  C_Q^{(i)} x_Q^{(i)}(t) + D_{Q_y}^{(i)}y(t)+ D_{Q_u}^{(i)}u(t) ,
\end{aligned}}
\]
where the residual signal $r^{(i)}(t)$ is a $q_i$-dimensional vector. For each system object \texttt{Q\{$i$\}}, two input groups \texttt{\bfseries 'outputs'} and \texttt{\bfseries 'controls'} are defined for $y(t)$ and $u(t)$, respectively, and the output group \texttt{\bfseries 'residuals'} is defined for the residual signal $r^{(i)}(t)$. \texttt{Q\{$i$\}} is empty if the index $i$ is not selected in \texttt{OPTIONS.FDSelect}.
\item
\texttt{R} an $n_b$-dimensional cell array, with \texttt{R\{$i$\}} containing  the resulting internal form of the $i$-th filter in a standard state-space representation
\[
{\begin{aligned}
\lambda x_R^{(i)}(t)  & =   A_Q^{(i)}x_R^{(i)}(t)+ B_{R_f}^{(i)}f(t)+ B_{R_w}^{(i)}w(t), \\
r^{(i)}(t) & =  C_Q^{(i)} x_R^{(i)}(t) + D_{R_f}^{(i)}f(t)+ D_{R_w}^{(i)}w(t).
\end{aligned}}
\]
The input groups \texttt{\bfseries 'faults'} and \texttt{\bfseries 'noise'} are defined for $f(t)$, and $w(t)$, respectively, and the output group \texttt{\bfseries 'residuals'} is defined for the residual signal $r^{(i)}(t)$. Note that the realizations of \texttt{Q\{$i$\}} and \texttt{R\{$i$\}} share the matrices $A_Q^{(i)}$ and $C_Q^{(i)}$.
\texttt{R\{$i$\}} is empty if the index $i$ is not selected in \texttt{OPTIONS.FDSelect}.
\item
\texttt{INFO} is a MATLAB structure containing additional information as follows:
\begin{center}
\begin{tabular}{|l|p{12cm}|} \hline \textbf{\texttt{INFO} fields} & \textbf{Description} \\ \hline
\texttt{tcond} & $n_b$-dimensional vector; \texttt{INFO.tcond}$(i)$ contains
                      the maximum of the condition numbers of the employed
                      non-orthogonal transformation matrices to determine
                      the $i$-th filter component \texttt{Q\{$i$\}}; a warning is
                      issued if any \texttt{INFO.tcond}$(i)$ $\geq$ \texttt{OPTIONS.tcond}.\\ \hline
\texttt{degs}     & $n_b$-dimensional cell array; if \texttt{OPTIONS.simple} = \texttt{true}, \texttt{INFO.degs\{$i$\}} contains the orders of the basis vectors of the employed simple nullspace basis for the synthesis of the $i$-th filter component \texttt{Q\{$i$\}};
if \texttt{OPTIONS.simple = false}, \texttt{INFO.degs\{$i$\}} contains the degrees of the basis vectors of an equivalent polynomial nullspace basis\\ \hline
 \texttt{HDesign}   & $n_b$-dimensional cell array; \texttt{INFO.HDesign\{$i$\}} is the $i$-th  design matrix actually employed for the
                      synthesis of the $i$-th fault detection filter \texttt{Q\{$i$\}}. \texttt{INFO.HDesin\{$i$\}} is empty if the index $i$ is not selected in \texttt{OPTIONS.FDSelect}. \\
                                        \hline
\end{tabular}
\end{center}
\end{description}

\subsubsection*{Method}

The \textbf{Procedure EFDI} from \cite[Sect.\ 5.4]{Varg17} is implemented, which
relies on the nullspace-based synthesis method proposed in \cite{Varg07d}. This method essentially determines each  filter $Q^{(i)}(\lambda)$ and its internal form $R^{(i)}(\lambda)$, by solving a suitably formulated EFDP  for a reduced system without control inputs, and with redefined disturbance and fault inputs. For this purpose, the function \texttt{efdisyn}  calls internally the function \texttt{efdsyn} to solve a suitably formulated EFDP for each specification (i.e., row) contained in the structure matrix $S_{FDI}$.
\index{M-functions!\texttt{\bfseries efdsyn}}

If the faulty system \texttt{SYSF} has the input-output form
\be\label{systemw1} {\mathbf{y}}(\lambda) =
G_u(\lambda){\mathbf{u}}(\lambda) +
G_d(\lambda){\mathbf{d}}(\lambda) +
G_f(\lambda){\mathbf{f}}(\lambda) +
G_w(\lambda){\mathbf{w}}(\lambda)
 \ee
and the $i$-th fault detection filter $Q^{(i)}(\lambda)$ contained in \texttt{Q\{$i$\}} has the input-output form
 \be\label{detec1i}
{\mathbf{r}}^{(i)}(\lambda) = Q^{(i)}(\lambda)\ba{c}
{\mathbf{y}}(\lambda)\\{\mathbf{u}}(\lambda)\ea  \, , \ee
then, taking into account the decoupling conditions (\ref{ens}), the resulting internal form of the $i$-th fault detection filter $R^{(i)}(\lambda)$, contained in \texttt{R\{$i$\}}, is

\be\label{resys1i} \hspace*{-5mm}{\mathbf{r}}^{(i)}(\lambda) = R^{(i)}(\lambda){\arraycolsep=1mm \ba{c}{\mathbf{f}}(\lambda)\\
{\mathbf{w}}(\lambda) \ea } =
R_f^{(i)}(\lambda){\mathbf{f}}(\lambda) +
R_w^{(i)}(\lambda){\mathbf{w}}(\lambda)  \, , \hspace*{-4mm}\ee
with $R^{(i)}(\lambda) = [\, R_f^{(i)}(\lambda)\; R_w^{(i)}(\lambda)\,]$ defined as
\be\label{resys2i} \ba{c|c} R_f^{(i)}(\lambda) & R_w^{(i)}(\lambda)  \ea :=
{\arraycolsep=1mm Q^{(i)}(\lambda)  \ba{c|c} G_f(\lambda) & G_w(\lambda)  \\
          0 & 0  \ea } \, .
         \ee
In accordance with (\ref{efdip}), the structure (row) vector corresponding to the zero and nonzero columns of the transfer function matrix  $R_f^{(i)}(\lambda)$ is equal to the $i$-th row of the specified $S_{FDI}$.


Each filter $Q^{(i)}(\lambda)$ is determined in the product form
\be\label{efdsyn:Qprodi} Q^{(i)}(\lambda) = \overline Q_1^{(i)}(\lambda)Q_1(\lambda) , \ee
where the factors are  determined as follows:
 \begin{itemize}
 \item[(a)] $Q_1(\lambda) = N_l(\lambda)$, with $N_l(\lambda)$ a $\big(p-r_d\big) \times (p+m_u)$ proper rational left nullspace basis satisfying $N_l(\lambda)\left[\begin{smallmatrix}G_u(\lambda) & G_d(\lambda)\\ I_{m_u} & 0 \end{smallmatrix}\right] = 0$, with $r_d := \text{rank}\, G_d(\lambda)$;
     \item[(b)] $\overline Q_1^{(i)}(\lambda)$ is the solution of a suitably formulated EFDP to achieve the specification contained in the $i$-th row of $S_{FDI}$. The function \texttt{efdsyn} is called for this purpose and internally uses a design matrix $H^{(i)}$, which can be specified in \texttt{OPTIONS.HDesign\{$i$\}} (see \textbf{Method} for \texttt{efdsyn}). The actually employed design matrix is returned in \texttt{INFO.HDesign\{$i$\}}.
 \end{itemize}
The computations to determine the  individual factors $\overline Q^{(i)}(\lambda)$ depend on the user's options (see \textbf{Method} for the function  \texttt{efdsyn}). In what follows, we only discuss shortly the computation of $Q_1(\lambda)$, performed only once as an initial reduction step.

If \texttt{OPTIONS.nullspace = true} or $m_d > 0$ or $E$ is singular, then $N_l(\lambda)$ is determined as a minimal proper nullspace basis. In this case, only orthogonal transformations are performed at this computational step.

If \texttt{OPTIONS.nullspace = false}, $m_d = 0$ and $E$ is invertible, then $N_l(\lambda) = [ \, I_{p} \; -G_u(\lambda)\,]$ is used, which corresponds to a full-order Luenberger observer. This option involves no numerical computations.

\subsubsection*{Examples}

\begin{example} \label{ex:Ex5.10}
This is \emph{Example} 5.10 from the book \cite{Varg17}, which considers a continuous-time system with triplex sensor redundancy on its measured scalar output, which we denote, respectively, by $y_1$, $y_2$ and $y_3$. Each output is related to the control and disturbance inputs by the input-output relation
\[ {\mathbf{y}}_i(s) = G_u(s){\mathbf{u}}(s) + G_d(s){\mathbf{d}}(s), \quad i = 1, 2, 3, \]
where $G_u(s)$ and $G_d(s)$ are $1\times m_u$ and $1\times m_d$ TFMs, respectively. We assume all three outputs are susceptible to additive sensor faults. Thus, the input-output model of the system with additive faults has the form
\[ {\mathbf{y}}(s) :=  \ba{c} {\mathbf{y}}_1(s)\\ {\mathbf{y}}_2(s)\\ {\mathbf{y}}_3(s) \ea = \ba{c} G_u(s)\\ G_u(s)\\ G_u(s) \ea{\mathbf{u}}(s) + \ba{c} G_d(s)\\ G_d(s)\\ G_d(s)  \ea{\mathbf{d}}(s)+
\ba{c} {\mathbf{f}}_1(s)\\ {\mathbf{f}}_2(s)\\ {\mathbf{f}}_3(s) \ea \, . 
\]
The maximal achievable structure matrix is
\[ S_{max} = \ba{ccc} 1 & 1 & 1 \\ 0 & 1 & 1 \\ 1 & 0 & 1 \\ 1 & 1 & 0 \ea \, .\]
If we assume that no simultaneous sensor faults occur, then we can target to solve an EFDIP for the structure matrix
\[ S_{FDI} = \ba{ccc} 0 & 1 & 1 \\ 1 & 0 & 1 \\ 1 & 1 & 0 \ea \, ,\]
where the columns of $S_{FDI}$ codify the desired fault signatures.

The resulting least order overall FDI filter has the generic form (i.e, independent of the numbers of control and disturbance inputs)
\be\label{voting} Q(s) = \ba{c} Q^{(1)}(s) \\ Q^{(2)}(s) \\ Q^{(3)}(s) \ea
= \ba{rrrccc} 0 & 1 & -1 & 0 & \cdots & 0\\-1 & 0 & 1 & 0 & \cdots & 0\\1 & -1 & 0 & 0& \cdots & 0\ea \ee
and the corresponding overall internal form is
\be\label{votingif} R_f(s) = \ba{c} R_f^{(1)}(s) \\ R_f^{(2)}(s) \\ R_f^{(3)}(s) \ea
= \ba{rrr} 0 & 1 & -1 \\-1 & 0 & 1 \\1 & -1 & 0 \ea \ee

\newpage
\vspace*{-10mm}
\begin{verbatim}
% Example - Solution of an EFDIP

p = 3; mf = 3;   % enter output and fault vector dimensions
% generate random dimensions for system order and input vectors
rng('default')
nu = floor(1+4*rand); mu = floor(1+4*rand);
nd = floor(1+4*rand); md = floor(1+4*rand);
% define random Gu(s) and Gd(s) with triplex sensor redundancy
% and Gf(s) = I for triplex sensor faults
Gu = ones(3,1)*rss(nu,1,mu); % enter Gu(s) in state-space form
Gd = ones(3,1)*rss(nd,1,md); % enter Gd(s) in state-space form

% build synthesis model with sensor faults
sysf = fdimodset([Gu Gd],struct('c',1:mu,'d',mu+(1:md),'fs',1:3));

SFDI = [ 0 1 1; 1 0 1; 1 1 0] > 0;  % enter structure matrix

% set options for least order synthesis with EFDISYN
options = struct('tol',1.e-7,'sdeg',-1,'rdim',1,'SFDI',SFDI);
[Qt,Rft] = efdisyn(sysf,options);

% normalize Q and Rf to match example
scale = sign([ Rft{1}.d(1,2) Rft{2}.d(1,3) Rft{3}.d(1,1)]);
for i = 1:3, Qt{i} = scale(i)*Qt{i}; Rft{i} = scale(i)*Rft{i}; end
Q = vertcat(Qt{:});  Rf = vertcat(Rft{:});
Q = set(Q,'InputName',['y1';'y2';'y3';strseq('u',1:mu)],...
          'OutputName',['r1';'r2';'r3'])
Rf = set(Rf,'InputName',['f1';'f2';'f3'],'OutputName',['r1';'r2';'r3'])

% check synthesis conditions: Q*[Gu Gd;I 0] = 0 and Q*[Gf; 0] = Rf
syse = [sysf;eye(mu,mu+md+mf)];  % form Ge = [Gu Gd Gf;I 0 0];
norm_Ru_Rd = norm(Q*syse(:,{'controls','disturbances'}),inf)
norm_rez = norm(Q*syse(:,'faults')-Rf,inf)

% check strong fault detectability
[S_strong,abs_dcgains] = fdisspec(Rft)

% determine achieved fault sensitivity conditions
FSCOND = fdifscond(Rft,0,SFDI)

% evaluate step responses
set(Rf,'InputName',strseq('f_',1:mf),'OutputName',strseq('r_',1:size(SFDI,1)));
step(Rf);
title('Step responses from the fault inputs')
ylabel('Residuals')
\end{verbatim}

\end{example}
\begin{example}\label{ex:Yuan3}
This is the  example of \cite{Yuan97}, already considered in Example~\ref{ex:Yuan}.  Using \texttt{efdisyn},   we can easily determine  a bank of least order fault detection filters, which achieve the computed maximal weak structure matrix $S_{weak}$. With the default least order synthesis option, we obtain a bank of 18 filters, each one of order one or two. The overall filters $Q(s)$ and $R_f(s)$ obtained by stacking the 18 component filters have state-space realizations of order 32, which are usually non-minimal. Typically, minimal realizations of orders about 20 can be computed for each of these filters. The bank of 12 component filters ensuring strong fault detection can easily be picked-out from the computed filters.

In this example, we show that using the pole assignment feature, the overall filters $Q(s)$ and $R_f(s)$ can be determined with minimal realizations of order 6, which is probably the least achievable global order. To arrive to this order, we enforce the same dynamics for all component filters by assigning, for example, all poles of the component filters to lie in the set $\{-1,-2\}$.
The resulting least order of the overall filter $Q(s)$ can be easily read-out from the plot of its Hankel-singular values shown in Fig.~\ref{fig:hsvQ}. The synthesis procedure also ensures that $R_f(s)$, and even of the joint overall filter $[\, Q(s)\;R_f(s)\,]$ have minimal realizations of order 6!. It is also straightforward to check that the resulting weak structure matrix of $R_f(s)$ and $S_{weak}$ coincide.

\begin{verbatim}
% Example of Yuan et al. IJC (1997)
rng('default');  % make results reproducible
p = 3; mu = 1; mf = 8;
A = [ -1 1 0 0; 1 -2 1 0; 0 1 -2 1; 0 0 1 -2  ];
Bu = [1 0 0 0]';
Bf = [ 1 0 0 0 1 0 0 0; 0 1 0 0 -1 1 0 0; 0 0 1 0 0 -1 1 0; 0 0 0 1 0 0 -1 1];
C = [ 1 0 0 0; 0 0 1 0; 0 0 0 1];
Du = zeros(p,mu); Df = zeros(p,mf);
% setup the model with additive faults
sysf = fdimodset(ss(A,[Bu Bf],C,[Du Df]),struct('c',1:mu,'f',mu+(1:mf)));

% compute the achievable weak specifications
opt = struct('tol',1.e-7,'FDTol',1.e-5);
S_weak = fdigenspec(sysf,opt);

% set options for least order synthesis with pole assignment
options = struct('tol',1.e-7,'sdeg',-5,'smarg',-5,'poles',[-1 -2],...
                    'FDTol',0.0001,'rdim',1,'simple',false,'SFDI',S_weak);
[Q,Rf] = efdisyn(sysf,options);

% minimal order of the overall filter is 6!
hsvd(vertcat(Q{:})) % only the first 6 Hankel singular values are nonzero

% check that the achieved structure matrix is the desired one
isequal(S_weak,fditspec(Rf))
\end{verbatim}

\begin{figure}[h]
\begin{center}
\includegraphics[height=10cm]{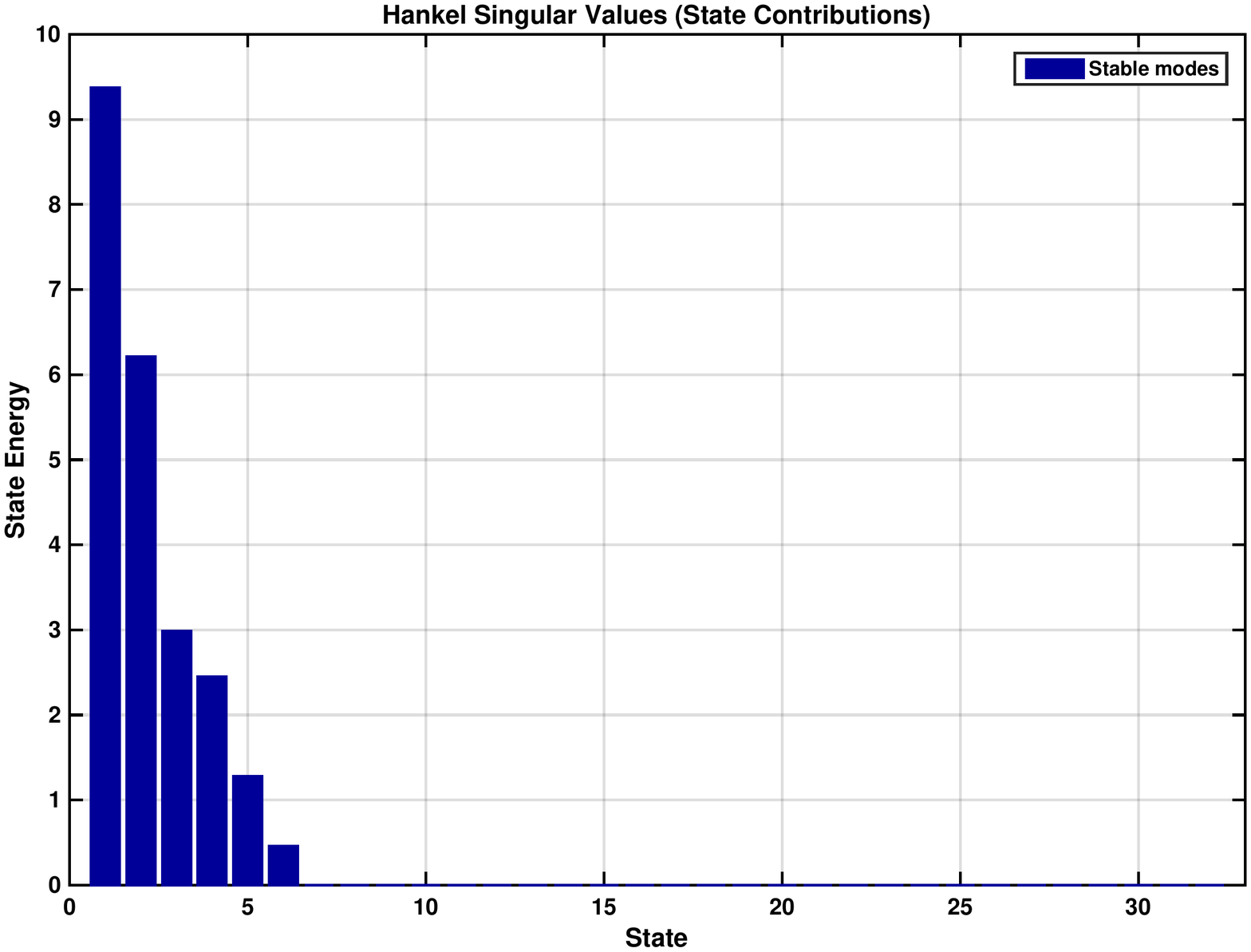}\vspace*{-5mm}
\caption{Hankel-singular values of the overall filter $Q(s)$.}
\label{fig:hsvQ}
 \end{center}
\end{figure}

\end{example}

\newpage
\subsubsection{\texttt{\bfseries afdisyn}}
\index{M-functions!\texttt{\bfseries afdisyn}}
\index{fault detection and isolation problem!approximate (AFDIP)}
\subsubsection*{Syntax}
\begin{verbatim}
[Q,R,INFO] = afdisyn(SYSF,OPTIONS)
\end{verbatim}

\subsubsection*{Description}
\texttt{\bfseries afdisyn} solves   the \emph{approximate fault detection and isolation problem} (AFDIP) (see Section \ref{sec:AFDIP}), for a given LTI system \texttt{SYSF} with additive faults and a given structure matrix $S_{FDI}$ (specified via the \texttt{OPTIONS} structure). Two banks of stable and proper filters are computed in the $n_b$-dimensional cell arrays \texttt{Q} and \texttt{R}, where $n_b$ is the number of specifications contained in $S_{FDI}$ (i.e., the number of rows of the structure matrix $S_{FDI}$).  \texttt{Q\{i\}} contains the $i$-th fault detection filter (\ref{ri_fdip}) in the overall solution (\ref{qbank}) of the AFDIP  and \texttt{R\{i\}} contains its internal form.

\subsubsection*{Input data}
\begin{description}
\item
\texttt{SYSF} is a  LTI system  in the state-space form
\be\label{afdisyn:sysss}
{\begin{aligned}
E\lambda x(t)  & =   Ax(t)+ B_u u(t)+ B_d d(t) + B_f f(t)+ B_w w(t)   , \\
y(t) & =  C x(t) + D_u u(t)+ D_d d(t) + D_f f(t) + D_w w(t) ,
\end{aligned}}
\ee
where any of the inputs components $u(t)$, $d(t)$, $f(t)$, or $w(t)$  can be void.  For the system \texttt{SYSF}, the input groups for $u(t)$, $d(t)$, $f(t)$, and $w(t)$ have the standard names \texttt{\bfseries 'controls'}, \texttt{\bfseries 'disturbances'}, \texttt{\bfseries 'faults'}, and \texttt{\bfseries 'noise'}, respectively.
\item
 \texttt{OPTIONS} is a MATLAB structure used to specify various synthesis  options and has the following fields:
{\setlength\LTleft{30pt}\begin{longtable}{|l|p{11.6cm}|} \hline \textbf{\texttt{OPTIONS} fields} & \textbf{Description} \\ \hline
       \texttt{SFDI}       & the desired  structure matrix $S_{FDI}$ with $n_b$ rows to solve the AFDIP\newline
                 (Default: \texttt{[1...1]}, i.e., solve an approximate fault
                                detection problem)\\ \hline
       \texttt{tol}       & relative tolerance for rank computations \newline
                 (Default: internally computed)\\ \hline
       \texttt{tolmin}       & absolute tolerance for observability tests \newline
                 (Default: internally computed)\\ \hline
       \texttt{FDTol}     & threshold for fault detectability checks
                 (Default: 0.0001)\\ \hline
       \texttt{FDGainTol} & threshold for strong fault detectability checks (Default: 0.01)\\ \hline
       \texttt{rdim}       & vector, whose $i$-th component $q_i$, specifies the desired number of residual outputs for the $i$-th component filters \texttt{Q\{$i$\}} and \texttt{R\{$i$\}}; if \texttt{OPTIONS.rdim} is a scalar $q$, then  a vector with all components $q_i = q$ is assumed.     \\
                 & (Default: \hspace*{-2.5mm}\begin{tabular}[t]{p{10cm}} \texttt{[ ]}, in which case $q_i = q_{i,1}+q_{i,2}$, with $q_{i,1}$ and $q_{i,2}$ selected   taking into account $r_w^{(i)}$, the rank of the transfer function matrix from the noise input to the reduced system output employed to determine \texttt{Q\{$i$\}},   as follows: \newline
                 if \texttt{OPTIONS.HDesign\{$i$\}} is empty, then \\ \hspace*{1em} $q_{i,1} = \min\big(1,r_w^{(i)}\big)$, if \texttt{OPTIONS.minimal} = \texttt{true}, or \\ \hspace*{1em} $q_{i,1} = r_w^{(i)}$ if \texttt{OPTIONS.minimal} = \texttt{false}; \\ if \texttt{OPTIONS.HDesign\{$i$\}} is nonempty, then $q_{i,1}$ is the row dimension of the design matrix contained in \texttt{OPTIONS.HDesign\{$i$\}}; \\
                               if \texttt{OPTIONS.HDesign2\{$i$\}} is empty, then \\ \hspace*{1em} $q_{i,2} = 1-\min\big(1,r_w^{(i)}\big)$, if \texttt{OPTIONS.minimal} = \texttt{true}, or \\ \hspace*{1em} $q_{i,2}$ is set to its maximum achievable value, \newline
                               \hspace*{2.8em} if \texttt{OPTIONS.minimal} = \texttt{false}; \\
                               if \texttt{OPTIONS.HDesign2\{$i$\}} is nonempty, then $q_{i,2}$ is the \\ row dimension of the design matrix contained in \\ \texttt{OPTIONS.HDesign2\{$i$\}}.)
                                \end{tabular}\\ \hline
       \texttt{FDFreq}  &  vector of real frequency values for strong detectability checks \newline (Default: \texttt{[ ]})  \\ \hline
       \texttt{smarg}   & stability margin for the poles of the component filters \texttt{Q\{$i$\}} and \texttt{R\{$i$\}}\\
                 &  (Default: \texttt{-sqrt(eps)} for a continuous-time system \texttt{SYSF}; \\
                 &  \hspace*{4.5em}\texttt{1-sqrt(eps)} for a discrete-time system \texttt{SYSF}).\\ \hline \pagebreak[4] \hline
      \texttt{sdeg}   & prescribed stability degree for the poles of the component filters \texttt{Q\{$i$\}} and \texttt{R\{$i$\}}\\
                 &  (Default: $-0.05$ for a continuous-time system \texttt{SYSF}; \\
                 &  \hspace*{4.8em} $0.95$ for a discrete-time system \texttt{SYSF}).\\ \hline
       \texttt{poles}   & complex vector containing a complex conjugate set of desired poles (within the stability domain) to be assigned for the component filters \texttt{Q\{$i$\}} and \texttt{R\{$i$\}} (Default: \texttt{[ ]})
                   \\\hline
 \texttt{nullspace}   & option to use a specific proper nullspace basis to be employed at the initial reduction step\\
                 & \hspace*{-.9mm}{\tabcolsep=0.7mm\begin{tabular}[t]{lcp{10cm}} \texttt{true } &--& use a minimal proper basis (default); \\
                 \texttt{false} &--& use a full-order observer based basis (see \textbf{Method}). \newline \emph{Note:} This option can  only be used if no disturbance inputs are present in (\ref{afdisyn:sysss}) and $E$ is invertible.
                 \end{tabular}} \\  \hline
       \texttt{simple} & option to employ simple proper bases for the synthesis of the component filters \texttt{Q\{$i$\}} and \texttt{R\{$i$\}} \\
                 &  \texttt{true }  -- use simple bases; \\
                 &  \texttt{false}  -- use non-simple bases (default)\\\hline
       \texttt{minimal} & option to perform least order synthesis of the component filters \\
                 &  \texttt{true }  -- perform least order synthesis (default); \\
                 &  \texttt{false}  -- perform full order synthesis \\[2mm] \hline
       \texttt{exact} & option to perform exact filter synthesis \\
                 &  \texttt{true }  -- perform exact synthesis (i.e., no optimization performed); \\
                 &  \texttt{false}  -- perform approximate synthesis (default). \\\hline
\texttt{freq}   & complex frequency value to be employed to check
                      the full row rank admissibility conditions within the function \texttt{afdsyn} (see \textbf{Method} for \texttt{afdsyn}) 
                      (Default:\texttt{[ ]}, i.e., a randomly generated frequency). \\
                                        \hline
       \texttt{tcond} & maximum alowed condition number of the employed non-orthogonal transformations (Default: $10^4$).\\ \hline
 \texttt{FDSelect}   & integer vector with increasing elements containing the indices of the desired filters to be designed
                     (Default: $[\,1, \ldots, n_b\,]$)\\
                                        \hline
 \texttt{HDesign}   & $n_b$-dimensional cell array; \texttt{OPTIONS.HDesign\{$i$\}}, if not empty,  is a
                      full row rank design matrix $H_1^{(i)}$ employed for the
                      synthesis of the $i$-th filter components $Q^{(1,i)}(\lambda)$ and  $R^{(1,i)}(\lambda)$ (see \textbf{Method}) \newline
                      (Default: \texttt{[ ]})\\
                                        \hline
 \texttt{HDesign2}   & $n_b$-dimensional cell array; \texttt{OPTIONS.HDesign2\{$i$\}}, if not empty,  is a
                      full row rank design matrix $H_2^{(i)}$ employed for the
                      synthesis of the $i$-th filter components $Q^{(2,i)}(\lambda)$ and  $R^{(2,i)}(\lambda)$ (see \textbf{Method}) \newline
                      (Default: \texttt{[ ]})\\
                                        \hline
\texttt{gamma}   & upper bound on the resulting $\|R_w^{(i)}(\lambda)\|_\infty$ (see \textbf{Method})  (Default: 1) \\ \hline
\texttt{epsreg}   & regularization parameter used in \textbf{\texttt{afdsyn}} (Default: 0.1) \\ \hline
\pagebreak[4]\hline
\texttt{sdegzer}   & prescribed stability degree for zeros shifting \\ &  (Default: $-0.05$ for a continuous-time system \texttt{SYSF}; \\
                 &  \hspace*{4.8em} $0.95$ for a discrete-time system \texttt{SYSF}).\\ \hline
\texttt{nonstd}   & option to handle nonstandard optimization problems used in  \textbf{\texttt{afdsyn}}:\\
                 &  1 -- use the quasi-co-outer--co-inner factorization (default); \\
                 &  2 -- use the modified co-outer--co-inner factorization
                          with the \\
                 & \hspace{1.7em}regularization parameter \texttt{OPTIONS.epsreg};  \\
                 &  3 -- use the Wiener-Hopf type co-outer--co-inner
                          factorization.  \\
                 &     4 -- use the Wiener-Hopf type co-outer-co-inner factorization with\\
                 & \hspace{1.7em}zero shifting of the  non-minimum phase factor using the\\
                 & \hspace{1.7em}stabilization parameter \texttt{OPTIONS.sdegzer} \\
                 &     5 -- use the Wiener-Hopf type co-outer-co-inner factorization with \\
                 & \hspace{1.7em}the regularization of the non-minimum phase factor using the \\
                 & \hspace{1.7em}regularization parameter \texttt{OPTIONS.epsreg}  \\                                       \hline
\end{longtable}}
\end{description}
\subsubsection*{Output data}
\begin{description}
\item
\texttt{Q} is an $n_b$-dimensional cell array, with \texttt{Q\{$i$\}} containing the resulting $i$-th filter in a standard state-space representation
\[
{\begin{aligned}
\lambda x_Q^{(i)}(t)  & =   A_Q^{(i)}x_Q^{(i)}(t)+ B_{Q_y}^{(i)}y(t)+ B_{Q_u}^{(i)}u(t) ,\\
r^{(i)}(t) & =  C_Q^{(i)} x_Q^{(i)}(t) + D_{Q_y}^{(i)}y(t)+ D_{Q_u}^{(i)}u(t) ,
\end{aligned}}
\]
where the residual signal $r^{(i)}(t)$ is a $q_i$-dimensional vector. For each system object \texttt{Q\{$i$\}}, two input groups \texttt{\bfseries 'outputs'} and \texttt{\bfseries 'controls'} are defined for $y(t)$ and $u(t)$, respectively, and the output group \texttt{\bfseries 'residuals'} is defined for the residual signal $r^{(i)}(t)$. \texttt{Q\{$i$\}} is empty if the index $i$ is not selected in \texttt{OPTIONS.FDSelect}.
\item
\texttt{R} is an $n_b$-dimensional cell array, with \texttt{R\{$i$\}} containing  the resulting internal form of the $i$-th filter in a standard state-space representation
\[
{\begin{aligned}
\lambda x_R^{(i)}(t)  & =   A_Q^{(i)}x_R^{(i)}(t)+ B_{R_f}^{(i)}f(t)+ B_{R_w}^{(i)}w(t), \\
r^{(i)}(t) & =  C_Q^{(i)} x_R^{(i)}(t) + D_{R_f}^{(i)}f(t)+ D_{R_w}^{(i)}w(t).
\end{aligned}}
\]
The input groups \texttt{\bfseries 'faults'} and \texttt{\bfseries 'noise'} are defined for $f(t)$, and $w(t)$, respectively, and the output group \texttt{\bfseries 'residuals'} is defined for the residual signal $r^{(i)}(t)$. Note that the realizations of \texttt{Q\{$i$\}} and \texttt{R\{$i$\}} share the matrices $A_Q^{(i)}$ and $C_Q^{(i)}$.
\texttt{R\{$i$\}} is empty if the index $i$ is not selected in \texttt{OPTIONS.FDSelect}.
\item
\texttt{INFO} is a MATLAB structure containing additional information as follows:
\begin{center}
\begin{tabular}{|l|p{12cm}|} \hline \textbf{\texttt{INFO} fields} & \textbf{Description} \\ \hline
\texttt{tcond} & $n_b$-dimensional vector; \texttt{INFO.tcond}$(i)$ contains
                      the maximum of the condition numbers of the employed
                      non-orthogonal transformation matrices to determine
                      the $i$-th filter component \texttt{Q\{$i$\}}; a warning is
                      issued if any \texttt{INFO.tcond}$(i)$ $\geq$ \texttt{OPTIONS.tcond}.\\ \hline
 \texttt{HDesign}   & $n_b$-dimensional cell array; \texttt{INFO.HDesign\{$i$\}} is the $i$-th  design matrix actually employed for the
                      synthesis of the filter component $Q_1^{(i)}$ of the $i$-th fault detection filter \texttt{Q\{$i$\}} (see \textbf{Method}). \texttt{INFO.HDesign\{$i$\}} is empty if the index $i$ is not selected in \texttt{OPTIONS.FDSelect}. \\
                                        \hline
 \texttt{HDesign2}   & $n_b$-dimensional cell array; \texttt{INFO.HDesign2\{$i$\}} is the $i$-th  design matrix actually employed for the
                      synthesis of the filter component $Q_2^{(i)}$ of the $i$-th fault detection filter \texttt{Q\{$i$\}} (see \textbf{Method}). \texttt{INFO.HDesign2\{$i$\}} is empty if the index $i$ is not selected in \texttt{OPTIONS.FDSelect}. \\
                                        \hline
\texttt{freq}   & complex frequency value employed to check the full row rank
                      admissibility conditions within the function \texttt{afdsyn} \\
                                        \hline
\texttt{gap}     & $n_b$-dimensional cell array; \texttt{INFO.gap$(i)$} contains the $i$-th gap $\eta_i$ resulted by calling \texttt{afdsyn} to
                      determine \texttt{Q\{$i$\}} (see \textbf{Method}).\\ \hline
\end{tabular}
\end{center}
\end{description}

\subsubsection*{Method}

The \textbf{Procedure AFDI} from \cite[Sect.\ 5.4]{Varg17} is implemented, which
relies on the optimization-based synthesis method proposed in \cite{Varg09b}. Each  filter $Q^{(i)}(\lambda)$ and its internal form $R^{(i)}(\lambda)$, are determined by solving a suitably formulated AFDP  for a reduced system without control inputs, and with redefined disturbance and fault inputs. For this purpose, the function \texttt{afdisyn}  calls internally the function \texttt{afdsyn} to solve a suitably formulated AFDP for each specification (i.e., row) contained in the structure matrix $S_{FDI}$.
\index{M-functions!\texttt{\bfseries afdsyn}}

If the faulty system \texttt{SYSF} has the input-output form
\be\label{afdi:systemw1} {\mathbf{y}}(\lambda) =
G_u(\lambda){\mathbf{u}}(\lambda) +
G_d(\lambda){\mathbf{d}}(\lambda) +
G_f(\lambda){\mathbf{f}}(\lambda) +
G_w(\lambda){\mathbf{w}}(\lambda)
 \ee
and the $i$-th fault detection filter $Q^{(i)}(\lambda)$ contained in \texttt{Q\{$i$\}} has the input-output form
 \be\label{afdi:detec1i}
{\mathbf{r}}^{(i)}(\lambda) = Q^{(i)}(\lambda)\ba{c}
{\mathbf{y}}(\lambda)\\{\mathbf{u}}(\lambda)\ea  \, , \ee
then, taking into account the decoupling conditions (\ref{ens}), the resulting internal form of the $i$-th fault detection filter $R^{(i)}(\lambda)$, contained in \texttt{R\{$i$\}}, is

\be\label{afdi:resys1i} \hspace*{-5mm}{\mathbf{r}}^{(i)}(\lambda) = R^{(i)}(\lambda){\arraycolsep=1mm \ba{c}{\mathbf{f}}(\lambda)\\
{\mathbf{w}}(\lambda) \ea } =
R_f^{(i)}(\lambda){\mathbf{f}}(\lambda) +
R_w^{(i)}(\lambda){\mathbf{w}}(\lambda)  \, , \hspace*{-4mm}\ee
with $R^{(i)}(\lambda) = [\, R_f^{(i)}(\lambda)\; R_w^{(i)}(\lambda)\,]$ defined as
\be\label{afdi:resys2i} \ba{c|c} R_f^{(i)}(\lambda) & R_w^{(i)}(\lambda)  \ea :=
{\arraycolsep=1mm Q^{(i)}(\lambda)  \ba{c|c} G_f(\lambda) & G_w(\lambda)  \\
          0 & 0  \ea } \, .
         \ee
When determining $Q^{(i)}(\lambda)$ it is first attempted to solve the \emph{strict} AFDIP (\ref{afdip-sr1}), such   that the structure vector corresponding to the zero and nonzero columns of $R_f^{(i)}(\lambda)$ is equal to the $i$-th row of the specified $S_{FDI}$. If this is not achievable, then the \emph{soft} AFDIP (\ref{afdip-sr}) is attempted to be solved.

\index{performance evaluation!fault detection and isolation!fault-to-noise gap}
The achieved gap $\eta_i$ is computed as $\eta_i = \big\|\overline R_f^{(i)}(\lambda)\big\|_{\infty -}/\big\|\big[\widetilde R_f^{(i)}(\lambda)\;R_w^{(i)}(\lambda)\big]\big\|_\infty$, where $\overline R_f^{(i)}(\lambda)$ is formed from the columns of $R_f^{(i)}(\lambda)$ corresponding to nonzero entries in the $i$-th row of $S_{FDI}$ and $\widetilde R_f^{(i)}(\lambda)$ is formed from the columns of $R_f^{(i)}(\lambda)$ corresponding to zero entries in the $i$-th row of $S_{FDI}$. If \texttt{OPTIONS.FDFreq} is nonempty, then $\big\|\overline R_f^{(i)}(\lambda)\big\|_{\infty -}$ is only evaluated over the frequency values contained in \texttt{OPTIONS.FDFreq}. The achieved gaps are returned in \texttt{INFO.gap}.

Each filter $Q^{(i)}(\lambda)$ is determined in the product form
\be\label{afdsyn:Qprodi} Q^{(i)}(\lambda) = \overline Q_1^{(i)}(\lambda)Q_1(\lambda) , \ee
where the factors are  determined as follows:
 \begin{itemize}
 \item[(a)] $Q_1(\lambda) = N_l(\lambda)$, with $N_l(\lambda)$ a $\big(p-r_d\big) \times (p+m_u)$ proper rational left nullspace basis satisfying $N_l(\lambda)\left[\begin{smallmatrix}G_u(\lambda) & G_d(\lambda)\\ I_{m_u} & 0 \end{smallmatrix}\right] = 0$, with $r_d := \text{rank}\, G_d(\lambda)$;
     \item[(b)] $\overline Q_1^{(i)}(\lambda)$ is the solution of a suitably formulated AFDP to achieve the specification contained in the $i$-th row of $S_{FDI}$. The function \texttt{afdsyn} is called for this purpose and internally uses the design matrices $H_1^{(i)}$, which can be specified in \texttt{OPTIONS.HDesign\{$i$\}}, and $H_2^{(i)}$, which can be specified in \texttt{OPTIONS.HDesign2\{$i$\}}  (see \textbf{Method} for \texttt{afdsyn}). The actually employed design matrices are returned in \texttt{INFO.HDesign\{$i$\}} and \texttt{INFO.HDesign2\{$i$\}}, respectively.
 \end{itemize}

Each factor $\overline Q_1^{(i)}(\lambda)$ is determined in the partitioned form
         \[ \overline Q_1^{(i)}(\lambda) = \ba{cc} \overline Q_1^{(i,1)}(\lambda) \\ \overline Q_1^{(i,2)}(\lambda) \ea , \]
where the computation of  individual factors $\overline Q_1^{(i,1)}(\lambda)$ and $\overline Q_1^{(i,2)}(\lambda)$ depends on the user's options (see \textbf{Method} for the function  \texttt{afdsyn}). In what follows, we only discuss shortly the computation of $Q_1(\lambda)$, performed only once as an initial reduction step.

If \texttt{OPTIONS.nullspace = true} or $m_d > 0$ or $E$ is singular, then $N_l(\lambda)$ is determined as a minimal proper nullspace basis. In this case, only orthogonal transformations are performed at this computational step.

If \texttt{OPTIONS.nullspace = false}, $m_d = 0$ and $E$ is invertible, then $N_l(\lambda) = [ \, I_{p} \; -G_u(\lambda)\,]$ is used, which corresponds to a full-order Luenberger observer. This option involves no numerical computations.

 \subsubsection*{Example}

\begin{example}
This is \emph{Example} 5.3 from the book \cite{Varg17}, with the disturbances redefined as noise inputs and by adding a sensor fault for the first output measurement. The TFMs of the system are:
\[ G_u(s) = {\arraycolsep=1mm\ba{c}  \displaystyle\frac{s+1}{s+2} \\ \\[-2mm] \displaystyle\frac{s+2}{s+3} \ea, \quad G_d(s) = 0, \quad G_f(s) = [\, G_u(s)\; I\,], \quad G_w(s) = \ba{c}  \displaystyle\frac{s-1}{s+2} \\ \\[-2mm]0 \ea} ,\]
where the fault input $f_1$ corresponds to an additive actuator fault, while the fault inputs $f_2$  and $f_3$ describe additive sensor faults in the outputs $y_1$ and $y_2$, respectively.
The transfer function matrix $G_w(s)$ is non-minimum phase, having an unstable zero at 1. Interestingly, the EFDIP formulated with $G_d(s) = G_w(s)$ is not solvable.

The maximal achievable structure matrix (for the EFDIP) is
\[ S_{max} = \ba{ccc} 1 & 1 & 1 \\ 0 & 1 & 1 \\ 1 & 0 & 1 \\ 1 & 1 & 0 \ea \, .\]
We assume that no simultaneous faults occur, and thus we can target to solve an AFDIP for the structure matrix
\[ S_{FDI} = \ba{ccc} 0 & 1 & 1 \\ 1 & 0 & 1 \\ 1 & 1 & 0 \ea \, ,\]
where the columns of $S_{FDI}$ codify the desired fault signatures.

We want to design a bank of three detection filters $Q^{(i)}(s)$, $i = 1, 2, 3$ which fulfill for each $i$:
\begin{itemize}
\item[--] the decoupling condition: $Q^{(i)}(s)\left[\begin{smallmatrix} G_u(s) & G_d(s) \\ I_{m_u} & 0  \end{smallmatrix}\right] = 0$;
\item[--] the fault isolability condition: $S_{R_f^{(i)}}$ is equal to the $i$-th row of $S_{FDI}$;
\item[--]  the maximization of the noise attenuation gap: $\eta_i := \|\overline R_f^{(i)}(s)\|_{\infty -}/\|R_w^{(i)}(s)\|_\infty = \max$, where $\overline R_f^{(i)}(s)$ is formed from the columns of $R_f^{(i)}(s)$ corresponding to nonzero entries in the $i$-th row of $S_{FDI}$.
\end{itemize}
The results computed with the following script are:\\
-- $\eta_1 = 1.5$ with
\[ Q^{(1)}(s) = {\ba{ccc} \displaystyle\frac{s+2}{s+1}  & -\displaystyle\frac{s+3}{s+2} &  0 \ea , \quad R_f^{(1)}(s) = \ba{ccc} 0 &  \displaystyle\frac{s+2}{s+1} & -\displaystyle\frac{s+3}{s+2} \ea }, \quad R_w^{(1)}(s) = \displaystyle\frac{s-1}{s+1}\; ; \]
-- $\eta_2 = \infty$ with
\[ Q^{(2)}(s) = {\ba{ccc} 0  & 1 &  -\displaystyle\frac{s+2}{s+3} \ea , \quad R_f^{(2)}(s) = \ba{ccc} \displaystyle\frac{s+2}{s+3} &  0 & 1 \ea }, \quad R_w^{(2)}(s) = 0\; ; \]
-- $\eta_3 = 1$ with
\[ Q^{(3)}(s) = {\ba{ccc} \displaystyle\frac{s+2}{s+1}  & 0 &  -1 \ea , \quad R_f^{(3)}(s) = \ba{ccc} 1 &  \displaystyle\frac{s+2}{s+1} & 0 \ea }, \quad R_w^{(3)}(s) = \displaystyle\frac{s-1}{s+1} \;. \]

\begin{verbatim}
% Example - Solution of an approximate fault detection problem (AFDIP)

% Example 5.3 of (V,2017) (modified)
s = tf('s'); % define the Laplace variable s
mu = 1; mw = 1; p = 2; mf = mu+p; % set dimensions
Gu = [(s+1)/(s+2); (s+2)/(s+3)];  % enter Gu(s)
Gw = [(s-1)/(s+2); 0];            % enter Gw(s)

% build the model with additive faults having Gf(s) = [Gu(s) eye(p)];
sysf = fdimodset(ss([Gu Gw]),struct('c',1,'f',1,'fs',1:2,'n',2));

% select SFDI
S = fdigenspec(sysf); SFDI = S(sum(S,2)==2,:)
nb = size(SFDI,1);

% perform synthesis with AFDISYN
options = struct('tol',1.e-7,'smarg',-3,...
                 'sdeg',-3,'SFDI',SFDI);
[Q,R,info] = afdisyn(sysf,options);

% display the implementation form Q{i}(s) and the internal forms
% Rf{i}(s) and Rw{i}(s) of the resulting fault detection filters
minreal(tf(Q{1})), minreal(tf(R{1}(:,'faults'))), minreal(tf(R{1}(:,'noise')))
minreal(tf(Q{2})), minreal(tf(R{2}(:,'faults'))), tf(R{2}(:,'noise'))
minreal(tf(Q{3})), minreal(tf(R{3}(:,'faults'))), tf(R{3}(:,'noise'))

% check the resulting gaps
format short e
info.gap
gap = fdif2ngap(R,[],SFDI)

% simulate step responses from fault and noise inputs
inpnames = {'f_1','f_2','f_3','w'};
outnames = {'r_1','r_2','r_3'};
Rtot = vertcat(R{:});
set(Rtot,'InputName',inpnames,'OutputName',outnames);
step(Rtot); ylabel('Residuals')
title('Step responses from fault and noise inputs')
\end{verbatim}

\end{example}

\pagebreak[4]
\subsubsection{\texttt{\bfseries emmsyn}}
\index{M-functions!\texttt{\bfseries emmsyn}}
\index{model-matching problem!a@exact (EMMP)}
\subsubsection*{Syntax}
\begin{verbatim}
[Q,R,INFO] = emmsyn(SYSF,SYSR,OPTIONS)
\end{verbatim}

\subsubsection*{Description}
\texttt{\bfseries emmsyn} solves the \emph{exact model matching problem} (EMMP) (as formulated in the more general form (\ref{emm:resys}) in Remark~\ref{rem:gemm}; see Section \ref{sec:EMMP}), for a given LTI system \texttt{SYSF} with additive faults and a given stable reference filter \texttt{SYSR}. Two stable and proper filters, \texttt{Q} and \texttt{R}, are computed, where \texttt{Q} contains the fault detection and isolation filter representing the solution of the EMMP,  and \texttt{R} contains its internal form.

\subsubsection*{Input data}
\begin{description}
\item
\texttt{SYSF} is a  LTI system  in the state-space form
\be\label{emmsyn:sysss}
{\begin{aligned}
E\lambda x(t)  & =   Ax(t)+ B_u u(t)+ B_d d(t) + B_f f(t)+ B_w w(t)   , \\
y(t) & =  C x(t) + D_u u(t)+ D_d d(t) + D_f f(t) + D_w w(t) ,
\end{aligned}}
\ee
where any of the inputs components $u(t)$, $d(t)$, $f(t)$, or $w(t)$  can be void.  For the system \texttt{SYSF}, the input groups for $u(t)$, $d(t)$, $f(t)$, and $w(t)$ have the standard names \texttt{\bfseries 'controls'}, \texttt{\bfseries 'disturbances'}, \texttt{\bfseries 'faults'}, and \texttt{\bfseries 'noise'}, respectively.

\item
\texttt{SYSR} is a proper and stable  LTI system  in the state-space form
\be\label{emmsyn:sysrss}
{\begin{aligned}
\lambda x_r(t)  & =   A_rx_r(t)+ B_{ru} u(t)+ B_{rd} d(t) + B_{rf} f(t)+ B_{rw} w(t)   , \\
y_r(t) & =  C_r x_r(t) + D_{ru} u(t)+ D_{rd} d(t) + D_{rf} f(t) + D_{rw} w(t) ,
\end{aligned}}
\ee
where the reference model output $ry_(t)$ is a $q$-dimensional vector and
any of the inputs components $u(t)$, $d(t)$, $f(t)$, or $w(t)$  can be void.  For the system \texttt{SYSR}, the input groups for $u(t)$, $d(t)$, $f(t)$, and $w(t)$ have the standard names \texttt{\bfseries 'controls'}, \texttt{\bfseries 'disturbances'}, \texttt{\bfseries 'faults'}, and \texttt{\bfseries 'noise'}, respectively.

\item
 \texttt{OPTIONS} is a MATLAB structure used to specify various synthesis  options and has the following fields:
{\setlength\LTleft{30pt}\begin{longtable}{|l|p{11.6cm}|}
\hline \textbf{\texttt{OPTIONS} fields} & \textbf{Description} \\ \hline
       \texttt{tol}       & relative tolerance for rank computations \newline
                 (Default: internally computed)\\ \hline
       \texttt{tolmin}       & absolute tolerance for observability tests \newline
                 (Default: internally computed)\\ \hline
       \texttt{smarg}   & stability margin for the poles of the filters \texttt{Q} and \texttt{R}\\
                 &  (Default: \texttt{-sqrt(eps)} for a continuous-time system \texttt{SYSF}; \\
                 &  \hspace*{4.5em}\texttt{1-sqrt(eps)} for a discrete-time system \texttt{SYSF}).\\ \hline
      \texttt{sdeg}   & prescribed stability degree for the poles of the filters \texttt{Q} and \texttt{R}\\
                 &  (Default: $-0.05$ for a continuous-time system \texttt{SYSF}; \\
                 &  \hspace*{4.8em} $0.95$ for a discrete-time system \texttt{SYSF}).\\ \hline
       \texttt{poles}   & complex vector containing a complex conjugate set of desired poles (within the stability domain) to be assigned for the filters \texttt{Q} and \texttt{R} (Default: \texttt{[ ]})
                   \\\hline
       \texttt{simple} & option to employ a simple proper basis for synthesis \\
                 &  \texttt{true }  -- use a simple basis; \\
                 &  \texttt{false}  -- use a non-simple basis (default)\\\hline
       \texttt{minimal} & option to perform least order synthesis of the filter \texttt{Q}  \\
                 &  \texttt{true }  -- perform least order synthesis (default); \\
                 &  \texttt{false}  -- perform full order synthesis \\\hline
       \texttt{regmin} & option to perform regularization selecting a least order left annihilator  \\
                 &  \texttt{true }  -- perform least order selection (default); \\
                 &  \texttt{false}  -- no least order selection performed \\\hline
       \texttt{tcond} & maximum allowed condition number of the employed non-orthogonal transformations (Default: $10^4$).\\ \hline
 \texttt{normalize}   & option for the normalization of the diagonal elements of the updating matrix $M(\lambda)$:\\
                 &  {\tabcolsep=1mm\begin{tabular}{llp{9.5cm}}\hspace*{-1.5mm}\texttt{'gain'}&--& scale with the gains of the  zero-pole-gain \newline representation (default)   \\
                   \hspace*{-1.5mm}\texttt{'dcgain'}&--& scale with the DC-gains \\
                   \hspace*{-1.5mm}\texttt{'infnorm'}&--& scale with the values of infinity-norms
                   \end{tabular}}  \\
                                        \hline
\texttt{freq}   & complex frequency value to be employed to check the
                      left-invertibility-based solvability condition (see \textbf{Method}) \newline
                      (Default:\texttt{[ ]}, i.e., a randomly generated frequency). \\
                                        \hline
\texttt{HDesign}   & full row rank design matrix $H$ employed for the
                      synthesis of the filter \texttt{Q}  (see \textbf{Method})
                      (Default: \texttt{[ ]}) \\
                     & \emph{Note.} This option can be only used in conjunction with the ``no least order synthesis'' option: \texttt{OPTIONS.minimal = false}. \\
                                        \hline
\end{longtable}}
\end{description}

\subsubsection*{Output data}
\begin{description}
\item
\texttt{Q} is the resulting fault detection filter in a standard state-space representation
\be\label{emmsyn:detss}
{\begin{aligned}
\lambda x_Q(t)  & =   A_Qx_Q(t)+ B_{Q_y}y(t)+ B_{Q_u}u(t) ,\\
r(t) & =  C_Q x_Q(t) + D_{Q_y}y(t)+ D_{Q_u}u(t) ,
\end{aligned}}
\ee
where the residual signal $r(t)$ is a $q$-dimensional vector. For the system object \texttt{Q}, two input groups \texttt{\bfseries 'outputs'} and \texttt{\bfseries 'controls'} are defined for $y(t)$ and $u(t)$, respectively, and the output group \texttt{\bfseries 'residuals'} is defined for the residual signal $r(t)$.

\item
\texttt{R} is the resulting internal form of the fault detection filter in a standard state-space representation
\be\label{emmsyn:detinss}
{\begin{aligned}
\lambda x_R(t)  & =   A_Rx_R(t)+ B_{R_u}u(t)+B_{R_d}d(t)+B_{R_f}f(t)+ B_{R_w}w(t) ,\\
r(t) & =  C_Rx_R(t)+ D_{R_u}u(t)+D_{R_d}d(t)+D_{R_f}f(t)+ D_{R_w}w(t)
\end{aligned}}
\ee
with the same input groups defined as for \texttt{SYSR} and the output group \texttt{\bfseries 'residuals'} defined for the residual signal $r(t)$.
\item
\texttt{INFO} is a MATLAB structure containing additional information as follows:
\begin{center}
\begin{longtable}{|l|p{12cm}|} \hline \textbf{\texttt{INFO} fields} & \textbf{Description} \\ \hline
\texttt{tcond} & the maximum of the condition numbers of the employed
                      non-orthogonal transformation matrices; a warning is
                      issued if \texttt{INFO.tcond} $\geq$ \texttt{OPTIONS.tcond}.\\ \hline
\texttt{degs}     & the left Kronecker indices of $G(\lambda) := \left[\begin{smallmatrix} G_u(\lambda) & G_u(\lambda) \\ I &                      0\end{smallmatrix}\right]$ (see \textbf{Metho}d); also the increasingly ordered degrees of a left minimal
                     polynomial nullspace basis of $G(\lambda)$; (\texttt{INFO.degs = [ ]} if no explicit left nullspace basis is computed)\\ \hline
\texttt{M} & state-space realization of the employed updating matrix $M(\lambda)$ (see \textbf{Method}) \\ \hline
\texttt{freq} & complex frequency value employed to check the left invertibility condition;
                      \texttt{INFO.freq = [ ]} if no frequency-based left invertibility
                      check was performed.  \\ \hline
\texttt{HDesign} & design matrix $H$ employed for the synthesis of
                      the fault detection filter; \texttt{INFO.HDesign = [ ]}  if no design  matrix was explicitly involved in the filter synthesis.
  \\ \hline
\end{longtable}
\end{center}
\end{description}

\subsubsection*{Method}

 Extensions of the \textbf{Procedure EMM} and \textbf{Procedure EMMS}  from \cite[Sect.\ 5.6]{Varg17} are implemented in the function \texttt{\bfseries emmsyn}. The \textbf{Procedure EMM}
relies on the model-matching synthesis method proposed in \cite{Varg04g}, while \textbf{Procedure EMMS} uses the inversion-based method proposed in \cite{Varg13a} in conjunction with the nullspace method. The
  \textbf{Procedure EMM} is employed to solve the general EMMP (see below), while  \textbf{Procedure EMMS} is employed to solve the more particular (but more practice relevant) strong exact fault
  detection and isolation problem (strong EFDIP).

Assume that the system \texttt{SYSF} has the input-output form
\be\label{emmsyn:systemw1} {\mathbf{y}}(\lambda) =
G_u(\lambda){\mathbf{u}}(\lambda) +
G_d(\lambda){\mathbf{d}}(\lambda) +
G_f(\lambda){\mathbf{f}}(\lambda) +
G_w(\lambda){\mathbf{w}}(\lambda)
 \ee
and the reference model \texttt{SYSR} has the input-output form
\be\label{emmsyn:systemw2} {\mathbf{y}}_r(\lambda) =
M_{ru}(\lambda){\mathbf{u}}(\lambda) +
M_{rd}(\lambda){\mathbf{d}}(\lambda) +
M_{rf}(\lambda){\mathbf{f}}(\lambda) +
M_{rw}(\lambda){\mathbf{w}}(\lambda) ,
 \ee
where the vectors $y$, $u$, $d$, $f$, $w$  and $y_r$ have dimensions $p$, $m_u$, $m_d$, $m_f$, $m_w$ and $q$, respectively. The resulting fault detection filter in (\ref{emmsyn:detss}) has the input-output form
 \be\label{emmsyn:detio}
{\mathbf{r}}(\lambda) = Q(\lambda)\ba{c}
{\mathbf{y}}(\lambda)\\{\mathbf{u}}(\lambda)\ea ,  \ee
where the resulting dimension of the residual vector $r$ is $q$.

The function \texttt{\bfseries emmsyn} determines $Q(\lambda)$ by solving the general exact model-matching problem
\be\label{gemmp}
{\arraycolsep=1mm Q(\lambda)  \ba{c|c|c|c} G_u(\lambda) & G_d(\lambda) & G_f(\lambda)  & G_w(\lambda) \\
         I_{m_u} & 0 & 0 & 0 \ea }  = M(\lambda) \big[\, M_{ru}(\lambda)\mid M_{rd}(\lambda) \mid M_{rf}(\lambda) \mid  M_{rw}(\lambda)\,\big] \, ,
         \ee
where $M(\lambda)$ is a stable, diagonal and invertible transfer function matrix chosen such that the resulting solution $Q(\lambda)$ of the linear rational matrix equation (\ref{gemmp}) is stable and proper. The resulting internal form $R(\lambda)$,  contained in \texttt{R}, is computed as
\be\label{Rtilde} R(\lambda) := \big[\, R_u(\lambda)\!\mid\! R_d(\lambda) \!\mid\! R_f(\lambda) \!\mid\!  R_w(\lambda)\,\big] := M(\lambda) \big[\, M_{ru}(\lambda)\!\mid\! M_{rd}(\lambda) \!\mid\! M_{rf}(\lambda) \!\mid\! M_{rw}(\lambda)\,\big].\ee
 Two cases are separately addressed, depending on the presence or absence of $M_{rw}(\lambda)$ in the reference model (\ref{emmsyn:systemw2}).

In the first case, when $M_{rw}(\lambda)$ is present, the general EMMP (\ref{gemmp}) is solved, with $M_{ru}(\lambda)$, or $M_{rd}(\lambda)$, or both of them, explicitly set to zero if not present in the reference model.
In the second case, when $M_{rw}(\lambda)$ is not present, then $Q(\lambda)$ is determined by solving the EMMP
\be\label{gemmp1}
{\arraycolsep=1mm Q(\lambda)  \ba{c|c|c} G_u(\lambda) & G_d(\lambda) & G_f(\lambda) \\
         I_{m_u} & 0 & 0 \ea }  = M(\lambda) \big[\, M_{ru}(\lambda)\mid M_{rd}(\lambda) \mid M_{rf}(\lambda)\,\big]
         \ee
with $M_{ru}(\lambda)$, or $M_{rd}(\lambda)$, or both of them, explicitly set to zero if not present in the reference model. In this case, if $G_w(\lambda)$ is present in the plant model (\ref{emmsyn:systemw1}), then $R_w(\lambda)$ is explicitly computed as
\[ R_w(\lambda) = Q(\lambda) \ba{c}G_w(\lambda) \\ 0 \ea . \]
The particular EMMP formulated in Section~\ref{sec:EMMP} corresponds to solve the EMMP (\ref{gemmp1}) with $M_{ru}(\lambda) = 0$, $M_{rd}(\lambda) = 0$.

If \texttt{OPTIONS.minimal = true}, then a least order solution $Q(\lambda)$ is determined in the form $Q(\lambda) = Q_2(\lambda)Q_1(\lambda)$, where $Q_1(\lambda)$ is a least McMillan degree  solution of the linear rational matrix equation
\be\label{emmp-direct}
{\arraycolsep=1mm Q_1(\lambda)  \ba{c|c|c|c} G_u(\lambda) & G_d(\lambda) & G_f(\lambda)  & G_w(\lambda) \\
         I_{m_u} & 0 & 0 & 0 \ea }  = \big[\, M_{ru}(\lambda)\mid M_{rd}(\lambda) \mid M_{rf}(\lambda) \mid  M_{rw}(\lambda)\,\big] \ee
and the diagonal updating factor $Q_2(\lambda) := M(\lambda)$ is determined to ensure that $Q(\lambda)$ is proper and stable.

If \texttt{OPTIONS.minimal = false} and either $M_{ru}(\lambda)$ or $M_{rd}(\lambda)$, or both, are present in the reference model (\ref{emmsyn:systemw2}), then $Q(\lambda)$ is determined in the form $Q(\lambda) = Q_2(\lambda)Q_1(\lambda)$, where $Q_1(\lambda)$ is a particular solution of the linear rational equation (\ref{emmp-direct}) and the diagonal updating factor $Q_2(\lambda) := M(\lambda)$ is determined to ensure that $Q(\lambda)$ is proper and stable.

If \texttt{OPTIONS.minimal = false} and if both $M_{ru}(\lambda)$ and $M_{rd}(\lambda)$ are absent in the reference model (\ref{emmsyn:systemw2}), then the nullspace method is employed as the first computational step of solving the EMMP (see \textbf{Procedure EMM} in \cite{Varg17}). The strong EFDIP arises if additionally $M_{rf}(\lambda)$ is diagonal and invertible, in which case, an extension of the  \textbf{Procedure EMMS} in \cite{Varg17} is employed. In fact, this procedure works for arbitrary invertible $M_{rf}(\lambda)$ and this case was considered for the implementation of \texttt{\bfseries emmsyn}. The solution of a fault estimation problem can be targeted by
  choosing $M_{rf}(\lambda) = I_{m_f}$ and checking that the resulting $M(\lambda) = I_{m_f}$. Recall that $M(\lambda)$ is provided in \texttt{INFO.M}. In what follows, we give some details of the implemented synthesis approach employed if $M_{ru}(\lambda)$, $M_{rd}(\lambda)$ and $M_{rw}(\lambda)$  are not present in reference model.

If \texttt{OPTIONS.regmin = false}, then $Q(\lambda)$ is determined in the form
\[  Q(\lambda)= M(\lambda)Q_2(\lambda)HN_l(\lambda), \]
where: $N_l(\lambda)$ is a $\big(p-r_d\big) \times (p+m_u)$ rational left nullspace basis satisfying
\[
N_l(\lambda)\ba{cc} G_u(\lambda) & G_d(\lambda)\\ I_{m_u} & 0 \ea  = 0 , \] with $r_d := \text{rank}\, G_d(\lambda)$; $H$ is a suitable full row rank design matrix used to build $q$ linear
combinations of the $p-r_d$ left nullspace basis vectors ($q$ is the number of outputs of \texttt{SYSR}); $Q_2(\lambda)$ is the solution of $Q_2(\lambda) H\overline G_f(\lambda) = M_{rf}(\lambda)$, where
\[ \overline G_f(\lambda) = N_l(\lambda) \ba{c} G_f(\lambda) \\ 0 \ea ; \]
and $M(\lambda)$ is a stable invertible transfer function matrix determined such that $Q(\lambda)$ and the corresponding $\widetilde R(\lambda)$ in (\ref{Rtilde}) have desired dynamics (specified via \texttt{OPTIONS.sdeg} and \texttt{OPTIONS.poles}). The internal form of the filter $Q(\lambda)$ is obtained as
\[   R(\lambda) = M(\lambda)Q_2(\lambda)H \overline G(\lambda), \]
where
\[ \overline G(\lambda) := \big[\, \overline G_f(\lambda) \mid \overline G_w(\lambda)  \,\big] = N_l(\lambda) \ba{c|c} G_f(\lambda) & G_w(\lambda) \\
          0 & 0  \ea  \, . \]
The solvability condition of the strong EFDIP is verified by checking the left invertibility condition
\be\label{leftinv} \rank H\overline G_{f}(\lambda_s) = m_f , \ee
where $\lambda_s$ is a suitable frequency value, which can be specified via the \texttt{OPTIONS.freq}.
The design parameter matrix $H$ is set as follows: if \texttt{OPTIONS.HDesign} is nonempty, then $H = \texttt{OPTIONS.HDesign}$;  if \texttt{OPTIONS.HDesign = [ ]}, then $H = I_{p-r_d}$, if $q = p-r_d$, or $H$ is a randomly generated $q \times \big(p-r_d\big)$ real matrix, if $q < p-r_d$. If \texttt{OPTIONS.simple = true}, then $N_l(\lambda)$ is determined as a simple rational basis. The orders of the basis vectors are provided in \texttt{INFO.degs}. These are also the degrees of the basis vectors of an equivalent polynomial nullspace basis.

If \texttt{OPTIONS.regmin = true}, then $[\, Q(\lambda) \; R_f(\lambda)\,]$ has the least McMillan degree, with $Q(\lambda)$  having $q$ outputs. $Q(\lambda)$ and $R_f(\lambda)$ are determined in the form
\[  [\, Q(\lambda) \mid R_f(\lambda)\,] = M(\lambda) Q_2(\lambda)\big[\, \overline Q(\lambda) \mid \overline R_f(\lambda)\,\big] \, , \]
where
\[  \big[\, \overline Q(\lambda) \mid \overline R_f(\lambda)\,\big] = H \big[\, N_l(\lambda) \mid \overline G_f(\lambda)\,\big] + Y(\lambda)\big[\, N_l(\lambda) \mid \overline G_f(\lambda)\,\big] \]
with
$[\, \overline Q(\lambda) \; \overline R_f(\lambda)\,]$ and $Y(\lambda)$  the least order solution of a left minimal cover problem \cite{Varg17g}; $Q_2(\lambda)$ is the solution of $Q_2(\lambda) \overline R_f(\lambda) = M_{rf}(\lambda)$; and $M(\lambda)$ is a stable invertible transfer function matrix determined such that $[\, Q(\lambda) \; R(\lambda)\,]$ has a desired dynamics. If \texttt{OPTIONS.HDesign} is nonempty, then $H = \texttt{OPTIONS.HDesign}$, and if \texttt{OPTIONS.HDesign = [ ]}, then a suitable randomly generated $H$ is employed, which fulfills the left invertibility condition (\ref{leftinv}).

The actually employed design matrix $H$ is provided in \texttt{INFO.HDesign}, and \texttt{INFO.HDesign = [ ]} if the solution of the EMMP is obtained by directly solving (\ref{gemmp}).

\subsubsection*{Examples}

\begin{example}\label{ex:5.12}
This is \emph{Example} 5.12 from the book \cite{Varg17} and was used in  \emph{Example}~\ref{ex:Ex5.10},
to solve an EFDIP for a system with triplex sensor redundancy. To solve the same problem by solving an EMMP, we use the resulting $R_f(s)$ to define the reference model
\[ M_{rf}(s) := R_f(s) = \ba{rrr} 0 & 1 & -1\\ -1 & 0 & 1 \\ 1 & -1 & 0 \ea  \, .\]
The resulting least order filter $Q(s)$, determined by employing \texttt{\bfseries emmsyn} with \textbf{Procedure EMM},  has the generic form
\[ Q(s) = \ba{rrrccc} 0 & 1 & -1 & 0 & \cdots & 0\\-1 & 0 & 1 & 0 & \cdots & 0\\1 & -1 & 0 & 0& \cdots & 0\ea  \, .\]

\begin{verbatim}
% Example - Solution of an EMMP

p = 3; mf = 3;   % enter output and fault vector dimensions
% generate random dimensions for system order and input vectors
rng('default')
nu = floor(1+4*rand); mu = floor(1+4*rand);
nd = floor(1+4*rand); md = floor(1+4*rand);
% define random Gu(s) and Gd(s) with triplex sensor redundancy
% and Gf(s) = I for triplex sensor faults
Gu = ones(3,1)*rss(nu,1,mu); % enter Gu(s) in state-space form
Gd = ones(3,1)*rss(nd,1,md); % enter Gd(s) in state-space form

% build synthesis model with sensor faults
sysf = fdimodset([Gu Gd],struct('c',1:mu,'d',mu+(1:md),'fs',1:3));

% enter reference model for the TFM from faults to residual
Mr = fdimodset(ss([ 0 1 -1; -1 0 1; 1 -1 0]),struct('f',1:mf));

% solve an exact model-matching problem using EMMSYN
[Q,R,info] = emmsyn(sysf,Mr);

% determine achieved fault sensitivity conditions
FSCOND = fdifscond(R,0,fdisspec(Mr))

% check the synthesis: Q*Ge = M*Me and R = M*Mr, where
% Ge = [Gu Gd Gf; I 0 0] and Me = [0 0 Mr ].
Ge = [sysf; eye(mu,mu+md+mf)]; Me = [zeros(p,mu+md) Mr];
norm(gminreal(Q*Ge)-info.M*Me,inf)
norm(R-info.M*Mr,inf)
\end{verbatim}

\end{example}
\begin{example}\label{ex:5.13}
This is \emph{Exampl}e 5.13 from the book \cite{Varg17}, with a continuous-time system with additive actuator faults, having the transfer-function matrices
\[
G_u(s) =
\left[\begin{array}{cc} \displaystyle\frac{s}{s^2 + 3\, s + 2} & \displaystyle\frac{1}{s + 2}\\ \\[-2mm] \displaystyle\frac{s}{s + 1} & 0\\ \\[-2mm]0 & \displaystyle\frac{1}{s + 2} \end{array}\right]
 , \quad G_d(s) = 0, \quad G_f(s) = G_u(s), \quad G_w(s) = 0 \, .
\]
We want to solve an EMMP with the reference model
\[  M_{rf}(s) = {\arraycolsep=2mm\ba{cc} 1 & 0\\0 & 1 \ea}  \, ,\]
which is equivalent to solve a strong EFDIP with the structure matrix
\[ S_{FDI} =  {\arraycolsep=2mm\ba{cc} 1 & 0\\0 & 1 \ea} .\]

A least order stable filter $Q(s)$, with poles assigned to $-1$, has been determined by employing \texttt{\bfseries emmsyn} with \textbf{Procedure EMMS}  of \cite{Varg17} (which is employed if \texttt{OPTIONS.minimal = false}).  The resulting $Q(s)$ is
\[ Q(s) = {\arraycolsep=2mm\ba{ccccc} 0 & 1  & 0 & -\displaystyle\frac{s}{s+1} & 0 \\ 
0 & 0 & \displaystyle\frac{s+2}{s+1} & 0 & -\displaystyle\frac{1}{s+1} \ea}   \]
and has the McMillan degree equal to 2. The resulting updating factor is
\[  M(s) = \ba{cc} \displaystyle\frac{s}{s+1} & 0 \\ 0 & \displaystyle\frac{1}{s+1} \ea \]
and has McMillan degree 2.
The presence of the zero at $s = 0$ in $R(s) := M(s)M_r(s)$ is unavoidable for the existence of a stable solution. It follows, that while a constant actuator fault $f_2$ is strongly detectable, a constant actuator fault $f_1$ is only  detectable during transients.

\begin{verbatim}
% Example - Solution of a strong EFDIP as an EMMP

% define s as an improper transfer function
s = tf('s');
% enter Gu(s)
Gu = [s/(s^2+3*s+2) 1/(s+2);
     s/(s+1) 0;
      0 1/(s+2)];
[p,mu] = size(Gu); mf = mu;

% build model with faults
sysf = fdimodset(ss(Gu),struct('c',1:mu,'f',1:mu));

% define Mr(s)
Mr = fdimodset(ss(eye(2)),struct('f',1:mf));

% solve a strong EFDIP using EMMSYN (for an invertible reference model)
opts_emmsyn = struct('tol',1.e-7,'sdeg',-1,'minimal',false);
[Q,R,info] = emmsyn(sysf,Mr,opts_emmsyn);

% check solution
G = [sysf;eye(mu,mu+mf)];
norm(Q*G-info.M*[zeros(mf,mu) Mr],inf)

% display results
minreal(tf(Q)), tf(info.M)
\end{verbatim}

\end{example}

\subsubsection{\texttt{\bfseries ammsyn}}
\index{M-functions!\texttt{\bfseries ammsyn}}
\index{model-matching problem!approximate (AMMP)}
\subsubsection*{Syntax}
\begin{verbatim}
[Q,R,INFO] = ammsyn(SYSF,SYSR,OPTIONS)
\end{verbatim}

\subsubsection*{Description}
\texttt{\bfseries ammsyn} solves the \emph{approximate model matching problem} (AMMP); (see Section \ref{sec:AMMP}), for a given LTI system \texttt{SYSF} with additive faults and a given stable reference filter \texttt{SYSR}. Two stable and proper filters, \texttt{Q} and \texttt{R}, are computed, where \texttt{Q} contains the fault detection and isolation filter representing the solution of the AMMP,  and \texttt{R} contains its internal form.

\subsubsection*{Input data}
\begin{description}
\item
\texttt{SYSF} is a  LTI system  in the state-space form
\be\label{ammsyn:sysss}
{\begin{aligned}
E\lambda x(t)  & =   Ax(t)+ B_u u(t)+ B_d d(t) + B_f f(t)+ B_w w(t)   , \\
y(t) & =  C x(t) + D_u u(t)+ D_d d(t) + D_f f(t) + D_w w(t) ,
\end{aligned}}
\ee
where any of the inputs components $u(t)$, $d(t)$, $f(t)$ or $w(t)$  can be void.  For the system \texttt{SYSF}, the input groups for $u(t)$, $d(t)$, $f(t)$, and $w(t)$ must have the standard names \texttt{\bfseries 'controls'}, \texttt{\bfseries 'disturbances'}, \texttt{\bfseries 'faults'}, and \texttt{\bfseries 'noise'}, respectively.

\item
\texttt{SYSR} is a proper and stable  LTI system  in the state-space form
\be\label{ammsyn:sysrss}
{\begin{aligned}
\lambda x_r(t)  & =   A_rx_r(t)+ B_{ru} u(t)+ B_{rd} d(t) + B_{rf} f(t)+ B_{rw} w(t)   , \\
y_r(t) & =  C_r x_r(t) + D_{ru} u(t)+ D_{rd} d(t) + D_{rf} f(t) + D_{rw} w(t) ,
\end{aligned}}
\ee
where the reference model output $y_r(t)$ is a $q$-dimensional vector and
any of the inputs components $u(t)$, $d(t)$, $f(t)$, or $w(t)$  can be void.  For the system \texttt{SYSR}, the input groups for $u(t)$, $d(t)$, $f(t)$, and $w(t)$ must have the standard names \texttt{\bfseries 'controls'}, \texttt{\bfseries 'disturbances'}, \texttt{\bfseries 'faults'}, and \texttt{\bfseries 'noise'}, respectively.


\item
 \texttt{OPTIONS} is a MATLAB structure used to specify various synthesis  options and has the following fields:
{\setlength\LTleft{30pt}\begin{longtable}{|l|p{11.6cm}|}
\hline \textbf{\texttt{OPTIONS} fields} & \textbf{Description} \\ \hline
       \texttt{tol}       & relative tolerance for rank computations \newline
                 (Default: internally computed)\\ \hline
       \texttt{tolmin}       & absolute tolerance for observability tests \newline
                 (Default: internally computed)\\ \hline
       \texttt{smarg}   & stability margin for the poles of the updating factor $M(\lambda)$ (see \textbf{Method})\\
                 &  (Default: \texttt{-sqrt(eps)} for a continuous-time system \texttt{SYSF}; \\
                 &  \hspace*{4.5em}\texttt{1-sqrt(eps)} for a discrete-time system \texttt{SYSF}).\\ \hline
      \texttt{sdeg}   & prescribed stability degree for the poles of the updating factor $M(\lambda)$ (see \textbf{Method})\\
                 &  (Default: $-0.05$ for a continuous-time system \texttt{SYSF}; \\
                 &  \hspace*{4.8em} $0.95$ for a discrete-time system \texttt{SYSF}).\\ \hline
       \texttt{poles}   & complex vector containing a complex conjugate set of desired poles (within the stability domain) to be assigned for the updating factor $M(\lambda)$ (see \textbf{Method}) (Default: \texttt{[ ]})
                   \\\hline
 \texttt{nullspace}   & option to use a specific proper nullspace basis to be employed in the nullspace-based synthesis step\\
                 & \hspace*{-.9mm}{\tabcolsep=0.7mm\begin{tabular}[t]{lcp{10cm}} \texttt{true } &--& use a minimal proper basis (default); \\
                 \texttt{false} &--& use a full-order observer based basis (see \textbf{Method}). \newline \emph{Note:} This option can  only be used if no disturbance inputs are present in (\ref{ammsyn:sysss}) and $E$ is invertible.
                 \end{tabular}} \\  \hline
       \texttt{simple} & option to employ a simple proper basis for the nullspace-based synthesis \\
                 &  \texttt{true }  -- use a simple basis; \\
                 &  \texttt{false}  -- use a non-simple basis (default)\\\hline
 \texttt{mindeg}   &option to compute a minimum degree solution\\
                 & \hspace*{-.9mm}{\tabcolsep=0.7mm\begin{tabular}[t]{lcp{10cm}}
                 \texttt{true} &--& determine, if possible, a  minimum order
                              stable solution  \\
                 \texttt{false} &--& determine a particular stable solution which
                              has possibly non-minimal (default).
                 \end{tabular}} \\  \hline
       \texttt{regmin} & regularization option with least order
                      left annihilator selection \\
                 &  \texttt{true }  -- perform least order selection  (default); \\
                 &  \texttt{false}  -- no least order selection to be performed  \\\hline
       \texttt{tcond} & maximum allowed condition number of the employed non-orthogonal transformations (Default: $10^4$).\\ \hline
 \texttt{normalize}   & option for the normalization of the diagonal elements of the updating matrix $M(\lambda)$ (see \textbf{Method}):\\
                 &  {\tabcolsep=1mm\begin{tabular}{llp{9.5cm}}\hspace*{-1.5mm}\texttt{'gain'}&--& scale with the gains of the  zero-pole-gain \newline representation (default)   \\
                   \hspace*{-1.5mm}\texttt{'dcgain'}&--& scale with the DC-gains \\
                   \hspace*{-1.5mm}\texttt{'infnorm'}&--& scale with the values of infinity-norms
                   \end{tabular}}  \\
                                        \hline
\texttt{freq}   & complex frequency value to be employed to check frequency response based (admissibility) rank conditions (see \textbf{Method}) \newline
                      (Default:\texttt{[ ]}, i.e., a randomly generated frequency). \\
                                        \hline
\texttt{HDesign}   & full row rank design matrix $H$ employed for the
                      synthesis of the filter \texttt{Q}  (see \textbf{Method})
                      (Default: \texttt{[ ]}) \\
                                        \hline
       \texttt{H2syn} & option to perform a $\mathcal{H}_2$-norm based synthesis \\
                 &  \texttt{true }  -- perform a $\mathcal{H}_2$-norm based synthesis; \\
                 &  \texttt{false}  -- perform a $\mathcal{H}_\infty$-norm based synthesis (default). \\\hline
\end{longtable}}
\end{description}

\subsubsection*{Output data}
\begin{description}
\item
\texttt{Q} is the resulting fault detection filter in a standard state-space representation
\be\label{ammsyn:detss}
{\begin{aligned}
\lambda x_Q(t)  & =   A_Qx_Q(t)+ B_{Q_y}y(t)+ B_{Q_u}u(t) ,\\
r(t) & =  C_Q x_Q(t) + D_{Q_y}y(t)+ D_{Q_u}u(t) ,
\end{aligned}}
\ee
where the residual signal $r(t)$ is a $q$-dimensional vector. For the system object \texttt{Q}, two input groups \texttt{\bfseries 'outputs'} and \texttt{\bfseries 'controls'} are defined for $y(t)$ and $u(t)$, respectively, and the output group \texttt{\bfseries 'residuals'} is defined for the residual signal $r(t)$.

\item
\texttt{R} is the resulting internal form of the fault detection filter in a standard state-space representation
\be\label{ammsyn:detinss}
{\begin{aligned}
\lambda x_R(t)  & =   A_Rx_R(t)+ B_{R_u}u(t)+B_{R_d}d(t)+B_{R_f}f(t)+ B_{R_w}w(t) ,\\
r(t) & =  C_Rx_R(t)+ D_{R_u}u(t)+D_{R_d}d(t)+D_{R_f}f(t)+ D_{R_w}w(t)
\end{aligned}}
\ee
with the same input groups defined as for \texttt{SYSR} and the output group \texttt{\bfseries 'residuals'} defined for the residual signal $r(t)$.
\item
\texttt{INFO} is a MATLAB structure containing additional information as follows:
\begin{center}
\begin{longtable}{|l|p{12cm}|} \hline \textbf{\texttt{INFO} fields} & \textbf{Description} \\ \hline
\texttt{tcond} & the maximum of the condition numbers of the employed
                      non-orthogonal transformation matrices; a warning is
                      issued if \texttt{INFO.tcond} $\geq$ \texttt{OPTIONS.tcond}.\\ \hline
\texttt{degs}     & the left Kronecker indices of $G(\lambda) := \left[\begin{smallmatrix} G_u(\lambda) & G_u(\lambda) \\ I &                      0\end{smallmatrix}\right]$ (see \textbf{Metho}d); also, if \texttt{OPTIONS.simple} = \texttt{true}, the orders of the basis vectors of the employed simple nullspace basis, or
if \texttt{OPTIONS.simple} = \texttt{false}, the degrees of the basis vectors of an equivalent polynomial nullspace basis. \texttt{INFO.degs = [ ]} if no strong FDI or fault detection oriented synthesis is performed. \\ \hline
\texttt{M} & state-space realization of the employed updating matrix $M(\lambda)$ (see \textbf{Method}) \\ \hline
\texttt{freq} & complex frequency value employed to check frequency response based (admissibility) rank conditions
                      \\ \hline
\texttt{HDesign} & design matrix $H$ employed for the synthesis of
                      the fault detection filter; \texttt{INFO.HDesign = [ ]}  if no design  matrix was explicitly involved in the filter synthesis.
  \\ \hline
 \texttt{nonstandard}   & logical value, which is set to \texttt{true} for a \emph{non-standard problem} (when $G_e(\lambda)$ in (\ref{ammsyn:Ge}) has zeros in $\partial\mathds{C}_s$), and to \texttt{false} for a \emph{standard problem} (when $G_e(\lambda)$ in (\ref{ammsyn:Ge}) has no zeros in $\partial\mathds{C}_s$) (see \textbf{Method}). \\ \hline
\texttt{gammaopt0} & resulting optimal model-matching performance value $\gamma_{opt,0}$ in (\ref{ammsyn:opthinf2})
(see \textbf{Method}).
  \\ \hline
\texttt{gammaopt} & resulting optimal model-matching performance value $\gamma_{opt,0}$ in (\ref{ammsyn:opthinf2}), in the \emph{standard case}, and $\gamma_{opt}$ in (\ref{ammsyn:modopthinf2}) in the \emph{non-standard case}
(see \textbf{Method}).
  \\ \hline
\texttt{gammasub} & resulting suboptimal model-matching performance value $\gamma_{sub}$ in (\ref{ammsyn:subopthinf2}) (see \textbf{Method}).
  \\ \hline
\end{longtable}
\end{center}
\end{description}

\subsubsection*{Method}

The function \texttt{\bfseries ammsyn} implements the  \textbf{Procedure AMMS}  from \cite[Sect.\ 5.7]{Varg17}, which relies on the approximate model-matching synthesis methods proposed in \cite{Varg11} (see also \cite{Varg12d} for more computational details). This procedure is primarily intended to approximately solve the particular (but more practice relevant) strong fault detection and isolation problem, which is addressed in the standard formulation of the AMMP in Section \ref{sec:AMMP}. However, the function  \texttt{\bfseries ammsyn} is also able to address the more general case of AMMP with an arbitrary stable reference model (see Remark \ref{rem:gamm} in Section \ref{sec:AMMP}). Therefore, in what follows, we consider the solution of the AMMP in this more general problem setting.

Assume that the system \texttt{SYSF} has the input-output form
\be\label{ammsyn:systemw1} {\mathbf{y}}(\lambda) =
G_u(\lambda){\mathbf{u}}(\lambda) +
G_d(\lambda){\mathbf{d}}(\lambda) +
G_f(\lambda){\mathbf{f}}(\lambda) +
G_w(\lambda){\mathbf{w}}(\lambda)
 \ee
and the reference model \texttt{SYSR} has the input-output form
\be\label{ammsyn:systemw2} {\mathbf{y}}_r(\lambda) =
M_{ru}(\lambda){\mathbf{u}}(\lambda) +
M_{rd}(\lambda){\mathbf{d}}(\lambda) +
M_{rf}(\lambda){\mathbf{f}}(\lambda) +
M_{rw}(\lambda){\mathbf{w}}(\lambda) ,
 \ee
where the vectors $y$, $u$, $d$, $f$, $w$  and $y_r$ have dimensions $p$, $m_u$, $m_d$, $m_f$, $m_w$ and $q$, respectively.
Furthermore, we assume that $M_r(\lambda) := \big[\, M_{ru}(\lambda)\; M_{rd}(\lambda) \; M_{rf}(\lambda) \;  M_{rw}(\lambda)\,\big] $ is stable.

The resulting fault detection filter in (\ref{ammsyn:detss}) has the input-output form
 \be\label{ammsyn:detio}
{\mathbf{r}}(\lambda) = Q(\lambda)\ba{c}
{\mathbf{y}}(\lambda)\\{\mathbf{u}}(\lambda)\ea ,  \ee
where the resulting dimension of the residual vector $r$ is $q$.

The function \texttt{\bfseries ammsyn} determines $Q(\lambda)$ by solving the approximate model-matching problem
\be\label{agemmp}
Q(\lambda) G_e(\lambda) \approx M(\lambda) M_r(\lambda \, ,
         \ee
where
\be\label{ammsyn:Ge} G_e(\lambda) := \ba{c|c|c|c} G_u(\lambda) & G_d(\lambda) & G_f(\lambda)  & G_w(\lambda) \\
         I_{m_u} & 0 & 0 & 0 \ea ,  \ee
$M(\lambda)$ is a stable, diagonal and invertible transfer function matrix chosen such that the resulting approximate solution $Q(\lambda)$ of (\ref{agemmp}) is stable and proper. The resulting internal form $R(\lambda)$,  contained in \texttt{R}, is computed as
\be\label{ammsyn:Rtilde} R(\lambda) := \big[\, R_u(\lambda)\!\mid\! R_d(\lambda) \!\mid\! R_f(\lambda) \!\mid\!  R_w(\lambda)\,\big] := Q(\lambda) G_e(\lambda) .\ee

\index{performance evaluation!fault detection and isolation!model-matching performance}
In the \emph{standard case}, $G_e(\lambda)$  has no zeros on the boundary of the
stability domain $\partial\mathds{C}_s$, and the resulting stable filter $Q(\lambda) = Q_{opt}(\lambda)$,  is the optimal solution of the $\mathcal{H}_\infty$- or $\mathcal{H}_2$-norm error minimization problem
\be\label{ammsyn:opthinf2} \gamma_{opt,0} := \|Q_{opt}(\lambda)G_e(\lambda)-M_0(\lambda)M_r(\lambda)\|_{\infty/2} = \min ,\ee
where $M_0(\lambda) = I$ in the case of $\mathcal{H}_\infty$-norm or of a discrete-time system. In the case of $\mathcal{H}_2$-norm and a continuous-time system,  $M_0(s)$ is determined as a stable, diagonal, and invertible transfer function matrix, which
ensures the existence of a finite $\mathcal{H}_2$-norm.

In the \emph{non-standard case}, $G_e(\lambda)$   has zeros on the boundary of the
stability domain $\partial\mathds{C}_s$, and the optimal solution $Q_{opt}(\lambda)$ of (\ref{ammsyn:opthinf2}) is possibly unstable or improper. In this case, $Q(\lambda)$ is chosen as $Q(\lambda) = M_1(\lambda)Q_{opt}(\lambda)$, with $M_1(\lambda)$ a stable, diagonal, and invertible transfer function matrix
  determined to ensure the stability of $Q(\lambda)$. In this case, $Q(\lambda)$ can be interpreted as a suboptimal solution of the updated $\mathcal{H}_\infty$- or $\mathcal{H}_2$-norm  error minimization problem
\be\label{ammsyn:modopthinf2}  \gamma_{opt} := \|\widetilde Q_{opt}(\lambda)G_e(\lambda) - M_1(\lambda)M_0(\lambda)M_r(\lambda)\|_{\infty/2} = \min ,\ee
whose optimal solution $\widetilde Q_{opt}(\lambda)$ is possibly unstable or improper, but for which $Q(\lambda)$ represents a stable and proper solution with the corresponding suboptimal model-matching error norm
\be\label{ammsyn:subopthinf2}   \gamma_{sub} := \|Q(\lambda)G_e(\lambda) - M_1(\lambda)M_0(\lambda)M_r(\lambda)\|_{\infty/2} . \ee
The value of $\gamma_{opt,0}$ in (\ref{ammsyn:opthinf2}) is returned in \texttt{INFO.gammaopt0}, and also in \texttt{INFO.gammaopt} and \texttt{INFO.gammasub} in the \emph{standard case}. In the \emph{non-standard case}, the value $\gamma_{opt}$ in (\ref{ammsyn:modopthinf2}) is returned in \texttt{INFO.gammaopt} and the value of $\gamma_{sub}$ in (\ref{ammsyn:subopthinf2}) is returned in \texttt{INFO.gammasub}. The updating matrix $M(\lambda) := M_0(\lambda)$, in the \emph{standard case}, or $M(\lambda) := M_1(\lambda)M_0(\lambda)$, in the \emph{non-standard case}, is returned in \texttt{INFO.M}.

Two cases are separately addressed in what follows, depending on the presence or absence of the terms $M_{ru}(\lambda)$, $M_{rd}(\lambda)$ and $M_{rw}(\lambda)$ in the reference model (\ref{emmsyn:systemw2}). In the first case, we consider the standard formulation of the AMMP of Section \ref{sec:AMMP}, with $M_{ru}(\lambda)=0$, $M_{rd}(\lambda)=0$ and $M_{rw}(\lambda)=0$, and  with $M_{rf}(\lambda)$ having all its columns nonzero (to enforce complete fault detectability). If $M_{rf}(\lambda)$ is invertible, we target the solution of a \emph{strong fault detection and isolation} (\emph{strong} FDI) problem. Additionally, we assume that $q \leq p-r_d$, where $r_d := \text{rank}\, G_d(\lambda)$. The second case, covers the rest of situations and is discussed separately.

To compute the solution $Q(\lambda)$ for the standard formulation of the AMMP, we employ a slight extension of the synthesis method which underlies \textbf{Procedure AMMD} in \cite{Varg17}. This procedure essentially determines the  filter $Q(\lambda)$ as a stable rational left annihilator of
\be\label{ammsyn:gtfm}  G(\lambda) := \ba{cc} G_u(\lambda) & G_d(\lambda) \\
 I_{m_u} & 0 \ea ,\ee
such that $Q(\lambda)$ also simultaneously solves the approximate model-matching problem
\be\label{ammsyn:gemmp}
{\arraycolsep=1mm Q(\lambda)  \ba{c|c} G_f(\lambda)  & G_w(\lambda) \\
          0 & 0 \ea }  \approx M(\lambda) \big[\, M_{rf}(\lambda) \mid  0 \,\big] \, ,
         \ee
where $M(\lambda)$ is a stable, diagonal and invertible transfer function matrix chosen such that the resulting solution $Q(\lambda)$  is stable and proper. The resulting internal form $R(\lambda)$,  contained in \texttt{R}, is computed as
\be\label{ammsyn:R} R(\lambda) = [\, R_f(\lambda) \mid R_w(\lambda) \,] := Q(\lambda)  \ba{c|c} G_f(\lambda)  & G_w(\lambda) \\
          0 & 0 \ea .\ee

If \texttt{OPTIONS.H2syn = false}, then a $\mathcal{H}_\infty$-norm based synthesis is performed by determining $Q(\lambda)$ and $M(\lambda)$ such that $\|\mathcal{E}(\lambda)\|_\infty$ is minimized, where
\be\label{ammsyn:Rez}  \mathcal{E}(\lambda) = [\, R_f(\lambda) \mid R_w(\lambda) \,] - M(\lambda) \big[\, M_{rf}(\lambda) \mid  0 \,\big] .\ee
If \texttt{OPTIONS.H2syn = true}, then a $\mathcal{H}_2$-norm based synthesis is performed by determining $Q(\lambda)$ and $M(\lambda)$ such that $\|\mathcal{E}(\lambda)\|_2$ is minimized.

The filter $Q(\lambda)$ is determined in the product form
\be\label{ammsym:Qprod} Q(\lambda) = Q_5(\lambda)Q_4(\lambda)Q_3(\lambda)Q_2(\lambda)Q_1(\lambda) , \ee
where the factors are  determined as follows:
 \begin{itemize}
 \item[(a)] $Q_1(\lambda) = N_l(\lambda)$, with $N_l(\lambda)$ a $\big(p-r_d\big) \times (p+m_u)$ proper rational left nullspace basis satisfying $N_l(\lambda)G(\lambda) = 0$ (recall that $r_d := \text{rank}\, G_d(\lambda)$);
     \item[(b)] $Q_2(\lambda)$ is a $q\times (p-r_d)$  admissible regularization factor guaranteeing complete fault detectability or strong fault isolability;
     \item[(c)] $Q_3(\lambda)$ is the inverse or left inverse of a quasi-co-outer factor;
     \item[(d)] $Q_4(\lambda)$ is the solution of a least distance problem (LDP);
     \item[(e)] $Q_5(\lambda)$ is a stable invertible factor determined  such that $Q(\lambda)$ has a desired dynamics.
 \end{itemize}
The computations of  individual factors depend on the user's options and specific choices are discussed in what follows.

\subsubsection*{Computation of $Q_1(\lambda)$}
If \texttt{OPTIONS.nullspace = true} or $m_d > 0$ or $E$ is singular, then $N_l(\lambda)$ is determined as a minimal proper nullspace basis. In this case, if \texttt{OPTIONS.simple = true}, then $N_l(\lambda)$ is determined as a simple rational basis and the orders of the basis vectors are provided in \texttt{INFO.degs}. If \texttt{OPTIONS.simple = false}, then $N_l(\lambda)$ is determined as a proper rational basis and \texttt{INFO.degs} contains the degrees of the basis vectors of an equivalent polynomial nullspace basis (see \cite[Section 9.1.3]{Varg17} for  definitions).

If \texttt{OPTIONS.nullspace = false}, $m_d = 0$ and $E$ is invertible, then $N_l(\lambda) = [ \, I_{p} \; -G_u(\lambda)\,]$ is used, which corresponds to a full-order Luenberger observer.

To check the solvability of the AMMP, the transfer function matrix  $\overline G_f(\lambda) :=  Q_1(\lambda)
\left[\begin{smallmatrix}  G_f(\lambda) \\ 0 \end{smallmatrix}\right]$ is determined and let $H$ be a full row rank design matrix, which is either specified in  a nonempty \texttt{OPTIONS.HDesign}, or otherwise $H = I_{p-r_d}$ is assumed. The AMMP with complete detectability requirement is solvable provided $H\overline G_f(\lambda_s)$ has all its columns nonzero, where $\lambda_s$ is a suitable frequency value, which can be specified via the \texttt{OPTIONS.freq}.
To check the solvability of the AMMP with a strong isolability condition, we check that $H\overline G_f(\lambda)$ has full column rank $m_f$.
This rank condition is verified by checking the left invertibility condition
\be\label{ammsyn:leftinv} \rank H\overline G_{f}(\lambda_s) = m_f . \ee
In the case when \texttt{OPTIONS.freq} is empty, the employed frequency value $\lambda_s$ is provided in \texttt{INFO.freq}.

\subsubsection*{Computation of $Q_2(\lambda)$}

If \texttt{OPTIONS.regmin = false}, then $Q_2(\lambda) = H$, where $H$ is a suitable $q\times \big(p-r_d\big)$ full row rank design matrix.  $H$ is set as follows. If \texttt{OPTIONS.HDesign} is nonempty, then $H = \texttt{OPTIONS.HDesign}$.
If \texttt{OPTIONS.HDesign} is empty, then the matrix $H$  is chosen such that, either: (1) $H\overline G_f(\lambda)$ is invertible, if the solution of a strong FDI problem is targeted; or (2)
$H\overline G_f(\lambda)$ has all its columns nonzero, if the focus is on guaranteeing complete fault detectability. In both cases, $H$ is built from an admissible set of $q$ rows of the identity matrix $I_{p-r_d}$.

If \texttt{OPTIONS.regmin = true}, then $Q_2(\lambda)$ is a $q\times \big(p-r_d\big)$ transfer function matrix determined in the form
\[  Q_2(\lambda) = H+Y_2(\lambda) \, , \]
where $Q_2(\lambda)Q_1(\lambda)$ $\big(\! = HN_l(\lambda)+Y_2(\lambda)N_l(\lambda)\big)$ and $Y_2(\lambda)$ are  the least order solution of a left minimal cover problem \cite{Varg17g}. If \texttt{OPTIONS.HDesign} is nonempty, then $H = \texttt{OPTIONS.HDesign}$, and if \texttt{OPTIONS.HDesign} is empty, then $H$ is chosen as above. This choice ensures that either: (1) $Q_2(\lambda)\overline G_f(\lambda)$ is invertible, if a strong FDI problem is solved; or (2)
$H\overline G_f(\lambda)$ has all its columns nonzero, if the focus is on guaranteeing complete fault detectability.

The structure field \texttt{INFO.HDesign} contains the employed value of the design matrix $H$.

\subsubsection*{Computation of $Q_3(\lambda)$}
Let  $\tilde r$ be the rank of $[\, \widetilde G_f(\lambda) \mid \widetilde G_w(\lambda) \,]:=  Q_2(\lambda)[\, \overline G_f(\lambda)\mid \overline G_w(\lambda)\,]$.  If a strong FDI problem is solved, then $\tilde r = m_f = q$.
The extended quasi-co-outer--co-inner factorization of $[\, \widetilde G_f(\lambda) \; \widetilde G_w(\lambda) \,]$ is computed as
\[ [\, \widetilde G_f(\lambda) \; \widetilde G_w(\lambda) \,] = [\, G_{o}(\lambda) \;  0 \, ] \ba{c} G_{i,1}(\lambda) \\ G_{i,2}(\lambda) \ea ,\]
where $G_{o}(\lambda)$ is a $q\times \tilde r$ full column rank quasi-co-outer factor and  $G_{i}(\lambda) = \left[\begin{smallmatrix} G_{i,1}(\lambda) \\ G_{i,2}(\lambda) \end{smallmatrix}\right] $ is a square inner factor. Note that the potential lack of stability of $G_{o}(\lambda)$ is not relevant at this stage for the employed solution method. If $\tilde r = q$ then,
$Q_3(\lambda) =  G_{o}^{-1}(\lambda)$, otherwise $Q_3(\lambda) =  G_{o}^{-L}(\lambda)$, with $G_{o}^{-L}(\lambda)$ a left inverse of $G_{o}(\lambda)$, determined such that all its free poles are assigned into the stable domain $\mathds{C}_s$. It follows, that in the \emph{standard case}, when $G_{o}(\lambda)$ has no zeros in $\partial\mathds{C}_s$, $Q_3(\lambda)$ results stable, while in the \emph{non-standard case}, when $G_{o}(\lambda)$ has zeros in $\partial\mathds{C}_s$, $Q_3(\lambda)$ results with poles which are either stable or lie in $\partial \mathds{C}_s$. These latter poles are precisely the unstable zeros of $G_{o}(\lambda)$ in $\partial\mathds{C}_s$.
The information on the type of the problem to be solved is returned in \texttt{INFO.nonstandard}, which is set equal to \texttt{false} for a standard problem, and \texttt{true} for a non-standard problem.

\subsubsection*{Computation of $Q_4(\lambda)$}
With
$\widetilde F_1(\lambda) = [\, M_r(\lambda) \; 0\,] G_{i,1}^\sim(\lambda)$ and  $\widetilde F_2(\lambda) = [\, M_r(\lambda) \; 0\,] G_{i,2}^\sim(\lambda)$, the $q\times \tilde r$ TFM $Q_4(\lambda)$ is determined as
the optimal solution  of the $\mathcal{H}_{\infty/2}$ least distance problem ($\mathcal{H}_{\infty/2}$-LDP)
\be\label{ammsyn:gammaopt} \gamma_{opt} = \min_{Q_4(\lambda) \in \mathcal{H}_\infty} \left\|{\arraycolsep=1mm\ba{cc} \widetilde F_1 (\lambda)- Q_4(\lambda) &  \widetilde F_2(\lambda)  \ea } \right\|_{\infty/2}. \ee
To ensure the existence of the solution of a $\mathcal{H}_{2}$-LDP in the continuous-time case with $\widetilde F_2(s)$ a non-strictly-proper transfer function matrix, a strictly proper and stable updating factor $\widetilde M(s) = \frac{k}{s+k} I$ is used, to form an updated reference model $\widetilde M(s)M_r(s)$ and the updated $\widetilde F_1(\lambda) = \widetilde M(\lambda)[\, M_r(\lambda) \; 0\,] G_{i,1}^\sim(\lambda)$ and  $\widetilde F_2(\lambda) = \widetilde M(\lambda[\, M_r(\lambda) \; 0\,] G_{i,2}^\sim(\lambda)$ are used. $\widetilde M(\lambda) = I$ in all other cases.

The value of $\gamma_{opt}$ in (\ref{ammsyn:gammaopt}) is returned in \texttt{INFO.gammaopt}.
For a standard $\mathcal{H}_\infty$-norm based synthesis, $\gamma_{opt}$ is the optimal approximation error $\|\mathcal{E}(\lambda)\|_\infty$, while for a
non-standard problem this value is the optimal $\mathcal{H}_\infty$
least distance which corresponds to an improper or
unstable filter.
For a standard $\mathcal{H}_2$-norm based synthesis, $\gamma_{opt}$  is the
optimal approximation error $\|\mathcal{E}(\lambda)\|_2$, while
for a non-standard problem this value is the optimal
$\mathcal{H}_2$ least distance which corresponds to an improper
or unstable filter.

\subsubsection*{Computation of $Q_5(\lambda)$}
In the standard case, $Q_5(\lambda) = I$. In the non-standard case,
 $Q_5(\lambda)$ is a stable, diagonal and invertible transfer function matrix determined such that $Q(\lambda)$ in (\ref{ammsym:Qprod}) has a desired dynamics (specified via \texttt{OPTIONS.sdeg} and \texttt{OPTIONS.poles}).
The overall updating factor used in (\ref{ammsyn:Rez}) is $M(\lambda) = Q_5(\lambda)\widetilde M(\lambda)$ and is provided in \texttt{INFO.M}.

In the second case, we solve model-matching problems which differ from the standard formulation of Section \ref{sec:AMMP}, as for example, having $M_{ru}(\lambda)\not=0$, $M_{rd}(\lambda)\not=0$ or $M_{rw}(\lambda)\not=0$, or   with a non-square $M_{rf}(\lambda)$, as well as other cases. To address the solution of these problems, we formulate general model-matching problems of the form (\ref{agemmp}) and solve these problems using general solvers as \texttt{glasol}, available in the Descriptor System Tools (DSTOOLS) \cite{Varg17a}.

The filter $Q(\lambda)$ is determined in the product form
\be\label{ammsym:Qprod2} Q(\lambda) = Q_3(\lambda)Q_2(\lambda)Q_1(\lambda) , \ee
where the factors are  determined as follows:
 \begin{itemize}
 \item[(a)] $Q_1(\lambda)$ is a proper rational left nullspace basis satisfying $Q_1(\lambda)\widetilde G(\lambda) = 0$ for a suitably defined $\widetilde G(\lambda)$ (see below);
     \item[(b)] $Q_2(\lambda)$ is the solution of a reduced or unreduced approximate model-matching problem;
     \item[(c)] $Q_3(\lambda)$ is a stable invertible factor determined  such that $Q(\lambda)$ has a desired dynamics.
 \end{itemize}
The computations of  individual factors depend on the user's options and specific choices are discussed in what follows.

\subsubsection*{Computation of $Q_1(\lambda)$}
Three cases are considered.
\begin{itemize}
\item[$(i)$] If both $M_{ru}(\lambda)\not=0$ and $M_{rd}(\lambda)\not=0$, then $Q_1(\lambda) = I_{p+m_u}$ and the corresponding reduced system and reference model are $\overline R(\lambda) := G_e(\lambda)$ and
$\overline M_r(\lambda) := M_r(\lambda)$. Note that $\widetilde G(\lambda)$ is an $(p+m_u)\times 0$ (empty) matrix.

\item[$(ii)$] If  $M_{ru}(\lambda)=0$, but $M_{rd}(\lambda)\not=0$, then $Q_1(\lambda)$ is a left proper rational
nullspace basis of $\widetilde G(\lambda) := \left[\begin{smallmatrix} G_u(\lambda) \\ I_{m_u} \end{smallmatrix}\right]$.

If \texttt{OPTIONS.nullspace = true} or $E$ is singular, then $Q_1(\lambda)$ is determined as a minimal proper nullspace basis. In this case, if \texttt{OPTIONS.simple = true}, then $Q_1(\lambda)$ is determined as a simple rational basis, while if \texttt{OPTIONS.simple = false}, then $Q_1(\lambda)$ is determined as a proper rational basis. The corresponding reduced system and reference model are
\[ \overline R(\lambda) := Q_1(\lambda)\ba{c|c|c} G_d(\lambda) & G_f(\lambda)  & G_w(\lambda) \\
         0 & 0 & 0 \ea , \qquad \overline M_r(\lambda) = \big[\, M_{rd}(\lambda) \mid M_{rf}(\lambda) \mid  M_{rw}(\lambda)\,\big] .\]

If \texttt{OPTIONS.nullspace = false} and $E$ is invertible, then $Q_1(\lambda) = [ \, I_{p} \; -G_u(\lambda)\,]$ is used, which corresponds to a full-order Luenberger observer. The corresponding reduced system and reference model are
\[ \overline R(\lambda) := \big[\, G_d(\lambda) \mid G_f(\lambda)  \mid G_w(\lambda)\,\big] , \qquad \overline M_r(\lambda) = \big[\, M_{rd}(\lambda) \mid M_{rf}(\lambda) \mid  M_{rw}(\lambda)\,\big] .\]
\item[$(iii)$] If  both $M_{ru}(\lambda)=0$ and $M_{rd}(\lambda)=0$, then $Q_1(\lambda)$  is a left proper rational
nullspace basis of $\widetilde G(\lambda) := \left[\begin{smallmatrix} G_u(\lambda) & G_d(\lambda) \\ I_{m_u} & 0 \end{smallmatrix}\right]$. If the resulting nullspace is empty, then the case $(ii)$ can be applied.

If \texttt{OPTIONS.nullspace = true} or $m_d > 0$ or $E$ is singular, then $Q_1(\lambda)$ is determined as a minimal proper nullspace basis. In this case, if \texttt{OPTIONS.simple = true}, then $Q_1(\lambda)$ is determined as a simple rational basis and if \texttt{OPTIONS.simple = false}, then $Q_1(\lambda)$ is determined as a proper rational basis. The corresponding reduced system and reference model are
\[ \overline R(\lambda) := Q_1(\lambda)\ba{c|c} G_f(\lambda)  & G_w(\lambda) \\
           0 & 0 \ea , \qquad \overline M_r(\lambda) = \big[\, M_{rf}(\lambda) \mid  M_{rw}(\lambda)\,\big] .\]

If \texttt{OPTIONS.nullspace = false}, $m_d = 0$ and $E$ is invertible, then $Q_1(\lambda) = [ \, I_{p} \; -G_u(\lambda)\,]$ is used, which corresponds to a full-order Luenberger observer. The corresponding reduced system and reference model are
\[ \overline R(\lambda) := \big[\, G_f(\lambda)  \mid G_w(\lambda)\,\big] , \qquad \overline M_r(\lambda) = \big[\, M_{rf}(\lambda) \mid  M_{rw}(\lambda)\,\big] .\]
\end{itemize}

\subsubsection*{Computation of $Q_2(\lambda)$}

$Q_2(\lambda)$ is computed as the optimal solution which minimizes the model-matching error norm such that
\be\label{ammsym:ammopt} \|Q_2(\lambda)\overline R_1(\lambda) - \overline M_r(\lambda)\|_{\infty/2} = \min .\ee
The standard-case corresponds to $\overline R_1(\lambda)$ without zeros in $\partial\mathds{C}_s$, while the non-standard case corresponds to $\overline R_1(\lambda)$ having zeros in $\partial\mathds{C}_s$.

\subsubsection*{Computation of $Q_3(\lambda)$}
In the standard case, $Q_3(\lambda) = I$. In the non-standard case,
 $Q_3(\lambda)$ is a stable, diagonal and invertible transfer function matrix determined such that $Q(\lambda)$ in (\ref{ammsym:Qprod2}) has a desired dynamics (specified via \texttt{OPTIONS.sdeg} and \texttt{OPTIONS.poles}).
The overall updating factor used in (\ref{gemmp}) is $M(\lambda) = Q_3(\lambda)$ and is provided in \texttt{INFO.M}.

In the second case, \texttt{INFO.degs = [ ]} and \texttt{INFO.HDesign = [ ]} are returned in the \texttt{INFO} structure.

\begin{example}\label{ex:5.11}
This is \emph{Example} 5.11 from the book \cite{Varg17}, with a continuous-time system with additive faults, having the transfer function matrices
\[ G_u(s) = \ba{c}  \displaystyle\frac{s+1}{s+2} \\ \\[-2mm] \displaystyle\frac{s+2}{s+3} \ea,  \quad G_d(s) = 0, \quad G_f(s) = \ba{cc} \ \displaystyle\frac{s+1}{s+2} & 0\\ \\[-2mm] 0 & 1 \ea  , \quad G_w(s) = \ba{c} \displaystyle \frac{1}{s+2} \\\\[-2mm] 0 \ea \, .\]
The maximally achievable structure matrix is
\[ S_{max} = \ba{cc}  0 & 1  \\ 1 & 0 \\  1 & 1 \ea  \]
and therefore we can target to solve an AMMP with strong fault isolability using the following reference model
\[  M_r(s) = {\arraycolsep=2mm\ba{cc} 1 & 0\\0 & 1 \ea}  \, .\]
This involves to determine a stable $Q(s)$, and possibly also an updating factor $M(s)$, which  fulfill
\[ \gamma_{opt} := \left\|Q(s)\ba{cccc} G_u(s) & G_d(s) & G_f(s) & G_w(s) \\ I_{m_u} & 0 & 0& 0 \ea - M(s)[\, 0 \; 0\; M_r(s) \; 0 \,] \right\|_\infty = \min .\]

A least order stable optimal filter $Q(s)$ has been determined by employing \texttt{\bfseries ammsyn} with $M(s) = I_2$. The resulting optimal performance is $\gamma_{opt} = \frac{\sqrt{2}}{2} = 0.7071$.  The resulting $Q(s)$ is
\[ Q(s)
         = \ba{ccc} 0.7072\displaystyle\frac{s+2}{s+\sqrt{2}} & 0 & -0.7072\displaystyle\frac{s+1}{s+\sqrt{2}} \\
         0 & 0.7072 & -0.7072\displaystyle\frac{s+2}{s+3} \ea  \]
and the resulting $R_f(s)$ and $R_w(s)$ are
\[ R_f(s)  = \ba{cc}  0.7072\displaystyle\frac{s+1}{s+\sqrt{2}} & 0  \\
         0 & 0.7072  \ea , \quad
R_w(s) =    \ba{c} 0.7072\displaystyle\frac{1}{s+\sqrt{2}}   \\
         0  \ea .
         \]
The fault-to-noise gaps can be computed using the function  \texttt{fdif2ngap}, by assuming as structure matrix $S_{FDI}$, the structure matrix of the reference model $M_r(s)$ (i.e., $S_{FDI} = I_2$). The resulting filter $Q(s)$ can be considered formed by column concatenating two separate filters $Q^{(1)}(s)$ and $Q^{(2)}(s)$, which aims to match the first and second rows of $M_r(s)$, respectively. The resulting fault-to-noise gaps are respectively, $\sqrt{2}$ and $\infty$, which indicate that the second filter solves, in fact, an EMMP.

The above results  have been computed with the following script.
\newpage
\begin{verbatim}
% Example - Solution of an AMMP

% define s as an improper transfer function
s = tf('s');
Gu = [(s+1)/(s+2); (s+2)/(s+3)];  % enter Gu(s)
Gf = [(s+1)/(s+2) 0; 0 1];        % enter Gf(s)
Gw = [1/(s+2); 0];                % enter Gw(s)
mu = 1; mf = 2; mw = 1; p = 2; % set dimensions

% build the synthesis model with additive faults
inputs = struct('c',1:mu,'f',mu+(1:mf),'n',mu+mf+(1:mw));
sysf = fdimodset(ss([Gu Gf Gw]),inputs);

% determine the maximally achievable structure matrix
Smax = fdigenspec(sysf,struct('tol',1.e-7))

% choose the targeted reference model
Mr = fdimodset(ss(eye(mf)),struct('faults',1:mf));

% solve the AMMP using AMMSYN
opts_ammsyn = struct('tol',1.e-7,'reltol',5.e-8);
[Q,R,info] = ammsyn(sysf,Mr,opts_ammsyn);

% display results
minreal(zpk(Q)),  tf(info.M)
Rf = minreal(zpk(R(:,'faults'))), Rw = minreal(zpk(R(:,'noise')))

% optimal and suboptimal performance, and achieved gaps
info
format short e
gap = fdif2ngap(R,[],fditspec(Mr))

% check synthesis performance
gammaopt = fdimmperf(R,Mr)

% check decoupling condition
Ge = [sysf;eye(mu,mu+mf+mw)];
norm_Ru = norm(Q*Ge(:,'controls'),inf)

% check synthesis results
norm_dif = norm(R-Q*Ge(:,{'faults','noise'}),inf)
\end{verbatim}

\end{example}

\newpage

\subsection{Functions for the Synthesis of Model Detection Filters }\label{fditools:mdsynthesis}

\subsubsection{\texttt{\bfseries emdsyn}}
\index{M-functions!\texttt{\bfseries emdsyn}}
\index{model detection problem!a@exact (EMDP)}
\subsubsection*{Syntax}
\begin{verbatim}
[Q,R,INFO] = emdsyn(SYSM,OPTIONS)
\end{verbatim}

\subsubsection*{Description}

\texttt{\bfseries emdsyn} solves   the \emph{exact model detection problem} (EMDP) (see Section \ref{sec:EMDP}), for a given stable LTI multiple model \texttt{SYSM} containing $N$ models. A bank of $N$ stable and proper residual generation filters $Q^{(i)}(\lambda)$, for $i = 1, \ldots, N$, is determined, in the form (\ref{detecmi}). For each filter $Q^{(i)}(\lambda)$, its associated  internal forms $R^{(i,j)}(\lambda)$, for $j = 1, \ldots, N$, are determined in accordance with  (\ref{ri_internal1}).

\subsubsection*{Input data}
\begin{description}
\item
\texttt{SYSM} is a multiple model which contains $N$ stable LTI systems in the state-space form
\be\label{emdsyn:sysiss}
\begin{array}{rcl}E^{(j)}\lambda x^{(j)}(t) &=& A^{(j)}x^{(j)}(t) + B^{(j)}_u u^{(j)}(t) + B^{(j)}_d d^{(j)}(t) + B^{(j)}_w w^{(j)}(t)  \, ,\\
y^{(j)}(t) &=& C^{(j)}x^{(j)}(t) + D^{(j)}_u u^{(j)}(t) + D^{(j)}_d d^{(j)}(t) + D^{(j)}_w w^{(j)}(t)   \, , \end{array} \ee
where $x^{(j)}(t) \in \mathds{R}^{n^{(j)}}$ is the state vector of the $j$-th system with control input $u^{(j)}(t) \in \mathds{R}^{m_u}$, disturbance input $d^{(j)}(t) \in \mathds{R}^{m_d^{(j)}}$ and noise input $w^{(j)}(t) \in \mathds{R}^{m_w^{(j)}}$, and
\index{faulty system model!physical}%
\index{faulty system model!multiple model}%
where any of the inputs components $u^{(j)}(t)$, $d^{(j)}(t)$, or $w^{(j)}(t)$  can be void. The multiple model \texttt{SYSM} is either an array of $N$ LTI systems of the form (\ref{emdsyn:sysiss}), in which case $m_d^{(j)} = m_d$ and $m_w^{(j)} = m_w$ for $j = 1, \ldots, N$,  or is a $1\times N$ cell array, with \texttt{SYSM\{$j$\}} containing the $j$-th component system in the form (\ref{emdsyn:sysiss}). The input groups for $u^{(j)}(t)$, $d^{(j)}(t)$, and $w^{(j)}(t)$  have the standard names \texttt{\bfseries 'controls'}, \texttt{\bfseries 'disturbances'}, and \texttt{\bfseries 'noise'}, respectively.

\item
 \texttt{OPTIONS} is a MATLAB structure used to specify various synthesis  options and has the following fields:
{\tabcolsep=1mm
\setlength\LTleft{30pt}\begin{longtable}{|l|lcp{9cm}|} \hline
\textbf{\texttt{OPTIONS} fields} & \multicolumn{3}{l|}{\textbf{Description}} \\ \hline
 \texttt{tol}   & \multicolumn{3}{p{9cm}|}{relative tolerance for rank computations \newline (Default: internally computed)} \\ \hline
 \texttt{tolmin}   & \multicolumn{3}{p{9cm}|}{absolute tolerance for observability tests
            \newline      (Default: internally computed)} \\ \hline
 \texttt{MDTol}   & \multicolumn{3}{p{9cm}|}{threshold for model detectability checks
        \newline         (Default: $10^{-4}$)} \\ \hline
 \texttt{MDGainTol}   & \multicolumn{3}{p{9cm}|}{threshold for strong model detectability checks \newline (Default: $10^{-2}$)} \\ \hline
 \texttt{rdim}   & \multicolumn{3}{p{12cm}|}{$N$-dimensional vector or a scalar; for a vector $q$, the $i$-th component $q_i$ specifies the desired number of residual outputs for the $i$-th filter  \texttt{Q\{$i$\}}; for a scalar value $\bar q$, a vector $q$ with all $N$ components $q_i = \bar q$ is assumed. \newline
(Default: \hspace*{-5.5mm}\begin{tabular}[t]{l} \hspace*{4.5mm}\texttt{[ ]}, in which case:\\ \hspace*{-2em} -- if \texttt{OPTIONS.HDesign\{$i$\}} is empty, then \\ $q_i = 1$, if \texttt{OPTIONS.minimal} = \texttt{true}, or \\ $q_i = n_v^{(i)}$, the dimension of the left nullspace which
                                underlies the \\ synthesis of \texttt{Q\{$i$\}}, if \texttt{OPTIONS.minimal} = \texttt{false} (see \textbf{Method}); \\ \hspace*{-2em} -- if \texttt{OPTIONS.HDesign\{$i$\}} is nonempty, then $q_i$ is the row dimension \\ of the design matrix contained in  \texttt{OPTIONS.HDesign\{$i$\}}.)
                                \end{tabular}}\\ \hline
 \texttt{MDFreq}  &  \multicolumn{3}{p{12cm}|}{real vector, which contains the frequency values $\omega_k$, $k = 1, \ldots, n_f$, to be used for strong model detectability checks. For each real frequency  $\omega_k$, there corresponds a complex frequency $\lambda_k$ which is used to evaluate the frequency-response gains. Depending on the system type, $\lambda_k = \mathrm{i}\omega_k$, in the continuous-time case, and $\lambda_k = \exp (\mathrm{i}\omega_k T)$, in the discrete-time case, where $T$ is the common sampling time of the component systems.  (Default: \texttt{[ ]}) } \\ \hline
 \texttt{emdtest}   & \multicolumn{3}{p{12cm}|}{option to perform extended model
                      detectability tests using both control and
                      disturbance input channels:}\\
                 &  \texttt{true} &--& use both control and disturbance input channels; \\
                 &  \texttt{false}&--& use only the control channel (default).  \\
                                        \hline
\texttt{smarg}   & \multicolumn{3}{p{11cm}|}{prescribed stability margin for the resulting filters \texttt{Q\{$i$\}} \newline
            (Default: \texttt{-sqrt(eps)} for continuous-time component systems; \newline
                   \hspace*{4.8em}\texttt{1-sqrt(eps)} for discrete-time component systems.
                   } \\ \hline
\texttt{sdeg}   & \multicolumn{3}{p{11cm}|}{prescribed stability degree for the resulting filters \texttt{Q\{$i$\}} \newline
            (Default: $-0.05$   for continuous-time component systems; \newline
                   \hspace*{4.8em} $0.95$ for discrete-time component systems.
                   } \\ \hline
 \texttt{poles}   & \multicolumn{3}{p{12cm}|}{complex vector containing a complex conjugate set of desired poles (within the stability margin) to be assigned for the resulting filters \texttt{Q\{$i$\}}
                     (Default: \texttt{[ ]})}\\
                                        \hline
 \texttt{nullspace}   & \multicolumn{3}{p{12cm}|}{option to use a specific type of proper nullspace bases:}\\
                 &  \texttt{true} &--& use minimal proper bases; \\
                 &  \texttt{false}&--& use full-order observer based bases (default)  \newline
                 \emph{Note:} This option can  only be used if no disturbance inputs are present in (\ref{emdsyn:sysiss}) and $\forall j$, $E^{(j)}$  is invertible.\\
                                        \hline
\texttt{simple}   & \multicolumn{3}{l|}{option to compute simple proper bases:}\\
                 &  \texttt{true} &--& compute simple bases; the orders of the
                            basis vectors are provided in \texttt{INFO.degs}; \\
                 &  \texttt{false}&--& no simple basis computed (default)  \\
                                        \hline
 \texttt{minimal}   & \multicolumn{3}{l|}{option to perform least order filter syntheses:}\\
                 &  \texttt{true} &--& perform least order syntheses (default); \\
                 &  \texttt{false}&--& perform full order syntheses.  \\
                                        \hline
 \texttt{tcond}   & \multicolumn{3}{l|}{maximum allowed value for the condition numbers of the  employed}\\
     & \multicolumn{3}{l|}{non-orthogonal transformation matrices (Default: $10^4$)}\\
    & \multicolumn{3}{l|}{(only used if \texttt{OPTIONS.simple = true}) } \\
                    \hline
 \texttt{MDSelect}   & \multicolumn{3}{p{12cm}|}{integer vector with increasing elements containing the indices of the desired filters to be designed
                     (Default: $[\,1, \ldots, N\,]$)}\\
                                        \hline
 \texttt{HDesign}   & \multicolumn{3}{p{12cm}|}{$N$-dimensional cell array; \texttt{OPTIONS.HDesign\{$i$\}}, if not empty,  is a
                      full row rank design matrix employed for the
                      synthesis of the $i$-th  filter.
                      If \texttt{OPTIONS.HDesign} is specified as a full row rank
                      design matrix $H$, then an $N$-dimensional cell array is assumed with  \texttt{OPTIONS.HDesign\{$i$\}} = $H$, for $i = 1, \ldots, N$.
                      (Default:~\texttt{[ ]}).}\\
                                        \hline
 \texttt{normalize}   & \multicolumn{3}{p{12cm}|}{option to normalize the the filters
                      \texttt{Q\{i\}} and \texttt{R\{i,j\}} such that the minimum gains of the
                      off-diagonal elements of \texttt{R\{i,j\}} are equal to one;
                      otherwise the standard normalization is  performed
                      to ensure equal gains for \texttt{R\{1,j\}} and \texttt{R\{j,1\}} :}\\
                 &  \texttt{true} &--& perform normalization to unit minimum gains; \\
                 &  \texttt{false}&--& perform standard normalization (default).  \\
                                        \hline
\end{longtable}}
\end{description}

\subsubsection*{Output data}
\begin{description}
\item
\texttt{Q} is an $N\times 1$ cell array of filters, where \texttt{Q\{$i$\}} contains the resulting $i$-th filter  in a standard state-space representation
\[
{\begin{aligned}
\lambda x_Q^{(i)}(t)  & =   A_Q^{(i)}x_Q^{(i)}(t)+ B_{Q_y}^{(i)}y(t)+ B_{Q_u}^{(i)}u(t) ,\\
r^{(i)}(t) & =  C_Q^{(i)} x_Q^{(i)}(t) + D_{Q_y}^{(i)}y(t)+ D_{Q_u}^{(i)}u(t) ,
\end{aligned}}
\]
where the residual signal $r^{(i)}(t)$ is a $q_i$-dimensional vector, with $q_i$ specified in \texttt{OPTIONS.rdim}.  For each system object \texttt{Q\{$i$\}}, two input groups \texttt{\bfseries 'outputs'} and \texttt{\bfseries 'controls'} are defined for $y(t)$ and $u(t)$, respectively, and the output group \texttt{\bfseries 'residuals'} is defined for the residual signal $r^{(i)}(t)$. \texttt{Q\{$i$\}} is empty for all $i$ which do not belong to the index set specified by \texttt{OPTIONS.MDSelect}.

\item \texttt{R} is an $N\times N$ cell array of filters, where the $(i,j)$-th filter \texttt{R\{$i,j$\}}, is the internal form of \texttt{Q\{$i$\}}
  acting on the $j$-th model.
The resulting \texttt{R\{$i,j$\}}  has a standard state-space representation
\[
{\begin{aligned}
\lambda x_R^{(i,j)}(t)  & =   A_R^{(i,j)}x_R^{(i,j)}(t)+ B_{R_u}^{(i,j)}u^{(j)}(t)+  B_{R_d}^{(i,j)}d^{(j)}(t)+ B_{R_w}^{(i,j)}w^{(j)}(t), \\
r^{(i,j)}(t) & =  C_R^{(i,j)} x_R^{(i,j)}(t) + D_{R_u}^{(i,j)}u^{(j)}(t)+ D_{R_d}^{(i,j)}d^{(j)}(t)+ D_{R_w}^{(i,j)}w^{(j)}(t)
\end{aligned}}
\]
and the input groups \texttt{\bfseries 'controls'}, \texttt{\bfseries 'disturbances'}  and \texttt{\bfseries 'noise'} are defined for $u^{(j)}(t)$, $d^{(j)}(t)$, and $w^{(j)}(t)$, respectively, and the output group \texttt{\bfseries 'residuals'} is defined for the residual signal $r^{(i,j)}(t)$. \texttt{R\{$i,j$\}}, $j = 1, \ldots, N$ are empty for all $i$ which do not belong to the index set specified by \texttt{OPTIONS.MDSelect}.
\pagebreak[4]
\item
 \texttt{INFO} is a MATLAB structure containing additional information, as follows:\\[-2mm]
{\setlength\LTleft{30pt}\begin{longtable}{|l|p{12cm}|} \hline \textbf{\texttt{INFO} fields} & \textbf{Description} \\ \hline
\texttt{tcond} & $N$-dimensional vector; \texttt{INFO.tcond}$(i)$ contains
                      the maximum of the condition numbers of the
                      non-orthogonal transformation matrices used to determine
                      the $i$-th filter \texttt{Q\{$i$\}}; a warning is
                      issued if any \texttt{INFO.tcond}$(i)$ $\geq$ \texttt{OPTIONS.tcond}.\\ \hline
\texttt{degs}     & $N$-dimensional cell array; if  \texttt{OPTIONS.simple} = \texttt{true}, then a nonempty \texttt{INFO.degs\{$i$\}} contains the orders of the basis vectors of the employed simple nullspace basis for the synthesis of the $i$-th filter component \texttt{Q\{$i$\}}; if  \texttt{OPTIONS.simple} = \texttt{false}, then a nonempty \texttt{INFO.degs\{$i$\}} contains the degrees of the basis vectors of an equivalent polynomial nullspace basis;
\texttt{INFO.degs\{$i$\} = [ ]} for all $i$ which do not belong to the index set specified by \texttt{OPTIONS.MDSelect}.
\\ \hline
\texttt{MDperf}     & $N\times N$-dimensional  array containing the achieved distance mapping
                      performance,  given as the peak gains associated
                      with the internal representations (see \textbf{Method}).
                       \newline
                      $\texttt{INFO.MDperf($i,j$)} = -1$, for $j = 1, \ldots, N$ and for all $i$ which do not belong to the index set specified by \texttt{OPTIONS.MDSelect}.
                      \\ \hline
\texttt{HDesign}     & $N$-dimensional cell array, where $\texttt{INFO.HDesign\{$i$\}}$ contains the $i$-th  design matrix $H^{(i)}$ employed for the synthesis of the $i$-th filter (see \textbf{Method}) \\ \hline
\end{longtable}
}
\end{description}

\subsubsection*{Method}

An extension of the \textbf{Procedure EMD} from \cite[Sect.\ 6.2]{Varg17} is implemented, which
relies on the nullspace-based synthesis method proposed in \cite{Varg09h}. Assume that the $j$-th model has the input-output form
\be\label{emd:systemi} {\mathbf{y}}^{(j)}(\lambda) =
G_u^{(j)}(\lambda){\mathbf{u}}^{(j)}(\lambda)
+ G_d^{(j)}(\lambda){\mathbf{d}}^{(j)}(\lambda)
+ G_w^{(j)}(\lambda){\mathbf{w}}^{(j)}(\lambda) \ee
and the resulting $i$-th filter $Q^{(i)}(\lambda)$ has the input-output form
 \be\label{emd:detec1i}
{\mathbf{r}}^{(i)}(\lambda) = Q^{(i)}(\lambda)\ba{c}
{\mathbf{y}}(\lambda)\\{\mathbf{u}}(\lambda)\ea  \, . \ee
The synthesis method, which underlies \textbf{Procedure EMD}, essentially determines each  filter $Q^{(i)}(\lambda)$ as a stable rational left annihilator of
\[ G^{(i)}(\lambda) := \ba{cc} G_u^{(i)}(\lambda) & G_d^{(i)}(\lambda) \\
 I_{m_u} & 0 \ea ,\]
such that for $i\neq j$ we have $[\, R_u^{(i,j)}(\lambda)\; R_d^{(i,j)}(\lambda) \,] \neq 0$, where
 \[ R^{(i,j)}(\lambda) := [\, R_u^{(i,j)}(\lambda)\; R_d^{(i,j)}(\lambda) \; R_w^{(i,j)}(\lambda)\,] = Q^{(i)}(\lambda) \ba{ccc} G_u^{(j)}(\lambda) & G_d^{(j)}(\lambda) & G_w^{(j)}(\lambda) \\
 I_{m_u} & 0 & 0\ea \]
is the internal form of $Q^{(i)}(\lambda)$ acting on the $j$-th model.

Each filter $Q^{(i)}(\lambda)$ is determined in the product form
\be\label{emdsym:Qprod} Q^{(i)}(\lambda) = Q_3^{(i)}(\lambda)Q_2^{(i)}(\lambda)Q_1^{(i)}(\lambda) , \ee
where the factors are  determined as follows:
 \begin{itemize}
 \item[(a)] $Q_1^{(i)}(\lambda) = N_l^{(i)}(\lambda)$, with $N_l^{(i)}(\lambda)$ a $\big(p-r_d^{(i)}\big) \times (p+m_u)$ proper rational left nullspace basis satisfying $N_l^{(i)}(\lambda)G^{(i)}(\lambda) = 0$, with $r_d^{(i)} := \text{rank}\, G_d^{(i)}(\lambda)$; ($n_v^{(i)} := p-r_d^{(i)}$ is the number of basis vectors)
     \item[(b)] $Q_2^{(i)}(\lambda)$ is an admissible factor (i.e., guaranteeing model detectability) to perform least order synthesis;
     \item[(c)] $Q_3^{(i)}(\lambda)$ is a stable invertible factor determined  such that $Q^{(i)}(\lambda)$ has a desired dynamics.
 \end{itemize}
The computations of  individual factors depend on the user's options. Specific choices are discussed in what follows.
\subsubsection*{Computation of $Q_1^{(i)}(\lambda)$}
If \texttt{OPTIONS.nullspace = true}, then the left nullspace basis $N_l^{(i)}(\lambda)$ is determined as a minimal proper rational basis, or, if \texttt{OPTIONS.nullspace = false} (the default option) and $m_d^{(i)} = 0$, then the simple observer based basis $N_l^{(i)}(\lambda) = [\, I \; -G_u^{(i)}(\lambda)\,]$ is employed. If $N_l^{(i)}(\lambda)$ is a minimal rational basis and
if \texttt{OPTIONS.simple = true}, then $N_l^{(i)}(\lambda)$ is determined as a simple rational basis and the orders of the basis vectors are provided in \texttt{INFO.degs}. These are also the degrees of the basis vectors of an equivalent polynomial nullspace basis. If \texttt{OPTIONS.minimal = false}, a stable basis is determined whose dynamics is specified via \texttt{OPTIONS.sdeg} and \texttt{OPTIONS.poles}.

\subsubsection*{Computation of $Q_2^{(i)}(\lambda)$}

If \texttt{OPTIONS.minimal = false}, then $Q_2^{(i)}(\lambda) =  H^{(i)}$, where $H^{(i)}$ is a suitable $q_i\times \big(p-r_d^{(i)}\big)$ full row rank design matrix. $H^{(i)}$ is set as follows. If \texttt{OPTIONS.HDesign}$\{i\}$ is nonempty, then $H^{(i)} = \texttt{OPTIONS.HDesign}\{i\}$.
If \texttt{OPTIONS.HDesign}$\{i\}$ is empty, then $q_i = \texttt{OPTIONS.rdim}$ if \texttt{OPTIONS.rdim} is nonempty and $q_i = p-r_d^{(i)}$ if \texttt{OPTIONS.rdim} is empty, and the matrix $H^{(i)}$  is chosen to build $q_i$ linear combinations of the $p-r_d^{(i)}$ left nullspace basis vectors, such that
$H^{(i)}Q_1^{(i)}(\lambda)$ has full row rank. If $q_i = p-r_d^{(i)}$ then the choice $H^{(i)} = I_{p-r_d^{(i)}}$ is used, otherwise $H^{(i)}$ is chosen a randomly generated $q_i \times \big(p-r_d^{(i)}\big)$ real matrix.

If \texttt{OPTIONS.minimal = true}, then $Q_2^{(i)}(\lambda)$ is a $q_i\times \big(p-r_d^{(i)}\big)$ transfer function matrix, with $q_i$ chosen as above.  $Q_2^{(i)}(\lambda)$ is determined in the form
\[  Q_2^{(i)}(\lambda) = H^{(i)}+Y_2^{(i)}(\lambda)\, , \]
such that $Q_2^{(i)}(\lambda)Q_1^{(i)}(\lambda)$ $\big(\!= H^{(i)}N_l^{(i)}(\lambda)+Y_2^{(i)}(\lambda)N_l^{(i)}(\lambda)\big)$ and $Y_2^{(i)}(\lambda)$ are  the least order solution of a left minimal cover problem \cite{Varg17g}.  If \texttt{OPTIONS.HDesign}$\{i\}$ is nonempty, then $H^{(i)} = \texttt{OPTIONS.HDesign}\{i\}$, and if \texttt{OPTIONS.HDesign}$\{i\}$ is empty, then a suitable randomly generated $H^{(i)}$ is employed (see above).

The structure field \texttt{INFO.HDesign}$\{i\}$ contains the employed value of the design matrix $H^{(i)}$.

\subsubsection*{Computation of $Q_3^{(i)}(\lambda)$}
 $Q_3^{(i)}(\lambda)$ is a stable invertible transfer function matrix determined such that $Q^{(i)}(\lambda)$ in (\ref{emdsym:Qprod}) has a desired dynamics (specified via \texttt{OPTIONS.sdeg} and \texttt{OPTIONS.poles}).
\vspace*{5mm}

\index{model detection!distance mapping}
\index{performance evaluation!model detection!distance mapping}
The resulting $N\times N$ matrix \texttt{INFO.MDperf} can be used for the assessment of the achieved distance mapping performance of the resulting model detection filters (see Section \ref{sec:mdperf} for definitions). If \texttt{OPTIONS.MDFreq} is empty, then \texttt{INFO.MDperf}$(i,j) = \big\| \big[\, R_u^{(i,j)}(\lambda)\; R_d^{(i,j)}(\lambda) \,\big] \big\|_\infty$ if \texttt{OPTIONS.emdtest = true} and \texttt{INFO.MDperf}$(i,j) = \big\|  R_u^{(i,j)}(\lambda) \big\|_\infty$ if \texttt{OPTIONS.emdtest = false}, and, ideally, represents a measure of the distance between the $i$-th and $j$-th component systems. If \texttt{OPTIONS.MDFreq} contains a set of $n_f$ real frequency values $\omega_k$, $k = 1, \ldots, n_f$ and $\lambda_k$, $k = 1, \ldots, n_f$ are the corresponding complex frequencies (see the description of \texttt{OPTIONS.MDFreq}), then \texttt{INFO.MDperf}$(i,j)$ $= \max_k\big\|\big[\, R_u^{(i,j)}(\lambda_k)\; R_d^{(i,j)}(\lambda_k) \,\big]\big\|_\infty$ if \texttt{OPTIONS.emdtest = true} and \texttt{INFO.MDperf}$(i,j) =$ $\max_k\big\|  R_u^{(i,j)}(\lambda_k) \big\|_\infty$ if \texttt{OPTIONS.emdtest = false}. In this case, the entry \texttt{INFO.MDperf}$(i,j)$ ideally represents a measure of the maximum distance between the frequency responses of the $i$-th and $j$-th component systems, evaluated in the selected set of frequency values.
If \texttt{OPTIONS.normalize = true}, then for each row $i$, the filters \texttt{Q\{i\}} and \texttt{R\{i,j\}} are scaled such that the least value of \texttt{INFO.MDperf}$(i,j)$ for $i\neq j$ is normalized to one. The standard normalization is performed if \texttt{OPTIONS.normalize = false}, in which case \texttt{INFO.MDperf}$(1,j)$ = \texttt{INFO.MDperf}$(j,1)$ for $j > 1$.

\subsubsection*{Example}

\begin{example} \label{ex:Ex6.1}
This is \emph{Example} 6.1 from the book \cite{Varg17}, which deals with a continuous-time state-space model, describing, in the fault-free case,  the lateral dynamics
of an F-16 aircraft with the matrices
\[ A^{(1)} = \left[\begin{array}{rrrr}
   -0.4492&    0.046&    0.0053&   -0.9926\\
         0&         0&    1.0000&    0.0067\\
   -50.8436&         0&   -5.2184&    0.7220\\
   ~16.4148&         0&    0.0026&   -0.6627
\end{array}\right]
 , \quad
 B_u^{(1)} = \left[\begin{array}{rr}
    0.0004&    0.0011\\
     0&        0\\
   -1.4161&    0.2621\\
   -0.0633&   -0.1205
\end{array}\right] ,\] \[ \;\, C^{(1)} = I_4, \;\qquad D_u^{(1)} = 0_{4\times 2} \, .\]
The four state variables are the sideslip angle, roll angle, roll rate and yaw rate,
and the two input variables are the aileron deflection and rudder
deflection. The model detection problem addresses the synthesis of model detection filters for the detection and identification of loss of efficiency of the two flight actuators, which control the deflections of the aileron and rudder.
The individual fault models correspond to different degrees of surface
efficiency degradation. A multiple model with  $N = 9$ component models is used, which correspond to a two-dimensional
 parameter grid for $N$ values of the parameter vector
 $\rho := [\rho_1,\rho_2]^T$.  For each component of $\rho$, we employ the three grid points
 $\{0,0.5,1\}$.
The component system matrices in (\ref{emdsyn:sysiss}) are defined
for $i = 1, 2, \ldots, N$ as: $E^{(i)} = I_4$,
$A^{(i)} = A^{(1)}$, $C^{(i)} = C^{(1)}$, and $B_u^{(i)} = B^{(1)}_u \Gamma^{(i)}$, where $\Gamma^{(i)} = \diag \big(1-\rho_1^{(i)},1-\rho_2^{(i)}\big)$
and $\big(\rho_1^{(i)},\rho_2^{(i)}\big)$ are the values of parameters $(\rho_1,\rho_2)$
on the chosen grid: \begin{center}
{\tabcolsep=2mm\begin{tabular}{|r|rrrrrrrrr|} \hline
 $\rho_1:$ &  0  &       0 &   0  &  0.5  &  0.5 &   0.5 &   1 &     1 &   1 \\
 $\rho_2:$ &  0  &     0.5 &   1  &    0  &  0.5 &     1 &   0 &   0.5 &   1 \\\hline
\end{tabular}  . }
\end{center} For example, $\big(\rho_1^{(1)},\rho_2^{(1)}\big) = (0,0)$ corresponds to
the fault-free situation, while $\big(\rho_1^{(9)},\rho_2^{(9)}\big) =
(1,1)$ corresponds to complete failure of both control surfaces. It follows, that the TFM  $G_u^{(i)}(s)$ of the $i$-th system can be expressed as
\be\label{Gui_ex} G_u^{(i)}(s) = G_u^{(1)}(s)\Gamma^{(i)}, \ee
where
\[ G_u^{(1)}(s) = C^{(1)}\big(sI-A^{(1)}\big)^{-1}B^{(1)}_u \]
is the TFM of the fault-free system. Note that $G_u^{(N)}(s) = 0$ describes the case of complete failure.

The distances between the $i$-th and $j$-th models can be evaluated---for example, as the $\mathcal{H}_\infty$-norm of $G_u^{(i)}(s)-G_u^{(j)}(s)$, for $i,j = 1, \ldots, N$ and are plotted in Fig.~\ref{MDexample}.
\begin{figure}[thpb]
  \begin{center}
    \includegraphics[width=14cm]{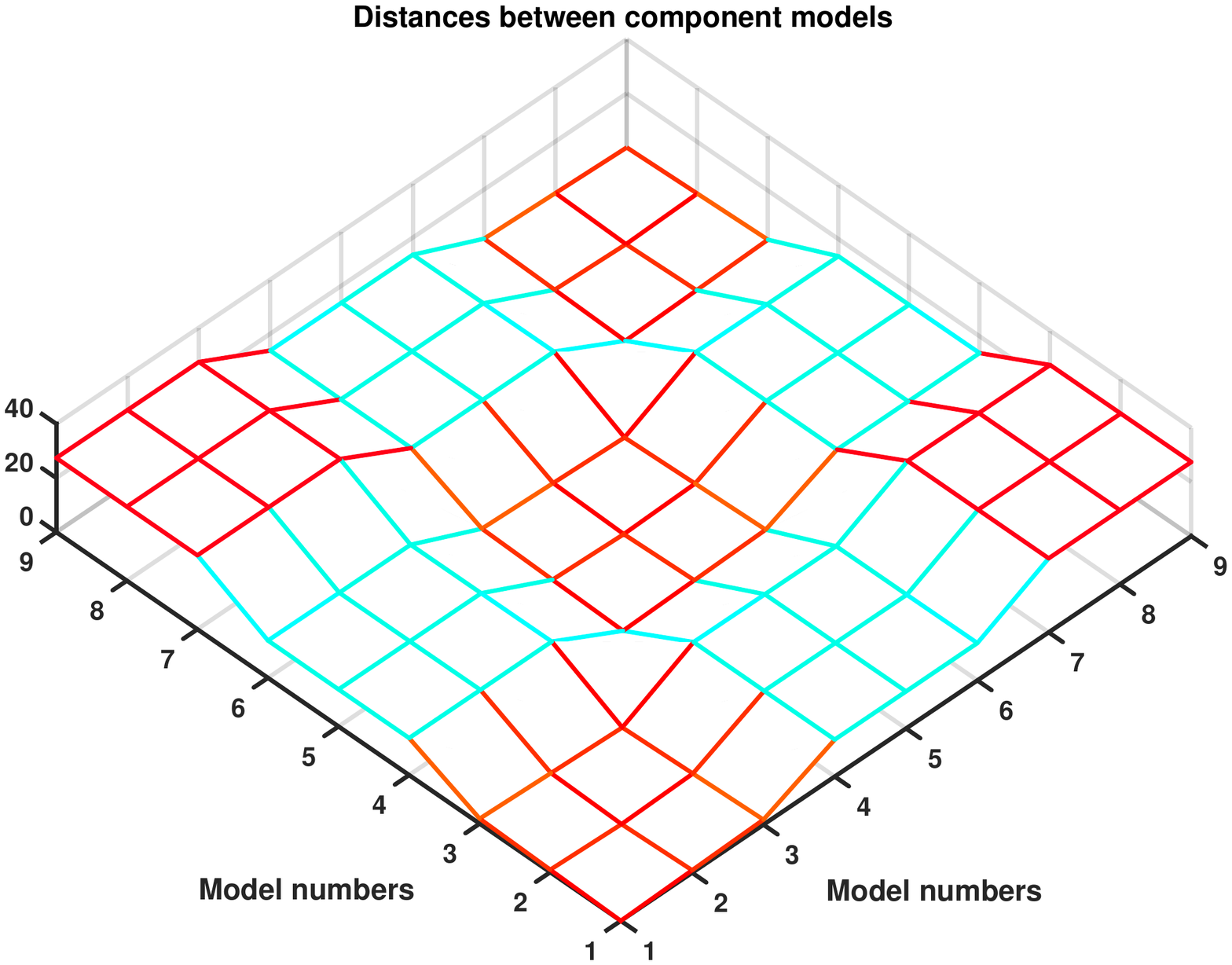} \vspace*{-12mm}
    \caption{Distances between component models in terms of $\big\|G_u^{(i)}(s)-G_u^{(j)}(s)\big\|_\infty$}
    \label{MDexample}
  \end{center}
\end{figure}

We aim to determine $N$ filters $Q^{(i)}(s)$, $i = 1, \ldots, N$, with scalar outputs, having least McMillan degrees and satisfactory dynamics,  which fulfill:
\begin{itemize}
\item[--] the decoupling conditions: $R_u^{(i,i)}(s) = 0$, $i = 1, \ldots, N$;
\item[--] the model detectability condition: $R_u^{(i,j)}(s) \neq 0, \quad \forall j\neq i, \;\;i,j = 1, \ldots, N$.
\end{itemize}
Additionally, the resulting model detection performance measures $\|R_u^{(i,j)}(s)\|_\infty$ should (ideally)  reproduce the shape of distances plotted in Fig.~\ref{MDexample}.

For the design of scalar filters, we used the same $1\times p$ design matrix $H$ for the synthesis of all filters, which has been chosen, after some trials with randomly generated values, as
\[ H = [ \,0.7645 \;\;  0.8848 \;\;  0.5778 \;\;  0.9026\,] \, .\]
The filter synthesis, performed by employing \texttt{emdsyn}, led to first order stable filters, which, as can be observed in Fig.~\ref{MDexampleres}, produces similar shapes of the model detection performance measure as those in Fig.~\ref{MDexample}.
\begin{figure}[thpb]
  \begin{center}
    \includegraphics[width=14cm]{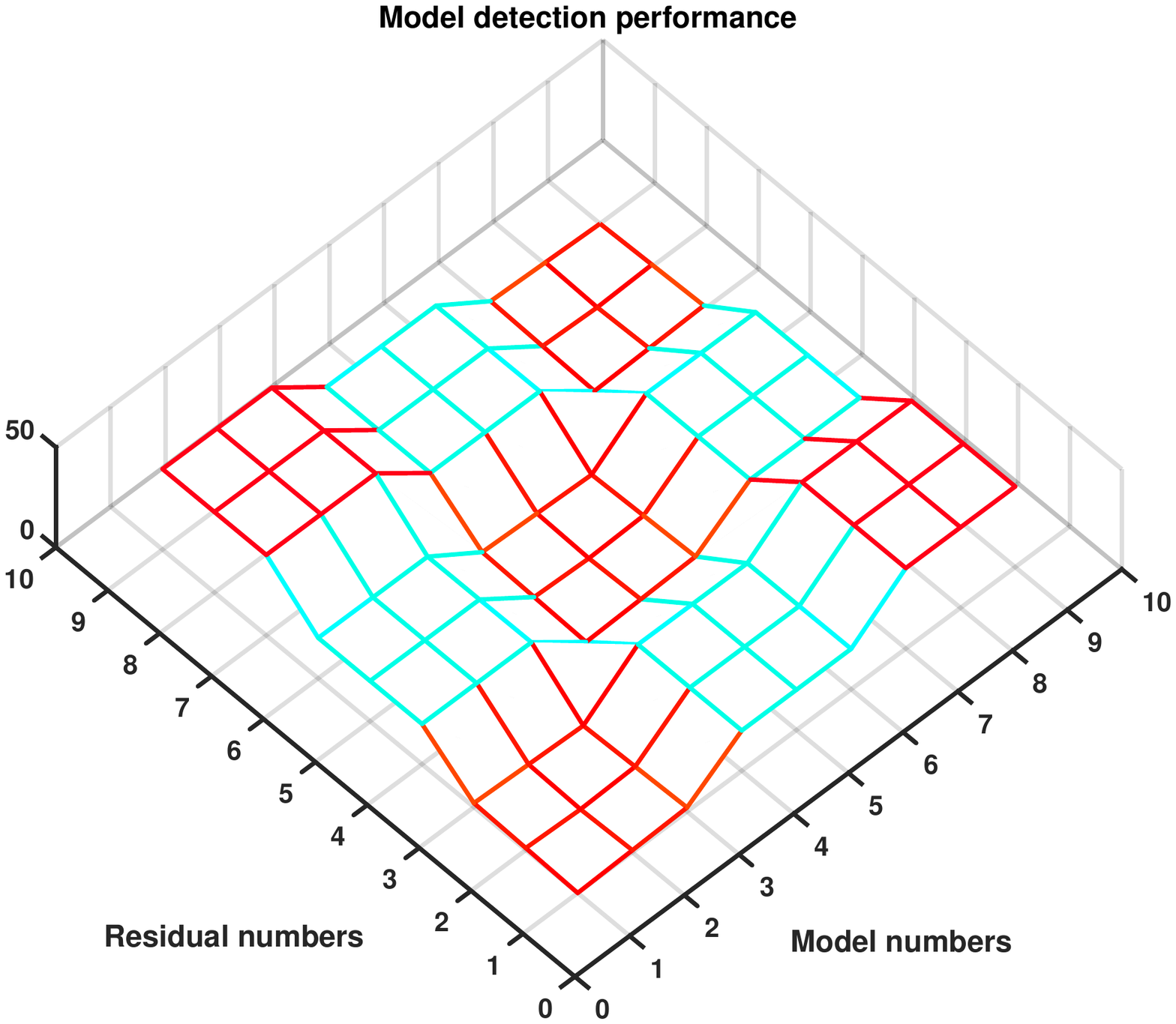}     \vspace*{-15mm}
  \end{center}
    \caption{Model detection performance in terms of $\big\|R_u^{(i,j)}(s)\big\|_\infty$}
    \label{MDexampleres}
\end{figure}

In Fig.~\ref{MDExampletimeresp} the step responses from $u_1$ (aileron) and $u_2$ (rudder) are presented for the $9\times 9$ block array, whose entries are the computed TFMs $R^{(i,j)}(s)$. Each column corresponds to a specific model for which the step responses of the $N$ residuals are computed.

\newpage
\begin{figure}[thpb]
  \begin{center}
    \hspace*{-1cm}
    \includegraphics[width=17.5cm]{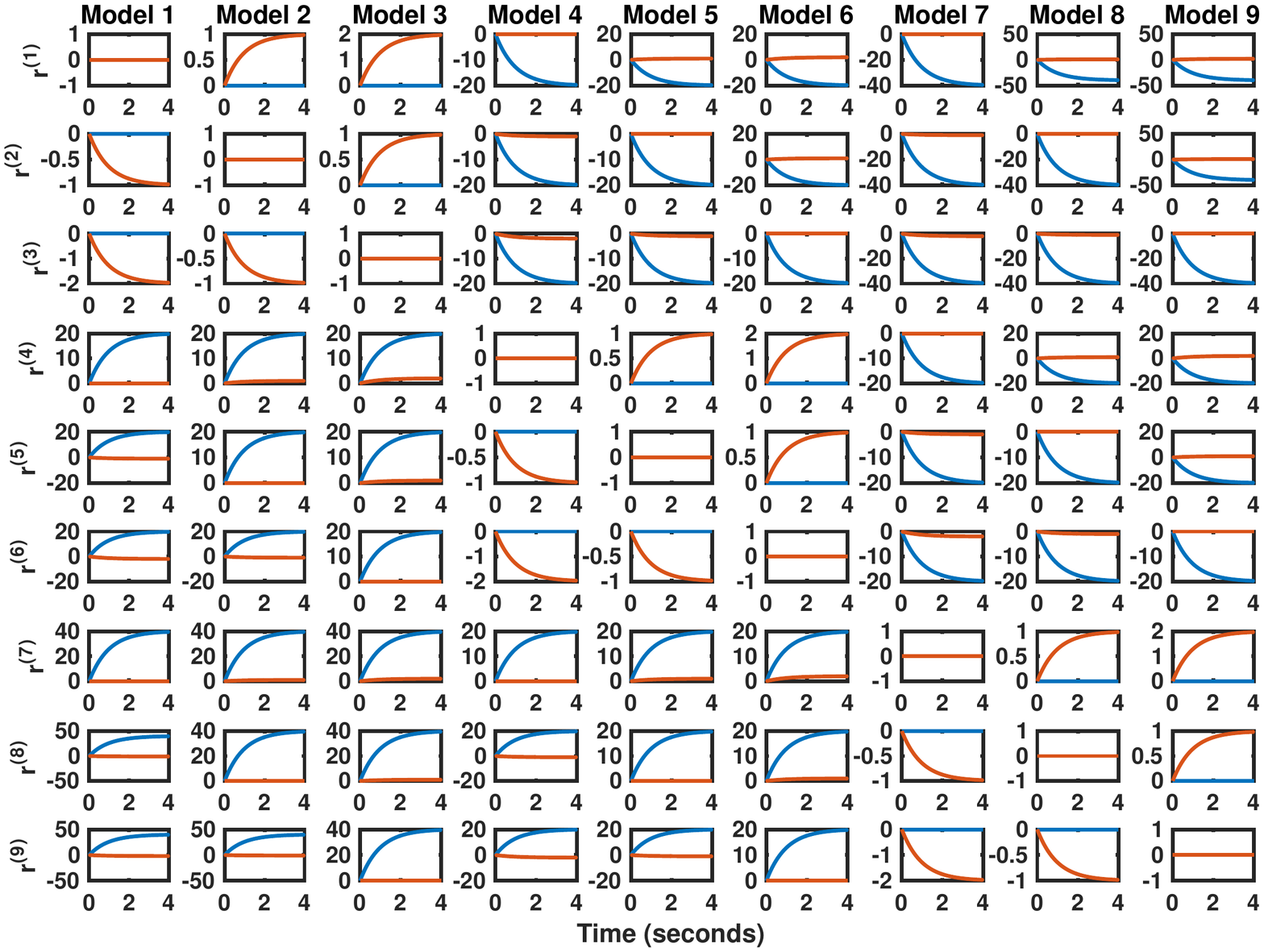} \vspace*{-.8cm}
    \caption[Step responses of $R^{(i,j)}(s)$ for least-order syntheses]{Step responses of $R^{(i,j)}(s)$ from $u_1$ (blue) and $u_2$ (red) for least order syntheses.}
    \label{MDExampletimeresp}
  \end{center}
\end{figure}

The following script implements the model building, filter synthesis and analysis steps.

\begin{verbatim}
% Example - Solution of an exact model detection problem (EMDP)

% Define a lateral aircraft dynamics model (without faults) with
% n  = 4 states
% mu = 2 control inputs
% p  = 4 measurable outputs
A = [-.4492 0.046 .0053 -.9926;
       0    0     1     0.0067;
   -50.8436 0   -5.2184  .722;
    16.4148 0     .0026 -.6627];
Bu = [0.0004 0.0011; 0 0; -1.4161 .2621; -0.0633 -0.1205];
C = eye(4); p = size(C,1); mu = size(Bu,2);

% define the loss of efficiency (LOE) faults as input scaling gains
% Gamma(i,:) = [ 1-rho1(i) 1-rho2(i) ]
Gamma = 1 - [ 0  0 0 .5 .5 .5 1  1 1;
              0 .5 1  0 .5  1 0 .5 1 ]';
N = size(Gamma,1);  % number of LOE cases

% define a multiple physical fault model Gui = Gu*diag(Gamma(i,:))
sysu = ss(zeros(p,mu,N,1));
for i=1:N
    sysu(:,:,i,1) = ss(A,Bu*diag(Gamma(i,:)),C,0);
end

% setup synthesis model
sysu = mdmodset(sysu,struct('controls',1:mu));

% nu-gap distance plots
nugapdist  = mddist(sysu);
figure, mesh(nugapdist), colormap hsv
title('\nu-gap distances between component models')
ylabel('Model numbers')
xlabel('Model numbers')

% H-infinity norm based distance plots
hinfdist  = mddist(sysu,struct('distance','Inf'));
figure, mesh(hinfdist), colormap hsv
title('H_\infty norm based distances between component models')
ylabel('Model numbers')
xlabel('Model numbers')


% call of EMDSYN with the options for stability degree -1 and pole -1 for
% the filters, tolerance and a design matrix H to form a linear combination
% of the left nullspace basis vectors
H = [ 0.7645   0.8848   0.5778   0.9026 ];
emdsyn_options = struct('sdeg',-1,'poles',-1,'HDesign',H);
[Q,R,info] = emdsyn(sysu,emdsyn_options);

% inspect achieved performance
figure, mesh(info.MDperf), colormap hsv
title('Distance mapping performance')
ylabel('Residual numbers')
xlabel('Model numbers')


% plot the step responses for the internal filter representations
figure
k1 = 0;
for j = 1:N,
  k1 = k1+1;
  k = k1;
  for i=1:N,
    subplot(N,N,k),
    [r,t] = step(R{j,i},4);
    plot(t,r(:,:,1),t,r(:,:,2)),
    if i == 1, title(['Model ',num2str(j)]), end
    if i == j, ylim([-1 1]), end
    if j == 1, ylabel(['r^(^', num2str(i),'^)'],'FontWeight','bold'), end
    if i == N && j == 5, xlabel('Time (seconds)','FontWeight','bold'), end
    k = k+N;
  end
end
\end{verbatim}
\end{example}

\subsubsection{\texttt{\bfseries amdsyn}}
\index{M-functions!\texttt{\bfseries amdsyn}}
\index{model detection problem!approximate (AMDP)}
\subsubsection*{Syntax}
\begin{verbatim}
[Q,R,INFO] = amdsyn(SYSM,OPTIONS)
\end{verbatim}

\subsubsection*{Description}

\texttt{\bfseries amdsyn} solves   the \emph{approximate model detection problem} (AMDP) (see Section \ref{sec:AMDP}), for a given stable LTI multiple model \texttt{SYSM} containing $N$ models. A bank of $N$ stable and proper residual generation filters $Q^{(i)}(\lambda)$, for $i = 1, \ldots, N$, is determined, in the form (\ref{detecmi}). For each filter $Q^{(i)}(\lambda)$, its associated  internal forms $R^{(i,j)}(\lambda)$, for $j = 1, \ldots, N$, are determined in accordance with  (\ref{ri_internal1}).

\subsubsection*{Input data}
\begin{description}
\item
\texttt{SYSM} is a multiple model which contains $N$ stable LTI systems in the state-space form
\be\label{amdsyn:sysiss}
\begin{array}{rcl}E^{(j)}\lambda x^{(j)}(t) &=& A^{(j)}x^{(j)}(t) + B^{(j)}_u u^{(j)}(t) + B^{(j)}_d d^{(j)}(t) + B^{(j)}_w w^{(j)}(t)  \, ,\\
y^{(j)}(t) &=& C^{(j)}x^{(j)}(t) + D^{(j)}_u u^{(j)}(t) + D^{(j)}_d d^{(j)}(t) + D^{(j)}_w w^{(j)}(t)   \, , \end{array} \ee
where $x^{(j)}(t) \in \mathds{R}^{n^{(j)}}$ is the state vector of the $j$-th system with control input $u^{(j)}(t) \in \mathds{R}^{m_u}$, disturbance input $d^{(j)}(t) \in \mathds{R}^{m_d^{(j)}}$ and noise input $w^{(j)}(t) \in \mathds{R}^{m_w^{(j)}}$, and
\index{faulty system model!physical}%
\index{faulty system model!multiple model}%
where any of the inputs components $u^{(j)}(t)$, $d^{(j)}(t)$, or $w^{(j)}(t)$  can be void. The multiple model \texttt{SYSM} is either an array of $N$ LTI systems of the form (\ref{amdsyn:sysiss}), in which case $m_d^{(j)} = m_d$ and $m_w^{(j)} = m_w$ for $j = 1, \ldots, N$,  or is a $1\times N$ cell array, with \texttt{SYSM\{$j$\}} containing the $j$-th component system in the form (\ref{amdsyn:sysiss}). The input groups for $u^{(j)}(t)$, $d^{(j)}(t)$, and $w^{(j)}(t)$  have the standard names \texttt{\bfseries 'controls'}, \texttt{\bfseries 'disturbances'}, and \texttt{\bfseries 'noise'}, respectively.

\item
 \texttt{OPTIONS} is a MATLAB structure used to specify various synthesis  options and has the following fields:
{\tabcolsep=1mm
\setlength\LTleft{30pt}\begin{longtable}{|l|lcp{10cm}|} \hline
\textbf{\texttt{OPTIONS} fields} & \multicolumn{3}{l|}{\textbf{Description}} \\ \hline
 \texttt{tol}   & \multicolumn{3}{p{9cm}|}{relative tolerance for rank computations \newline (Default: internally computed)} \\ \hline
 \texttt{tolmin}   & \multicolumn{3}{p{9cm}|}{absolute tolerance for observability tests
            \newline      (Default: internally computed)} \\ \hline
 \texttt{MDTol}   & \multicolumn{3}{l|}{threshold for model detectability checks
                 (Default: $10^{-4}$)} \\ \hline
 \texttt{MDGainTol}   & \multicolumn{3}{l|}{threshold for strong model detectability checks (Default: $10^{-2}$)} \\ \hline
 \texttt{rdim}   & \multicolumn{3}{p{12cm}|}{$N$-dimensional vector or a scalar; for a vector $q$, the $i$-th component $q_i$ specifies the desired number of residual outputs for the $i$-th filter  \texttt{Q\{$i$\}}; for a scalar value $\bar q$, a vector $q$ with all $N$ components $q_i = \bar q$ is assumed. \newline
(Default: \hspace*{-5.5mm}\begin{tabular}[t]{l} \hspace*{4.5mm}\texttt{[ ]}, in which case:\\ \hspace*{-2em} -- if \texttt{OPTIONS.HDesign\{$i$\}} is empty, then \\ $q_i = 1$, if \texttt{OPTIONS.minimal} = \texttt{true}, or \\ $q_i = n_v^{(i)}$, the dimension of the left nullspace which
                                underlies the \\ synthesis of \texttt{Q\{$i$\}}, if \texttt{OPTIONS.minimal} = \texttt{false} and $r_w^{(i)} = 0$ \\ (see \textbf{Method}); \\ $q_i = r_w^{(i)}$,  if \texttt{OPTIONS.minimal} = \texttt{false} and $r_w^{(i)} > 0$ \\ (see \textbf{Method}); \\ \hspace*{-2em} -- if \texttt{OPTIONS.HDesign\{$i$\}} is nonempty, then $q_i$ is the row dimension \\ of the design matrix contained in  \texttt{OPTIONS.HDesign\{$i$\}}.)
                                \end{tabular}}\\ \hline
 \texttt{emdtest}   & \multicolumn{3}{p{12cm}|}{option to perform extended model
                      detectability tests using both control and
                      disturbance input channels:}\\
                 &  \texttt{true} &--& use both control and disturbance input channels; \\
                 &  \texttt{false}&--& use only the control channel (default).  \\
                                        \hline
\texttt{smarg}   & \multicolumn{3}{p{11cm}|}{prescribed stability margin for the resulting filters \texttt{Q\{$i$\}} \newline
            (Default: \texttt{-sqrt(eps)} for continuous-time component systems; \newline
                   \hspace*{4.8em}\texttt{1-sqrt(eps)} for discrete-time component systems.
                   } \\ \hline
\texttt{sdeg}   & \multicolumn{3}{p{11cm}|}{prescribed stability degree for the resulting filters \texttt{Q\{$i$\}} \newline
            (Default: $-0.05$   for continuous-time component systems; \newline
                   \hspace*{4.8em} $0.95$ for discrete-time component systems.
                   } \\ \hline
 \texttt{poles}   & \multicolumn{3}{p{12cm}|}{complex vector containing a complex conjugate set of desired poles (within the stability margin) to be assigned for the resulting filters \texttt{Q\{$i$\}}
                     (Default: \texttt{[ ]})}\\
                                        \hline
 \texttt{MDFreq}  &  \multicolumn{3}{p{12cm}|}{real vector, which contains the frequency values $\omega_k$, $k = 1, \ldots, n_f$, to be used for strong model detectability checks. For each real frequency  $\omega_k$, there corresponds a complex frequency $\lambda_k$ which is used to evaluate the frequency-response gains. Depending on the system type, $\lambda_k = \mathrm{i}\omega_k$, in the continuous-time case, and $\lambda_k = \exp (\mathrm{i}\omega_k T)$, in the discrete-time case, where $T$ is the common sampling time of the component systems.  (Default: \texttt{[ ]}) } \\ \hline
 \texttt{nullspace}   & \multicolumn{3}{p{12cm}|}{option to use a specific type of proper nullspace bases:}\\
                 &  \texttt{true} &--& use minimal proper bases; \\
                 &  \texttt{false}&--& use full-order observer based bases (default)  \newline
                 \emph{Note:} This option can  only be used if no disturbance inputs are present in (\ref{amdsyn:sysiss}) and $\forall j$, $E^{(j)}$  is invertible.\\ \hline
\texttt{simple}   & \multicolumn{3}{l|}{option to compute simple proper bases:}\\
                 &  \texttt{true} &--& compute simple bases; the orders of the
                            basis vectors are provided in \texttt{INFO.degs}; \\
                 &  \texttt{false}&--& no simple basis computed (default)  \\
                                        \hline
 \texttt{minimal}   & \multicolumn{3}{l|}{option to perform least order filter syntheses:}\\
                 &  \texttt{true} &--& perform least order syntheses (default); \\
                 &  \texttt{false}&--& perform full order syntheses.  \\
                                        \hline
 \texttt{tcond}   & \multicolumn{3}{l|}{maximum allowed value for the condition numbers of the  employed}\\
     & \multicolumn{3}{l|}{non-orthogonal transformation matrices (Default: $10^4$)}\\
    & \multicolumn{3}{l|}{(only used if \texttt{OPTIONS.simple = true}) } \\
                    \hline
 \texttt{MDSelect}   & \multicolumn{3}{p{12cm}|}{integer vector with increasing elements containing the indices of the desired filters to be designed
                     (Default: $[\,1, \ldots, N\,]$)}\\
                                        \hline
 \texttt{HDesign}   & \multicolumn{3}{p{12cm}|}{$N$-dimensional cell array; \texttt{OPTIONS.HDesign\{$i$\}}, if not empty,  is a
                      full row rank design matrix employed for the
                      synthesis of the $i$-th  filter.
                      If \texttt{OPTIONS.HDesign} is specified as a full row rank
                      design matrix $H$, then an $N$-dimensional cell array is assumed with  \texttt{OPTIONS.HDesign\{$i$\}} = $H$, for $i = 1, \ldots, N$.
                      (Default:~\texttt{[ ]}).}\\
                                        \hline
\texttt{epsreg}   & \multicolumn{3}{p{11cm}|}{regularization parameter (Default: 0.1)} \\ \hline
\texttt{sdegzer}   & \multicolumn{3}{p{11cm}|}{prescribed stability degree for zeros shifting \newline (Default: $-0.05$ for a continuous-time system \texttt{SYSF}; \newline
                 \hspace*{4.8em} $0.95$ for a discrete-time system \texttt{SYSF}).}\\ \hline
\texttt{nonstd}   & \multicolumn{3}{l|}{option to handle nonstandard optimization problems (see \textbf{Method}):}\\
                 &  ~~~1 &--& use the quasi-co-outer--co-inner factorization (default); \\
                 &  ~~~2 &--& use the modified co-outer--co-inner factorization
                          with the regularization parameter \texttt{OPTIONS.epsreg};  \\
                 &  ~~~3 &--& use the Wiener-Hopf type co-outer--co-inner
                          factorization.  \\
                 &  ~~~4 &--& use the Wiener-Hopf type co-outer-co-inner factorization with zero shifting of the  non-minimum phase factor using the stabilization parameter \texttt{OPTIONS.sdegzer} \\
                 & ~~~5 &--& use the Wiener-Hopf type co-outer-co-inner factorization with the regularization of the non-minimum phase factor using the regularization parameter \texttt{OPTIONS.epsreg}  \\                                                                                                \hline
\end{longtable}}
\end{description}

\subsubsection*{Output data}
\begin{description}
\item
\texttt{Q} is an $N\times 1$ cell array of filters, where \texttt{Q\{$i$\}} contains the resulting $i$-th filter  in a standard state-space representation
\[
{\begin{aligned}
\lambda x_Q^{(i)}(t)  & =   A_Q^{(i)}x_Q^{(i)}(t)+ B_{Q_y}^{(i)}y(t)+ B_{Q_u}^{(i)}u(t) ,\\
r^{(i)}(t) & =  C_Q^{(i)} x_Q^{(i)}(t) + D_{Q_y}^{(i)}y(t)+ D_{Q_u}^{(i)}u(t) ,
\end{aligned}}
\]
where the residual signal $r^{(i)}(t)$ is a $q_i$-dimensional vector, with $q_i$ specified in \texttt{OPTIONS.rdim}. For each system object \texttt{Q\{$i$\}}, two input groups \texttt{\bfseries 'outputs'} and \texttt{\bfseries 'controls'} are defined for $y(t)$ and $u(t)$, respectively, and the output group \texttt{\bfseries 'residuals'} is defined for the residual signal $r^{(i)}(t)$. \texttt{Q\{$i$\}} is empty for all $i$ which do not belong to the index set specified by \texttt{OPTIONS.MDSelect}.

\item \texttt{R} is an $N\times N$ cell array of filters, where the $(i,j)$-th filter \texttt{R\{$i,j$\}}, is the internal form of \texttt{Q\{$i$\}}
  acting on the $j$-th model.
The resulting \texttt{R\{$i,j$\}}  has a standard state-space representation
\[
{\begin{aligned}
\lambda x_R^{(i,j)}(t)  & =   A_R^{(i,j)}x_R^{(i,j)}(t)+ B_{R_u}^{(i,j)}u^{(j)}(t)+  B_{R_d}^{(i,j)}d^{(j)}(t)+ B_{R_w}^{(i,j)}w^{(j)}(t), \\
r^{(i,j)}(t) & =  C_R^{(i,j)} x_R^{(i,j)}(t) + D_{R_u}^{(i,j)}u^{(j)}(t)+ D_{R_d}^{(i,j)}d^{(j)}(t)+ D_{R_w}^{(i,j)}w^{(j)}(t)
\end{aligned}}
\]
and the input groups \texttt{\bfseries 'controls'}, \texttt{\bfseries 'disturbances'}  and \texttt{\bfseries 'noise'} are defined for $u^{(j)}(t)$, $d^{(j)}(t)$, and $w^{(j)}(t)$, respectively, and the output group \texttt{\bfseries 'residuals'} is defined for the residual signal $r^{(i,j)}(t)$. \texttt{R\{$i,j$\}}, $j = 1, \ldots, N$ are empty for all $i$ which do not belong to the index set specified by \texttt{OPTIONS.MDSelect}.

\item
 \texttt{INFO} is a MATLAB structure containing additional information, as follows:\\
{\setlength\LTleft{30pt}\begin{longtable}{|l|p{12cm}|} \hline \textbf{\texttt{INFO} fields} & \textbf{Description} \\ \hline
\texttt{tcond} & $N$-dimensional vector; \texttt{INFO.tcond}$(i)$ contains
                      the maximum of the condition numbers of the
                      non-orthogonal transformation matrices used to determine
                      the $i$-th filter \texttt{Q\{$i$\}}; a warning is
                      issued if any \texttt{INFO.tcond}$(i)$ $\geq$ \texttt{OPTIONS.tcond}.\\ \hline
\texttt{degs}     & $N$-dimensional cell array; if  \texttt{OPTIONS.simple} = \texttt{true}, then a nonempty \texttt{INFO.degs\{$i$\}} contains the orders of the basis vectors of the employed simple nullspace basis for the synthesis of the $i$-th filter component \texttt{Q\{$i$\}}; if  \texttt{OPTIONS.simple} = \texttt{false}, then a nonempty \texttt{INFO.degs\{$i$\}} contains the degrees of the basis vectors of an equivalent polynomial nullspace basis;
\texttt{INFO.degs\{$i$\} = [ ]} for all $i$ which do not belong to the index set specified by \texttt{OPTIONS.MDSelect}.
\\ \hline
\texttt{MDperf}     & $N\times N$-dimensional  array containing the achieved model
                      detection performance measure,  given as the gains associated
                      with the internal representations (see \textbf{Method}).
                       \newline
                      $\texttt{INFO.MDperf($i,j$)} = -1$, for $j = 1, \ldots, N$ and for all $i$ which do not belong to the index set specified by \texttt{OPTIONS.MDSelect}.
                      \\ \hline
\texttt{HDesign}     & $N$-dimensional cell array, where $\texttt{INFO.HDesign\{$i$\}}$ contains the $i$-th  design matrix $H^{(i)}$ employed for the synthesis of the $i$-th filter (see \textbf{Method}) \\ \hline
\texttt{MDgap}     & $N$-dimensional vector, which contains the achieved noise gaps. \texttt{INFO.MDgap}$(i)$ contains the $i$-th gap $\eta_i$ achieved by the $i$-th filter (see \textbf{Method}). \\ \hline
\end{longtable}
}

\end{description}

\subsubsection*{Method}

An extension of the \textbf{Procedure AMD} from \cite[Sect.\ 6.3]{Varg17} is implemented, which
relies on the nullspace-based synthesis method proposed in \cite{Varg09h}. Assume that the $j$-th model has the input-output form
\be\label{amd:systemi} {\mathbf{y}}^{(j)}(\lambda) =
G_u^{(j)}(\lambda){\mathbf{u}}^{(j)}(\lambda)
+ G_d^{(j)}(\lambda){\mathbf{d}}^{(j)}(\lambda)
+ G_w^{(j)}(\lambda){\mathbf{w}}^{(j)}(\lambda) \ee
and the resulting $i$-th filter $Q^{(i)}(\lambda)$ has the input-output form
 \be\label{amd:detec1i}
{\mathbf{r}}^{(i)}(\lambda) = Q^{(i)}(\lambda)\ba{c}
{\mathbf{y}}(\lambda)\\{\mathbf{u}}(\lambda)\ea  \, . \ee
The synthesis method, which underlies \textbf{Procedure AMD}, essentially determines each  filter $Q^{(i)}(\lambda)$ as a stable rational left annihilator of
\[ G^{(i)}(\lambda) := \ba{cc} G_u^{(i)}(\lambda) & G_d^{(i)}(\lambda) \\
 I_{m_u} & 0 \ea ,\]
such that for $i\neq j$ we have $[\, R_u^{(i,j)}(\lambda)\; R_d^{(i,j)}(\lambda) \,] \neq 0$, where
 \[ R^{(i,j)}(\lambda) := [\, R_u^{(i,j)}(\lambda)\; R_d^{(i,j)}(\lambda) \; R_w^{(i,j)}(\lambda)\,] = Q^{(i)}(\lambda) \ba{ccc} G_u^{(j)}(\lambda) & G_d^{(j)}(\lambda) & G_w^{(j)}(\lambda) \\
 I_{m_u} & 0 & 0\ea \]
is the internal form of $Q^{(i)}(\lambda)$ with respect to the $j$-th model. Additionally, the gap $\eta_i$ achieved by the $i$-th filter is maximized.

\index{model detection!distance mapping}
\index{performance evaluation!model detection!distance mapping}
The resulting $N\times N$ matrix \texttt{INFO.MDperf} can be used for the assessment of the achieved distance mapping performance of the resulting model detection filters (see Section \ref{sec:mdperf} for definitions). If \texttt{OPTIONS.MDFreq} is empty, then \texttt{INFO.MDperf}$(i,j) = \big\| \big[\, R_u^{(i,j)}(\lambda)\; R_d^{(i,j)}(\lambda) \,\big] \big\|_\infty$ if \texttt{OPTIONS.emdtest = true} and \texttt{INFO.MDperf}$(i,j) = \big\|  R_u^{(i,j)}(\lambda) \big\|_\infty$ if \texttt{OPTIONS.emdtest = false}, and, ideally, represents a measure of the distance between the $i$-th and $j$-th component systems. If \texttt{OPTIONS.MDFreq} contains a set of $n_f$ real frequency values $\omega_k$, $k = 1, \ldots, n_f$ and $\lambda_k$, $k = 1, \ldots, n_f$ are the corresponding complex frequencies (see the description of \texttt{OPTIONS.MDFreq}), then \texttt{INFO.MDperf}$(i,j)$ $= \max_k\big\|\big[\, R_u^{(i,j)}(\lambda_k)\; R_d^{(i,j)}(\lambda_k) \,\big]\big\|_\infty$ if \texttt{OPTIONS.emdtest = true} and \texttt{INFO.MDperf}$(i,j) =$ $\max_k\big\|  R_u^{(i,j)}(\lambda_k) \big\|_\infty$ if \texttt{OPTIONS.emdtest = false}. In this case, the entry \texttt{INFO.MDperf}$(i,j)$ ideally represents a measure of the maximum distance between the frequency responses of the $i$-th and $j$-th component systems, evaluated in the selected set of frequency values.
If \texttt{OPTIONS.normalize = true}, then for each row $i$, the filters \texttt{Q\{i\}} and \texttt{R\{i,j\}} are scaled such that the least value of \texttt{INFO.MDperf}$(i,j)$ for $i\neq j$ is normalized to one. The standard normalization is performed if \texttt{OPTIONS.normalize = false}, in which case \texttt{INFO.MDperf}$(1,j)$ = \texttt{INFO.MDperf}$(j,1)$ for $j > 1$.

\index{performance evaluation!model detection!noise gap}
The $N$-dimensional vector \texttt{INFO.MDgap}, contains the resulting noise gaps $\eta_i$, for $i = 1, \ldots, N$ (see Section \ref{sec:mdgap} for definitions). If \texttt{OPTIONS.MDFreq} is empty,  then $\eta_i$ is evaluated as
\be\label{gapi_ext} \eta_i := \min_{j\neq i} \big\|\big[\,R_u^{(i,j)}(\lambda)\; R_d^{(i,j)}(\lambda)\,\big]\big\|_\infty/\big\|R_w^{(i,i)}(\lambda)\big\|_\infty , \ee
if \texttt{OPTIONS.emdtest = true} and
\be\label{gapi} \eta_i := \min_{j\neq i} \big\|R_u^{(i,j)}(\lambda)\big\|_\infty/\big\|R_w^{(i,i)}(\lambda)\big\|_\infty , \ee
if \texttt{OPTIONS.emdtest = false}.
If \texttt{OPTIONS.MDFreq} contains a set of $n_f$ real frequency values $\omega_k$, $k = 1, \ldots, n_f$ and $\lambda_k$, $k = 1, \ldots, n_f$ are the corresponding complex frequencies (see the description of \texttt{OPTIONS.MDFreq}), then
\be\label{gapifr-ext} \eta_i := \min_{j\neq i}\max_k \big\|\big[\,R_u^{(i,j)}(\lambda_k)\; R_d^{(i,j)}(\lambda_k)\,\big]\big\|_2/\big\|R_w^{(i,i)}(\lambda)\big\|_\infty , \ee
if \texttt{OPTIONS.emdtest = true} and
\be\label{gapifr} \eta_i := \min_{j\neq i}\max_k \big\|R_u^{(i,j)}(\lambda_k)\big\|_2/\big\|R_w^{(i,i)}(\lambda)\big\|_\infty , \ee
if \texttt{OPTIONS.emdtest = false}.

Each filter $Q^{(i)}(\lambda)$ is determined in the product form
\be\label{amdsym:Qprod} Q^{(i)}(\lambda) = Q_4^{(i)}(\lambda)Q_3^{(i)}(\lambda)Q_2^{(i)}(\lambda)Q_1^{(i)}(\lambda) , \ee
where the factors are  determined as follows:
 \begin{itemize}
 \item[(a)] $Q_1^{(i)}(\lambda) = N_l^{(i)}(\lambda)$, with $N_l^{(i)}(\lambda)$ a $\big(p-r_d^{(i)}\big) \times (p+m_u)$ proper rational left nullspace basis satisfying $N_l^{(i)}(\lambda)G^{(i)}(\lambda) = 0$, with $r_d^{(i)} := \text{rank}\, G_d^{(i)}(\lambda)$; ($n_v^{(i)} := p-r_d^{(i)}$ is the number of basis vectors)
     \item[(b)] $Q_2^{(i)}(\lambda)$ is an admissible regularization factor guaranteeing model detectability;
     \item[(c)] $Q_3^{(i)}(\lambda)$ represents an optimal choice to maximize the $i$-th gap $\eta_i$ in (\ref{gapi_ext}) -- (\ref{gapifr});
     \item[(d)] $Q_4^{(i)}(\lambda)$ is a stable invertible factor determined  such that $Q^{(i)}(\lambda)$ has a desired dynamics.
 \end{itemize}
The computations of  individual factors depend on the user's options and the optimization problem features. Specific choices are discussed in what follows.
\subsubsection*{Computation of $Q_1^{(i)}(\lambda)$}
If \texttt{OPTIONS.nullspace = true}, then the left nullspace basis $N_l^{(i)}(\lambda)$ is determined as a stable minimal proper rational basis, or, if \texttt{OPTIONS.nullspace = false} (the default option) and $m_d^{(i)} = 0$, then the simple observer based basis $N_l^{(i)}(\lambda) = [\, I \; -G_u^{(i)}(\lambda)\,]$ is employed. If $N_l^{(i)}(\lambda)$ is a minimal rational basis and
if \texttt{OPTIONS.simple = true}, then $N_l^{(i)}(\lambda)$ is determined as a simple rational basis and the orders of the basis vectors are provided in \texttt{INFO.degs}. These are also the degrees of the basis vectors of an equivalent polynomial nullspace basis. The dynamics of $Q_1^{(i)}(\lambda)$ is specified via \texttt{OPTIONS.sdeg} and \texttt{OPTIONS.poles}.

\subsubsection*{Computation of $Q_2^{(i)}(\lambda)$}
Let $r_w^{(i)}$ be the rank of $\overline G_w^{(i,i)}(\lambda) :=  Q_1^{(i)}(\lambda)
\left[\begin{smallmatrix}  G_w^{(i)}(\lambda) \\ 0 \end{smallmatrix}\right]$ and let $q_i$ be the desired number of residual outputs for the $i$-th filter.  If \texttt{OPTIONS.rdim} is nonempty, then $q_i = \texttt{OPTIONS.rdim}\{i\}$, while if \texttt{OPTIONS.rdim} is empty a default value of $q_i$ is defined depending on the setting of $\texttt{OPTIONS.HDesign}\{i\}$ and \texttt{OPTIONS.minimal}. If $\texttt{OPTIONS.HDesign}\{i\}$ is nonempty, then $q_i$ is the row dimension of the design matrix contained in $\texttt{OPTIONS.HDesign}\{i\}$.  If  $\texttt{OPTIONS.HDesign}\{i\}$ is empty, then $q_i = 1$ if \texttt{OPTIONS.minimal = true}. If \texttt{OPTIONS.minimal = false} and $r_w^{(i)} = 0$, then $q_i = p-r_d^{(i)}$, while if $r_w^{(i)} > 0$, then $q_i = r_w^{(i)}$.

If \texttt{OPTIONS.minimal = false}, then $Q_2^{(i)}(\lambda) = M^{(i)}(\lambda) H^{(i)}$, where $H^{(i)}$ is a suitable $q_i\times \big(p-r_d^{(i)}\big)$ full row rank design matrix
and $M^{(i)}(\lambda)$ is a stable invertible transfer function matrix determined such that $Q_2^{(i)}(\lambda)Q_1^{(i)}(\lambda)$ has a desired dynamics (specified via \texttt{OPTIONS.sdeg} and \texttt{OPTIONS.poles}). $H^{(i)}$ is set as follows. If \texttt{OPTIONS.HDesign}$\{i\}$ is nonempty, then $H^{(i)} = \texttt{OPTIONS.HDesign}\{i\}$.
If \texttt{OPTIONS.HDesign}$\{i\}$ is empty, the matrix $H^{(i)}$  is chosen to build $q_i$ linear combinations of the $p-r_d^{(i)}$ left nullspace basis vectors, such that
$H^{(i)}Q_1^{(i)}(\lambda)$ has full row rank. If $q_i = p-r_d^{(i)}$ then the choice $H^{(i)} = I_{p-r_d^{(i)}}$ is used, otherwise $H^{(i)}$ is chosen a randomly generated $q_i \times \big(p-r_d^{(i)}\big)$ real matrix.

If \texttt{OPTIONS.minimal = true}, then $Q_2^{(i)}(\lambda)$ is a $q_i\times \big(p-r_d^{(i)}\big)$ transfer function matrix, with $q_i$ chosen as above.  $Q_2^{(i)}(\lambda)$ is determined in the form
\[  Q_2^{(i)}(\lambda) = M^{(i)}(\lambda)\widetilde Q_2^{(i)}(\lambda) \, , \]
where $\widetilde Q_2^{(i)}(\lambda) := H^{(i)}+Y_2^{(i)}(\lambda)$, $\widetilde Q_2^{(i)}(\lambda)Q_1^{(i)}(\lambda)$ $\big(\! = H^{(i)}N_l^{(i)}(\lambda)+Y_2^{(i)}(\lambda)N_l^{(i)}(\lambda)\big)$ and $Y_2^{(i)}(\lambda)$ are  the least order solution of a left minimal cover problem \cite{Varg17g}, and $M^{(i)}(\lambda)$, a stable invertible transfer function matrix determined such that $M^{(i)}(\lambda)\widetilde Q_2^{(i)}(\lambda)Q_1^{(i)}(\lambda)$ has a desired dynamics.  If \texttt{OPTIONS.HDesign}$\{i\}$ is nonempty, then $H^{(i)} = \texttt{OPTIONS.HDesign}\{i\}$, and if \texttt{OPTIONS.HDesign}$\{i\}$ is empty, then a suitable randomly generated $H^{(i)}$ is employed (see above).

\emph{Note:} The stabilization with $M^{(i)}(\lambda)$ is only performed if $r_w^{(i)} = 0$ (i.e.,  to also cover the case of an exact synthesis).

The structure field \texttt{INFO.HDesign}$\{i\}$ contains the employed value of the design matrix $H^{(i)}$.

\subsubsection*{Computation of $Q_3^{(i)}(\lambda)$}
Let define $\widetilde G_w^{(i,i)}(\lambda) :=  Q_2^{(i)}(\lambda)\overline G_w^{(i,i)}(\lambda)$ and let $\tilde r_w^{(i)} = \rank \widetilde G_w^{(i,i)}(\lambda)$, which satisfies $\tilde r_w^{(i)} \leq q_i$.
If  $\tilde r_w^{(i)} = 0$ then $Q_3^{(i)}(\lambda) = I$. If $\tilde r_w^{(i)} > 0$, then the quasi-co-outer--co-inner factorization of $\widetilde G_w^{(i,i)}(\lambda)$ is computed as
\[ \widetilde G_w^{(i,i)}(\lambda) = R_{wo}^{(i)}(\lambda)R_{wi}^{(i)}(\lambda) ,\]
where $R_{wo}^{(i)}(\lambda)$ is a (full column rank) quasi-co-outer factor and  $R_{wi}^{(i)}(\lambda)$ is a (full row rank) co-inner factor. In the \emph{standard case} $R_{wo}^{(i)}(\lambda)$ is outer (i.e., has no zeros on the boundary of the stability domain $\partial\mathds{C}_s$) and thus, there exists a stable left inverse $\big(R_{wo}^{(i)}(\lambda)\big)^{-L}$ such that $\big(R_{wo}^{(i)}(\lambda)\big)^{-L}R_{wo}^{(i)}(\lambda) = I$. In this case, we choose $Q_3^{(i)}(\lambda) = \big(R_{wo}^{(i)}(\lambda)\big)^{-L}$.
This is an optimal choice which ensures that the maximal gap $\eta_i$ is achieved by the $i$-th filter (see below). If $\tilde r_w = q_i$ (the usual case), then $Q_3^{(i)}(\lambda)$ is simply $Q_3^{(i)}(\lambda) = \big(R_{wo}^{(i)}(\lambda)\big)^{-1}$.

In the \emph{non-standard case} $R_{wo}^{(i)}(\lambda)$ is only quasi-outer and thus,  has zeros on the boundary of the stability domain $\partial\mathds{C}_s$. Depending on the selected option to  handle nonstandard optimization problems \texttt{OPTIONS.nonstd}, several choices are possible for $Q_3^{(i)}(\lambda)$ in this case:
 \begin{itemize}
 \item If \texttt{OPTIONS.nonstd} = 1, then $Q_3^{(i)}(\lambda) = \big(R_{wo}^{(i)}(\lambda)\big)^{-L}$ is used, where all spurious poles of the left inverse are assigned to values specified via \texttt{OPTIONS.sdeg} and \texttt{OPTIONS.poles}.
 \item If \texttt{OPTIONS.nonstd} = 2, then a modified co-outer--co-inner factorization of $[\,R_{wo}^{(i)}(\lambda)\;\epsilon I\,]$ is computed, whose co-outer factor $R_{wo,\epsilon}^{(i)}(\lambda)$ satisfies
\[ R_{wo,\epsilon}^{(i)}(\lambda)\big(R_{wo,\epsilon}^{(i)}(\lambda)\big)^\sim = \epsilon^2 I + R_{wo}^{(i)}(\lambda)\big(R_{wo}^{(i)}(\lambda)\big)^\sim . \]
Then $Q_3^{(i)}(\lambda) = \big(R_{wo,\epsilon}^{(i)}(\lambda)\big)^{-L}$ is used. The value of the regularization parameter $\epsilon$ is specified via \texttt{OPTIONS.epsreg}.
\item If \texttt{OPTIONS.nonstd} = 3, then a Wiener-Hopf type co-outer--co-inner factorization is computed in the form
\be\label{WHfact-md} \widetilde G_w^{(i,i)}(\lambda) = R_{wo}^{(i)}(\lambda)R_{wb}^{(i)}(\lambda)R_{wi}^{(i)}(\lambda) ,\ee
where $R_{wo}^{(i)}(\lambda)$ is co-outer, $R_{wi}^{(i)}(\lambda)$ is co-inner,  and $R_{wb}^{(i)}(\lambda)$ is a square stable factor whose zeros are precisely the zeros of $\widetilde G_w^{(i,i)}(\lambda)$ in $\partial\mathds{C}_s$.
$Q_3^{(i)}(\lambda)$ is determined as before  $Q_3^{(i)}(\lambda) = \big(R_{wo}^{(i)}(\lambda)\big)^{-L}$.
\item If \texttt{OPTIONS.nonstd} = 4, then the Wiener-Hopf type co-outer--co-inner factorization (\ref{WHfact-md}) is computed and
$Q_3^{(i)}(\lambda)$ is determined as  $Q_3^{(i)}(\lambda) = \big(R_{wb}^{(i)}(\tilde{\lambda})\big)^{-1}\big(R_{wo}^{(i)}(\lambda)\big)^{-1}$, where $\tilde{\lambda}$ is a small perturbation of $\lambda$ to move all zeros of $R_{wb}^{(i)}(\lambda)$ into the stable domain. In the continuous-time case $\tilde s = \frac{s-\beta_z}{1-\beta_zs}$, while in the discrete-time case $\tilde z = z/\beta_z$, where the zero shifting parameter $\beta_z$ is the prescribed stability degree for the zeros specified in  \texttt{OPTIONS.sdegzer}. For the evaluation of $R_{wb}^{(i)}(\tilde\lambda)$, a suitable bilinear transformation is performed.
\item If \texttt{OPTIONS.nonstd} = 5, then the Wiener-Hopf type co-outer--co-inner factorization (\ref{WHfact-md}) is computed and
$Q_3^{(i)}(\lambda)$ is determined as  $Q_3^{(i)}(\lambda) = \big(R_{wb,\epsilon}^{(i)}(\lambda)\big)^{-1}\big(R_{wo}^{(i)}(\lambda)\big)^{-1}$, where $R_{wb,\epsilon}^{(i)}(\lambda)$ is the co-outer factor of the co-outer--co-inner factorization of $\big[\, R_{wb}^{(i)}(\lambda) \; \epsilon I\,\big]$ and satisfies
\[ R_{wb,\epsilon}^{(i)}(\lambda)\big(R_{wb,\epsilon}^{(i)}(\lambda)\big)^\sim = \epsilon^2 I + R_{wb}^{(i)}(\lambda)\big(R_{wb}^{(i)}(\lambda)\big)^\sim . \]
The value of the regularization parameter $\epsilon$ is specified via \texttt{OPTIONS.epsreg}.
\end{itemize}

A typical feature of the non-standard case is that, with the exception of using the option \texttt{OPTIONS.nonstd} = 3, all other choices of \texttt{OPTIONS.nonstd} lead to a poor dynamical performance of the resulting filter, albeit arbitrary large noise gaps $\eta_i$ can be occasionally achieved.

\subsubsection*{Computation of $Q_4^{(i)}(\lambda)$}
In the standard case, $Q_4^{(i)}(\lambda) = I$. In the non-standard case,
 $Q_4^{(i)}(\lambda)$ is a stable invertible transfer function matrix determined such that $Q^{(i)}(\lambda)$ in (\ref{amdsym:Qprod}) has a desired dynamics (specified via \texttt{OPTIONS.sdeg} and \texttt{OPTIONS.poles}).

\subsubsection*{Example}

\begin{example} \label{ex:Ex6.2}
This is Example 6.2 from the book \cite{Varg17}, which deals with a continuous-time state-space model, describing, in the fault-free case,  the lateral dynamics
of an F-16 aircraft with the matrices\\[-2mm]
\[ {\arraycolsep=1mm A^{(1)}\! =\! \ba{rrrr}
   -0.4492&    0.046&    0.0053&   -0.9926\\
         0&         0&    1.0000&    0.0067\\
  -50.8436&         0&   -5.2184&    0.7220\\
   16.4148&         0&    0.0026&   -0.6627
\ea,
\;\;
 B_u^{(1)} \!=\! \ba{rr}
    0.0004&    0.0011\\
         0&         0\\
   -1.4161&    0.2621\\
   -0.0633&   -0.1205
\ea,} \;\, B_w^{(1)} \!=\! [\, I_4 \; 0_{4\times 2}\,] \; , \]
\[ C^{(1)} = \ba{cccc} 57.2958 & 0 & 0 & 0\\ 0 & 57.2958 & 0 & 0 \ea, \;\, D_u^{(1)} = 0_{2\times 2} , \;\, D_w^{(1)} = [\, 0_{2\times 4} \;\; I_2\,] \, .\]
The four state variables are the sideslip angle, roll angle, roll rate and yaw rate,
and the two input variables are the aileron deflection and rudder
deflection. The two measured outputs are the sideslip angle and roll angle, and, additionally input noise and output noise are included in the model.
The component system matrices in (\ref{amdsyn:sysiss}) are defined
for $i = 1, 2, \ldots, N$ as: $E^{(i)} = I_4$,
$A^{(i)} = A^{(1)}$, $C^{(i)} = C^{(1)}$, $B_w^{(i)} = B^{(1)}_w$, $D_w^{(i)} = D^{(1)}_w$, and $B_u^{(i)} = B^{(1)}_u \Gamma^{(i)}$, where $\Gamma^{(i)} = \diag \big(1-\rho_1^{(i)},1-\rho_2^{(i)}\big)$
and $\big(\rho_1^{(i)},\rho_2^{(i)}\big)$ are the values of parameters $(\rho_1,\rho_2)$
on the chosen grid points:\\[-6mm]
\begin{center}
{\tabcolsep=2mm\begin{tabular}{|r|rrrrrrrrr|} \hline
 $\rho_1:$ &  0  &       0 &   0  &  0.5  &  0.5 &   0.5 &   1 &     1 &   1 \\
 $\rho_2:$ &  0  &     0.5 &   1  &    0  &  0.5 &     1 &   0 &   0.5 &   1 \\\hline
\end{tabular}  . }
\end{center}
The TFMs  $G_u^{(i)}(s)$ and $G_w^{(i)}(s)$ of the $i$-th system can be expressed as
\be\label{Gui_ex2} G_u^{(i)}(s) = G_u^{(1)}(s)\Gamma^{(i)}, \quad G_w^{(i)}(s) = G_w^{(1)}(s) \; ,\ee
where\\[-6mm]
\[ G_u^{(1)}(s) = C^{(1)}\big(sI-A^{(1)}\big)^{-1}B^{(1)}_u, \quad G_w^{(1)}(s) = C^{(1)}\big(sI-A^{(1)}\big)^{-1}B^{(1)}_w + D^{(1)}_w \, .\]
The individual fault models correspond to different degrees of surface
efficiency degradation.
The values  $\big(\rho_1^{(1)},\rho_2^{(1)}\big) = (0,0)$ correspond to
the fault-free situation, while $G_u^{(N)}(s) = 0$ describes the case of complete failure.

The approximate model detection problem addresses the synthesis of model detection filters for the detection and identification of loss of efficiency of the two flight actuators, which control the deflections of the aileron and rudder, in the presence of noise inputs.
For the design of the  model detection system, we aim to determine the $N$ filters $Q^{(i)}(s)$, $i = 1, \ldots, N$ having satisfactory dynamic responses and exhibiting the maximally achievable gaps.

In Fig.~\ref{AMDExampletimeresp} the time responses of the residual evaluation signals $\theta_i(t)$ are presented, where $\theta_i(t)$ are computed using a Narendra-type evaluation filter \cite{Nare97} with input $\|r^{(i)}(t)\|_2^2$ and parameters $\alpha = 0.9$, $\beta = 0.1$, $\gamma=10$ (see also Remark 3.13 in \cite{Varg17}). The control inputs have been chosen as follows: $u_1(t)$ is a step of amplitude 0.3 added to a square wave of period $2\pi$, and $u_2(t)$ is a step of amplitude 1.5 added to a sinus function of unity amplitude and period $\pi$. The noise inputs are zero mean white noise of amplitude 0.01 for the input noise and 0.03 for the measurement noise. Each column corresponds to a specific model for which the time responses of the $N$ residual evaluation signals are computed. The achieved typical  structure matrix for model detection (with zeros down the diagonal) can easily be read out from this signal based assessment, even in presence of noise.

\begin{figure}[thpb]
  \begin{center}
    \hspace*{-1cm}
    \includegraphics[width=13.5cm]{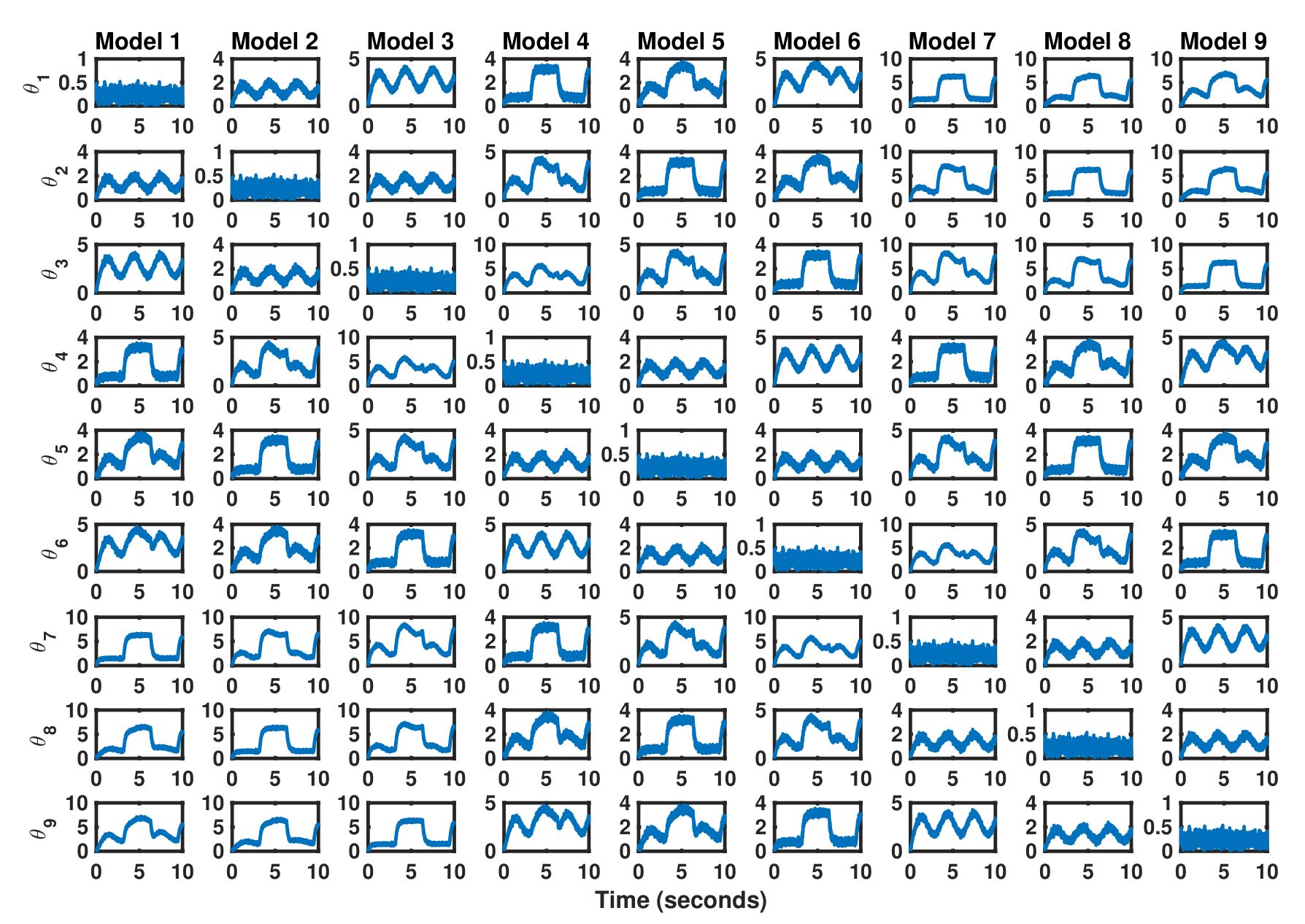} \vspace*{-.4cm}
    \caption{Time responses of evaluation signals for optimal syntheses}
    \label{AMDExampletimeresp}
  \end{center}
\end{figure}
The following script implements the model building, synthesis and analysis steps.

\begin{verbatim}
% Example - Solution of an approximate model detection problem (AMDP)

% Define a lateral aircraft dynamics model (without faults) with
% n  = 4 states
% mu = 2 control inputs
% p  = 2 measurable outputs
% mw = 6 noise inputs

A = [-.4492 .046  .0053 -.9926;
       0    0     1      .0067;
   -50.8436 0   -5.2184  .722;
    16.4148 0     .0026 -.6627];
Bu = [.0004 .0011; 0 0; -1.4161 .2621; -.0633 -.1205];
[n,mu] = size(Bu); p = 2; mw = n+p; m = mu+mw;
Bw = eye(n,mw);
C = 180/pi*eye(p,n);  Du = zeros(p,mu); Dw = [zeros(p,n) eye(p)];

% define the loss of efficiency (LOE) faults as input scaling gains
% Gamma(i,:) = [ 1-rho1(i) 1-rho2(i) ]
Gamma = 1 - [ 0  0 0 .5 .5 .5 1  1 1;
              0 .5 1  0 .5  1 0 .5 1 ]';
N = size(Gamma,1);  % number of LOE cases

% define a multiple physical fault model Gui = Gu*diag(Gamma(i,:))
sysuw = ss(zeros(p,mu+mw,N,1));
for j = 1:N
    sysuw(:,:,j,1) = ss(A,[Bu*diag(Gamma(j,:)) Bw],C,[Du Dw]);
end

% setup synthesis model
sysuw = mdmodset(sysuw,struct('controls',1:mu,'noise',mu+(1:mw)));

% use nonminimal design with  AMDSYN with stability degree -1 and poles  at -1
opt_amdsyn = struct('sdeg',-1,'poles',-1,'minimal',false);
[Q,R,info] = amdsyn(sysuw,opt_amdsyn);
info.MDperf, info.MDgap

% inspect achieved performance
figure, mesh(info.MDperf), colormap hsv
title('Model detection performance')
ylabel('Residual numbers')
xlabel('Model numbers')

% generate input signals for Ex. 6.2
d = diag([ 1 1 0.01 0.01 0.01 0.01 0.03 0.03]);
t = (0:0.01:10)';  ns = length(t);
usin = gensig('sin',pi,t(end),0.01)+1.5;
usquare = gensig('square',pi*2,t(end),0.01)+0.3;
u = [ usquare usin (rand(ns,mw)-0.5)]*d;

% plot the step responses for the internal filter representations
figure
k1 = 0; alpha = 0.9; beta = 0.1; gamma = 10;
for j = 1:N,
  k1 = k1+1; k = k1;
  for i=1:N,
    subplot(N,N,k),
    [r,t] = lsim(R{j,i},u,t);
    % use a Narendra filter with (alpha,beta,gamma) = (0.9,0.1,10)
    theta = alpha*sqrt(r(:,1).^2+r(:,2).^2)+...
        beta*sqrt(lsim(tf(1,[1 gamma]),r(:,1).^2+r(:,2).^2,t));
    plot(t,theta),
    if i == 1, title(['Model ',num2str(j)]), end
    if i == j, ylim([0 1]), end
    if j == 1, ylabel(['\theta_', num2str(i)],'FontWeight','bold'), end
    if i == N && j == 5, xlabel('Time (seconds)','FontWeight','bold'), end
    k = k+N;
  end
end
\end{verbatim}
\end{example}

\bookmarksetup{startatroot}
\cleardoublepage
\phantomsection

\newpage
\pdfbookmark[1]{References}{Refs}

\appendix
\newpage
\section{Installing FDITOOLS} \label{appA}

\textbf{FDITOOLS} runs  with MATLAB R2015b (or later versions) under 64-bit Windows 7 (or later). Additionally, the \emph{Control System Toolbox} (Version 9.10 or later) and the \emph{Descriptor Systems Tools} (\textbf{DSTOOLS}) collection (Version 0.64 or later) are necessary to be installed.
To install \textbf{FDITOOLS},  perform the following steps:
\begin{itemize}
\item  download \textbf{FDITOOLS} and \textbf{DSTOOLS} as zip files from Bitbucket\footnote{Download \textbf{FDITOOLS} from \url{https://bitbucket.org/DSVarga/fditools}, and \textbf{DSTOOLS} from \url{https://bitbucket.org/DSVarga/dstools}}
\item create on your computer the directories \texttt{fditools} and \texttt{dstools}
\item  extract, using any unzip utility, the functions of the \textbf{FDITOOLS} and \textbf{DSTOOLS} collections in the corresponding directories  \texttt{fditools} and \texttt{dstools}, respectively
\item  start MATLAB and put the directories \texttt{fditools}  and \texttt{dstools} on the MATLAB path, by using the \texttt{pathtool} command; for repeated use, save the new MATLAB search path, or alternatively, use the \texttt{addpath} command to set new path entries in \texttt{startup.m}
\item try out the installation by running the demonstration script \texttt{FDIToolsdemo.m}
\end{itemize}

\noindent\emph{Note: }The software accompanying the book \cite{Varg17} can be also downloaded as a zip file,\footnote{\url{https://sites.google.com/site/andreasvargacontact/home/book/matlab}} which also includes \textbf{FDITOOLS} V0.2 and \textbf{DSTOOLS} V0.5.   To install and execute the example and case-study scripts listed in the book, follow the steps indicated in the web page. An updated collection  of MATLAB scripts with examples is also available from Bitbucket.\footnote{\url{https://bitbucket.org/DSVarga/fdibook_examples}}

\newpage
\section{Current \texttt{Contents.m} File }\label{app:contents}

The M-functions available in the current version of \textbf{FDITOOLS}  are listed in the current version of the \texttt{Contents.m} file, given below:

\begin{verbatim}
% FDITOOLS - Fault detection and isolation filter synthesis tools.
% Version 1.0           30-November-2018
% Copyright 2016-2018 A. Varga
%
% Demonstration.
%   FDIToolsdemo - Demonstration of FDITOOLS.
%
% Setup of synthesis models.
%   fdimodset  - Setup of models for solving FDI synthesis problems.
%   mdmodset   - Setup of models for solving model detection synthesis problems.
%
% Analysis.
%   fdigenspec - Generation of achievable FDI specifications.
%   fdichkspec - Checking the feasability of a set of specifications.
%   mddist     - Computation of distances between component models.
%   mddist2c   - Computation of distances to a set of component models.
%
% Performance evaluation of FDI filters
%   fditspec   - Computation of the weak or strong structure matrix.
%   fdisspec   - Computation of the strong structure matrix.
%   fdifscond  - Fault sensitivity condition of FDI filters.
%   fdif2ngap  - Fault-to-noise gap of FDI filters.
%   fdimmperf  - Model-matching performance of FDI filters.
%
% Performance evaluation of model detection filters
%   mdperf     - Distance mapping performance of model detection filters.
%   mdmatch    - Distance matching performance of model detection filters.
%   mdgap      - Noise gaps of model detection filters.
%
% Synthesis of FDI filters.
%   efdsyn     - Exact synthesis of fault detection filters.
%   afdsyn     - Approximate synthesis of fault detection filters.
%   efdisyn    - Exact synthesis of fault detection and isolation filters.
%   afdisyn    - Approximate synthesis of fault detection and isolation filters.
%   emmsyn     - Exact model matching based synthesis of FDI filters.
%   ammsyn     - Approximate model matching based synthesis of FDI filters.
%
% Synthesis of model detection filters.
%   emdsyn     - Exact synthesis of model detection filters.
%   amdsyn     - Approximate synthesis of model detection filters.
%
% Miscellaneous.
%   hinfminus  - H-(infinity-) index of a stable transfer function matrix.
%   hinfmax    - Maximum of H-inf norms of columns of a transfer function matrix.
%   efdbasesel - Selection of admissible basis vectors to solve the EFDP.
%   afdbasesel - Selection of admissible basis vectors to solve the AFDP.
%   emmbasesel - Selection of admissible basis vectors to solve the strong EFDIP.
%   ammbasesel - Selection of admissible basis vectors to solve the strong AFDIP.
%   emdbasesel - Selection of admissible basis vectors to solve the EMDP.
 \end{verbatim}

\newpage
\section{FDITOOLS Release Notes} \label{appB}

The \textbf{FDITOOLS} Release Notes describe the changes introduced in the successive versions of the \textbf{FDITOOLS} collection, as new features, enhancements to functions, or major bug fixes. The following versions of \textbf{FDITOOLS} have been released:\\

\begin{tabular}{llp{9cm}}
Version & Release date & Comments\\ \hline
V0.2 &  December 31, 2016 & Initial version accompanying the book \cite{Varg17}.\\
V0.21 &  February 15, 2017 & Enhanced user interface of function \texttt{genspec}. \\
V0.3 &  April 7, 2017 &  New function for the exact model-matching based synthesis.\\
V0.4 &   August 31, 2017 & New function for the exact synthesis of model detection filters and substantial enhancements of most functions. The option to use ``design matrices'' has been added to all synthesis functions.  \\
V0.8 &  February 28, 2018 & Complete set of functions for the approximate synthesis of fault detection and model detection filters implemented. \\
V0.85 &   July 7, 2018 & New functions for performance evaluation of fault detection filters have been implemented together with notable enhancements of existing analysis functions. New synthesis options have been implemented in almost all synthesis functions (e.g., using an observer-based nullspace basis, option to perform exact synthesis for approximate synthesis functions, etc.)  \\
V0.87 &    August 15, 2018 & Enhancements of the approximate model-matching
synthesis function performed and a new function implemented for model-matching performance evaluation.\\
V1.0 &    November 30, 2018 & Many enhancements of the approximate synthesis  functions performed, by using the version V0.71 of \textbf{DSTOOLS}. A set of new functions have been implemented for the analysis of multiple models and performance evaluation of model detection filters.   \\
\hline
\end{tabular}

\subsection{Release Notes V0.2 }
This is the initial version of the \textbf{FDITOOLS} collection of M-functions, which accompanies the book \cite{Varg17}. All numerical results presented in this book have been obtained using this version of \textbf{FDITOOLS}.

\subsubsection{New Features}
The M-functions available in the Version 0.2 of \textbf{FDITOOLS} are listed below:

\begin{verbatim}
% FDITools - Fault detection and isolation filter synthesis tools.
% Version 0.2            31-Dec-2016
% Copyright 2017 A. Varga
%
% Demonstration.
%   FDIToolsdemo - Demonstration of FDITools.
%
% Analysis functions.
%   fditspec   - Computation of the structure matrix of a system.
%   fdisspec   - Computation of the strong structure matrix of a system.
%   genspec    - Generation of achievable fault detection specifications.
%
% Synthesis functions.
%   efdsyn     - Exact synthesis of fault detection filters.
%   efdisyn    - Exact synthesis of fault detection and isolation filters.
%
% Miscellaneous.
%   efdbasesel - Selection of admissible basis vectors to solve the EFDP.
%
\end{verbatim}

\subsection{Release Notes V0.21 }
Version 0.21, dated February 15, 2017, is a minor improvement over version 0.2 of \textbf{FDITOOLS}.

\subsubsection{New Features}
A new version of the function \texttt{\bfseries genspec} is provided, with an enhanced user interface. The new calling syntax, which also covers the previously used calling syntax, is similar to that used by the synthesis functions and allows the direct handling of systems having control, disturbance and additive fault inputs. \index{M-functions!\texttt{\bfseries genspec}}

\subsection{Release Notes V0.3 }
Version 0.3, dated April 7, 2017, provides a complete set of functions implementing the exact synthesis approaches of fault detection and isolation filters.

\subsubsection{New Features}
Two new functions have been implemented:
\begin{itemize}
\item
The function \texttt{\bfseries emmsyn} is provided for the exact synthesis of fault detection and isolation filters, by using an exact model-matching approach. \index{M-functions!\texttt{\bfseries emmsyn}}
\item The function \texttt{\bfseries emmbasesel} is provided for the selection of admissible left nullspace basis vectors to solve the strong exact fault detection and isolation problem, using an exact model-matching approach.
\end{itemize}

\subsection{Release Notes V0.4 }
Version 0.4, dated August 31, 2017, is a major new release, including substantial revisions of most functions, by adding exhaustive input parameter checks, new user options and several enhancements and simplifications of the implemented codes. Besides these modifications, two new functions have been implemented for the exact synthesis of model detection filters.

\subsubsection{New Features}
Two new functions have been implemented:
\begin{itemize}
\item
The function \texttt{\bfseries emdsyn} is provided for the exact synthesis of model detection filters, by using a nullspace-based synthesis approach. \index{M-functions!\texttt{\bfseries emdsyn}}
\item The function \texttt{\bfseries emdbasesel} is provided for the selection of admissible left nullspace basis vectors for solving the exact model detection problem.
\end{itemize}

\noindent Several new features have been implemented:
\begin{itemize}
\item The functionality of \texttt{\bfseries fditspec} has been enhanced, by generating a three-dimensional structure matrix in the case of several frequency values specified at input. In this case, the pages (along the third dimension) of this array contain the strong structure matrices at different frequency values.
\item The functionality of \texttt{\bfseries fdisspec} has been enhanced, by generating a three-dimensional structure matrix in the case of several frequency values, whose pages (along the third dimension) contain the strong structure matrices at different frequency values. An error message is issued if any of the specified frequencies is a system pole.
\item The functionality of \texttt{\bfseries genspec} has been restricted, by only allowing a single frequency value to generate strong specifications.
\item
In the function \texttt{\bfseries efdsyn} a new  option to specify a design parameter has been implemented and the internally employed/generated value of this parameter is returned in the \texttt{INFO} structure. This allows, among others, the reproducibility of the computed results. This feature also can serve for optimization purposes (e.g., minimizing the sensitivity conditions; see \cite[Remark 5.6]{Varg17}).
\item
In the function \texttt{\bfseries efdisyn} a new  option has been implemented to specify design parameters for the synthesis of individual filters. The internally employed/generated values of these design parameters are returned in the \texttt{INFO} structure. This allows, among others, the reproducibility of the computed results. This feature also can serve for optimization purposes (e.g., minimizing the sensitivity conditions of the individual filters; see \cite[Remark 5.6]{Varg17}).
\item
In the function \texttt{\bfseries efdisyn} a new  option has been implemented to specify a subset of indices of the filters to be designed. This allows, for example, to design separately individual filters of the overall bank of filters.
\item
In the function \texttt{\bfseries emmsyn} several enhancements of the algorithm implementation have been performed, new  options have been implemented for the normalization of the diagonal elements of the updating factor, for the specification of a regularization option, for the specification of a design parameter and for the specification of a complex frequency value to be used for solvability checks. The actually employed design matrix and frequency are returned in two fields of the \texttt{INFO} structure.
\item A new function \texttt{\bfseries emdsyn} is provided for the exact synthesis of model detection filters, by using a nullspace-based synthesis approach. \index{M-functions!\texttt{\bfseries emdsyn}}
\item The implementation of \texttt{\bfseries efdbasesel} has been simplified and its functionality has been enhanced, by allowing as input, a three-dimensional structure matrix as computed by \texttt{\bfseries fdisspec}. Also, the selection of admissible vectors is performed, regardless the degree information for the basis vectors is provided or not.
\item The implementation of \texttt{\bfseries emmbasesel} has been simplified and the selection of admissible vectors is performed, regardless the degree information for the basis vectors is provided or not.
\end{itemize}
\subsubsection{Bug Fixes}

The function \texttt{\bfseries efdsyn} has been updated by fixing a bug in the strong fault detectability test, in the case of more than one frequency values, or in the case when one of the specified frequency values coincides with a system pole.

\subsubsection{Compatibility Issues}

The function \texttt{\bfseries emmsyn} has been updated to comply with the version V0.6 of \textbf{DSTOOLS}.

\subsection{Release Notes V0.8 }
Version 0.8, dated February 28, 2018, is a major new release, including four new functions for the approximate synthesis of fault and model detection filters. Besides this, a few modifications have been performed by adding new user options and fixing some bugs.

\subsubsection{New Features}
Several new functions have been implemented:
\begin{itemize}
\item The function \texttt{\bfseries afdsyn} is provided for the approximate synthesis of fault detection filters, by using a nullspace-based synthesis approach. \index{M-functions!\texttt{\bfseries afdsyn}}
\item The function \texttt{\bfseries afdisyn} is provided for the approximate synthesis of fault detection and isolation filters, by using a nullspace-based synthesis approach. \index{M-functions!\texttt{\bfseries afdisyn}}
\item The function \texttt{\bfseries ammsyn} is provided for the approximate synthesis of fault detection and isolation filters, by using an approximate model-matching approach. \index{M-functions!\texttt{\bfseries ammsyn}}
\item The function \texttt{\bfseries amdsyn} is provided for the approximate synthesis of model detection filters, by using a nullspace-based synthesis approach. \index{M-functions!\texttt{\bfseries amdsyn}}
\item The function \texttt{\bfseries ammbasesel} is provided for the selection of admissible left nullspace basis vectors to solve the strong fault detection and isolation problem, using an approximate model-matching approach.
\end{itemize}
Several new features have been implemented:
\begin{itemize}
\item In the function \texttt{\bfseries emdsyn}, the following modifications have been performed in specifying user options: (1) the option for a design matrix has been enhanced, by allowing to specify, besides a cell array of matrices, also a unique design matrix; (2) a new option allows to choose either an observer-based nullspace or a minimal rational nullspace; (3) the option specifying the maximum number of residual outputs has been enhanced.
\end{itemize}

\subsubsection{Bug Fixes}

Several bug fixes has been performed:
\begin{itemize}
\item
The function \texttt{\bfseries efdsyn} has been updated to correctly handle the case of no fault inputs and to fix several bugs related to handling the optional design matrix.
\item
The function \texttt{\bfseries emmsyn} has been updated by fixing several bugs related to handling the optional design matrix.
\item
The function \texttt{\bfseries emdsyn} has been updated by fixing several bugs (e.g., in selecting a preliminary design, in handling the optional design matrix, fixing names of variables).
\end{itemize}

\subsection{Release Notes V0.85 }
Version 0.85, dated July 7, 2018, provides notable enhancements of several functions and includes a number of new functions as well.

\subsubsection{New Features}
The following new functions have been implemented:
\begin{itemize}
\item The function \texttt{\bfseries fdimodset} is provided for an easy setup of synthesis models with additive faults for solving FDI synthesis problems. \index{M-functions!\texttt{\bfseries fdimodset}}
\item The function \texttt{\bfseries mdmodset} is provided for an easy setup of multiple synthesis models for solving model detection synthesis problems. \index{M-functions!\texttt{\bfseries mdmodset}}
\item The function \texttt{\bfseries fdifscond} is provided to compute the  fault sensitivity condition of a stable system or of a collection of systems with additive faults. \index{M-functions!\texttt{\bfseries fdif2ngap}}
\item The function \texttt{\bfseries fdif2ngap} is provided to compute the fault-to-noise gap of a stable system or of a collection of systems. \index{M-functions!\texttt{\bfseries fdif2ngap}}
\item The function \texttt{\bfseries hinfminus} is provided for the computation of the $\mathcal{H}_{\infty -}$-index of the transfer function matrix of a stable LTI system. \index{M-functions!\texttt{\bfseries hinfminus}}
\item The function \texttt{\bfseries hinfmax} is provided for the computation of the maximum of the $\mathcal{H}_{\infty}$-norm of the columns of the transfer function matrix of a stable LTI system. \index{M-functions!\texttt{\bfseries hinfmax}}
\item The function \texttt{\bfseries afdbasesel} is provided for the selection of admissible left nullspace basis vectors to solve the approximate fault detection problem.
\end{itemize}
Several new features have been implemented:
\begin{itemize}
\item In the function \texttt{\bfseries fditspec}, the following enhancements have been implemented: (1) an option has been added to perform a block-structure based evaluation of the structure matrix (see (\ref{structure_matrix}) in Section \ref{isolability}); (2) the applicability  has been extended to cell arrays of systems sharing the same inputs or the same input group \texttt{{'faults'}}, to allow a block-structure based evaluation of the joint structure matrix (see (\ref{structure_matrix}) in Section \ref{isolability}). This extension permits the evaluation of the joint structure matrix of the resulting batch of internal filter representations, as computed by the functions  \texttt{\bfseries efdisyn} and \texttt{\bfseries afdisyn}.
\item In the function \texttt{\bfseries fdisspec}, the following enhancements have been implemented: (1) an option has been added to perform a block-structure based evaluation of the structure matrix (see (\ref{structure_matrix}) in Section \ref{isolability}); (2) the applicability  has been extended to cell arrays of systems sharing the same inputs or the same input group \texttt{{'faults'}}, to allow a block-structure based evaluation of the joint structure matrix (see (\ref{structure_matrix}) in Section \ref{isolability}). This extension permits the evaluation of the joint structure matrix of the resulting batch of internal filter representations, as computed by the functions  \texttt{\bfseries efdisyn} and \texttt{\bfseries afdisyn}.
\item In the function \texttt{\bfseries efdsyn}, a new option allows to choose  in the initial synthesis step either an observer-based (possibly non-minimal) nullspace basis (only in the case of lack of disturbance inputs)  or a minimal rational nullspace basis (the default option).
\item In the function \texttt{\bfseries afdsyn}, the following enhancements have been performed: (1) the resulting fault detection filter has, in general, a two-block structure (see \cite[Remark 5.10]{Varg17}); (2) accordingly, the option to specify the number of residual outputs has been enhanced; (3) a new user option allows to specify separate design matrices for the two components of the filter; (4) a new option allows to choose in the initial synthesis step either an observer-based (possibly non-minimal) nullspace basis or a minimal rational nullspace; (5) a new option to perform \emph{exact} synthesis is provided (this functionality equivalent to that of the function \texttt{\bfseries efdsyn}); (6) a new option is available to specify a test frequency value to be employed to check the  rank-based admissibility condition; (7) an extended set of additional information is returned to the user.
\item In the function \texttt{\bfseries efdisyn}, the following enhancements have been performed: (1) a new option allows to choose  in the initial synthesis step either an observer-based (possibly non-minimal) nullspace basis (only in the case of lack of disturbance inputs)  or a minimal rational nullspace basis (the default option); (2) the option to specify the numbers of residual outputs of each filter has been implemented.
\item In the function \texttt{\bfseries afdisyn}, the following enhancements have been performed: (1) each component of the resulting bank of fault detection filters has, in general, a two-block structure (see \cite[Remark 5.10]{Varg17}); (2) accordingly, the option to individually specify the numbers of residual outputs of each filter has been implemented; (3) a new user option allows to specify separate design matrices for the two components of the filters; (4) a new option allows to choose  in the initial synthesis step either an observer-based (possibly non-minimal) nullspace basis (only in the case of lack of disturbance inputs) or a minimal rational nullspace basis; (5) a new option to perform \emph{exact} synthesis is provided (this functionality equivalent to that of the function \texttt{\bfseries efdisyn}); (6) a new option is available to specify a test frequency value to be employed to check the  rank-based admissibility conditions; (7) the additional information returned to the user has been updated.
\item In the function \texttt{\bfseries emdsyn},  the option to individually specify the numbers of residual outputs of each filter has been implemented.
\item In the function \texttt{\bfseries amdsyn},  the option to individually specify the numbers of residual outputs of each filter has been implemented.
\item The demonstration script \texttt{\bfseries FDIToolsdemo} has been updated to use calls to the newly implemented confort functions to setup models and evaluate suitable performance measures.
\end{itemize}

\noindent The checking of solvability conditions related to the existence of left nullspace bases has been enhanced for all synthesis functions, by issuing appropriate error messages in the case of empty nullspace bases.

\subsubsection{Bug Fixes}

Several minor bug fixes have been performed in the functions \texttt{\bfseries efdsyn} and \texttt{\bfseries afdsyn}.

\subsection{Release Notes V0.87 }
Version 0.87, dated August 15, 2018, provides enhancements of the approximate model-matching synthesis function \texttt{ammsyn} and includes the new function \texttt{\bfseries fdimmperf} for model-matching performance evaluation.

\subsubsection{New Features}
The following new function has been implemented:
\begin{itemize}
\item The function \texttt{\bfseries fdimmperf} is provided to compute the  model-matching performance of a stable system or of a collection of stable systems. \index{M-functions!\texttt{\bfseries fdimmperf}}
\end{itemize}
Several new features have been implemented:
\begin{itemize}
\item In the function \texttt{\bfseries ammsyn}, the following enhancements have been performed: (1) extension to handle arbitrary reference models with control, disturbance, fault and noise inputs; (2) enhancing the nullspace-based synthesis, by explicitly considering strong FDI problems and fault detection problems; (3) a new option allows to choose  in the initial synthesis step either an observer-based (possibly non-minimal) nullspace basis (only in the case of lack of disturbance inputs)  or a minimal rational nullspace basis (the default option); (4) support for void fault inputs provided; (5) providing both optimal and suboptimal performance values in the \texttt{INFO} structure.
\item The function \texttt{\bfseries ammbasesel} has been extended to handle both strong FDI related selection as well as fault detection oriented selection of basis vectors.
\end{itemize}

\subsubsection{Bug Fixes}

Several minor bug fixes have been performed in the functions \texttt{\bfseries fdifscond} and \texttt{\bfseries fdif2ngap}.

\subsection{Release Notes V1.0 }
Version 1.0, dated November 30, 2018, concludes the development of a fairly complete collection of  tools
to address the main computational aspects of the synthesis of fault detection and model detection filters using linear synthesis techniques. This version provides several new functions for the analysis of model detection problems and the evaluation of the performance of model detection filters.
Besides this, several enhancements of the approximate synthesis functions have been performed related to the handling of non-standard cases, as well as of the synthesis functions of model detection filters. This version of \textbf{FDTOOLS} relies on the version V0.71 of the Descriptor Systems Tools \textbf{DSTOOLS}.

\subsubsection{New Features}
The following new function has been implemented:
\begin{itemize}
\item The function \texttt{\bfseries fdichkspec} is provided to check the feasibility of a set of FDI specifications.\index{M-functions!\texttt{\bfseries fdichkspec}}
\item The function \texttt{\bfseries mdperf} is provided to compute the  distance mapping performance of model detection filters. \index{M-functions!\texttt{\bfseries mdperf}}
\item The function \texttt{\bfseries mdmatch} is provided to compute the  distance matching performance of model detection filters. \index{M-functions!\texttt{\bfseries mdmatch}}
\item The function \texttt{\bfseries mdgap} is provided to compute the  noise gap performance of model detection filters. \index{M-functions!\texttt{\bfseries mdgap}}
\end{itemize}
The following new features have been added:
\begin{itemize}
\item  The function \textbf{\texttt{fdigenspec}} replaces \textbf{\texttt{genspec}} and can now handle multiple frequency values to determine strong FDI specifications.  For compatibility purposes, the obsolete function \textbf{\texttt{genspec}} can be still used, but it will be removed in a future version of \textbf{FDITOOLS}.
\item In the functions \texttt{\bfseries afdsyn} and \texttt{\bfseries afdisyn}, new options have been added to the \texttt{OPTIONS} structure, to handle non-standard problems.  Also, the detection of non-standard problems has been simplified, being done without additional computations, using the new information provided by \texttt{goifac} in the version V0.7 (and later) of \textbf{DSTOOLS}.
\item In the function \texttt{\bfseries ammsyn}, the optimal performance value for the original problem is provided in the \texttt{INFO} structure, jointly with the optimal and suboptimal values of the updated problem. Also, the detection of non-standard problems has been simplified, being done without additional computations, using the new information provided by \texttt{goifac} in the version V0.7  (and later) of \textbf{DSTOOLS}.
\item In the function \texttt{\bfseries emdsyn}, a new option has been added to the \texttt{OPTIONS} structure, to handle the extended model detectability. Also, an new option has been added to perform specific normalizations of the resulting model detection filters.
\item In the function \texttt{\bfseries amdsyn}, a new option has been added to the \texttt{OPTIONS} structure, to handle the extended model detectability. Also, new options have been added, to handle non-standard problems.  In this context, the detection of non-standard problems has been simplified, being done without additional computations, using the new information provided by \texttt{goifac} in the version V0.7  (and later) of \textbf{DSTOOLS}.
\end{itemize}

\subsubsection{Bug Fixes and Minor Updates}

\begin{itemize}
\item
The functions \texttt{\bfseries fditspec}, \texttt{\bfseries fdisspec}, \texttt{\bfseries fdifscond}, \texttt{\bfseries fdif2ngap} and \texttt{\bfseries fdimmperf} and their documentations have been updated by using a new notation for the underlying internal forms of the FDI filters.

\item  Several minor bug fixes have been performed in the function \texttt{\bfseries efdisyn}.

\item The functions \texttt{\bfseries emdsyn} and \texttt{\bfseries amdsyn} have been updated to comply with the new definitions of model detectability and extended model detectability introduced in Section \ref{sec:mdetectability}.
\end{itemize}

\pdfbookmark[1]{Index}{Ind}
\begin{theindex}

  \item EFDP, \hyperindexformat{\see{fault detection problem}}{20}
  \item EFDIP,
		\hyperindexformat{\see{fault detection and isolation problem}}{20}
  \item AFDP, \hyperindexformat{\see{fault detection problem}}{20}
  \item AFDIP,
		\hyperindexformat{\see{fault detection and isolation problem}}{20}
  \item EMMP, \hyperindexformat{\see{model-matching problem}}{20}
  \item EFEP, \hyperindexformat{\see{model-matching problem}}{20}
  \item AMMP, \hyperindexformat{\see{model-matching problem}}{20}
  \item EMDP, \hyperindexformat{\see{model detection problem}}{20}
  \item AMDP, \hyperindexformat{\see{model detection problem}}{20}

  \indexspace

  \item FDD, \hyperindexformat{\see{fault detection and diagnosis}}{12}
  \item FDI, \hyperindexformat{\see{fault detection and isolation}}{12}
  \item fault detectability, \hyperpage{13}, \hyperpage{16}
    \subitem complete, \hyperpage{16}
    \subitem complete, strong, \hyperpage{17}
    \subitem strong, \hyperpage{16, 17}
  \item fault detection and diagnosis (FDD)
    \subitem fault detection, \hyperpage{13}
    \subitem fault isolation, \hyperpage{13}
    \subitem strong fault isolation, \hyperpage{14}
    \subitem weak fault isolation, \hyperpage{14}
  \item fault detection problem
    \subitem exact (EFDP), \hyperindexformat{\ii}{21}, \hyperpage{89}
      \subsubitem solvability, \hyperindexformat{\ii}{21},
		\hyperpage{22, 23}
    \subitem exact (EFDP) with strong detectability,
		\hyperindexformat{\ii}{21}
      \subsubitem solvability, \hyperindexformat{\ii}{21}
    \subitem approximate (AFDP), \hyperindexformat{\ii}{21},
		\hyperpage{22}, \hyperpage{95}
      \subsubitem solvability, \hyperindexformat{\ii}{22}
  \item fault detection and isolation (FDI), \hyperpage{12}
  \item fault detection and isolation problem
    \subitem exact (EFDIP), \hyperindexformat{\ii}{22}, \hyperpage{107}
      \subsubitem solvability, \hyperindexformat{\ii}{22}
    \subitem exact (EFDIP) with strong isolability,
		\hyperindexformat{\ii}{22}
      \subsubitem solvability, \hyperindexformat{\ii}{22}
    \subitem approximate (AFDIP), \hyperindexformat{\ii}{22},
		\hyperpage{23}, \hyperpage{114}
      \subsubitem solvability, \hyperindexformat{\ii}{23}
  \item fault isolability, \hyperpage{18, 19}
    \subitem strong, \hyperpage{19}
    \subitem structure matrix, \hyperpage{18}, \hyperpage{65},
		\hyperpage{69}
      \subsubitem fault signature, \hyperpage{18}
      \subsubitem specification, \hyperpage{18}
    \subitem weak, \hyperpage{19}
  \item faulty system model
    \subitem additive, \hyperpage{14}
      \subsubitem input-output, \hyperpage{14}
      \subsubitem state-space, \hyperpage{15}
    \subitem multiple model, \hyperpage{30}, \hyperpage{57},
		\hyperpage{60}, \hyperpage{141}, \hyperpage{151}
    \subitem physical, \hyperpage{30}, \hyperpage{57}, \hyperpage{60},
		\hyperpage{141}, \hyperpage{151}

  \indexspace

  \item M-functions
    \subitem \texttt{\bfseries afdisyn}, \hyperpage{114},
		\hyperpage{171}
    \subitem \texttt{\bfseries afdsyn}, \hyperpage{95}, \hyperpage{118},
		\hyperpage{171}
    \subitem \texttt{\bfseries amdsyn}, \hyperpage{151},
		\hyperpage{172}
    \subitem \texttt{\bfseries ammsyn}, \hyperpage{129},
		\hyperpage{171}
    \subitem \texttt{\bfseries efdisyn}, \hyperpage{107}
    \subitem \texttt{\bfseries efdsyn}, \hyperpage{89}, \hyperpage{110}
    \subitem \texttt{\bfseries emdsyn}, \hyperpage{141},
		\hyperpage{170, 171}
    \subitem \texttt{\bfseries emmsyn}, \hyperpage{122},
		\hyperpage{169}
    \subitem \texttt{\bfseries fdichkspec}, \hyperpage{54},
		\hyperpage{175}
    \subitem \texttt{\bfseries fdif2ngap}, \hyperpage{75},
		\hyperpage{172}
    \subitem \texttt{\bfseries fdifscond}, \hyperpage{72}
    \subitem \texttt{\bfseries fdigenspec}, \hyperpage{49}
    \subitem \texttt{\bfseries fdimmperf}, \hyperpage{79},
		\hyperpage{174}
    \subitem \texttt{\bfseries fdimodset}, \hyperpage{44},
		\hyperpage{172}
    \subitem \texttt{\bfseries fdisspec}, \hyperpage{69}
    \subitem \texttt{\bfseries fditspec}, \hyperpage{65}
    \subitem \texttt{\bfseries genspec}, \hyperpage{169}
    \subitem \texttt{\bfseries hinfmax}, \hyperpage{173}
    \subitem \texttt{\bfseries hinfminus}, \hyperpage{173}
    \subitem \texttt{\bfseries mddist2c}, \hyperpage{59}
    \subitem \texttt{\bfseries mddist}, \hyperpage{57}
    \subitem \texttt{\bfseries mdgap}, \hyperpage{86}, \hyperpage{175}
    \subitem \texttt{\bfseries mdmatch}, \hyperpage{84},
		\hyperpage{175}
    \subitem \texttt{\bfseries mdmodset}, \hyperpage{46},
		\hyperpage{172}
    \subitem \texttt{\bfseries mdperf}, \hyperpage{82}, \hyperpage{175}
  \item model detectability, \hyperpage{31}
    \subitem extended, \hyperpage{32}
    \subitem strong, \hyperpage{33}
  \item model detection, \hyperpage{29}
    \subitem $\mathcal{H}_2$-norm distances, \hyperpage{36},
		\hyperpage{59}
    \subitem $\mathcal{H}_\infty$-norm distances, \hyperpage{36},
		\hyperpage{59}
    \subitem $\nu$-gap distances, \hyperpage{35}, \hyperpage{37},
		\hyperpage{57}, \hyperpage{59}
    \subitem distance mapping, \hyperpage{38}, \hyperpage{82},
		\hyperpage{146}, \hyperpage{155}
    \subitem distance matching, \hyperpage{39}, \hyperpage{84}
    \subitem noise gap, \hyperpage{39}, \hyperpage{86}
  \item model detection problem
    \subitem exact (EMDP), \hyperindexformat{\ii}{34}, \hyperpage{141}
      \subsubitem solvability, \hyperindexformat{\ii}{34, 35}
    \subitem exact (EMDP) with strong model detectability,
		\hyperindexformat{\ii}{34}
      \subsubitem solvability, \hyperindexformat{\ii}{34}
    \subitem approximate (AMDP), \hyperindexformat{\ii}{35},
		\hyperpage{151}
      \subsubitem solvability, \hyperindexformat{\ii}{35}
  \item model-matching problem
    \subitem exact (EMMP), \hyperindexformat{\ii}{23},
		\hyperpage{24, 25}, \hyperpage{122}
      \subsubitem solvability, \hyperindexformat{\ii}{24},
		\hyperpage{25}
    \subitem exact fault estimation (EFEP), \hyperindexformat{\ii}{23}
      \subsubitem solvability, \hyperindexformat{\ii}{24}
    \subitem approximate (AMMP), \hyperindexformat{\ii}{25},
		\hyperpage{129}
      \subsubitem solvability, \hyperindexformat{\ii}{25}

  \indexspace

  \item performance evaluation
    \subitem fault detection and isolation
      \subsubitem fault sensitivity condition, \hyperpage{26},
		\hyperpage{72}
      \subsubitem fault-to-noise gap, \hyperpage{27}, \hyperpage{75},
		\hyperpage{101}, \hyperpage{119}
      \subsubitem model-matching performance, \hyperpage{28},
		\hyperpage{79}, \hyperpage{133}
    \subitem model detection
      \subsubitem distance mapping, \hyperpage{38}, \hyperpage{82},
		\hyperpage{146}, \hyperpage{155}
      \subsubitem distance matching, \hyperpage{39}, \hyperpage{84}
      \subsubitem noise gap, \hyperpage{39}, \hyperpage{86},
		\hyperpage{156}

  \indexspace

  \item residual generation
    \subitem for fault detection and isolation, \hyperpage{15},
		\hyperpage{25}
    \subitem for model detection, \hyperpage{30}
  \item residual generator
    \subitem implementation form, \hyperpage{15}, \hyperpage{31}
    \subitem internal form, \hyperpage{15}, \hyperpage{25},
		\hyperpage{31}

  \indexspace

  \item structure matrix, \hyperindexformat{\ii}{18}, \hyperpage{49},
		\hyperpage{54}, \hyperpage{65}, \hyperpage{69}
    \subitem fault signature, \hyperpage{18}
    \subitem specification, \hyperpage{18}

  \indexspace

  \item transfer function
    \subitem winding number, \hyperpage{35}

\end{theindex}


\end{document}